\documentclass[a4paper,11pt]{article}
\pdfoutput=1 

\usepackage{jheppub} 

\usepackage[T1]{fontenc} 

\usepackage{bm}
\usepackage{dsfont}
\usepackage{mathtools}
\usepackage{stackengine}
\usepackage{subcaption}
\usepackage{tensor}

\usepackage{changes}
\definechangesauthor[name=Vincent]{hzc}

\newcommand{\naturals}{\mathbb{N}}
\newcommand{\integers}{\mathbb{Z}}

\newcommand{\AdS}{\mathrm{AdS}}
\newcommand{\anom}{\mathrm{an}}

\newcommand{\aux}{\mathrm{aux}}
\newcommand{\bare}{\mathrm{bare}}
\newcommand{\bc}{b.c.}
\newcommand{\BH}{\mathrm{BH}}
\newcommand{\bkg}{\mathrm{bkg}}
\newcommand{\cf}{\textit{c.f.}\ }
\newcommand{\Cf}{\textit{C.f.}\ }
\newcommand{\CFT}{\mathrm{CFT}}
\newcommand{\elec}{\mathrm{e}}

\newcommand{\diff}{\mathrm{diff}}
\newcommand{\diver}{\mathrm{div}}
\newcommand{\Dong}{\mathrm{D}}
\newcommand{\eg}{\textit{e.g.}\ }
\newcommand{\Eg}{\textit{E.g.}\ }

\newcommand{\gen}{\mathrm{gen}}
\newcommand{\GH}{\mathrm{GH}}
\newcommand{\grav}{\mathrm{grav}}
\newcommand{\ie}{\textit{i.e.}\ }

\newcommand{\jt}{\mathrm{jt}}
\newcommand{\mat}{\mathrm{mat}}
\newcommand{\Max}{\mathrm{Max}}
\newcommand{\newton}{\mathrm{N}}
\newcommand{\perse}{\textit{per se}}

\newcommand{\QG}{\mathrm{QG}}
\newcommand{\ren}{\mathrm{ren}}
\newcommand{\Wald}{\mathrm{W}}

\newcommand{\state}{\rho}
\newcommand{\unState}{\hat{\rho}}
\newcommand{\stateMat}{\rho^\mat}
\newcommand{\stateMatPurified}{\varphi^{\mat,\aux}}

\newcommand{\unStateMat}{\unState^\mat}
\newcommand{\unStateMatR}[1]{\unStateMat_{#1}}

\newcommand{\pureState}{\Psi}

\newcommand{\ent}{S}
\newcommand{\entR}[1]{S_{#1}}

\newcommand{\entDong}[1]{\ent_{#1}^{\Wald\Dong}}

\newcommand{\entDongMat}[1]{\ent_{#1}^{\mat,\Wald\Dong}}

\newcommand{\entDongMatRen}[1]{\ent_{\ren,#1}^{\mat,\Wald\Dong}}

\newcommand{\entDongGrav}[1]{\ent_{#1}^{\grav,\Wald\Dong}}
\newcommand{\entDongGravBare}[1]{\ent_{\bare,#1}^{\grav,\Wald\Dong}}
\newcommand{\entDongGravRen}[1]{\ent_{\ren,#1}^{\grav,\Wald\Dong}}

\newcommand{\entDongAnom}[1]{\ent_{#1}^{\Wald\Dong,\anom}}
\newcommand{\entGen}[1]{\ent_{#1}^\gen}

\newcommand{\area}[1]{A_{#1}}

\newcommand{\bdyMan}{\mathcal{M}}
\newcommand{\bdyManR}[1]{\bdyMan_{#1}}
\newcommand{\bdyReg}{\mathcal{R}}
\newcommand{\entSurf}{\sigma}
\newcommand{\entSurfR}[1]{\entSurf_{#1}}
\newcommand{\bdyEntSurf}{\varsigma}
\newcommand{\entReg}{\Sigma}
\newcommand{\egReg}{\Xi}
\newcommand{\egPatch}{\mathcal{S}}

\newcommand{\NelPatchPart}[1]{{\egPatch_{#1}}}

\newcommand{\neigh}[2]{\mathcal{N}_{#1}(#2)}
\newcommand{\neighSize}{a}

\newcommand{\metric}{g}
\newcommand{\metricR}[1]{\metric_{#1}}

\newcommand{\metricCFT}{g^{\CFT}}
\newcommand{\entRegMetric}{\gamma_{\entReg}}
\newcommand{\bgFieldsOther}{\phi}
\newcommand{\bgFieldsOtherR}[2][\bgFieldsOther]{{#1}_{#2}}

\newcommand{\graviton}{h}
\newcommand{\gravitonR}[1]{\graviton_{#1}}
\newcommand{\qFieldsOther}{\psi}
\newcommand{\qFieldsOtherR}[1]{\qFieldsOther_{#1}}
\newcommand{\fieldMax}{A}
\newcommand{\fieldMaxP}{\fieldMax^{+}}
\newcommand{\fieldMaxM}{\fieldMax^{-}}

\newcommand{\bgFieldMax}{\bar{\fieldMax}}

\newcommand{\qFieldMax}{\tilde{\fieldMax}}

\newcommand{\formMax}{\mathbf{\fieldMax}}
\newcommand{\formMaxP}{\formMax^{+}}
\newcommand{\formMaxM}{\formMax^{-}}
\newcommand{\formMaxBH}{\formMax_{\BH}}
\newcommand{\formMaxBHP}{\formMaxBH^{+}}
\newcommand{\bgFormMax}{\bar{\formMax}}

\newcommand{\qFormMax}{\tilde{\formMax}}
\newcommand{\qFormMaxP}{\qFormMax^{+}}
\newcommand{\qFormMaxM}{\qFormMax^{-}}
\newcommand{\fieldMaxGauge}{\alpha}
\newcommand{\fieldMaxGaugeBH}{\fieldMaxGauge_{\BH}}

\newcommand{\qFieldMaxGauge}{\tilde{\alpha}}
\newcommand{\fieldStrMax}{F}
\newcommand{\fieldStrMaxP}{\fieldStrMax^{+}}
\newcommand{\fieldStrMaxM}{\fieldStrMax^{-}}
\newcommand{\fieldStrMaxBH}{\fieldStrMax_{\BH}}
\newcommand{\fieldStrMaxBHP}{\fieldStrMaxBH^{+}}
\newcommand{\qFieldStrMax}{\tilde{\fieldStrMax}}
\newcommand{\qFieldStrMaxP}{\qFieldStrMax^{+}}

\newcommand{\formStrMax}{\mathbf{\fieldStrMax}}
\newcommand{\formStrMaxP}{\formStrMax^{+}}
\newcommand{\formStrMaxM}{\formStrMax^{-}}
\newcommand{\formStrMaxBH}{\formStrMax_{\BH}}
\newcommand{\formStrMaxBHP}{\formStrMaxBH^{+}}

\newcommand{\blkOp}{\mathcal{O}}
\newcommand{\blkOpInit}{\blkOp_{\entReg}}
\newcommand{\unitary}{U}
\newcommand{\id}{\mathds{1}}
\newcommand{\testFunc}{\iota}
\newcommand{\egField}{\gamma}
\newcommand{\timeOrder}{\mathcal{T}}

\newcommand{\act}{I}
\newcommand{\actQG}{\act^\QG}
\newcommand{\actGrav}{\act^\grav}
\newcommand{\actGravBare}{\act^\grav_{\bare}}
\newcommand{\actGravRen}{\act^\grav_{\ren}}

\newcommand{\actMat}{\act^\mat}

\newcommand{\actMatDiver}{\act_\diver^\mat}

\newcommand{\eActCon}{E}
\newcommand{\eActConR}[1]{\eActCon_{#1}}

\newcommand{\eActMix}{W}

\newcommand{\eActMixMat}{\eActMix^\mat}

\newcommand{\eActMixMatRen}{\eActMix_{\ren}^{\mat}}
\newcommand{\eActMixMatBare}{\eActMix_{\bare}^{\mat}}
\newcommand{\actEntSurf}[1]{\act_{#1}}

\newcommand{\actEntSurfMat}[1]{\actEntSurf{#1}^\mat}

\newcommand{\actTimeClosedInterval}[2]{\act_{\timeClosedInterval{#1}{#2}}}
\newcommand{\actTimeClosedIntervalMat}[2]{\actTimeClosedInterval{#1}{#2}^{\mat}}
\newcommand{\LorActTimeClosedInterval}[2]{\Lor{\act}_{\timeClosedInterval{#1}{#2}}}
\newcommand{\LorActTimeClosedIntervalMat}[2]{\LorActTimeClosedInterval{#1}{#2}^{\mat}}
\newcommand{\rangleQG}{\rangle^\QG}

\newcommand{\GNewton}{G_{\newton}}
\newcommand{\potElec}{\mu}
\newcommand{\potElecBH}{\mu_{\BH}}
\newcommand{\qPotElec}{\tilde{\potElec}}
\newcommand{\chargeElec}{\charge{\elec}}
\newcommand{\chargeElecBH}{\charge{\elec,\BH}}
\newcommand{\qChargeElec}{\tilde{\mathbf{Q}}_{\elec}}
\newcommand{\totalChargeElec}{\mathcal{Q}_{\elec}}

\newcommand{\qTotalChargeElec}{\tilde{\mathcal{Q}}_{\elec}}

\newcommand{\srcGrav}{\mathbf{t}}

\newcommand{\current}[1]{\mathbf{J}_{#1}}
\newcommand{\EucCurrent}[1]{\Euc{\mathbf{J}}_{#1}}
\newcommand{\LorCurrent}[1]{\Lor{\mathbf{J}}_{#1}}
\newcommand{\currentGrav}[1]{\mathbf{J}_{#1}^{\grav}}
\newcommand{\currentMat}[1]{\mathbf{J}_{#1}^{\mat}}
\newcommand{\currentAnom}[1]{\mathbf{J}_{#1}^{\anom}}
\newcommand{\EucCurrentMat}[1]{\Euc{\mathbf{J}}_{#1}^{\mat}}
\newcommand{\LorCurrentMat}[1]{\Lor{\mathbf{J}}_{#1}^{\mat}}
\newcommand{\charge}[1]{\mathbf{Q}_{#1}}
\newcommand{\EucCharge}[1]{\Euc{\mathbf{Q}}_{#1}}
\newcommand{\LorCharge}[1]{\Lor{\mathbf{Q}}_{#1}}
\newcommand{\chargeGrav}[1]{\mathbf{Q}_{#1}^{\grav}}
\newcommand{\chargeMat}[1]{\mathbf{Q}_{#1}^{\mat}}
\newcommand{\chargeMax}[1]{\mathbf{Q}_{#1}^{\Max}}
\newcommand{\chargeAnom}[1]{\mathbf{Q}_{#1}^{\anom}}
\newcommand{\symp}{\bm{\omega}}
\newcommand{\sympGrav}{\bm{\omega}^{\grav}}
\newcommand{\sympMat}{\bm{\omega}^{\mat}}
\newcommand{\sympPot}{\bm{\theta}}

\newcommand{\sympPotGrav}{\sympPot^{\grav}}
\newcommand{\sympPotMat}{\sympPot^{\mat}}

\newcommand{\sympPotAnom}{\sympPot^{\anom}}
\newcommand{\sympPotAnomQ}[1]{\sympPot^{\anom}_{#1}}
\newcommand{\sympPotMax}{\sympPot^{\Max}}
\newcommand{\eom}[1]{E_{#1}}

\newcommand{\eomGrav}[1]{\eom{#1}^\grav}
\newcommand{\eomAnom}[1]{\eom{#1}^\anom}
\newcommand{\eomDensity}[1]{\mathbf{E}_{#1}}
\newcommand{\EucEomDensity}[1]{\Euc{\mathbf{E}}_{#1}}
\newcommand{\LorEomDensity}[1]{\Lor{\mathbf{E}}_{#1}}
\newcommand{\eomDensityMat}[1]{\eomDensity{#1}^\mat}

\newcommand{\eomDensityGrav}[1]{\eomDensity{#1}^\grav}
\newcommand{\eomDensityAnom}[1]{\eomDensity{#1}^\anom}
\newcommand{\Lag}{L}

\newcommand{\LagDensity}{\mathbf{L}}
\newcommand{\LagDensityGrav}{\LagDensity^\grav}
\newcommand{\LagDensityMat}{\LagDensity^\mat}

\newcommand{\LagDensityMax}{\LagDensity^{\Max}}
\newcommand{\LagDensityJt}{\mathbf{L}^{\jt}}

\newcommand{\LagDensityJtMat}{\LagDensity^{\jt,\mat}}
\newcommand{\LagDensityJtAnom}{\LagDensity^{\jt,\anom}}

\newcommand{\LagDensityGH}{\mathbf{L}^{\GH}}

\newcommand{\LagDensityGHMat}{\LagDensity^{\GH,\mat}}
\newcommand{\LagDensityGHAnom}{\LagDensity^{\GH,\anom}}

\newcommand{\LagDensityBdy}{\mathbf{L}^{\bdyMan}}
\newcommand{\LagDensityBdyGrav}{\LagDensity^{\bdyMan,\grav}}
\newcommand{\LagDensityBdyMat}{\LagDensity^{\bdyMan,\mat}}
\newcommand{\LagDensityBdyAnom}{\LagDensity^{\bdyMan,\anom}}
\newcommand{\LagDensityBdyMax}{\LagDensity^{\bdyMan,\Max}}

\newcommand{\EucLagDensityBdyMat}{\Euc{\LagDensity}^{\bdyMan,\mat}}

\newcommand{\LorLagDensityBdyMat}{\Lor{\LagDensity}^{\bdyMan,\mat}}
\newcommand{\FaulknerA}{\mathbf{A}}
\newcommand{\FaulknerAAnom}{\FaulknerA^{\anom}}

\newcommand{\FaulknerB}{\mathbf{B}}
\newcommand{\FaulknerBOf}[1]{\FaulknerB_{#1}}
\newcommand{\FaulknerBAnomOf}[1]{\FaulknerBOf{#1}^\anom}
\newcommand{\FaulknerC}{\mathbf{C}}
\newcommand{\FaulknerCGrav}{\mathbf{C}^{\grav}}
\newcommand{\FaulknerCMat}{\mathbf{C}^{\mat}}
\newcommand{\FaulknerCAnom}{\mathbf{C}^{\anom}}
\newcommand{\FaulknerCOf}[1]{\FaulknerC_{#1}}
\newcommand{\FaulknerCMatOf}[1]{\mathbf{C}^{\mat}_{#1}}
\newcommand{\FaulknerCAnomOf}[1]{\FaulknerC_{#1}^\anom}

\newcommand{\modHam}{K}
\newcommand{\modHamMat}{\modHam^{\mat}}
\newcommand{\modHamNel}[1]{\mathcal{K}_{#1}}
\newcommand{\modHamNelDiff}[1]{\modHamNel{#1}^{\diff}}
\newcommand{\modHamNelMat}[1]{\modHamNel{#1}^{\mat}}
\newcommand{\modHamNelMatDiff}[1]{\modHamNel{#1}^{\mat,\diff}}
\newcommand{\modHamNelS}[1]{\tilde{\mathcal{K}}_{#1}}
\newcommand{\modHamNelSDiff}[1]{\modHamNelS{#1}^{\diff}}
\newcommand{\Ham}[1]{H_{#1}}
\newcommand{\HamBdy}[1]{\Ham{#1}^{\bdyMan}}

\newcommand{\HamBdyMax}[1]{\Ham{#1}^{\bdyMan,\Max}}
\newcommand{\HamDensity}[1]{\mathbf{H}_{#1}}
\newcommand{\HamDensityMat}[1]{\HamDensity{#1}^{\mat}}
\newcommand{\HamDensityAnom}[1]{\HamDensity{#1}^{\anom}}
\newcommand{\HamDensitySimple}[1]{\tilde{\mathbf{H}}_{#1}}
\newcommand{\HamDensitySimpleMat}[1]{\HamDensitySimple{#1}^{\mat}}
\newcommand{\HamDensitySimpleAnom}[1]{\HamDensitySimple{#1}^{\anom}}
\newcommand{\EucHamDensityMat}[1]{\Euc{\mathbf{H}}_{#1}^{\mat}}
\newcommand{\LorHamDensityMat}[1]{\Lor{\mathbf{H}}_{#1}^{\mat}}
\newcommand{\HamDensityBdyMat}[1]{\HamDensity{#1}^{\bdyMan,\mat}}
\newcommand{\HamDensityBdyAnom}[1]{\HamDensity{#1}^{\bdyMan,\anom}}
\newcommand{\EucHamDensityBdyMat}[1]{\Euc{\mathbf{H}}_{#1}^{\bdyMan,\mat}}
\newcommand{\LorHamDensityBdyMat}[1]{\Lor{\mathbf{H}}_{#1}^{\bdyMan,\mat}}
\newcommand{\stress}{T}
\newcommand{\stressBare}{T_{\bare}}
\newcommand{\stressRen}{T_{\ren}}

\newcommand{\stressDensity}{\mathbf{\stress}}
\newcommand{\stressBareDensity}{\mathbf{\stress}_\bare}
\newcommand{\stressRenDensity}{\mathbf{\stress}_\ren}

\newcommand{\volComp}{\epsilon}
\newcommand{\volForm}{\bm{\volComp}}
\newcommand{\egForm}{\bm{\egField}}
\newcommand{\Lor}[1]{\tensor[^{\mathrm{L}}]{#1}{}}
\newcommand{\Euc}[1]{\tensor[^{\mathrm{E}}]{#1}{}}
\newcommand{\extrK}{\kappa}
\newcommand{\WaldB}{\mathbf{B}}

\newcommand{\WaldX}{\mathbf{X}}
\newcommand{\WaldY}{\mathbf{Y}}
\newcommand{\WaldZ}{\mathbf{Z}}

\newcommand{\thTime}{\tau}
\newcommand{\timeClosedInterval}[2]{[#1,#2]}
\newcommand{\timeOpenInterval}[2]{(#1,#2)}
\newcommand{\thTimeNelPatch}[1]{\thTime_{#1}}

\newcommand{\LorTime}{t}
\newcommand{\thVec}{\xi}

\newcommand{\NelFunc}{\kappa}
\newcommand{\NelVec}{\zeta}
\newcommand{\egVec}{\chi}
\newcommand{\normalVec}{n}
\newcommand{\normalVecOf}[1]{\normalVec_{(#1)}}

\newcommand{\Lie}[1]{\mathcal{L}_{#1}}
\newcommand{\vary}{\delta}
\newcommand{\varyWRT}[1]{\vary_{#1}}

\DeclareMathOperator{\tr}{tr}

\stackMath
\newcommand{\rep}[1]{\stackunder[0pt]{\scriptstyle\times}{\scriptscriptstyle #1}}
\newcommand{\repS}[1]{\stackunder[0pt]{\scriptstyle\tilde{\times}}{\scriptscriptstyle #1}}
\newcommand{\orb}[1]{\stackunder[0pt]{\scriptstyle\div}{\scriptscriptstyle #1}}
\newcommand{\orbS}[1]{\stackunder[0pt]{\scriptstyle\tilde{\div}}{\scriptscriptstyle #1}}
\newcommand{\diffRel}{\stackrel{\diff}{\sim}}

\newcommand{\eqDubious}{\stackrel{?}{=}}

\newcommand{\DongStressBkg}{\stress^{\bkg}}
\newcommand{\DongActOrbR}[1]{\hat{\act}_{#1}}

\title{\boldmath The generalized first law for more general matter}

\author[a,b]{Hong Zhe Chen}

\affiliation[a]{
  Perimeter Institute for Theoretical Physics,\\
  31 Caroline St. N. Waterloo, Ontario N2L 2Y5, Canada } \affiliation[b]{
  Department of Physics and Astronomy, University of Waterloo,\\
  200 University Avenue West, Waterloo, Ontario, N2L 3G1, Canada }

\emailAdd{hchen2@perimeterinstitute.ca}

\abstract{In previous work, a first law of generalized entropy was derived from
  semiclassical gravitational dynamics around thermal setups using an assumed
  relation between the matter modular Hamiltonian and the gravitational stress
  tensor. Allowing for non-minimal coupling between curvature and any tensor
  matter fields, we show however, that the modular Hamiltonian of thermal states
  is given by the integrated bulk Noether current associated to time translation
  plus a spacetime boundary term. One generally cannot express this in terms of
  gravitational stress tensor components. Still, working with the correct
  expression for the modular Hamiltonian, we are able to recover a first law of
  generalized entropy, with added benefits over the previous result. Firstly,
  any Wald-Dong contributions to generalized entropy resulting from non-minimal
  coupling between matter and curvature are included. Secondly, in gravitational
  equations of motion, we allow for a non-vanishing stress tensor expectation
  value in the unperturbed background and state, and account for background
  field perturbations as part of its variation. Finally, the quantum matter is
  allowed to contribute nontrivially to asymptotic energy, \eg{}as is necessary,
  even for a minimally coupled Maxwell field, to recover the expected
  thermodynamic first law of charged black holes.}

\begin{document}
\maketitle
\flushbottom

\section{Introduction}
\label{sec:rockBottom}

Holographic duality provides an equivalence between theories of quantum gravity
in asymptotically anti-de Sitter (AdS) spacetimes and strongly-coupled conformal
field theories (CFTs) of a large number (\ie{}$N^2$-many) degrees of freedom
residing on the spacetimes' boundaries. In recent years, the AdS/CFT
correspondence has fuelled a surge of interest in connecting geometric
quantities to quantum information measures of entanglement. Most notable in
these developments is the proposal by Ryu and Takayanagi
\cite{Ryu:2006bv,Ryu:2006ef} relating the entanglement entropy of a region
$\bdyReg$ in the boundary CFT to the area of an extremal surface in the bulk
spacetime. This can be understood as the leading contribution in a saddle-point
approximation for Einstein gravity \cite{Lewkowycz:2013nqa}. Corrections have
since been found, both in the case of higher-curvature gravity
\cite{Jacobson:1993xs,Hung:2011xb,Dong:2013qoa} and at next-to-leading order in
$1/N^2$, or equivalently in Newton's constant $\GNewton$,
\cite{Faulkner:2013ana,Engelhardt:2014gca} giving the bulk formula for
generalized entropy:
\begin{align}
  \entGen{\entSurf}
  =& \left\langle
     \entDong{\entSurf}
     \right\rangle
     + \ent[\stateMat]
     \;,
     \label{eq:resentment}
\end{align}
where $\entSurf$ is a codimension-two `quantum extremal' surface
\cite{Engelhardt:2014gca,Dong:2017xht}
\footnote{Being quantum extremal means that $\entSurf$ extremizes
  $\entGen{\entSurf}$ among surfaces homologous to $\bdyReg$ \cite{Dong:2017xht}
  --- towards the end of appendix \ref{sec:d4f5g3nf6bg2e6nf3be7o-oo-oc4d5}, we
  briefly review the derivation of quantum extremality in holography. If
  multiple quantum extrema exist, the one which gives the smallest
  $\entGen{\entSurf}$ is the one for which $\entGen{\entSurf}$ gives the entropy
  of the boundary region $\bdyReg$. \label{foot:d4nf6c4e6nf3b6}}, the Wald-Dong
entropy $\entDong{\entSurf}$ \cite{Wald:1993nt,Iyer:1994ys,Dong:2013qoa} is a
local quantity evaluated on $\entSurf$, reducing to the area $\area{\entSurf}$
\begin{align}
  \entDong{\entSurf}
  =& \frac{\area{\entSurf}}{4\GNewton}
     \qquad
     (\text{Einstein gravity})
\end{align} 
in the Einstein case, and $\ent[\stateMat]$ is the (von Neumann) entropy of the
bulk matter state $\stateMat$ in the region $\entReg$ bounded by $\entSurf$ and
$\bdyReg$.

The name ``generalized entropy'' for the quantity \eqref{eq:resentment} harks
back to its origins in black hole thermodynamics, where the non-decreasing
properties of \eqref{eq:resentment}, with $\entSurf$ now the black hole horizon
on evolving time slices, has been interpreted as a generalized second law for
black holes \cite{Bekenstein:1973ur}. Indeed, \eqref{eq:resentment} seems to be
the natural generalization of horizon area, promoting classical laws of black
hole thermodynamics to allow for the inclusion of quantum matter. For some more
recent discussions of the dynamics of generalized entropy evaluated on or near
horizons, see \eg{}\cite{Wall:2011hj,Bousso:2015eda,Bousso:2015mna}.

\begin{figure}
  \centering
  \begin{subfigure}[t]{0.48\textwidth}
    \centering \includegraphics[width=0.9\textwidth]{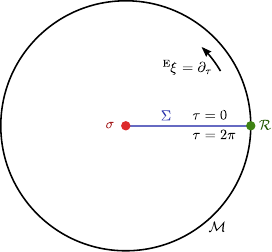}
    \caption{Euclidean path integral preparation of a thermal state on a
      rotation- (\ie{}$U(1)$-) symmetric Euclidean geometry.}
    \label{fig:e4e5ke2ke7}
  \end{subfigure}
  \hfill
  \begin{subfigure}[t]{0.48\textwidth}
    \centering \includegraphics[width=0.9\textwidth]{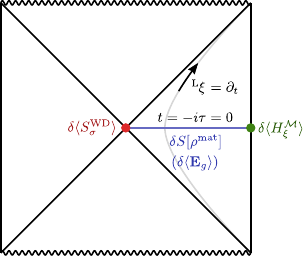}
    \caption{An eternal AdS black hole, exemplifying the physical evolution of a
      thermal setup prepared by panel \subref{fig:e4e5ke2ke7}.}
    \label{fig:d4d5c4c6}
  \end{subfigure}
  \caption{Euclidean preparation (\subref{fig:e4e5ke2ke7}) and Lorentzian
    evolution (\subref{fig:d4d5c4c6}) of a thermal setup. The Euclidean
    background in panel \subref{fig:e4e5ke2ke7} is symmetric with respect to the
    flow $\Euc{\thVec}$ of thermal time $\thTime$. The state is prepared on the
    surface $\entReg$, bounded in the bulk by the entangling surface $\entSurf$,
    around which $\Euc{\thVec}$ generates rotations, and on the spacetime
    boundary by $\bdyReg$. In panel \subref{fig:d4d5c4c6}, the Lorentzian time
    $\LorTime=-i\thTime$ evolution corresponds to a boost $\Lor{\thVec}$ that
    generates horizons. The $\entReg$, $\entSurf$, and $\bdyReg$ surfaces are
    embedded as shown (in the same colours as in panel \subref{fig:e4e5ke2ke7})
    in this Lorentzian geometry, with $\entSurf$ giving the bifurcation surface
    of the horizons. Elements of the first law of generalized entropy are
    written in the colours of the surfaces on which they respectively reside:
    the variations of the Dong entropy $\langle \entDong{\entSurf} \rangle$, the
    matter entropy $\ent[\stateMat]$, and the asymptotic energy $\langle
    \HamBdy{\thVec} \rangle$. Sufficient conditions of the first law are for the
    unperturbed and linearized gravitational equations of motion, $\langle
    \eomDensity{\metric} \rangle=0$ and
    $\vary\langle\eomDensity{\metric}\rangle=0$, to hold on $\entReg$.}
  \label{fig:d4f5e4}
\end{figure}

Among the laws of black hole thermodynamics, of particular focus in this note is
the first law\footnote{This first law in fact applies to any thermal setup, not
  just black holes. For instance, one may consider a Rindler wedge, which is
  thermal with respect to the boost direction (this would be generated by the
  $\thVec$ mentioned next in the text).} which relates the variation of entropy
to the variation of mass. The first law considers a setup that is thermal with
respect to a flow $\thVec=\partial_{\LorTime}$ in a certain time $\LorTime$ ---
by this, we mean that the background and state have a Euclidean preparation that
is symmetric under rotation in a thermal time direction $\thTime=i\LorTime$,
generated by (the Wick rotation\footnote{Elsewhere in this paper, when the
  distinction is important, we will use superscripts $\Euc{\bullet}$ and
  $\Lor{\bullet}$ to mark Euclidean and Lorentzian objects.} of) $\thVec$ ---
see Figure \ref{fig:e4e5ke2ke7}. An example would be an eternal black hole in
the Hartle-Hawking vacuum, whose Lorentzian evolution is illustrated in Figure
\ref{fig:d4d5c4c6}. In such cases where $\thVec$ is bifurcate, the bifurcation
surface $\entSurf$ bipartitions a Cauchy surface into two pieces, with each
piece corresponding to a region exterior to horizons generated by $\thVec$. The
generalized entropy $\entGen{\entSurf}$ evaluated for the entangling surface
$\entSurf$ is then interpreted as the entropy of one such half-space $\entReg$.

At the classical level, $\entGen{\entSurf}$ reduces to the Wald-Dong entropy
$\entDong{\entSurf}$ and the first law of black hole thermodynamics is perhaps
most elegantly stated using Wald's Noether charge formalism
\cite{Wald:1993nt,Iyer:1994ys}. Due to the symmetry of the metric generated by
$\thVec$, the Wald-Dong entropy $\entDong{\entSurf}$ can be related
\cite{Dong:2013qoa} to a Noether charge (density) $\charge{\thVec}$
\cite{Wald:1993nt,Iyer:1994ys}\footnote{The Noether charge entropy
  \eqref{eq:e4c5nf3d6d4cxd4nxd4nf6nc3a6} is often simply referred to as the Wald
  entropy.}, corresponding to $\thVec$, integrated along $\entSurf$:
\begin{align}
  \entDong{\entSurf}
  =& 2\pi \int_\entSurf \charge{\thVec}
     \;,
     \label{eq:e4c5nf3d6d4cxd4nxd4nf6nc3a6}
\end{align}
In fact, \eqref{eq:e4c5nf3d6d4cxd4nxd4nf6nc3a6} continues to hold for linear
perturbations away from thermal setups\footnote{ Some details regarding how one
  can see \eqref{eq:e4c5nf3d6d4cxd4nxd4nf6nc3a6} and the fact that it continues
  to hold at linear order in perturbations are explained later in this paper in
  footnote \ref{foot:keg}, below \eqref{eq:everyone}, and in footnotes
  \ref{foot:sequester} and \ref{foot:evoke}.
}. It is then shown that classically, for perturbations between nearby on-shell
configurations,
\begin{align}
  \vary\entDong{\entSurf}
  =& 2\pi \int_{\entSurf} \vary\charge{\thVec}
     = \vary\HamBdy{\thVec}
     \label{eq:hunger}
\end{align}
where $\HamBdy{\thVec}$ is an asymptotic energy, \eg{}including the ADM mass and
angular momentum for Einstein gravity \cite{Iyer:1994ys} and, in holographic
setups, giving the energy in the boundary CFT corresponding to the $\thVec$ flow
\cite{Faulkner:2013ica}. Thus, variations of entropy and energy are related in a
manner reminiscent of the first law of thermodynamics. In fact,
\cite{Faulkner:2013ica} found that for arbitrary perturbations, potentially to
off-shell configurations, \eqref{eq:hunger} simply receives a contribution given
by an integral of linearized gravitational equations of motion
$\vary\eomDensity{\metric}$ for the perturbation (where
$\eomDensity{\metric}^{ab}=\frac{\delta \act}{\delta \metric_{ab}}$ is the
functional derivative of the action $\act$ in the metric $\metric$). Moreover,
in the vacuum of AdS, sufficient symmetries exists such that
\cite{Faulkner:2013ica} were able to argue the converse direction, that the
first law \eqref{eq:hunger} implies the linearized gravitational equations of
motion are locally satisfied in the bulk.

In \cite{Swingle:2014uza}\footnote{See also
  \cite{Jacobson:2015hqa,Jacobson:2018ahi} for similar discussions in the
  context of causal diamonds.}, these results were lifted to the semiclassical
level, allowing for the presence of quantum matter in the bulk. It was presumed
that the modular Hamiltonian $\modHamMat$ corresponding to the bulk matter state
$\stateMat\propto e^{-2\pi\modHamMat}$ is given by the integral of components of
a stress tensor $\stress^{ab}$:
\begin{align}
  \modHamMat
  \eqDubious& -\int_{\entReg} \volForm_a \, \thVec_b\, \stress^{ab} \;,
              \label{eq:e4e6d4d5e5}
\end{align} 
where $\volForm_a$ gives the volume form on codimension-one
surfaces\footnote{Note that our default convention for orienting constant time
  $\LorTime$ surfaces such as $\entReg$ is similar to that of
  \cite{Iyer:1994ys}, taking the orientation to be given by
  $\normalVecOf{\LorTime}\cdot\volForm=\normalVecOf{\LorTime}^a \,\volForm_a$
  with $\volForm$ the spacetime volume form and $\normalVecOf{\LorTime}$ the
  normal vector directed towards increasing $\LorTime$. This seems to disagree
  with \cite{Swingle:2014uza}, as seen by comparison of our
  \eqref{eq:e4e6d4d5e5} and their (3.6). (It is possible that
  \cite{Swingle:2014uza} inherited a sign error from \cite{Faulkner:2013ica} ---
  see footnote \ref{foot:frail}.) At any rate, writing out \eqref{eq:e4e6d4d5e5}
  explicitly, one has
  $ -\int_{\entReg} \volForm_a \, \thVec_b\, \stress^{ab} = \int_{\entReg}
  d^dx\, \sqrt{\entRegMetric} \, \normalVecOf{t}^a \, \thVec^b \, \stress_{ab} $
  with $\entRegMetric$ the induced metric on $\entReg$. Note also that it is the
  Lorentzian stress tensor that is written here --- see section
  \ref{sec:albatross}.
  \label{foot:jailhouse}
}, in this case $\entReg$. The idea was then to leverage the first law of von
Neumann entropy
\begin{align}
  \vary\ent[\stateMat]
  =& 2\pi \varyWRT{\stateMat} \langle \modHamMat \rangle
     \label{eq:e4e6}
\end{align}
relating the variations of matter entropy $\ent[\stateMat]$ and the expectation
value of $\langle \modHamMat \rangle$. Here, $\varyWRT{\stateMat}$ means the
variation of an expectation value due to the matter state perturbation
$\vary\stateMat$, fixing the operator. As $\langle \stress^{ab} \rangle$
conveniently sources the gravitational equations of motion, again one naively
finds that linearized equations of motion $\vary\langle \eomDensity{\metric}
\rangle$ imply a first law
\begin{align}
  \vary\entGen{\entSurf}
  =& \vary\langle \HamBdy{\thVec}  \rangle
     \label{eq:fake}
\end{align}
at linear order in the background and matter state perturbations. Additionally,
as before, the converse direction can also be argued in the Poincar\'e vacuum.

There are, however, some inconvenient subtleties in this narrative.
Firstly, 
the modular Hamiltonian is not, in general, related to the gravitational stress
tensor as in \eqref{eq:e4e6d4d5e5}\footnote{To put it another way, for
  \eqref{eq:e4e6d4d5e5} to hold, the $\stress^{ab}$ there is in general not the
  \emph{gravitational} stress tensor appearing in gravitational equations of
  motion. In maximally-symmetric spacetimes for example, \eqref{eq:e4e6d4d5e5}
  typically holds (or at least gives the bulk part of the modular Hamiltonian)
  with the \emph{canonical} stress tensor, whose components correspond to those
  of Noether currents. For the remainder of this paper, when we say ``stress
  tensor'', we mean the gravitational stress tensor.}, especially in the
presence of non-minimal couplings between gravity and matter. For the
non-minimally coupled free scalar, for example, the modular Hamiltonian is
always the canonical generator for the $\thVec$ time flow
\cite{Casini:2014yca,Casini:2019qst}. Secondly, in the presence of non-minimal
coupling, there is also potential for the matter sector to make additional
contributions $\langle \entDongMat{\entSurf} \rangle$ to the Wald-Dong entropy
$\langle \entDong{\entSurf} \rangle$ --- these have been called ``Wald-like''
terms in \cite{Faulkner:2013ana}. Such terms are not accounted for in the
derivation of \cite{Swingle:2014uza}. Drawing inspiration from the calculations
of non-minimally coupled scalars, \cite{Jafferis:2015del} suggests that these
two aforementioned subtleties resolve each other --- namely that the variations
of Wald-like terms cancel against the incongruities between the modular
Hamiltonian and the stress tensor integral. In this note, we shall bear out
these claims in far greater generality than scalar field theory.

Yet, there are still more issues related to the size of background
perturbations, \eg{}perturbations $\vary\metric$ of the metric, and what
``linearized'' means for gravitational equations of motion and the first law of
generalized entropy \eqref{eq:fake}. In previous work,
\eg{}\cite{Swingle:2014uza,Jafferis:2015del}, the authors were content with
perturbations $\vary\metric\sim O(\GNewton)$, say resulting from back-reaction
of the matter state perturbation $\vary\stateMat$. Then, part of the meaning of
linearization, in their results, is that the gravitational equations of motion
for the perturbation $\vary\metric\sim O(\GNewton)$ are written to leading order
in $\GNewton$ and the generalized first law \eqref{eq:fake} is obtained to
zeroth order in $\GNewton$. This, for example, is what allows, in the argument
summarized around \eqref{eq:fake}, for the transition from
$\varyWRT{\stateMat}\langle \modHamMat \rangle$ in \eqref{eq:e4e6} to the full
variation of the stress tensor $\vary\langle \stress^{ab} \rangle$ required to
source the linearized gravitational equations $\vary\langle \eom{\metric}^{ab}
\rangle$ --- note that, in general, there should also be a contribution to
$\vary \langle\stress^{ab} \rangle$ resulting from the variation of the operator
$\stress^{ab}$ itself due to perturbations in the background fields. The
assumption of $\vary\metric$ being perturbatively small in $\GNewton$ is also
needed in previous work, if quantum fields make nontrivial contributions to the
matter stress tensor in the unperturbed thermal setup. (Of course, in the
Poincar\'e vacuum on which \cite{Swingle:2014uza} is focused, the matter stress
tensor vanishes.) This is because, implicit in the derivation of the first law
\eqref{eq:fake} (even in its classical statement \eqref{eq:hunger}) is the
assumption that the unperturbed setup satisfies gravitational equations of
motion --- otherwise, these equations appear, multiplied by $\vary\metric$, as
an extra term in the first law. To add quantum matter, one should therefore
provide a term like $\langle \stress^{ab} \rangle \vary\metric_{ab}$ ---
something which the derivation described around \eqref{eq:fake} seems to miss.
Of course, if $\vary\metric \sim O(\GNewton)$ and one is only interested in the
first law up to $O(\GNewton)$ corrections, then this is a nonissue.

In contrast, we shall allow for background variations that are not
perturbatively small in $\GNewton$ --- for example, one might envision turning
on a classical gravitational wave background that is not suppressed by
$\GNewton$. We shall find that, assuming gravitational equations linearized in
the perturbations of the background fields and the matter state $\stateMat$, the
generalized first law \eqref{eq:fake} holds, also at linearized order in the
perturbations. In fact, there will be no expansion in $\GNewton$, except
possibly in justifying the semiclassical theory in the first place to frame it
in the context of a full quantum gravity theory (see appendix
\ref{sec:d4f5g3nf6bg2e6nf3be7o-oo-oc4d5})\footnote{If we pretend the
  semiclassical theory is a self-contained theory, and we take the generalized
  entropy formula \eqref{eq:resentment} to be an exact notion of entropy in this
  theory, then we will find no further need to expand in $\GNewton$.}. The
issues mentioned in the preceding paragraph simply resolve each other.

Altogether, we shall resolve all the issues mentioned in the preceding
paragraphs very generally, considering arbitrary theories of gravity possibly
non-minimally coupled to quantum matter, which may consist of arbitrary tensor
fields. We shall show how linearized gravitational equations of motion imply
that the generalized entropy \eqref{eq:resentment} satisfies the first law
\eqref{eq:fake} relating variations of entropy and energy away from an arbitrary
thermal setup. This will be in spite of the fact that, as we will find, the
modular Hamiltonian of the matter state $\stateMat$ corresponds to the integral
of a Noether current plus a boundary term, as opposed to a gravitational stress
tensor $\stress^{ab}$
component. 
Moreover, we allow for background perturbations unsuppressed by $\GNewton$. As
an added bonus for working carefully with the correct expression for the modular
Hamiltonian, we will also find that contributions of the matter sector to the
asymptotic energy $\HamBdy{\thVec}$ automatically appear in the first law of
generalized entropy. In \cite{Swingle:2014uza}, matter is taken to fall off
sufficiently quickly at the spacetime boundary so as to not contribute new terms
to the asymptotic energy. However, as we summarize in the discussion section
\ref{sec:d4f5g3nf6bg2e6nf3o-oo-oc4} and explicitly show in appendix
\ref{sec:garnish}, such terms, for the example of a minimally coupled quantum
Maxwell field propagating on a black hole background, are important for
recovering the expected thermodynamic first law of electrically charged black
holes.

The remainder of this paper is organized as follows. We begin by establishing
notation in section \ref{sec:helium} for our semiclassical theory and re-express
generalized entropy in terms of the theory's effective action in a form bearing
resemblance to Callan and Wilczek's entropy formula \cite{Callan:1994py}. With
this, we shall show in section \ref{sec:fluorine} that the matter modular
Hamiltonian is in general given by the integral of the Noether current
associated with the thermal time flow $\thVec$ plus a boundary term. In section
\ref{sec:neon} we proceed to show that the first law \eqref{eq:fake} of
generalized entropy nonetheless holds, fully accounting for the variation of
Wald-like contributions $\langle \entDongMat{\entSurf} \rangle$ resulting from
non-minimal coupling. Our results and possible future directions are discussed
in section \ref{sec:d4f5g3nf6bg2e6nf3o-oo-oc4}. Details relating our
semiclassical approach, involving the effective action, to perhaps more
conventional holographic entropy calculations, featuring quantum gravity
partition functions, are included in appendix
\ref{sec:d4f5g3nf6bg2e6nf3be7o-oo-oc4d5}\footnote{Our approach is not entirely
  new, as we draw many connections to the use of effective actions in
  \cite{Dong:2017xht}, but we have aimed to be more explicit than perhaps
  previous discussions existing in literature.}. Some intermediate results are
derived in appendices \ref{sec:immorally} and \ref{sec:commodore}, including a
generalization of our expression for the thermal modular Hamiltonian to give the
instantaneous generator of time evolution on time-dependent backgrounds. In
appendix \ref{sec:garnish} we consider the contribution that a Maxwell field
makes to the asymptotic energy. Finally, in appendix \ref{sec:deflate} we
describe how UV-divergences are renormalized in generalized entropy, and how the
variation of the matter path integration measure might correct various
quantities encountered in this paper.


\section{Preliminaries}

\label{sec:helium}
Let us begin by introducing some notation. In this note, we are motivated by the
consideration of a quantum gravitational theory on a spacetime\footnote{Our
  calculations here and in the remainder of the paper carry through just as well
  in spacetimes without boundaries, \eg{}in de Sitter spacetime, simply by
  ignoring spacetime boundary terms. It is interesting to note that, in those
  cases, the RHS of first law \eqref{eq:fake}, which we shall derive in section
  \ref{sec:neon}, is zero as there is no asymptotic energy.} with boundary
$\bdyMan$. The theory will have a metric $\metric$ and other background fields
$\bgFieldsOther$. On top of this background will live a graviton $\graviton$ and
other quantum fields $\qFieldsOther$. The action $\act$ will consist of a
classical gravitational piece $\actGrav$ and a matter piece $\actMat$ (with
gravitons included in the latter, as discussed shortly):
\begin{align}
  \act[\metric,\graviton,\bgFieldsOther+\qFieldsOther]
  =& \actGrav[\metric]
     + \actMat[\metric,\graviton,\bgFieldsOther+\qFieldsOther] \;.
     \label{eq:alone}
\end{align}
More generally, we shall attach superscripts $\bullet^{\grav}$ and
$\bullet^{\mat}$ to symbols when we wish to refer to the purely gravitational or
the matter parts of some object\footnote{Note that some objects associated with
  $\graviton$, $\bgFieldsOther$, and $\qFieldsOther$ can only come from
  $\actMat$, so we will omit $\bullet^{\mat}$ superscripts on those objects. For
  example, we write for the equations of motion and symplectic potential terms
  to be introduced in \eqref{eq:copier} below:
  $\eomDensity{\graviton}=\eomDensityMat{\graviton}$,
  $\eomDensity{\bgFieldsOther}=\eomDensityMat{\bgFieldsOther}=\eomDensity{\qFieldsOther}=\eomDensityMat{\qFieldsOther}$,
  and
  $\sympPot[\vary\graviton,\vary\bgFieldsOther+\vary\qFieldsOther]=\sympPotMat[\vary\graviton,\vary\bgFieldsOther+\vary\qFieldsOther]$.}.
We shall clarify the appearance of the combination of arguments
$(\metric,\graviton,\bgFieldsOther+\qFieldsOther)$ in these actions further
below. The action is given by the integrals
\begin{align}
  \act =& \int \LagDensity + \int_{\bdyMan} \LagDensityBdy\;,
          \label{eq:treadmill}
\end{align} 
of Lagrangian densities
\begin{align}
  \LagDensity
  =& \LagDensityGrav
     + \LagDensityMat
     = \volForm \Lag
     \;,
   &
     \LagDensityBdy
     =& \LagDensityBdyGrav
        + \LagDensityBdyMat
\end{align} 
over the bulk spacetime and its boundary $\bdyMan$ respectively, where
$\volForm$ is the bulk volume form. As stated previously, in this paper, we
shall allow for arbitrary tensor matter fields $\bgFieldsOther+\qFieldsOther$
(unless required, we most often will notationally suppress tensor indices). As
we explain in the following paragraph, our reason for allowing a boundary
contribution $\int_{\bdyMan}\LagDensityBdy$ to the action is not to introduce
extra degrees of freedom or dynamics on the boundary $\bdyMan$, but rather to
ensure the bulk theory has a good variational principle. The effective action
$\eActMix[\metric]$ is then given by the logarithm of a path integral:
\begin{align}
  \eActMix[\metric]
  =& \actGrav[\metric] + \eActMixMat[\metric]
     \;,
   &
     \eActMixMat[\metric]
     =&
        -\log \int [d\graviton] [d\qFieldsOther]
        e^{-\actMat[\metric,\graviton,\bgFieldsOther+\qFieldsOther]}
        \;.
        \label{eq:gadolinium}
\end{align}
As suggested by the above notation, we begin here by considering a Euclidean
theory.

In this paper, we shall consider the situation where the fields
$\metric+\graviton$ and $\bgFieldsOther+\qFieldsOther$ are required to satisfy
certain boundary conditions at the spacetime boundary; we take the background
fields $\metric$ and $\bgFieldsOther$ to satisfy these and we shall suppose the
boundary conditions then require the quantum fields $\graviton$ and
$\qFieldsOther$ to have certain asymptotics on approach to the spacetime
boundary $\bdyMan$. For example, in holography, $\metric$ and $\bgFieldsOther$
would have the asymptotics of non-normalizable modes while $\graviton$ and
$\qFieldsOther$ would behave like normalizable modes. Note that, even if
$\graviton$ and $\qFieldsOther$ are required to vanish at the spacetime boundary
$\bdyMan$, this does not necessarily imply that the boundary action
$\int_{\bdyMan} \LagDensityBdy$ is independent of the quantum fields. One should
think of this action as evaluated on the boundary limit of some IR cutoff
surface --- terms involving the quantum fields may yet be finite if they also
contain compensating factors, say of the background fields, which become large
in the boundary limit. In fact, the boundary action is sometimes required for
the bulk theory to have a well-defined variational principle. Specifically, let
us write the field variation of the bulk Lagrangian $\LagDensity$ as
\begin{align}
  \varyWRT{\egField}\LagDensity
  =& \eomDensity{\egField} \vary\egField + d\sympPot[\vary\egField] \;,
  & (\egField\in\{\metric,\bgFieldsOther,\graviton,\qFieldsOther\})
    \label{eq:copier}
\end{align}
where
\begin{align}
  \frac{\delta \act}{\delta \egField}
  =& \eomDensity{\egField}
     = \eom{\egField} \volForm
  & (\egField\in\{\metric,\bgFieldsOther,\graviton,\qFieldsOther\},\; \text{$\vary\egField$ away from $\bdyMan$})
\end{align}
denotes equations of motion (multiplied by the spacetime volume form $\volForm$)
and the symplectic potential $\sympPot$ accounts for boundary terms that arise
when extracting the equations of motion. Then, in order for the equations of
motion alone to be sufficient for extremizing the action $\act$ with respect to
the fields (and thus for there to be a sensible classical limit), and for the
path integral \eqref{eq:gadolinium} to be sensible\footnote{To see what can go
  wrong in the path integral if one has non-vanishing $\sympPot|_{\bdyMan}$ but
  does not include a compensating boundary action, say taking
  $\LagDensityBdy=0$, consider the argument leading to \eqref{eq:uranium} below.
  The argument for $\varyWRT{\bgFieldsOther}\eActMix[\metric]=0$ actually
  generally applies to any variation $\vary\bgFieldsOther$ which preserves the
  boundary conditions on the spacetime boundary $\bdyMan$ and thus can be
  absorbed as a shift of the quantum fields. For such variations, we have
  \begin{align}
    0
    =& \varyWRT{\bgFieldsOther} \eActMix[\metric]
       = \int_{\bdyMan} \langle \sympPot[\vary\bgFieldsOther] \rangle\;,
    &
      (\text{$\LagDensityBdy=0$ and $\vary\bgFieldsOther$ preserving boundary conditions})
  \end{align}
  where we have made use of \eqref{eq:uranium} in the second equality.
  (Actually, there may also be contributions due to variations of the path
  integral measure, but we leave such issues for appendix \ref{sec:deflate}.)
  However, this is obviously untrue in certain examples. For example, for a
  scalar field in holography, when one considers quantum fluctuations in the
  faster falling off normalizable modes, $\sympPot[\vary\bgFieldsOther]$, with
  $\vary\bgFieldsOther$ having the asymptotics of such a mode, is actually a
  c-number determined by the fixed non-normalizable modes. The resolution is to
  recognize that one must include in $\LagDensityBdy$ the holographic
  renormalization counterterm for the scalar field --- see (5.106) in
  \cite{Ammon:2015wua} and (2.11) in \cite{Marolf:2006nd} --- which cancels
  against the symplectic potential as in \eqref{eq:wiring}. 
  A similar exercise can also be carried out for the metric $\metric$ and
  graviton $\graviton$ in the quantum gravity path integral, prior to
  Legendre-transforming to the effective action as described in appendix
  \ref{sec:d4f5g3nf6bg2e6nf3be7o-oo-oc4d5}. One finds, in Einstein gravity, that
  including the usual Gibbons-Hawking boundary term as well as holographic
  renormalization counterterms --- see (5.123) in \cite{Ammon:2015wua} --- in
  $\LagDensityBdy$ allows \eqref{eq:wiring} to be satisfied.
  Legendre-transforming to the semiclassical theory at one-loop-in-$\GNewton$,
  one expects \eqref{eq:wiring} to continue to hold as one expands in the
  graviton and discards all but the zeroth and second order terms.
}, one must require\footnote{When we write, \eg{}$\bullet_{\bdyMan}$ or
  $\bullet|_{\bdyMan}$ for tensors or forms, we mean not only evaluation at
  $\bdyMan$, but also projection to components along $\bdyMan$, unless indices
  are otherwise specified. \label{foot:kinfolk}}
\begin{align}
  \left. \sympPot[\vary\egField] \right|_{\bdyMan}
  =& -\varyWRT{\egField}\LagDensityBdy|_{\bdyMan} \;,
  &
    (
    \egField\in\{\metric,\graviton,\bgFieldsOther,\qFieldsOther\},
    \text{$\vary\egField$ preserves b.c.}
    )
    \label{eq:wiring}
\end{align}
for any field variations which preserve the boundary conditions on the spacetime
boundary $\bdyMan$. One can verify that \eqref{eq:wiring} is satisfied in
holography, where $\int_{\bdyMan}\LagDensityBdy$ contains Gibbons-Hawking-like
actions and boundary counterterm actions\footnote{For Einstein gravity and
  scalar fields where the modes falling off faster at the AdS boundary are
  path-integrated over while slower-falloff modes are fixed,
  $\int_\bdyMan\LagDensityBdy$ is precisely the sum of the Gibbons-Hawking
  action and counterterms needed for holographic renormalization --- for the
  latter see (5.106) and (5.123) of \cite{Ammon:2015wua}. In certain mass ranges
  of the scalar field, however, one can instead choose to swap the modes which
  are integrated over and those which are fixed by boundary conditions
  \cite{Klebanov:1999tb,Marolf:2006nd}; then, one must adjust the boundary
  action $\int_{\bdyMan}\LagDensityBdy$ appropriately --- see (2.16) in
  \cite{Marolf:2006nd}. As described in \cite{Marolf:2006nd} (and we discuss
  briefly in appendix \ref{sec:garnish}) an analogous choice exists for vector
  fields. In any case, we expect \eqref{eq:wiring} to hold for a given choice of
  boundary conditions and the corresponding boundary action $\int_\bdyMan
  \LagDensityBdy$.}. (As stated below however, when the entangling surface
$\entSurf$ stretches to the spacetime boundary $\bdyMan$, in this paper, we do
not allow $\int_{\bdyMan} \LagDensityBdy$ to contribute to Wald-Dong entropy ---
such contributions would be interpreted as counterterms to bulk generalized
entropy that renormalize its IR divergence. Rather, we consider $\int_{\bdyMan}
\LagDensityBdy$ primarily to ensure that the bulk dynamics and path integral are
well-defined.)

Let us now discuss the interpretation of the effective action. The effective
action \eqref{eq:gadolinium} can be either thought of as the
(one-loop-in-$\GNewton$ approximation of) the effective action for the metric or
simply the logarithm of the partition function in a semiclassical theory where
$\graviton$ is just an extra matter field --- we leave the details of these
descriptions to appendix \ref{sec:d4f5g3nf6bg2e6nf3be7o-oo-oc4d5}, but highlight
here some key points. We recall that the first variations of effective actions
provide nontrivial equations of motion. (This is described around
\eqref{eq:tavern}.) Here, a nontrivial equation of motion for $\metric$ is
obtained from the first derivative of $\eActMix[\metric]$:
\begin{align}
  0
  =& \frac{\delta \eActMix[\metric]}{\delta\metric}
     = \eomDensityGrav{\metric}
     +\frac{1}{2}\left\langle \stressDensity \right\rangle \;,
  &
  &(\text{$\vary\metric$ away from $\bdyMan$, $\metric$ on-shell})
    \label{eq:lonely}
\end{align}
where
\begin{align}
  \frac{\delta \actGrav[\metric]}{\delta \metric_{ab}}
  \equiv& (\eomDensityGrav{\metric})^{ab}
          \equiv (\eomGrav{\metric})^{ab} \volForm
  &
  &(\text{$\vary\metric$ away from $\bdyMan$})
    \label{eq:earthly}
\end{align} 
gives the gravitational part of the metric's equation of motion and
\begin{align}
  2\frac{\delta \eActMixMat[\metric]}{\delta \metric_{ab}}
  =& \langle \stressDensity^{ab} \rangle
     \;,
  &
    2(\eomDensityMat{\metric})^{ab}
    =& \stressDensity^{ab}=\stress^{ab} \volForm \;.
  &
  &(\text{$\vary\metric$ away from $\bdyMan$})
    \label{eq:friendless}
\end{align}
is the matter stress tensor\footnote{As we shall show in section
  \ref{sec:albatross}, the Euclidean stress tensor $\stress^{ab}$ considered
  here is the negative of the Lorentzian stress tensor, \eg{}as appears in
  \eqref{eq:e4e6d4d5e5}.}. (Actually, in the second equality, the stress tensor
should generally also receive a contribution from the variation of the path
integral measure with respect to the metric. We leave consideration of such
contributions to appendix \ref{sec:deflate}, where we collectively discuss all
anomalous corrections to quantities in this paper resulting from variations of
the path integral measure.) The meaning of \eqref{eq:lonely} is that the
background metric $\metric$ satisfying this equation of motion correctly
captures the expectation value of the quantum field $\langle \metric+\graviton
\rangleQG = \metric$, \ie $\langle \graviton \rangleQG = 0$, in the quantum
gravitational (QG) theory --- see \eqref{eq:sequence}. Recall that the
calculation of an effective action requires the removal of tadpoles. To get the
generating functional at one-loop-order-in-$\GNewton$, this means the removal of
terms in the action linear in $\graviton$. The quadratic terms, on the other
hand, are included in $\actMat[\metric,\graviton,\bgFieldsOther+\qFieldsOther]$
--- see \eqref{eq:covenant}\footnote{To go to higher orders in $\GNewton$, one
  should keep interaction terms involving $\graviton$ but remove the resulting
  loop diagrams contributing to $\graviton$ tadpoles. As emphasized in appendix
  \ref{sec:d4f5g3nf6bg2e6nf3be7o-oo-oc4d5}, the latter step presents an obstacle
  to expressing the effective action in the form \eqref{eq:gadolinium}, since it
  is not clear that a local modification to the action is sufficient to
  implement the prescription of killing loop tadpoles. Of course, gravity is
  infamously nonrenormalizable, so this higher-loop story is sketchy at best
  anyway. At any rate, we will be satisfied in this note to stay at
  one-loop-order-in-$\GNewton$}. By construction then, the expectation value
$\langle \graviton \rangle$ evaluated using the semiclassical action
$\actMat[\metric,\graviton,\bgFieldsOther+\qFieldsOther]$ always vanishes:
\begin{align}
  \langle \graviton \rangle
  =& 0 \;,
     \label{eq:jeep}
\end{align}
for any background $\metric$ (see discussion above \eqref{eq:trifocal}).
However, the graviton two-point function is nontrivial and, in particular, the
matter stress tensor \eqref{eq:friendless} receives contributions from the
graviton owing to the terms in
$\actMat[\metric,\graviton,\bgFieldsOther+\qFieldsOther]$ quadratic in
$\graviton$.

In order to consider nontrivial gravitational equations of motion
\eqref{eq:lonely}, we have thus distinguished the treatment of the metric and
graviton from other background and quantum fields. These other fields, in
contrast, only appear in the combination $\bgFieldsOther+\qFieldsOther$ so that
the background fields $\bgFieldsOther$ merely shift the quantum fields
$\qFieldsOther$. Note, in particular that, due to the path integration of
$\qFieldsOther$, \eqref{eq:gadolinium} is independent of the bulk profile of
$\bgFieldsOther$. (One, might however, imagine that the near-boundary profile of
$\bgFieldsOther$ specifies the boundary conditions imposed on
$\bgFieldsOther+\qFieldsOther$, on which the path-integral implicitly depends.)
Hence, the first variation of \eqref{eq:gadolinium} in $\bgFieldsOther$ vanishes
trivially:
\begin{align}
  0
  =& \frac{\delta \eActMix[\metric]}{\delta \bgFieldsOther}
     = \langle \eomDensity{\bgFieldsOther} \rangle \;.
  &(\text{$\vary\bgFieldsOther$ away from $\bdyMan$})
    \label{eq:uranium}
\end{align}
This should be contrasted with the nontrivial gravitational equations of motion
\eqref{eq:lonely}; indeed, one motivation for us to consider the gravitational
effective action, as opposed to a partition function where $\metric$ merely
shifts $\graviton$, is so that one can sensibly ask whether a given $\metric$ is
on-shell, as defined by \eqref{eq:lonely}.

From here on, we shall simply take the form of the action \eqref{eq:alone}
as-is, with the path integral of \eqref{eq:gadolinium} describing a
semiclassical theory of quantum matter $\graviton, \qFieldsOther$ on a classical
background $\metric,\bgFieldsOther$. In this theory, the path integral of
\eqref{eq:gadolinium} provides a definition of an expectation value in the
semiclassical theory: for a given operator $\blkOp$,
\begin{align}
  \langle  \blkOp \rangle
  =& \frac{
     \int [d\graviton] [d\qFieldsOther]\;
     e^{-\actMat[\metric,\graviton,\bgFieldsOther+\qFieldsOther]}\; \blkOp
     }{
     e^{-\eActMixMat[\metric]}
     }
     \;.
     \label{eq:desperate}
\end{align}
In quantum field theory, one often thinks of a Euclidean path integral as
preparing a state of quantum matter on a codimension-one time slice. One can
then consider a subregion $\entReg$ on this slice%
\footnote{In the present discussion, we aim to be quite general and independent
  of holography. In particular, $\entReg$ need not be bounded in the bulk by a
  quantum extremal entangling surface $\entSurf$. We recognize, however, there
  is some discussion to be had about gauging diffeomorphisms in the graviton
  path integral; one may ask, for example, how the region $\entReg$, or
  equivalently the location of $\entSurf$, is chosen in each configuration
  $\metric+\graviton$ included in the graviton path integral computing
  $\unStateMat$. We will leave such subtleties related to gauge-fixing to future
  work. In the discussion section \ref{sec:d4f5g3nf6bg2e6nf3o-oo-oc4}, however,
  we speculate on how the formula \eqref{eq:astatine} to be introduced below may
  provide a reasonable definition of generalized entropy for arbitrary regions,
  even in gauge theories.
  \label{foot:d4nf6c4e6nf3b6g3}} with reduced state $\stateMat$. Computing the
expectation value of an operator $\blkOpInit$ on $\entReg$ using
\eqref{eq:desperate} is then thought of as tracing against the state
$\stateMat$. Namely,
\begin{align}
  \tr(\stateMat \blkOpInit)
  =& \langle \blkOpInit \rangle \;,
\end{align}
or introducing the unnormalized state $\unStateMat$,
\begin{align}
  \stateMat =& \frac{\unStateMat}{\tr\unStateMat}
               \;,
                 &
                   \tr\unStateMat
                   =& e^{-\eActMixMat[\metric]}
                      \;,
                      \label{eq:panic}
\end{align} 
we have
\begin{align}
  \tr(\unStateMat \blkOpInit)
  =& \int [d\graviton] [d\qFieldsOther]\;
     e^{-\actMat[\metric,\graviton,\bgFieldsOther+\qFieldsOther]}\; \blkOpInit
     \;.
     \label{eq:aspirin}
\end{align}

Given the background $(\metric,\bgFieldsOther)$ and state $\stateMat$ of quantum
fields $(\graviton,\qFieldsOther)$, we may calculate the generalized entropy of
the region $\entReg$ --- that is, the generalized entropy across the
codimension-two entangling surface $\entSurf$ which bi-partitions the initial
time slice into $\entReg$ and its complement. This generalized entropy is given
by
\begin{align}
  \entGen{\entSurf}[\metric]
  =& \left\langle
     \entDong{\entSurf}[\metric,\graviton,\bgFieldsOther+\qFieldsOther]
     \right\rangle
     + \ent[\stateMat]
     \;,
     \label{eq:solitude}
\end{align}
where the Wald-Dong entropy $\entDong{\entSurf}$ is a local contribution given
by the integral of some covariant local quantity along $\entSurf$
\cite{Dong:2013qoa} --- to be described further below --- and $\ent[\stateMat]$
is the von Neumann entropy of the reduced state $\stateMat$ in the region
$\entReg$:
\begin{align}
  \ent[\stateMat]
  =& -\tr\left( \stateMat \log \stateMat \right).
     \label{eq:depression}
\end{align}
An alternative expression
\begin{align}
  \ent[\stateMat]
  =& 2\pi \left\langle \modHamMat \right\rangle
     +\log\tr\unStateMat
     \label{eq:sadness}
\end{align}
also exists in terms of the trace of the unnormalized state $\unStateMat$ and
the modular Hamiltonian $\modHamMat$, defined by
\begin{align}
  \unStateMat
  =& e^{-2\pi\modHamMat} \;.
\end{align} 
In practice, it is often helpful to think of \eqref{eq:depression} as the $n\to
1$ limit of R\'enyi entropies
\begin{align}
  \entR{n}[\stateMat]
  =& -\frac{1}{n-1}\log\tr [(\stateMat)^n] \;,
     \label{eq:erasable}
  \\
  \ent[\stateMat]
  =& \lim_{n\to 1}\entR{n}[\stateMat]
     = - \partial_n \left(
     \log \tr[(\stateMat)^n]
     \right)_{n=1}
     \;.
     \label{eq:lost}
\end{align}

The discussion in the previous paragraph generally extends beyond setups that
have simple Euclidean preparations --- the background and quantum fields in
\eqref{eq:solitude} can be those that describe a physical Lorentzian background
and $\unStateMat$ can be an arbitrary state on a spatial region $\entReg$ with
entangling surface $\entSurf$ in that Lorentzian system. The utility of
\eqref{eq:lost}, however, is most evident when a Euclidean preparation is
possible and the R\'enyi entropies can be calculated using the replica
trick\footnote{One may also consider setups prepared by more complicated
  Schwinger-Keldysh path integrals, see \eg{}\cite{Dong:2016hjy}. In those
  cases, in the replica procedure described in the following paragraph, one
  should glue together copies of the Lorentzian and Euclidean geometries through
  which the Schwinger-Keldysh path integral runs.} --- to discuss this, it is
helpful now to introduce some notation useful for gluing together replicas of
manifolds and fields.

Given a codimension-two surface $\entSurf$ in a (Euclidean) manifold with metric
$\metric$, we shall use $\metricR{\rep{\entSurf}n}$ to denote the metric of the
replicated manifold obtained by a `branched' stitching together of $n$ replicas
of the original geometry, such that one moves from replica to replica as one
goes around $\entSurf$. We will define similar procedures
$\bullet_{\rep{\entSurf}n}$ for other fields. The inverse operation, which we
shall denote $\bullet_{\orb{\entSurf}n}$, corresponds to taking the orbifold
around $\entSurf$. The replicated manifold with other replicated background
fields obtained from the $\bullet_{\rep{\entSurf}n}$ operation are precisely the
background over which one expects to path-integrate in order to obtain the trace
in the R\'enyi entropy \eqref{eq:erasable}. There is, however a subtlety that we
must clarify: replication generally introduces or modifies conical singularities
on $\entSurf$; \eg{}if $\metric$ is smooth on $\entSurf$, then
$\metricR{\rep{\entSurf}n}$ will have a conical singularity with total angle
$2\pi n$ around $\entSurf$. When we write an action involving fields replicated
or orbifolded with $\bullet_{\rep{\entSurf}n}$ or $\bullet_{\orb{\entSurf}n}$,
we shall \emph{exclude} any modification to conical singularities as
contributions to the action localized on $\entSurf$ resulting from the
$\bullet_{\rep{\entSurf}n}$ or $\bullet_{\orb{\entSurf}n}$ operation. That is,
if we define $\actEntSurf{\entSurf}$ as the restriction of the integral defining
$\act$ to a small $a$-radius neighbourhood $\neigh{\neighSize}{\entSurf}$ of
$\entSurf$, then we write
\begin{align}
  \actEntSurf{\entSurf}[
  \metricR{\rep{\entSurf}n},
  \gravitonR{\rep{\entSurf}n},
  (\bgFieldsOther+\qFieldsOther)_{\rep{\entSurf}n}
  ]
  =& n\actEntSurf{\entSurf}[
     \metric,
     \graviton,
     \bgFieldsOther
     +\qFieldsOther
     ]
     \label{eq:trustless}
\end{align} 
(which vanishes in the $a\to 0$ limit when
$(\metric,\graviton,\bgFieldsOther+\qFieldsOther)$ are smooth at $\entSurf$)
and, moreover, the action on the replicated configuration is precisely $n$ times
the original:
\begin{align}
  \act[
  \metricR{\rep{\entSurf}n},
  \gravitonR{\rep{\entSurf}n},
  (\bgFieldsOther+\qFieldsOther)_{\rep{\entSurf}n}
  ]
  =& n\act[
     \metric,
     \graviton,
     \bgFieldsOther
     +\qFieldsOther
     ] \;.
     \label{eq:carwash}
\end{align}
The path integral evaluation of $\tr[(\stateMat)^n]$ in the R\'enyi entropy
\eqref{eq:erasable}, and more generally the trace $\tr[(\stateMat)^n
\blkOpInit]$ against an arbitrary operator $\blkOpInit$ on $\entReg$, should
similarly \emph{not} acquire extra contributions to its action\footnote{This is
  perhaps most obvious when starting with the path integral preparation of the
  state $\unStateMat$ --- here, the path integral is done on the unreplicated
  geometry and there is obviously no extra $n$-dependent contribution on
  $\entSurf$. The prescription to get $\tr[(\unStateMat)^n]$ is then merely to
  take $n$ copies of this path integral and perform a branched-identification of
  the quantum fields along $\entReg$ between the different
  copies. \label{foot:f3}} localized on $\entSurf$ with increasing $n$. Thus,
our notation allows us to simply write
\begin{align}
  \tr[(\unStateMat)^n \blkOpInit]
  =& \int [d\graviton][d\qFieldsOther]\;
     \exp\left\{
     - \actMat[
     \metricR{\rep{\entSurf}n},
     \graviton,
     \bgFieldsOtherR{\rep{\entSurf}n}
     + \qFieldsOther
     ]
     \right\}
     \blkOpInit
     \;.
     \label{eq:rename}
\end{align}
While the RHS only has an obvious definition for $n\in\naturals$, the LHS
suggests that analytic continuation to real $n>0$ is possible. Then, applying
\eqref{eq:panic} and \eqref{eq:lost} to \eqref{eq:rename} with $\blkOpInit=\id$
allows for a calculation of the matter von Neumann entropy $\ent[\stateMat]$.

To define the Wald-Dong entropy $\entDong{\entSurf}$ appearing in the
generalized entropy \eqref{eq:solitude} however, it is also helpful to introduce
the deformed replication operation $\bullet_{\repS{\entSurf}n}$ which involves
an additional deformation in the tiny $a$-radius neighbourhood
$\neigh{\neighSize}{\entSurf}$ of $\entSurf$ such that the conical properties of
the field exactly at $\entSurf$ remain invariant under
$\bullet_{\repS{\entSurf}n}$ --- in particular, the opening angle around
$\entSurf$ must not change. We shall also require the deformation respect the
$\integers_n$-symmetry of the replicated field. For example, if $\metric$ is
smooth at $\entSurf$, then $\metricR{\repS{\entSurf}n}$ is a
$\integers_n$-symmetric geometry with a smoothed, \ie{}regulated, conical
singularity with large curvature in the neighbourhood
$\neigh{\neighSize}{\entSurf}$ of $\entSurf$, giving extra contributions to the
action that become localized on $\entSurf$ in the limit $a\to 0$ of vanishing
neighbourhood size. Thus, in contrast to \eqref{eq:trustless}, the
$\bullet_{\repS{\entSurf}n}$ operation nontrivially modifies contributions to
the action localized in $\neigh{\neighSize}{\entSurf}$; let us quantify this
with
\begin{align}
  \actEntSurf{\entSurf,n}[\metric,\graviton,\bgFieldsOther+\qFieldsOther]
  \equiv& \frac{1}{n} \act[
          \metricR{\repS{\entSurf}n},
          \gravitonR{\repS{\entSurf}n},
          (\bgFieldsOther + \qFieldsOther)_{\repS{\entSurf}n}
          ]
          - \act[\metric,\graviton,\bgFieldsOther+\qFieldsOther]
          \label{eq:catering}
  \\
  =&
     \frac{1}{n} \actEntSurf{\entSurf}[
     \metricR{\repS{\entSurf}n},
     \gravitonR{\repS{\entSurf}n},
     (\bgFieldsOther + \qFieldsOther)_{\repS{\entSurf}n}
     ]
     - \actEntSurf{\entSurf}[\metric,\graviton,\bgFieldsOther+\qFieldsOther] \;.
     \label{eq:molybdenum}
\end{align}
Analogous to $\bullet_{\orb{\entSurf}n}$, it is helpful also to denote by
$\bullet_{\orbS{\entSurf}n}$ the deformed orbifolding that is the inverse
operation to $\bullet_{\repS{\entSurf}n}$. Then, for any configuration
$(\metric,\graviton,\bgFieldsOther+\qFieldsOther)$ which is
$\integers_{n'}$-symmetric around $\entSurf$, we can also define, for any $n$
dividing $n'$,
\begin{align}
  \actEntSurf{\entSurf,\frac{1}{n}}[\metric,\graviton,\bgFieldsOther+\qFieldsOther]
  \equiv& n \act[
          \metricR{\orbS{\entSurf}n},
          \gravitonR{\orbS{\entSurf}n},
          (\bgFieldsOther + \qFieldsOther)_{\orbS{\entSurf}n}
          ]
          - \act[\metric,\graviton,\bgFieldsOther+\qFieldsOther]
          \label{eq:turkey}
  \\
  =&
     n \actEntSurf{\entSurf}[
     \metricR{\orbS{\entSurf}n},
     \gravitonR{\orbS{\entSurf}n},
     (\bgFieldsOther + \qFieldsOther)_{\orbS{\entSurf}n}
     ]
     - \actEntSurf{\entSurf}[\metric,\graviton,\bgFieldsOther+\qFieldsOther] \;.
     \label{eq:pony}
\end{align}
While we have constructed $\actEntSurf{\entSurf,n}$ above in \eqref{eq:catering}
and \eqref{eq:turkey} for some discrete set of $n$, one can argue that an
analytic continuation to real $n>0$ is possible for any configuration
$(\metric,\graviton,\bgFieldsOther+\qFieldsOther)$ (which need not have some
$\integers_{n'}$-symmetry). In particular, applying \eqref{eq:trustless} and
\eqref{eq:carwash}, we can recast \eqref{eq:catering} and \eqref{eq:molybdenum}
(as well as \eqref{eq:turkey} and \eqref{eq:pony}) in the form
\begin{align}
  \actEntSurf{\entSurf,n}[\metric,\graviton,\bgFieldsOther+\qFieldsOther]
  =&
     \act[
     \metricR{\repS{\entSurf}n\orb{\entSurf}n},
     \gravitonR{\repS{\entSurf}n\orb{\entSurf}n},
     (\bgFieldsOther + \qFieldsOther)_{\repS{\entSurf}n\orb{\entSurf}n}
     ]
     - \act[\metric,\graviton,\bgFieldsOther+\qFieldsOther]
  \\
  =&
     \actEntSurf{\entSurf}[
     \metricR{\repS{\entSurf}n\orb{\entSurf}n},
     \gravitonR{\repS{\entSurf}n\orb{\entSurf}n},
     (\bgFieldsOther + \qFieldsOther)_{\repS{\entSurf}n\orb{\entSurf}n}
     ]
     - \actEntSurf{\entSurf}[\metric,\graviton,\bgFieldsOther+\qFieldsOther] \;,
     \label{eq:fondue}
\end{align}
where
$\bullet_{\repS{\entSurf}n\orb{\entSurf}n}=\bullet_{\orb{\entSurf}n\repS{\entSurf}n}$
has a natural meaning extended to real $n>0$ as a deformation localized in the
small neighbourhood $\neigh{\neighSize}{\entSurf}$ of $\entSurf$ such that the
opening angle around exactly $\entSurf$ is divided by $n$. With the definition
of $\actEntSurf{\entSurf,n}$ established, the Wald-Dong entropy
\cite{Dong:2013qoa}\footnote{
  In particular, focusing on the bulk metric field in holography,
  \cite{Dong:2013qoa} considers a sequence $\metricR{n}$ of smooth
  configurations which are $\integers_n$-symmetric around codimension-two
  surfaces $\entSurfR{n}$, with $\metricR{1}=\metric$ being the configuration on
  which the Wald-Dong entropy of $\entSurfR{1}=\entSurf$ is to be calculated. As
  described around \eqref{eq:cardiac} in our appendix
  \ref{sec:d4f5g3nf6bg2e6nf3be7o-oo-oc4d5}, these metrics are bulk solutions
  which arise naturally from the boundary replica trick. Then, (3.6) in
  \cite{Dong:2013qoa} reads
  \begin{align}
    \entDong{\entSurf}[\metric]
    =& -\partial_n \left(
       \act[\metricR{n\orbS{\entSurfR{n}}n}]
       - \act[\metricR{n\orb{\entSurfR{n}}n}]
       \right)_{n=1} \;.
       \label{eq:twig}
  \end{align} 
  The ``$S_{\mathrm{total}}$'' and ``$S_{\mathrm{outside}}$'' of
  \cite{Dong:2013qoa} respectively correspond to the two terms written in
  \eqref{eq:twig}. (Contrary to what Figure 2 therein might suggest, one should
  evaluate ``$S_{\mathrm{outside}}$'' as the action
  $\act[\metricR{n\orb{\entSurfR{n}}n}]$ \emph{exactly up to}, but not
  including, the conical singularity of $\metricR{n\orb{\entSurfR{n}}n}$ at
  $\entSurfR{n}$. For example, to recover (3.21) therein, for the singular cone
  ``$a\to0$'' term of (3.20), it is important to take $\rho\to 0$ as the lower
  bound of the integral after making this cone singular but before evaluating
  $\partial_n|_{n=1}=\partial_{\epsilon}|_{\epsilon=0}$ and hence before sending
  $a\to 0$ in the regularized cone, \ie{}finite $a$, term of (3.20).) The
  expression \eqref{eq:twig} can be seen to agree with \eqref{eq:unglue}:
  \begin{align}
    -\partial_n \left(
    \act[\metricR{n\orbS{\entSurfR{n}}n}]
    - \act[\metricR{n\orb{\entSurfR{n}}n}]
    \right)_{n=1}
    =&
       -\partial_n \left(
       \act[\metricR{\orbS{\entSurf}n}]
       - \act[\metricR{\orb{\entSurf}n}]
       \right)_{n=1}
    =
         -\partial_n \left.
         \act[\metricR{\orbS{\entSurf}n}]
         \right|_{n=1}
         -\act[\metric]
  \end{align}
  where we have used \eqref{eq:carwash} 
  in the last 
  equality.
  \label{foot:numbing}} appearing in the generalized entropy formula
\eqref{eq:solitude} can now be defined as:
\begin{align}
  \entDong{\entSurf}[\metric,\graviton,\bgFieldsOther+\qFieldsOther]
  \equiv& \partial_n \left.
          \actEntSurf{\entSurf,n}[
          \metric,\graviton,\bgFieldsOther+\qFieldsOther
          ]
          \right|_{n=1}
          = -\partial_n \left.
          \actEntSurf{\entSurf,\frac{1}{n}}[
          \metric,\graviton,\bgFieldsOther+\qFieldsOther
          ]
          \right|_{n=1}
          \label{eq:handwork}
  \\
  =& \partial_n \left(
     \act[
     \metricR{\repS{\entSurf}n},
     \gravitonR{\repS{\entSurf}n},
     (\bgFieldsOther+\qFieldsOther)_{\repS{\entSurf}n}
     ]
     -n\act[\metric,\graviton,\bgFieldsOther+\qFieldsOther]
     \right)_{n=1}
     \label{eq:bodacious}
  \\
  =& \partial_n \left(
     -\act[
     \metricR{\orbS{\entSurf}n},
     \gravitonR{\orbS{\entSurf}n},
     (\bgFieldsOther+\qFieldsOther)_{\orbS{\entSurf}n}
     ]
     -n\act[\metric,\graviton,\bgFieldsOther+\qFieldsOther]
     \right)_{n=1}
     \;,
     \label{eq:unglue}
\end{align}
where, to obtain the expression \eqref{eq:bodacious}, or equivalently
\eqref{eq:unglue}, we have used \eqref{eq:catering}. We shall see later in this
section that the generalized entropy $\entGen{\entSurf}[\metric]$ defined in
\eqref{eq:solitude} has an expression very similar to \eqref{eq:bodacious}, with
the action $\act$ replaced by the effective action $\eActMix$.

In this paper, the parameter $\neighSize$ describing the radius of the
neighbourhood $\neigh{\neighSize}{\entSurf}$ of $\entSurf$, in which
$\bullet_{\repS{\entSurf}n}$ replicas and $\bullet_{\orbS{\entSurf}n}$ orbifolds
are deformed, will always be a small value, whose vanishing limit is implicitly
understood to be taken \emph{last}, for those quantities that have finite $a\to
0$ limits. For instance, in higher-curvature theories, the action
$\actEntSurf{\entSurf}[\metricR{\repS{\entSurf}n},\ldots]$ in the neighbourhood
$\neigh{\neighSize}{\entSurf}$ of $\entSurf$ typically diverges for smooth
$\metric$ and $n\ne 1$ in the $a\to 0$ limit, so, for
$\actEntSurf{\entSurf}[\metricR{\repS{\entSurf}n},\ldots]$ and
$\actEntSurf{\entSurf,n}[\metric,\ldots]$, $a$ is understood to be small but
finite. However, it has been shown \cite{Dong:2013qoa} that this regulator can
be safely removed for the Wald-Dong entropy \eqref{eq:handwork} --- after taking
the $n$ derivative and setting $n=1$ in \eqref{eq:handwork}, one can take the
$a\to 0$ limit and obtain a finite answer\footnote{At the end of appendix
  \ref{sec:moonrise}, we give an argument for the finiteness of generalized
  entropy in the $\neighSize\to 0$ limit. Given the parallels between
  \eqref{eq:bodacious} and the below \eqref{eq:astatine}, a similar argument
  would seem to apply to Wald-Dong entropy, with the effective action replaced
  by the action.}.

To be clear, in this paper, we shall not allow the boundary part
$\int_{\bdyMan}\LagDensityBdy$ of the action \eqref{eq:treadmill} to contribute
to Wald-Dong entropy\footnote{A way to make this consistent with
  \eqref{eq:catering} and \eqref{eq:handwork}, where the full action is used, is
  to view $\int_{\bdyMan}\LagDensityBdy$ as actually being evaluated on the
  boundary limit of IR cutoff surfaces which always remain at large proper
  separation from $\entSurf$ --- the $\bullet_{\rep{\entSurf}n}$ and
  $\bullet_{\repS{\entSurf}n}$ operations would then be indistinguishable to
  $\int_{\bdyMan} \LagDensityBdy$. The cutoff surfaces would thus have parts
  which run off to infinity at $\entSurf\cap\bdyMan$. For example, this is the
  picture that one naturally has for a Euclidean $\AdS_3$ planar black hole: in
  the standard black hole picture, one views the bifurcation surface $\entSurf$
  and the IR cutoff surface as being parallel and thus never meeting; however,
  upon redrawing the spacetime as a global AdS ball, one finds that the two run
  off together to the AdS boundary at two
  points.
}, even if the entangling surface $\entSurf$ stretches to the spacetime boundary
$\bdyMan$. Our reasons for considering this part of the action in this paper are
primarily dynamical, as described around \eqref{eq:wiring}. In holography
however, apart from its role in dynamics, $\int_{\bdyMan}\LagDensityBdy$ also
includes counterterms which implement holographic renormalization of the
effective action. If we were to allow these counterterms to contribute to
Wald-Dong entropy, then we would be renormalizing IR divergences of generalized
entropy in the bulk and UV divergences of the CFT entropy. Thus, allowing
Wald-Dong contributions of $\int_{\bdyMan}\LagDensityBdy$ would seem to give a
bulk generalized entropy that is dual to a UV-renormalized\footnote{For a
  description of how UV divergences are renormalized in generalized entropy, see
  section \ref{sec:moonrise}. The discussion there is written with the bulk
  generalized entropy in mind, but is expected to extend generically to any
  theory of quantum fields on fixed backgrounds. In the present discussion, the
  divergent parts of the holographic renormalization counterterms would play the
  part of ``$-\actMatDiver[\metric]$'' for the CFT.} `CFT generalized entropy'
which includes Wald-Dong terms of the CFT theory needed to cancel UV
divergences. In this paper, we are interested in studying the bulk first law of
generalized entropy, which according to the intuition of previous work
\cite{Faulkner:2013ica,Swingle:2014uza}, should be equivalent in holography to a
first law of the CFT. While the bare von Neumann entropy of the CFT ought to
satisfy the standard first law \eqref{eq:e4e6} of any von Neumann entropy, no
sensible first law for the CFT generalized entropy is expected to hold as the
CFT metric is nondynamical. This motivates us to exclude
$\int_{\bdyMan}\LagDensityBdy$ from contributing to the bulk Wald-Dong entropy
so that, in holography, the bulk generalized entropy we consider is dual to a
bare von Neumann entropy and thus should possess a first law.

Before moving on, let us comment also briefly on how \eqref{eq:handwork} relates
to alternative expressions for Wald-Dong entropy (see
\eg{}\cite{Faulkner:2013ana}\footnote{The authors of \cite{Faulkner:2013ana},
  similar to \cite{Dong:2013qoa}, consider the Wald-Dong entropy \eqref{eq:twig}
  constructed using smooth bulk backgrounds $\metricR{n}$, as described in
  footnote \ref{foot:numbing}. From \eqref{eq:twig}, one can propose an
  expression for the Wald-Dong entropy analogous to \eqref{eq:escalator}:
  \begin{align}
    \entDong{\entSurf}[\metric]
    \eqDubious&
                -
                \int_{\partial\neigh{0}{\entSurf}}
                \sympPot[\partial_n \metricR{n\orb{\entSurf{n}} n}|_{n=1}]
                \;,
                \label{eq:entrap}
  \end{align} 
  where, in taking the $n$ derivative, an analytic continuation of
  $\metricR{n\orb{\entSurfR{n}}n}$ to real $n$ near $n=1$ is understood. The
  argument for \eqref{eq:entrap} is quite similar to that for
  \eqref{eq:escalator}, following from naively evaluating the action variations
  of \eqref{eq:twig} using \eqref{eq:copier}. Since the variations
  $\partial_n\metricR{n\orbS{\entSurfR{n}}n}|_{n=1}$ and
  $\partial_n\metricR{n\orb{\entSurfR{n}}n}|_{n=1}$ agree except in
  $\neigh{\neighSize}{\entSurf}$, the equations of motion contributions arising
  from the two terms in \eqref{eq:twig} should cancel in the $\neighSize\to 0$
  limit. What remains is the symplectic potential contribution \eqref{eq:entrap}
  from the second term.}) given in terms of the symplectic potential $\sympPot$
defined in \eqref{eq:copier}:
\begin{align}
  \entDong{\entSurf}[\metric,\graviton,\bgFieldsOther+\qFieldsOther]
  =& \partial_n \left.
     \act[
     \metricR{\repS{\entSurf}n\orb{\entSurf}n},
     \gravitonR{\repS{\entSurf}n\orb{\entSurf}n},
     (\bgFieldsOther + \qFieldsOther)_{\repS{\entSurf}n\orb{\entSurf}n}
     ]
     \right|_{n=1}
     \label{eq:throat}
  \\
  \eqDubious& -\int_{\partial\neigh{0}{\entSurf}} \sympPot[
              \partial_n \metricR{\repS{\entSurf}n\orb{\entSurf}n} |_{n=1} ,
              \partial_n \gravitonR{\repS{\entSurf}n\orb{\entSurf}n} |_{n=1} ,
              \partial_n (\bgFieldsOther+\qFieldsOther)_{\repS{\entSurf}n\orb{\entSurf}n} |_{n=1}
              ] \;,
              \label{eq:escalator}
\end{align}
or, alternatively,
\begin{align}
  \entDong{\entSurf}[\metric,\graviton,\bgFieldsOther+\qFieldsOther]
  =& -\partial_n \left.
     \act[
     \metricR{\rep{\entSurf}n\orbS{\entSurf}n},
     \gravitonR{\rep{\entSurf}n\orbS{\entSurf}n},
     (\bgFieldsOther + \qFieldsOther)_{\rep{\entSurf}n\orbS{\entSurf}n}
     ]
     \right|_{n=1}
     \label{eq:scone}
  \\
  \eqDubious& \int_{\partial\neigh{0}{\entSurf}} \sympPot[
              \partial_n \metricR{\rep{\entSurf}n\orbS{\entSurf}n} |_{n=1} ,
              \partial_n \gravitonR{\rep{\entSurf}n\orbS{\entSurf}n} |_{n=1} ,
              \partial_n (\bgFieldsOther+\qFieldsOther)_{\rep{\entSurf}n\orbS{\entSurf}n} |_{n=1}
              ] \;,
              \label{eq:crummiest}
\end{align}
where $\neigh{0}{\entSurf}$ denotes a zero-sized neighbourhood of $\entSurf$
(not to be confused with the $\neigh{\neighSize}{\entSurf}$ associated with
$\bullet_{\repS{\entSurf}n}$, where $a\to 0$ should be taken only at the very
end). To obtain the second lines \eqref{eq:escalator} and \eqref{eq:crummiest},
we note that, as described below \eqref{eq:fondue},
$\bullet_{\repS{\entSurf}n\orb{\entSurf}n}$ and, analogously,
$\bullet_{\rep{\entSurf}n\orbS{\entSurf}n}$ are deformations localized in the
neighbourhood $\neigh{\neighSize}{\entSurf}$ of $\entSurf$ such that the opening
angle around $\entSurf$ is respectively divided and multiplied by $n$. Since
these act nontrivially only in the tiny neighbourhood
$\neigh{\neighSize}{\entSurf}$, one expects the only sizable contribution to the
variation of the action to come from the symplectic potential $\sympPot$
contribution --- see \eqref{eq:copier} --- surrounding the zero-sized
neighbourhood $\neigh{0}{\entSurf}$ of $\entSurf$. (Recall our convention is for
the actions in \eqref{eq:throat} and \eqref{eq:scone} to exclude contributions
from the strict conical singularities exactly at
$\entSurf$.) 
Implicit in all this, however, is the assumption that the $\partial_n|_{n=1}$
derivative can be taken inside the action to act on the Lagrangian. As
\cite{Dong:2013qoa}\footnote{In particular, the finite $a$ and ``$a\to0$'' terms
  in (3.18)-(3.20) of \cite{Dong:2013qoa} respectively correspond to
  contributions to the two terms in \eqref{eq:twig}; the offending (3.21)
  correction of \cite{Dong:2013qoa} to \eqref{eq:entrap} is part of the latter.
  This correction arises because the smooth $\integers_n$-symmetric backgrounds
  $\metricR{n}$ have orbifolds $\metricR{n\orb{\entSurfR{n}}n}$ which, once
  analytically continued near $n=1$, give large curvature contributions that
  diverge \emph{as one approaches} $\entSurfR{n}$ in
  $\metricR{n\orb{\entSurfR{n}}n}$ such that certain $O[(n-1)^2]$ terms of the
  Lagrangian are promoted to $O(n-1)$ terms in
  $\act[\metricR{n\orb{\entSurfR{n}}n}]$. (This is somewhat surprising since,
  for $n\in\naturals$, the orbifold $\metricR{n\orb{\entSurfR{n}}n}$ is smooth
  except \emph{exactly at} the conically singular $\entSurfR{n}$, which is
  excluded from $\act[\metricR{n\orb{\entSurfR{n}}n}]$.)} finds, this is not
generally the case in higher-curvature theories when the entangling surface
$\entSurf$ has non-vanishing extrinsic curvature --- here, taking $n$ close to
$1$ gives rise to curvature contributions to the actions in \eqref{eq:throat}
and \eqref{eq:scone} concentrated near the very centre of
$\neigh{\neighSize}{\entSurf}$ which blow up as one approaches the entangling
surface $\entSurf$ (even at finite $\neighSize$), promoting certain $O[(n-1)^2]$
terms of the Lagrangian to $O(n-1)$ contributions to the action, containing
factors of the extrinsic curvature of $\entSurf$. Nonetheless, it will be
interesting to later compare \eqref{eq:crummiest} with a very similar expression
derived from Noether charge, in the thermal case where the extrinsic curvatures
of $\entSurf$ vanish.

To end this section, let us now use the deformed replication operation
$\bullet_{\repS{\entSurf}n}$ defined above to derive a simple expression for
generalized entropy in terms of the effective action $\eActMix$. Using
\eqref{eq:carwash} and \eqref{eq:turkey}, we can rewrite \eqref{eq:rename} in
terms of the deformed replication operation $\bullet_{\repS{\entSurf}n}$:
\begin{align}
  \tr[(\unStateMat)^n \blkOpInit]
  =& \int [d\graviton][d\qFieldsOther]\;
     \exp\left\{
     - (\actMat+\actEntSurfMat{\entSurf,\frac{1}{n}})[
     \metricR{\repS{\entSurf}n},
     \graviton,
     \bgFieldsOtherR{\repS{\entSurf}n}
     + \qFieldsOther
     ]
     \right\}
     \blkOpInit
     \;,
     \label{eq:berkelium}
\end{align}
where we have made a change of path integration variables from the $\graviton$
and $\qFieldsOther$ appearing in \eqref{eq:rename} to
$\gravitonR{\orb{\entSurf}n\repS{\entSurf}n}=\gravitonR{\repS{\entSurf}n\orb{\entSurf}n}$
and
$\qFieldsOtherR{\orb{\entSurf}n\repS{\entSurf}n}=\qFieldsOtherR{\repS{\entSurf}n\orb{\entSurf}n}$
which we have renamed back to $\graviton$ and $\qFieldsOther$ above\footnote{It
  is natural to select boundary conditions for the quantum fields at $\entSurf$
  such that they share the same conical properties as the background. Assuming
  the original background $(\metric,\bgFieldsOther)$ is smooth at $\entSurf$,
  then in \eqref{eq:rename}, the background
  $(\metricR{\rep{\entSurf}n},\bgFieldsOtherR{\rep{\entSurf}n})$ is conically
  singular with opening angle $2\pi n$ around $\entSurf$, while the
  $(\metricR{\repS{\entSurf}n},\bgFieldsOtherR{\repS{\entSurf}n})$ of
  \eqref{eq:berkelium} are regulated to be smooth at $\entSurf$. Note that
  $\bullet_{\orb{\entSurf}n\repS{\entSurf}n}=\bullet_{\repS{\entSurf}n\orb{\entSurf}n}$
  has a natural definition, even acting on fields $(\graviton,\qFieldsOther)$
  which do not necessarily have $\integers_n$-symmetry around $\entSurf$, as
  deforming, in the neighbourhood $\neigh{\neighSize}{\entSurf}$ of $\entSurf$,
  conically singular configurations with opening angle $2\pi n$ at $\entSurf$,
  so that they become smooth.}. Actually, due to transformations of the path
integration measure resulting from this change of variables and the change in
the background from
$(\metricR{\rep{\entSurf}n},\bgFieldsOtherR{\rep{\entSurf}n})$ to
$(\metricR{\repS{\entSurf}n},\bgFieldsOtherR{\repS{\entSurf}n})$, we generally
expect an anomalous correction to \eqref{eq:berkelium}; however, in this paper,
we will relegate all variations of the path integral measure to the
comprehensive discussion of appendix \ref{sec:deflate}. Setting this to the
side, combining \eqref{eq:panic}, \eqref{eq:lost}, \eqref{eq:handwork}, and
\eqref{eq:berkelium}, a simple expression for the generalized entropy
\eqref{eq:solitude} is then given in terms of the effective action
\eqref{eq:gadolinium} evaluated on replicated backgrounds\footnote{Note that,
  even though replicated fields are written in \eqref{eq:berkelium}, since
  $\actEntSurfMat{\entSurf,1}=0$, we see the pertinent quantity to consider for
  the difference between the generalized and matter von Neumann entropies is
  just $-\partial_n
  \actEntSurf{\entSurf,\frac{1}{n}}[\metric,\graviton,\bgFieldsOther+\qFieldsOther]|_{n=1}$,
  as appears in the Wald-Dong entropy \eqref{eq:handwork}.}:
\begin{align}
  \entGen{\entSurf}[\metric]
  =& \partial_n \left(
     \eActMix[\metricR{\repS{\entSurf}n}] - n\eActMix[\metric]
     \right)_{n=1}
     \;.
     \label{eq:astatine}
\end{align}
(In appendix \ref{sec:affix}, we shall argue that the anomalous corrections to
\eqref{eq:berkelium} mentioned above are absorbed into a corrected definition of
Wald-Dong entropy, so that \eqref{eq:astatine} still remains exactly correct.)
This is just like the expression \eqref{eq:bodacious} for Wald-Dong entropy, but
with the action $\act$ replaced by the effective action $\eActMix$. This lends
weight to the intuition that the generalized entropy \eqref{eq:astatine} of a
semiclassical theory is the natural analogue of the Wald-Dong entropy of a
classical theory. Moreover, \eqref{eq:astatine} makes a clear connection between
UV divergences in generalized entropy and the effective action
$\eActMix[\metric]$; in particular, for an effective action $\eActMix[\metric]$
that is rendered UV finite from the renormalization of UV divergences between
$\actGrav[\metric]$ and $\eActMixMat[\metric]$, we also expect the generalized
entropy given by \eqref{eq:astatine} to be UV finite. This is discussed further
in appendix \ref{sec:moonrise}. We also explain there why we expect
\eqref{eq:astatine} to remain finite as we take the zero-size limit of the
regulator $\neighSize$ for the smoothed conical singularities produced by the
action of $\bullet_{\repS{\entSurf}n}$ on smooth backgrounds.

Our result \eqref{eq:astatine} is essentially the Callan-Wilczek equation
\cite{Callan:1994py}\footnote{ In early work,
  \eg{}\cite{Solodukhin:1995ak,Cooperman:2013iqr}, the point that the
  calculation of $\tr[(\unStateMat)^n]$ should exclude extra contributions from
  conical singularities produced by replication is often missed --- see
  (2.5)-(2.6) and the discussion around (4.2) in \cite{Solodukhin:1995ak} and
  (2.14)-(2.15) and section 3.1 in \cite{Cooperman:2013iqr} --- or the
  calculation of $\tr[(\unStateMat)^n]$ is thought to be renormalized by the
  gravitational theory in a way that includes the contributions of conical
  singularities. (\Cf{}the discussion in our appendix \ref{sec:moonrise}, where
  we however identify the renormalized quantity as a generalized, rather than
  von Neumann, entropy.) Consequently, the authors associate the entirety of
  \eqref{eq:astatine}'s RHS with an entanglement \ie{}von Neumann entropy.
  Instead, we maintain that the Wald-Dong entropy \eqref{eq:handwork} should be
  distinguished from the matter von Neumann entropy, with the latter being truly
  given by \eqref{eq:depression} or \eqref{eq:lost} for some density matrix
  $\stateMat$ --- see footnote \ref{foot:f3}. \label{foot:elm}}, except that we
have (not yet) specialized to thermal setups, characterized by a $U(1)$-symmetry
around the entangling surface $\entSurf$ corresponding to a rotation in thermal
time (recall our discussion of Figure \ref{fig:e4e5ke2ke7}). In general, it
seems practically unclear how one might compute
$\eActMix[\metric_{\repS{\entSurf}n}]$ for real $n$ near $1$, in order to take
the derivative \eqref{eq:astatine}\footnote{This is to be contrasted with the
  action $\act[ \metricR{\repS{\entSurf}n}, \gravitonR{\repS{\entSurf}n},
  (\bgFieldsOther + \qFieldsOther)_{\repS{\entSurf}n} ]$, which can be expressed
  in terms of
  $\actEntSurf{\entSurf,n}[\metric,\graviton,\bgFieldsOther+\qFieldsOther]$ as
  in \eqref{eq:catering} and thus be readily computed for real $n>0$ as
  described around \eqref{eq:fondue}. One can, however provide a formal
  definition of $\eActMix[\metricR{\repS{\entSurf}n}]$ for real $n>0$ as
  follows. The classical action $\actGrav[\metricR{\repS{\entSurf}n}]$ can be
  analytically continued as described above, so it is sufficient to consider
  just the quantum matter piece $\eActMixMat[\metricR{\repS{\entSurf}n}]$ (see
  \eqref{eq:gadolinium}). As discussed around \eqref{eq:fondue}, the background
  $(\metricR{\repS{\entSurf}n\orb{\entSurf}n},\bgFieldsOtherR{\repS{\entSurf}n\orb{\entSurf}n})$
  has a natural analytic continuation to non-integer $n$ as a deformation of
  $(\metric,\bgFieldsOther)$ in the neighbourhood $\neigh{\neighSize}{\entSurf}$
  of $\entSurf$ such that the opening angle around exactly $\entSurf$ is divided
  by $n$. On
  $(\metricR{\repS{\entSurf}n\orb{\entSurf}n},\bgFieldsOtherR{\repS{\entSurf}n\orb{\entSurf}n})$,
  one can evaluate the path integral over quantum fluctuations
  $(\graviton,\qFieldsOther)$ which preserve the conical properties of the
  background fields at $\entSurf$ and further have boundary conditions fixed on
  the upper and lower faces of the codimension-one surface $\entReg$ --- this
  path integral can be thought of as defining unnormalized density matrices of
  the matter on $\entReg$ that we call
  $\unStateMatR{\repS{\entSurf}n\orb{\entSurf}n}\equiv(\unStateMatR{\repS{\entSurf}n})^{1/n}$.
  Then, analogous to \eqref{eq:panic}, note that we have
  $\tr\unStateMatR{\repS{\entSurf}n}=e^{-\eActMixMat[\metricR{\repS{\entSurf}n}]}$.
  Since our definitions for $\unStateMatR{\repS{\entSurf}n\orb{\entSurf}n}$ and
  $\unStateMatR{\repS{\entSurf}n}$ are valid for non-integer $n$, this gives a
  formal definition for the analytic continuation of
  $\eActMixMat[\metricR{\repS{\entSurf}n}]$. \label{foot:pretext}}. However, in
thermal setups, there is an obvious analytic continuation of the replicated
background $(\metricR{\repS{\entSurf}n},\bgFieldsOtherR{\repS{\entSurf}n})$,
obtained by varying the period of thermal time $\thTime$ to $2\pi n$ (from its
original period of $2\pi$ in $\metric$). With this, our goal in section
\ref{sec:fluorine} will be to establish a connection between the generalized
entropy formulas \eqref{eq:solitude} and \eqref{eq:astatine} and Wald's Noether
charge formalism \cite{Wald:1993nt,Iyer:1994ys} for computing entropy in thermal
cases.

It is perhaps worth pausing to remark that \eqref{eq:astatine}, despite first
appearances, differs from the standard holographic derivation of the generalized
entropy formula from the boundary von Neumann entropy \cite{Faulkner:2013ana}
--- in appendix \ref{sec:d4f5g3nf6bg2e6nf3be7o-oo-oc4d5} we compare and contrast
these approaches in detail. For the calculation of R\'enyi entropies in that
context, the replicated boundary manifold forms a shell within which one is
instructed to fill with a bulk path integral that produces smooth expectation
values for background plus quantum fields. This can be shown to lead to the
generalized entropy formula \eqref{eq:solitude} \cite{Faulkner:2013ana} on a
bulk solution $\metric$ with an additional quantum extremal condition for
$\entSurf$ \cite{Dong:2017xht}, recovering the expected RT formula. In contrast,
the replicated background $\metricR{\repS{\entSurf}n}$ of \eqref{eq:astatine}
has a regulated conical singularity around $\entSurf$. As we have derived,
\eqref{eq:astatine} gives a formula for the generalized entropy
\eqref{eq:solitude} of an arbitrary entangling surface $\entSurf$ on an
arbitrary background $\metric$. Of course, if evaluated on the same background
solution with the same RT surface $\entSurf$, \eqref{eq:astatine} and the
holographic results must agree to give \eqref{eq:solitude}. In appendix
\ref{sec:d4f5g3nf6bg2e6nf3be7o-oo-oc4d5}, we show this agreement more directly,
by relating our Callan-Wilczek formula \eqref{eq:astatine} to the holographic
calculation.



\section{Generalized entropy in thermal setups}

\label{sec:fluorine}
Here, we wish to relate the generalized entropy formulas \eqref{eq:solitude} and
\eqref{eq:astatine} to Iyer and Wald's Noether charge formalism
\cite{Wald:1993nt,Iyer:1994ys} in a thermal setting where, as described above
\eqref{eq:e4c5nf3d6d4cxd4nxd4nf6nc3a6}, there is a Euclidean $U(1)$ rotation
symmetry around the bulk entangling surface $\entSurf$. Specifically, we seek to
write the generalized entropy formula \eqref{eq:solitude}, with von Neumann
entropy $\ent[\stateMat]$ given by \eqref{eq:sadness}, as the sum of a Wald
Noether charge (identified with the Wald-Dong entropy $\entDong{\entSurf}$) plus
the expectation value of a modular Hamiltonian (and an effective action giving
the second term of \eqref{eq:sadness} by \eqref{eq:panic}). We shall find that
the matter modular Hamiltonian density is given by the canonical form of the
$U(1)$ rotation generator, \ie{}the integral of a Noether current plus a
boundary term, as opposed to some component of the gravitational stress tensor
obtained by varying the matter action with respect to the metric. This is in
agreement with existing work --- see \cite{Jafferis:2015del} section 4.1,
\cite{Casini:2014yca}, and \cite{Casini:2019qst} --- which focus primarily on
scalar field theory non-minimally coupled to curvature. We will aim to be more
general, allowing for arbitrary vector and tensor fields.

We organize this section as follows. We shall start in section \ref{sec:velcro}
by introducing notation relevant to computing generalized entropy using our
Callan-Wilczek formula \eqref{eq:astatine} in a thermal setup. In particular, we
review a trick inspired by \cite{Nelson:1994na,Wong:2013gua} for extending the
proper period of thermal time, \ie{}producing the replicated background
$(\metricR{\repS{\entSurf}n}, \bgFieldsOtherR{\repS{\entSurf}n})$ appearing in
\eqref{eq:astatine}. In section \ref{sec:obliged}, we employ this trick to
obtain an expression for generalized entropy in terms of integrals of Noether
charge and current in the Iyer-Wald formalism \cite{Wald:1993nt,Iyer:1994ys}.
This result motivates equating the matter modular Hamiltonian $\modHamMat$ with
the matter Noether current integral plus a boundary term, at least as expectation values. In section
\ref{sec:kebab}, we argue that this is in fact an operator equality by
considering expectation values with other operators; we also take a closer look
at the role played by terms previously set aside in the derivation of section
\ref{sec:obliged}. Finally, in section \ref{sec:albatross}, we move from
Euclidean to Lorentzian signature, rewriting our main results in the latter; in
particular, we explain how our conventions lead to the appearance of overall
signs in the Noether current and charge when moving between the two signatures.
This will prepare us for section \ref{sec:neon}, where we consider the first law
of generalized entropy in Lorentzian signature involving state and background
variations which may not necessarily permit simple Euclidean descriptions.

\subsection{Extending the proper thermal period}
\label{sec:velcro}
We are interested here in (Euclidean) spacetimes $\metric$ with a $U(1)$
symmetry --- see Figure \ref{fig:e4e5ke2ke7}. Let $\thTime\sim\thTime+2\pi$ be
the thermal time parameterizing the $U(1)$ direction, with $\thTime=0$ on the
bulk surface $\entReg$, the half-space for which generalized entropy is being
computed. The bulk part of $\partial\entReg$ is given by the entangling surface
$\entSurf$. The spacetime may or may not have a boundary; in case it does, let
us call it $\bdyMan$ and write the intersection of $\partial\entReg$ with
$\bdyMan$ as $\bdyReg$. In a holographic setting, $\bdyReg$ would be the
boundary region whose von Neumann entropy is computed by the bulk generalized
entropy across $\entSurf$. Let
\begin{align}
  \thVec =& \partial_\thTime
\end{align}
be the generator of the $U(1)$ symmetry in the thermal time direction.

\begin{figure}
  \centering
  \begin{subfigure}[b]{0.48\textwidth}
    \centering \includegraphics[width=0.9\textwidth]{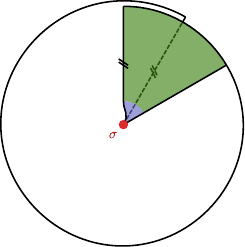}
    \caption{Extending the period of thermal time beyond $2\pi$ with locally
      fixed metric.}
    \label{fig:e4e5f4exf4nf3}
  \end{subfigure}
  \hfill
  \begin{subfigure}[b]{0.48\textwidth}
    \centering \includegraphics[width=0.9\textwidth]{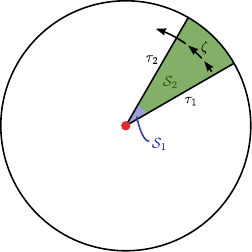}
    \caption{A diffeomorphism of panel \subref{fig:e4e5f4exf4nf3} such that the
      coordinate period of thermal time is back to $2\pi$.}
    \label{fig:e4e5f4exf4bc4}
  \end{subfigure}
  \caption{A background $(\metric,\bgFieldsOther)$ that is rotation-
    (\ie{}$U(1)$-) symmetric around a codimension-two surface $\entSurf$ has
    obvious analytic continuations for the replicated backgrounds
    $(\metricR{\rep{\entSurf}n},\bgFieldsOtherR{\rep{\entSurf}n})$ and
    $(\metricR{\repS{\entSurf}n},\bgFieldsOtherR{\repS{\entSurf}n})$ (with the
    latter illustrated here) to real $n$ near $n=1$. This is obtained by
    extending the period of thermal time $\thTime$ to $2\pi n$, as illustrated
    in panel \ref{fig:e4e5f4exf4nf3}. Here, the metric is locally kept fixed
    (except perhaps in a tiny neighbourhood $\neigh{\neighSize}{\entSurf}$ of
    $\entSurf$) but the period of $\thTime$ has been extended such that the two
    constant $\thTime$ surfaces marked by slashes are identified. Alternatively,
    one can apply a diffeomorphism to the coloured regions such that thermal
    time $\thTime$ retains its original period of $2\pi$, resulting in panel
    \ref{fig:e4e5f4exf4bc4}. For $n$ close to $1$, this involves perturbing the
    background by $((n-1)\Lie{\NelVec}\metric,(n-1)\Lie{\NelVec}\bgFieldsOther)$
    within the region marked as $\NelPatchPart{2}\subset
    \timeOpenInterval{\thTimeNelPatch{1}}{\thTimeNelPatch{2}}$ --- a qualitative
    sketch of the vector field $\NelVec$ is shown. Within the infinitesimal
    region
    $\NelPatchPart{1}=\neigh{\neighSize}{\entSurf}\cap\timeOpenInterval{\thTimeNelPatch{1}}{\thTimeNelPatch{2}}$,
    one should smoothly fill in the background fields.}
  \label{fig:e4e5f4exf4}
\end{figure}

Our starting point for obtaining an expression for the generalized entropy in
terms of Wald's Noether charge formalism is \eqref{eq:astatine}. Thus, to get
started, let us now introduce a diffeomorphism which has the effect of
replication $\bullet_{\rep{\entSurf}n}$ around the bulk entangling surface
$\entSurf$ for $U(1)$ symmetric backgrounds --- this trick is inspired by
\cite{Nelson:1994na,Wong:2013gua} and explained in Figure \ref{fig:e4e5f4exf4}.
Choose two thermal times $\thTimeNelPatch{1}$ and $\thTimeNelPatch{2}$ and
consider\footnote{When we write
  $\timeOpenInterval{\thTimeNelPatch{1}}{\thTimeNelPatch{2}}$, we always mean
  the thermal time interval going in the positive $\thTime$ direction from
  $\thTimeNelPatch{1}$ to $\thTimeNelPatch{2}$, even if
  $\thTimeNelPatch{1}>\thTimeNelPatch{2}$ --- recall the periodicity of
  $\thTime$.} the thermal time interval
$\timeOpenInterval{\thTimeNelPatch{1}}{\thTimeNelPatch{2}}$. Let
$\NelFunc(\tau)$ be a smooth function interpolating from $0$ to $2\pi$ within
$\timeOpenInterval{\thTimeNelPatch{1}}{\thTimeNelPatch{2}}$, with all
derivatives vanishing at $\thTimeNelPatch{1}$ and $\thTimeNelPatch{2}$:
\begin{align}
  \NelFunc(\thTimeNelPatch{1}^+) =& 0 \;
  &
    \NelFunc(\thTimeNelPatch{2}^-) =& 2\pi \;
  &
    \NelFunc'(\thTimeNelPatch{1}^+)
    =& \NelFunc'(\thTimeNelPatch{2}^-)
       = \NelFunc''(\thTimeNelPatch{1}^+)
       = \NelFunc''(\thTimeNelPatch{2}^-)
       = \cdots
       = 0 \;,
       \label{eq:finished}
\end{align} 
where we have used superscripts $\bullet^-$ and $\bullet^+$ to respectively mean
limits from below and above. We may then define
\begin{align}
  \NelVec
  =& \NelFunc \thVec
     \;,
     \label{eq:abamectine}
\end{align} 
which generates a transformation that just about captures the effect of
replication of $U(1)$-symmetric fields around $\entSurf$, up to a
diffeomorphism:
\begin{align}
  \partial_n \metricR{\rep{\entSurf}n}|_{n=1}
  \diffRel& \Lie{\NelVec} \metric = 2 (d\NelFunc)_{(a} \xi_{b)} \;,
  &
    \partial_n \bgFieldsOtherR{\rep{\entSurf}n}|_{n=1}
    \diffRel& \Lie{\NelVec} \bgFieldsOther
              \;,
              \label{eq:samarium}
\end{align}
where the Lie derivative of the metric was evaluated using
\begin{align}
  \nabla_{(a} \xi_{b)}
  =& 0
     \;,
  &
  \nabla_{(a} \NelVec_{b)}
  =& (d\NelFunc)_{(a} \xi_{b)}
     \;.
\end{align}
For simplicity, we shall, until the discussion in section \ref{sec:kebab},
choose coordinates for $\metricR{\rep{\entSurf}n}$ such that the $\diffRel$ in
\eqref{eq:samarium} become equalities (that is, we take the perspective of
Figure \ref{fig:e4e5f4exf4bc4} as opposed to Figure \ref{fig:e4e5f4exf4nf3}).
Note that the transformations \eqref{eq:samarium} interpolate to doing nothing
on $\thTimeNelPatch{1}$ and $\thTimeNelPatch{2}$. So far, the choices of
$\thTimeNelPatch{1}$, $\thTimeNelPatch{2}$, and $\NelFunc$, have been fairly
arbitrary; we will soon leverage this arbitrariness in calculations below.

As indicated in \eqref{eq:samarium} the transformations defined above describe
the replication operation $\bullet_{\rep{\entSurf}n}$ applied to the background
fields; as described by \eqref{eq:trustless} and \eqref{eq:carwash}, recall that
our convention is for functionals of such fields is to neglect any extra
contributions arising from the introduction or modification of conical
singularities produced by $\bullet_{\rep{\entSurf}n}$. In contrast, as made
explicit therein, our Callan-Wilczek formula \eqref{eq:astatine} for generalized
entropy involves backgrounds produced by the deformed replication operation
$\bullet_{\repS{\entSurf}n}$ which generates regulated conical singularities
that should be included in the evaluation of the effective action. In
particular, $\bullet_{\repS{\entSurf}n}$ is defined with an additional
deformation in a tiny neighbourhood $\neigh{\neighSize}{\entSurf}$ of $\entSurf$
such that the opening angle ($2\pi$ for smooth backgrounds) exactly at
$\entSurf$ remains unchanged by $\bullet_{\repS{\entSurf}n}$. Let us phrase this
in terms of augmenting our transformation \eqref{eq:samarium}. We divide the
spacetime in the thermal time interval
$\timeOpenInterval{\thTimeNelPatch{1}}{\thTimeNelPatch{2}}$ into two pieces:
$\NelPatchPart{1}=\neigh{\neighSize}{\entSurf}\cap\timeOpenInterval{\thTimeNelPatch{1}}{\thTimeNelPatch{2}}$,
given by a tiny neighbourhood of $\entSurf$ within
$\timeOpenInterval{\thTimeNelPatch{1}}{\thTimeNelPatch{2}}$, and
$\NelPatchPart{2}$, the complement of $\NelPatchPart{1}$ in
$\timeOpenInterval{\thTimeNelPatch{1}}{\thTimeNelPatch{2}}$. Within
$\NelPatchPart{2}$, we apply the transformation generated by $\NelVec$, as
written in \eqref{eq:samarium}. Within $\NelPatchPart{1}$, we shall apply a
deformation which smoothly interpolates to doing nothing near $\entSurf$; as
with \eqref{eq:samarium}, we require that the transformation here is also
trivial on $\thTimeNelPatch{1,2}$. The complete picture is illustrated in figure
\ref{fig:e4e5f4exf4bc4}.

\subsection{Generalized entropy as Noether charge and current}
\label{sec:obliged}
Let us now consider the generalized entropy functional \eqref{eq:astatine}. Let
us begin with the first term. Notice here that, to evaluate
$\eActMix[\metricR{\repS{\entSurf}n}]$, in addition to replicating the metric
$\metric$, we should also replicate other background fields $\bgFieldsOther$ ---
if, for example, the background $\bgFieldsOther$ is chosen so as to satisfy
certain boundary conditions on $\bdyMan$, then
$\bgFieldsOtherR{\repS{\entSurf}n}$ will satisfy the requisite boundary
conditions on the replicated spacetime boundary. (In this paper, we have tended
to suppress $\bgFieldsOther$ as an argument of $\eActMix$ because $\eActMix$ is
independent of the \emph{bulk} profile of $\bgFieldsOther$.) To proceed, we
shall assume that a field configuration
$(\metric,\graviton,\bgFieldsOther+\qFieldsOther)$ satisfies the required
boundary conditions on the boundary of the spacetime if and only if
$(\metricR{\repS{\entSurf}n},\graviton,\bgFieldsOtherR{\repS{\entSurf}n}+\qFieldsOther)$
satisfies the corresponding boundary conditions on the $n$-replicated spacetime
boundary
. We expect this to be the case, for example, when boundary conditions take the
form of requiring $\graviton$ and $\qFieldsOther$ to scale at certain asymptotic
rates on approach to the boundary. This allows us to write
\begin{align}
  \eActMix[\metric_{\repS{\entSurf}n}]
  =&
     -\log \int [d\graviton] [d\qFieldsOther]
     e^{-\act[\metricR{\repS{\entSurf}n},\graviton,\bgFieldsOtherR{\repS{\entSurf}n}+\qFieldsOther]},
     \label{eq:ytterbium}
\end{align}
where the boundary conditions for $(\graviton,\qFieldsOther)$ in the path
integral at the spacetime boundary are independent of $n$. Thus, the $n$
dependence is mostly accounted for by the background fields appearing in the
action
$\act[\metricR{\repS{\entSurf}n},\graviton,\bgFieldsOtherR{\repS{\entSurf}n}+\qFieldsOther]$.
We say ``mostly'' because it is conceivable that there may be an anomalous
dependence of the integration measure on the background.

Reserving further discussion of such anomalous contributions to appendix
\ref{sec:deflate}, using \eqref{eq:samarium}\footnote{Note that we are assuming
  a diffeomorphism invariance of the action and hence effective action in
  utilizing the trick described in section \ref{sec:velcro} to implement
  $\bullet_{\repS{\entSurf}n}$. In holography, the boundary part
  $\int_{\bdyMan}\LagDensityBdy$ of the action, containing counterterms for
  holographic renormalization, is not invariant under diffeomorphisms carrying
  along fields and the cutoff surface in the transverse direction when the
  boundary spacetime dimension $d$ is even --- this is precisely the origin of
  the CFT Weyl anomaly. (A review is given in section 6.3.2 of
  \cite{Ammon:2015wua}.) In this paper, we shall take cutoff surfaces to be
  invariant under rotations in the thermal time direction and only assume that
  $\int_{\bdyMan}\LagDensityBdy$ is invariant under diffeomorphisms parallel to
  cutoff surfaces, such as that appearing in \eqref{eq:samarium}.} and
\eqref{eq:ytterbium}, we write\footnote{Note that this is essentially the
  calculation in section II of \cite{Wong:2013gua}, except they seem to neglect
  $\sympPot$ terms as well as any $\LagDensityBdy$ (see (II.4)-(II.7) therein).
  In the following calculation here, we shall find that a Dong-like or Noether
  charge contribution to the entropy (a contribution which should be
  distinguished from the bulk von Neumann entropy) can be extracted from
  \eqref{eq:francium} by appropriately adding and subtracting $\sympPot$ terms
  around the entangling surface. Given that \cite{Wong:2013gua} never introduces
  such terms, the modular Hamiltonian \cite{Wong:2013gua} derives (see (II.9),
  (II.12), and (II.13) therein), an integral of the stress tensor components,
  likely does in general correspond to a valid von Neumann entropy. In appendix
  \ref{sec:immorally}, we present a calculation on time-dependent backgrounds
  that even more closely resembles \cite{Wong:2013gua} and point out there also
  how certain boundary terms were neglected by \cite{Wong:2013gua}.}
\begin{align}
  \partial_n \eActMix[\metric_{\repS{\entSurf}n}] |_{n=1}
  =& \left\langle \modHamNelS{\NelVec} \right\rangle
     \label{eq:thorium}
  \\
  \modHamNelS{\NelVec}
  \equiv&
          (
          \varyWRT{\metric,\Lie{\NelVec}\metric}
          + \varyWRT{\bgFieldsOther,\Lie{\NelVec}\bgFieldsOther}
          )\act[\metric,\graviton,\bgFieldsOther+\qFieldsOther]
          \label{eq:sprain}
  \\
  \begin{split}
    =& \int_{\NelPatchPart{2}} \left(
       \eomDensity{\metric}
       \cdot \Lie{\NelVec} \metric + \eomDensity{\bgFieldsOther} \cdot
       \Lie{\NelVec} \bgFieldsOther \right)
    \\
     &+ \int_{\partial\NelPatchPart{2} \cap \bdyMan} \left\{
       \sympPot[
       \Lie{\NelVec} \metric, \Lie{\NelVec} \bgFieldsOther
       ]
       + (
       \varyWRT{\metric,\Lie{\NelVec}\metric} +
       \varyWRT{\bgFieldsOther,\Lie{\NelVec}\bgFieldsOther}
       ) \LagDensityBdy
       \right\}\;,
       \label{eq:francium}
  \end{split}
\end{align}
with the equations of motion $\eomDensity{}$ and symplectic potential $\sympPot$
as defined in \eqref{eq:copier}. In writing \eqref{eq:francium}, we have dropped
\begin{align}
  \int_{\NelPatchPart{1}}
  \left( 
  \eomDensity{\metric}
  \cdot \partial_n \metricR{\repS{\entSurf}n}|_{n=1}
  + \eomDensity{\bgFieldsOther}
  \cdot \partial_n \bgFieldsOtherR{\repS{\entSurf}n}|_{n=1}
  \right)
  =& 0
     \label{eq:terbium}
\end{align}
which vanishes due to the vanishing size\footnote{Recall that the expectation
  value in \eqref{eq:thorium} is to be evaluated on the unperturbed background
  which is smooth at $\entSurf$. For states that are not particularly
  pathological exactly at $\entSurf$, there is no reason for $\langle
  \eomDensity{\metric} \rangle$ or $\langle \eomDensity{\bgFieldsOther} \rangle$
  to become infinite at $\entSurf$. Further, recall that
  $\partial_n\metricR{\repS{\entSurf}n}|_{n=1}$ and $\partial_n
  \bgFieldsOtherR{\orbS{\entSurf}n}|_{n=1}$ in $\NelPatchPart{1}$ smoothly
  interpolates from \eqref{eq:samarium} (which is finite) to zero at
  $\entSurf$.} of $\NelPatchPart{1}$ in the $\neighSize\to 0$ limit. In fact,
due to \eqref{eq:uranium}, we could have also omitted the
$\eomDensity{\bgFieldsOther}$ term from the RHS of \eqref{eq:thorium}, but we
have kept it around as it will be helpful for later discussion to consider the
operator $\modHamNelS{\NelVec}$ itself, rather than just its expectation value.
In writing \eqref{eq:francium}, we have also used the fact that
$\partial_n\metric_{\repS{\entSurf}n}|_{n=1}$ and
$\partial_n\bgFieldsOther_{\repS{\entSurf}n}|_{n=1}$ vanish on
$\thTimeNelPatch{1}$ and $\thTimeNelPatch{2}$.


For clarity, let us state that, in this paper, we shall take the orientation of
$m$-dimensional surfaces, written explicitly as the boundary $\partial\egReg$ of
some $(m+1)$-dimensional region $\egReg$, to be the usual one induced from that
of $\egReg$ such that Stokes' theorem is written
\begin{align}
  \int_{\egReg} d\egForm
  =& \int_{\partial\egReg} \egForm
     \label{eq:gallon}
\end{align}
for any $m$-form $\egForm$. We shall also take the spacetime boundary $\bdyMan$
to have the orientation induced in this way from the orientation of the bulk
spacetime. Furthermore, when we take the intersection of two $m$-dimensional
surfaces such that the intersection is itself also $m$-dimensional, we shall
implicitly take the intersection to inherit the orientation of the first written
surface. For example, $\partial\NelPatchPart{1}\cap\partial\NelPatchPart{2}$ and
$\partial\NelPatchPart{2}\cap\partial\NelPatchPart{1}$ shall implicitly inherit
the orientations of $\partial\NelPatchPart{1}$ and $\partial\NelPatchPart{2}$.
As we proceed, we shall state conventions for orienting other surfaces as we
need them.

Continuing with our calculation, let us now take advantage of the arbitrariness
of $\thTimeNelPatch{1}$, $\thTimeNelPatch{2}$, and $\NelFunc$ in order to take
the limit $\thTimeNelPatch{1}\to\thTimeNelPatch{2}^-$ where the thermal time
interval $\timeOpenInterval{\thTimeNelPatch{1}}{\thTimeNelPatch{2}}$ becomes
vanishingly small. Note that \eqref{eq:francium} does not vanish in this limit
because the Lie derivative $\Lie{\NelVec}\egField$ of a general tensor field
$\egField$ becomes sizable owing to the derivative of $\NelFunc$ in
$\NelVec=\NelFunc\thVec$ becoming large:
\begin{align}
  \Lie{\NelVec} \egField^{c_1\cdots c_r}_{b_1 \cdots b_s}
  =& \NelVec^a \nabla_a \egField^{c_1\cdots c_r}_{b_1 \cdots b_s}
     - \sum_{i=1}^r
     \egField^{c_1 \cdots a \cdots c_r}_{b_1 \cdots b_s} \nabla_a \NelVec^{c_i}
     + \sum_{i=1}^s
     \egField^{c_1 \cdots c_r}_{b_1 \cdots a \cdots b_s} \nabla_{b_i} \NelVec^{a}
  \\
  \sim&
        - \sum_{i=1}^r
        \egField^{c_1 \cdots a \cdots c_r}_{b_1 \cdots b_s} (d\kappa)_a \thVec^{c_i}
        + \sum_{i=1}^s
        \egField^{c_1 \cdots c_r}_{b_1 \cdots a \cdots b_s} (d\kappa)_{b_i} \thVec^{a}
      &
        (\thTimeNelPatch{1}\to\thTimeNelPatch{2}^{-})
        \label{eq:doze}
\end{align} 
with the sums being over the positions $i$ of vector indices $\bullet^{c_1
  \cdots c_i \cdots c_r}$ and dual-vector indices $\bullet_{b_1 \cdots b_i
  \cdots b_s}$ of the field $\egField$ in question. Thus, it is helpful to
introduce the quantities \cite{Faulkner:2013ica}
\begin{align}
  (\FaulknerCOf{\egField})_a
  \equiv& \sum_{i=1}^r
          (\eom{\egField})^{b_1 \cdots b_s}_{c_1 \cdots a \cdots c_r}
          \egField^{c_1 \cdots c_i \cdots c_r}_{b_1 \cdots b_s}
          \volForm_{c_i}
          - \sum_{i=1}^s
          (\eom{\egField})^{b_1 \cdots b_i \cdots b_s}_{c_1 \cdots c_r}
          \egField^{c_1 \cdots c_r}_{b_1 \cdots a \cdots b_s}
          \volForm_{b_i}
          \;,
          \label{eq:roentgenium}
\end{align}
so that
\begin{align}
  \lim_{\thTimeNelPatch{1}\to\thTimeNelPatch{2}^-}
  \int_{\NelPatchPart{2}} \eomDensity{\egField} \cdot \Lie{\NelVec} \egField
  =& -2\pi \int_{\thTimeNelPatch{2}} \thVec^a(\FaulknerCOf{\egField})_a \,.
     \label{eq:feisty}
\end{align} 
where $\thTimeNelPatch{2}$ is shorthand for the $\thTime=\thTimeNelPatch{2}$
surface.

To be clear, here and elsewhere in this paper, we shall take the orientation of
constant time surfaces, \eg{}the $\thTimeNelPatch{2}$ surface of
\eqref{eq:feisty}, to be given by the interior product of the normal vector
$\normalVecOf{\thTime}$ of constant time slices, pointed in the direction of
increasing $\thTime$, and the spacetime volume form
\begin{align}
  & \normalVecOf{\thTime} \cdot \volForm \;.
  &(\text{orientation of constant-time surfaces})
    \label{eq:stucco}
\end{align} 
Of course, if an $d$-dimensional surface is explicitly written as the boundary
$\partial\egReg$ of some $(d+1)$-dimensional region $\egReg$, then we will still
take the orientation on $\partial\egReg$ to be the one induced by that of
$\egReg$, even if part of $\partial\egReg$ lies on a constant-time surface. That
is, the rules stated previously around \eqref{eq:gallon} take priority.

Now, applying \eqref{eq:feisty} to \eqref{eq:francium}, we then find
\begin{align}
  \lim_{\thTimeNelPatch{1}\to\thTimeNelPatch{2}^-} \modHamNelS{\NelVec}
  &= 
    -2\pi \int_{\thTimeNelPatch{2}}
    \thVec^a \left(  
    \FaulknerCOf{\metric}
    + \FaulknerCOf{\bgFieldsOther}
    \right)_a
    + \lim_{\thTimeNelPatch{1}\to\thTimeNelPatch{2}^-}\int_{\partial\NelPatchPart{2} \cap \bdyMan} \left\{ \sympPot[
    \Lie{\NelVec} \metric, \Lie{\NelVec} \bgFieldsOther ] + (
    \varyWRT{\metric,\Lie{\NelVec}\metric} +
    \varyWRT{\bgFieldsOther,\Lie{\NelVec}\bgFieldsOther} ) \LagDensityBdy
    \right\}
    \;.
    \label{eq:guzzler}
\end{align}
In particular, from the definition \eqref{eq:roentgenium} applied to the metric
$\metric$,
\begin{align}
  \FaulknerCOf{\metric}^a
  =& -2\volForm_b \eom{\metric}^{ab}
     = -\volForm_b \left(2\eomGrav{\metric} + \stress\right)^{ab} \;,
     \label{eq:e4c5}
\end{align}
where $\stress^{ab}$ is the stress tensor \eqref{eq:friendless}\footnote{As
  remarked below \eqref{eq:friendless}, the stress tensor should also receive a
  contribution from the variation of the path integral measure --- this is
  precisely one of the anomalous terms we dropped below \eqref{eq:ytterbium}.},
we see in \eqref{eq:guzzler} the appearance of the term
\begin{align}
  2\pi \int_{\thTimeNelPatch{2}} \volForm_a \thVec_b \stress^{ab}
  \label{eq:botch}
\end{align}
which, if $\thTimeNelPatch{2}=0$ is chosen to place the integral on $\entReg$,
matches $2\pi$ times the often-written expression \eqref{eq:e4e6d4d5e5} for the
matter modular Hamiltonian\footnote{The relative sign between \eqref{eq:botch}
  and \eqref{eq:e4e6d4d5e5} is a result of the fact that \eqref{eq:botch} is
  written with the Euclidean stress tensor, while \eqref{eq:e4e6d4d5e5} is
  written with the Lorentzian stress tensor. As we shall show in section
  \ref{sec:albatross}, the two stress tensors are off by a sign.}. But, we shall
soon see that the matter modular Hamiltonian $\modHamMat$ should instead be
identified with the integral $- \int_{\thTimeNelPatch{2}} \currentMat{\thVec}$
of a Noether current $\currentMat{\thVec}$ plus a boundary term. Roughly
speaking, our calculations below will more generally bear out
\cite{Jafferis:2015del}'s observations (made there in scalar field theory) that
\eqref{eq:botch} should contribute to both the matter modular Hamiltonian
$2\pi\modHamMat$ and the matter part of the Wald-Dong entropy. In particular,
rewriting the $(\FaulknerCOf{\metric}+\FaulknerCOf{\bgFieldsOther})_a$ term of
\eqref{eq:guzzler}, we will be able to pick out: a Noether current integral $-
2\pi \int_{\thTimeNelPatch{2}} \current{\thVec}$ whose matter part $- 2\pi
\int_{\thTimeNelPatch{2}} \currentMat{\thVec}$ contributes to $2\pi\modHamMat$ ;
and a total derivative $2\pi \int_{\thTimeNelPatch{2}} d\charge{\thVec}$ whose
contribution $-2\pi\int_{\entSurf} \charge{\thVec}$ at the entangling surface is
identified as the Wald-Dong part $\entDong{\entSurf}$ of generalized entropy
\eqref{eq:solitude}.

To separate out the Wald-Dong and modular Hamiltonian contributions to
\eqref{eq:guzzler}, let us now write down the definition of Noether currents and
charges. For any vector field $\egVec$, the Noether current $\current{\egVec}$
is defined by \cite{Wald:1993nt,Iyer:1994ys}
\begin{align}
  \current{\egVec}
  \equiv&\, \sympPot[
          \Lie{\egVec} \metric,\Lie{\egVec} \graviton,\Lie{\egVec} (\bgFieldsOther+\qFieldsOther)
          ]
          - \egVec \cdot \LagDensity \,.
          \label{eq:d4d5c4e6}
\end{align}
To define the Noether charge (up to a closed form),
\cite{Faulkner:2013ica}\footnote{See (5.11)-(5.12) in therein.} showed that one
can generally express the Noether current in the form
\begin{align}
  \current{\egVec}
  \equiv&\, d\charge{\egVec} + \egVec^a \FaulknerC_a \,,
  &
    \FaulknerC_a
    \equiv& \sum_{\egField\in\{\metric,\graviton,\bgFieldsOther,\qFieldsOther\}}
            (\FaulknerCOf{\egField})_a
            \label{eq:lutetium}
\end{align}
where $\charge{\egVec}$ is the Noether charge and $\FaulknerC_a$ is the sum of
the $(\FaulknerCOf{\egField})_a$, defined above in \eqref{eq:roentgenium}, over
all fields $\egField\in\{g,\bgFieldsOther,\graviton,\qFieldsOther\}$. (Note
that, for classically on-shell configurations, $\current{\egVec}$ becomes exact
and thus closed --- this is the expression of current conservation in Iyer-Wald
formalism \cite{Wald:1993nt,Iyer:1994ys}.) Now, applying \eqref{eq:lutetium}
with $\egVec=\thVec$ to the first integral of \eqref{eq:guzzler}, we find
\begin{align}
  \begin{split}
    \lim_{\thTimeNelPatch{1}\to\thTimeNelPatch{2}^-} \modHamNelS{\NelVec}
    =&
       2\pi \left\{
       \int_{\thTimeNelPatch{2}\cap\bdyMan} \charge{\thVec}
       -\int_{\entSurf} \charge{\thVec}
       + \int_{\thTimeNelPatch{2}}
       \left[ \thVec^a\left( \FaulknerCOf{\graviton} +
       \FaulknerCOf{\qFieldsOther} \right)_a - \current{\thVec} \right]
       \right\}
    \\
     &+\lim_{\thTimeNelPatch{1}\to\thTimeNelPatch{2}^-}\int_{\partial\NelPatchPart{2}
       \cap \bdyMan}
       \left\{
       \sympPot[
       \Lie{\NelVec} \metric, \Lie{\NelVec} \bgFieldsOther ] + (
       \varyWRT{\metric,\Lie{\NelVec}\metric} +
       \varyWRT{\bgFieldsOther,\Lie{\NelVec}\bgFieldsOther} ) \LagDensityBdy
       \right\} \,.
  \end{split}
       \label{eq:campsite}
\end{align}

Here and elsewhere, we shall take the orientation of the codimension-two
surfaces (\eg{}the entangling surface $\entSurf$, the boundary region
$\bdyReg=\entReg\cap\bdyMan$ on the initial time slice, and the intersection
$\thTimeNelPatch{2}\cap\bdyMan$ of the $\thTime=\thTimeNelPatch{2}$ surface with
the spacetime boundary $\bdyMan$) residing on the boundaries of constant-time
surfaces to all have the orientation given by
\begin{align}
  &\normalVecOf{r} \cdot (\normalVecOf{\thTime} \cdot \volForm) \;,
  &(\text{orientation of codimension-two surfaces})
    \label{eq:defame}
\end{align}
where $\normalVecOf{\thTime}$ is the normal vector to the constant time surface,
pointed in the direction of increasing $\thTime$ (as in \eqref{eq:stucco}), and
$\normalVecOf{r}$ is the other normal of the codimension-two surface in
question, tangent to the constant time surface and pointed away from $\entSurf$
towards the spacetime boundary $\bdyMan$. (On the spacetime boundary,
$\normalVecOf{r}$ would correspond to the outward-directed normal vector of
$\bdyMan$.) Again, the rules stated around \eqref{eq:gallon} take priority for
any surfaces explicitly written as the boundary $\partial\egReg$ of some higher
dimensional region $\egReg$.

Let us now massage the spacetime boundary terms of \eqref{eq:campsite} a bit.
Using the definitions \eqref{eq:d4d5c4e6} and \eqref{eq:lutetium} for Noether
current and charge, we have the relation
\begin{align}
  2\pi\int_{\thTimeNelPatch{2}\cap\bdyMan} \charge{\thVec}
  =& -\lim_{\thTimeNelPatch{1}\to\thTimeNelPatch{2}^-}
     \int_{\partial\NelPatchPart{2}\cap\bdyMan} d\charge{\NelVec}
     = - \lim_{\thTimeNelPatch{1}\to\thTimeNelPatch{2}^-}
     \int_{\partial\NelPatchPart{2}\cap\bdyMan} \sympPot[
     \Lie{\NelVec}\metric,
     \Lie{\NelVec}\graviton,
     \Lie{\NelVec}(\bgFieldsOther+\qFieldsOther)
     ] \;,
     \label{eq:lilac}
\end{align}
where, in the second equality, we have noted that the $\NelVec\cdot\LagDensity$
and $\NelVec^a\FaulknerC_a$ terms arising from \eqref{eq:d4d5c4e6} and
\eqref{eq:lutetium} do not involve any derivatives of $\NelVec$ and thus have a
negligible integral over $\partial\NelPatchPart{2}\cap\bdyMan$ in the
$\thTimeNelPatch{1}\to\thTimeNelPatch{2}^-$ limit. Additionally, using the
relation \eqref{eq:wiring} between the symplectic potential and quantum field
variations of the boundary Lagrangian $\LagDensityBdy$, we have
\begin{align}
  (
  \varyWRT{\metric,\Lie{\NelVec}\metric}
  +\varyWRT{\bgFieldsOther,\Lie{\NelVec}\bgFieldsOther}
  ) \LagDensityBdy |_{\bdyMan}
  =& \Lie{\NelVec}\LagDensityBdy |_{\bdyMan}
     - (
     \varyWRT{\graviton,\Lie{\NelVec}\graviton}
     +\varyWRT{\qFieldsOther,\Lie{\NelVec}\qFieldsOther}
     ) \LagDensityBdy |_{\bdyMan}
  \\
  =& \Lie{\NelVec}\LagDensityBdy |_{\bdyMan}
     + \sympPot[\Lie{\NelVec}\graviton,\Lie{\NelVec}\qFieldsOther] |_{\bdyMan}\;.
     \label{eq:jigsaw}
\end{align}
One can use the general identity
\begin{align}
  \Lie{\egVec} \egForm
  =& \egVec \cdot d\egForm
     + d(\egVec\cdot\egForm)
     \label{eq:struggle}
\end{align}
for the Lie derivative of an arbitrary form field $\egForm$ with respect to an
arbitrary vector field $\egVec$ to integrate the $\Lie{\NelVec}\LagDensityBdy$
appearing in \eqref{eq:jigsaw}:
\begin{align}
  \int_{\partial\NelPatchPart{2}\cap\bdyMan} \Lie{\NelVec} \LagDensityBdy
  =& \int_{\partial\NelPatchPart{2}\cap\bdyMan} d(\NelVec\cdot\LagDensityBdy)
     = - 2\pi\int_{\thTimeNelPatch{2}\cap\bdyMan} \thVec\cdot\LagDensityBdy
     \label{eq:yarn}
\end{align} 
Altogether, applying \eqref{eq:lilac}, \eqref{eq:jigsaw}, and \eqref{eq:yarn} to
the spacetime boundary terms of \eqref{eq:campsite} simplifies the expression to
\begin{align}
  \begin{split}
    \lim_{\thTimeNelPatch{1}\to\thTimeNelPatch{2}^-} \modHamNelS{\NelVec}
    =&
       2\pi \left\{ 
       -\int_{\entSurf} \charge{\thVec} + \int_{\thTimeNelPatch{2}}
       \left[ \thVec^a\left( \FaulknerCOf{\graviton} +
       \FaulknerCOf{\qFieldsOther} \right)_a - \current{\thVec} \right]
       -\int_{\thTimeNelPatch{2}\cap\bdyMan} \thVec\cdot\LagDensityBdy
       \right\}\,.
  \end{split}
       \label{eq:canal}
\end{align}

Shortly in section \ref{sec:kebab}, we will argue that the terms in
\eqref{eq:canal} involving $\FaulknerCOf{\graviton}$ and
$\FaulknerCOf{\qFieldsOther}$ not only have a vanishing expectation value, but
are artifacts of the diffeomorphism implicit in using \eqref{eq:samarium} to
model replication. Setting these to the side, we see in \eqref{eq:canal} that we
have extracted Noether charge and current terms as we had hoped. Let us use
superscripts $\bullet^\grav$ and $\bullet^\mat$ to denote quantities, \eg{}the
Noether current, derived respectively from the gravitational and matter actions
$\actGrav$ and $\actMat$. Then, using the equality \eqref{eq:thorium} between
$\partial_n \eActMix[\metricR{\repS{\entSurf}n}] |_{n=1}$ and $\langle
\modHamNelS{\NelVec} \rangle$ (holding for any choice of $\thTimeNelPatch{1}$,
$\thTimeNelPatch{2}$, and $\NelFunc$) we obtain, from the Callan-Wilczek
expression \eqref{eq:astatine} for generalized entropy,
\begin{align}
  \entGen{\entSurf}[\metric]
  =& -2\pi \left\langle
     \int_{\entSurf} \charge{\thVec}
     + \int_{\thTimeNelPatch{2}} \currentMat{\thVec}
     + \int_{\thTimeNelPatch{2}\cap\bdyMan} \thVec\cdot \LagDensityBdyMat
     \right\rangle
     - \eActMixMat[\metric]
     \;,
     \label{eq:radon}
\end{align}
where we have used \eqref{eq:gadolinium} and
\begin{align}
  \current{\thVec}
  =& \currentGrav{\thVec} + \currentMat{\thVec}
     = -\thVec \cdot \LagDensityGrav + \currentMat{\thVec} \;,
     \label{eq:famished}
  \\
  \actGrav[\metric]
  =& 2\pi \left(
     \int_{\thTimeNelPatch{2}} \thVec\cdot\LagDensityGrav
     - \int_{\thTimeNelPatch{2}\cap\bdyMan} \thVec\cdot\LagDensityBdyGrav
     \right)
     \;,
     \label{eq:taps}
\end{align}
following from \eqref{eq:treadmill}, \eqref{eq:d4d5c4e6}, and the $\thVec$
rotation symmetry of the background metric $\metric$.

We can compare \eqref{eq:radon} with \eqref{eq:panic}, \eqref{eq:solitude}, and
\eqref{eq:sadness}, which read
\begin{align}
  \entGen{\entSurf}[\metric]
  =& \left\langle
     \entDong{\entSurf}[\metric,\graviton,\bgFieldsOther+\qFieldsOther]
     \right\rangle
     + 2\pi \langle \modHamMat \rangle
     - \eActMixMat[\metric]
     \;.
     \label{eq:parish}
\end{align}
The peculiar minus sign of the first terms in \eqref{eq:radon} is an expected
result of working in Euclidean signature%
\footnote{ In this paper, we have decided to stay consistent with the way we
  define Noether current and charge (always according to \eqref{eq:d4d5c4e6} and
  \eqref{eq:lutetium}) as well as orientation (always according to the rules
  stated around \eqref{eq:gallon}, \eqref{eq:stucco}, and \eqref{eq:defame})
  when moving between Euclidean and Lorentzian signatures. Alternatively, one
  can decide to move, in Euclidean signature, the sign on the Noether charge and
  current in \eqref{eq:radon} over to \eqref{eq:d4d5c4e6} or to the orientations
  of the constant-time surfaces and the entangling surface $\entSurf$.
  \label{foot:cartridge}
} --- we will see in section \ref{sec:albatross} how these signs arise when
moving between Lorentzian and Euclidean signatures. As hinted in
\eqref{eq:e4c5nf3d6d4cxd4nxd4nf6nc3a6} (which acquires a sign on the RHS in
Euclidean signature), in thermal setups, the Wald-Dong entropy can be equated to
the integral of Noether charge on $\entSurf$. Obviously, one can just compare
the formulas for Noether charge entropy \cite{Iyer:1994ys} and Wald-Dong entropy
\cite{Dong:2013qoa} to see that they are equal\footnote{To see
  \eqref{eq:e4c5nf3d6d4cxd4nxd4nf6nc3a6}, compare Proposition 4.1 in
  \cite{Iyer:1994ys} describing Noether charge with the general expression (1.5)
  for Wald-Dong entropy in \cite{Dong:2013qoa}. (The extrinsic curvatures of the
  entangling surface $\entSurf$ appearing in the latter vanish in the thermal
  setup.) Notice what is sufficient for \eqref{eq:e4c5nf3d6d4cxd4nxd4nf6nc3a6}
  to hold is for one to choose $\WaldZ=0$ for the Noether charge in Proposition
  4.1 and either all fields to be $\thVec$-symmetric, or for merely the metric
  $\metric$ to be $\thVec$-symmetric and additionally $\WaldY=0$ to be chosen.
  (Through our derivation, it is evident that \eqref{eq:radon} is independent of
  the ambiguities $\WaldY$ and $\WaldZ$ in Noether charge.) It will be
  convenient for us to take $\WaldY=0$ and $\WaldZ=0$, seeing as quantum
  fluctuations in matter fields may not be
  $\thVec$-symmetric. \label{foot:keg}}, but let us also give a more intuitive
albeit rough explanation. Using \eqref{eq:d4d5c4e6} and \eqref{eq:lutetium}
which relate Noether charge $\charge{\thVec}$ to the symplectic potential
$\sympPot$, we can put the Noether charge integrated over $\entSurf$ in a form
\begin{align}
  -2\pi\int_{\entSurf} \charge{\thVec}
  =& 
     \int_{\partial\NelPatchPart{1}\cap\partial\NelPatchPart{2}} d\charge{\NelVec}
     = 
     \int_{\partial\NelPatchPart{1}\cap\partial\NelPatchPart{2}} \sympPot[
     \Lie{\NelVec}\metric,
     \Lie{\NelVec}\graviton,
     \Lie{\NelVec}(\bgFieldsOther+\qFieldsOther)
     ] \;
     \label{eq:everyone}
\end{align}
which closely resembles the formula \eqref{eq:crummiest} for the Wald-Dong
entropy in terms of the symplectic potential. Let $\testFunc$ be a smoothed
indicator function that vanishes outside $\neigh{\neighSize}{\entSurf}$ and is
mostly identically $1$ in $\neigh{\neighSize}{\entSurf}$ except for a fixed
fraction of $\neigh{\neighSize}{\entSurf}$ near its boundary. Then, the action
of $\testFunc\Lie{\NelVec}$ on fields is a deformation in
$\neigh{\neighSize}{\entSurf}$ that has roughly the effect of producing the
variation of a conical excess at exactly $\entSurf$, just like the $\partial_n(
\bullet_{\rep{\entSurf}n\orbS{\entSurf}n})_{n=1}$ operation seen in
\eqref{eq:crummiest}. Replacing the $\partial_n(
\bullet_{\rep{\entSurf}n\orbS{\entSurf}n})_{n=1}$ field variations in
\eqref{eq:crummiest} with this variation $\testFunc\Lie{\NelVec}\,\bullet$ and
integrating the symplectic potential on $\partial\neigh{0}{\entSurf}$ where
$\testFunc=1$, we see that we recover\footnote{Our discussion here is obviously
  quite rough. For example, it is unable to resolve the $\WaldY$- and
  $\WaldZ$-type ambiguities in Noether charge appearing in Proposition 4.1 of
  \cite{Iyer:1994ys}, specifically (52) therein, mentioned in footnote
  \ref{foot:keg}. To deal with the $\WaldY$-type ambiguities which arise as
  exact contributions to the symplectic potential $\sympPot$, we would need to
  be more careful in relating $\testFunc\Lie{\NelVec}$ to $\partial_n(
  \bullet_{\rep{\entSurf}n\orbS{\entSurf}n})_{n=1}$; \eg{}the former generates a
  transformation that is discontinuous at $\thTimeNelPatch{2}$ for the matter
  fields which might not be $\thVec$-symmetric. On the other hand, $\WaldZ$-type
  ambiguities give exact contributions to Noether charge $\charge{\thVec}$. To
  resolve $\WaldY$- and $\WaldZ$-type ambiguities, when the entangling surface
  $\entSurf$ is not compact, \eg{}reaching the spacetime boundary $\bdyMan$, one
  must also be cautious in writing the first equality of \eqref{eq:everyone} as
  there may be extra contributions at
  $\partial\NelPatchPart{1}\cap\partial\NelPatchPart{2}\cap\bdyMan$. As noted in
  footnote \ref{foot:keg} we should make the choice $\WaldY=0$ and $\WaldZ=0$ in
  order to equate Wald-Dong and Noether charge
  entropies. 
} \eqref{eq:everyone} with the limit of zero size for $\NelPatchPart{1}$. As
noted below \eqref{eq:crummiest}, the subtle corrections found by
\cite{Dong:2013qoa} that might modify that equation in higher curvature theories
involve factors of extrinsic curvature which indeed vanish in the thermal setup
considered here. We can also give a rough argument for this in the thermal case.
Recall, as summarized below \eqref{eq:crummiest}, the corrections to that
equation are expected to come from curvature contributions to the action $\act[
\metricR{\rep{\entSurf}n\orbS{\entSurf}n},
\gravitonR{\rep{\entSurf}n\orbS{\entSurf}n}, (\bgFieldsOther +
\qFieldsOther)_{\rep{\entSurf}n\orbS{\entSurf}n} ]$ concentrated at the centre
of $\neigh{\neighSize}{\entSurf}$ that become divergent on approach to
$\entSurf$ for $n$ close to $1$ at finite $\neighSize$. For thermal setups,
however, the metric $\metric$ is symmetric in $\thVec$ and the statement
$\partial_n\metricR{\rep{\entSurf}n\orbS{\entSurf}n}|_{n=1}\diffRel
\testFunc\Lie{\NelVec}\metric$ becomes precise (\cf{}section \ref{sec:velcro}).
Sufficiently close to $\entSurf$, this is just a Lie derivative
$\Lie{\NelVec}\metric$ in the $\NelVec\propto\thVec$ direction. Thus, the
curvatures of $\metricR{\rep{\entSurf}n\orbS{\entSurf}n}$ should give
contributions to the action that, at $n$ close to $1$ and finite $\neighSize$,
are no more divergent as one approaches $\entSurf$ than on the original manifold
$\metric$. (This is a statement about curvatures near, but not exactly at
$\entSurf$; recall that our convention, stated around \eqref{eq:carwash}, is to
exclude from $\act[ \metricR{\rep{\entSurf}n\orbS{\entSurf}n},
\gravitonR{\rep{\entSurf}n\orbS{\entSurf}n}, (\bgFieldsOther +
\qFieldsOther)_{\rep{\entSurf}n\orbS{\entSurf}n} ]$ the strict conical
singularity exactly at $\entSurf$.)

Finally, moving onto the latter terms of \eqref{eq:radon} and \eqref{eq:parish},
we see that equating these two expressions suggests
\begin{align}
  \modHamMat
  =& - \int_{\thTimeNelPatch{2}} \currentMat{\thVec}
     - \int_{\thTimeNelPatch{2}\cap\bdyMan} \thVec\cdot\LagDensityBdyMat
     \;,
     \label{eq:mendelevium}
\end{align}
at least as expectation values. In the below section \ref{sec:kebab}, we shall
argue that this is in fact correct as an operator equation, by considering
correlators with other operators.

\subsection{Modular Hamiltonian as Noether current}
\label{sec:kebab}
Our goal now is to show that the equality \eqref{eq:mendelevium} between the
matter modular Hamiltonian and the Noether current integral
$-\int_{\thTimeNelPatch{2}} \currentMat{\thVec}$ does indeed hold as an operator
equation. We shall do this by considering correlators of $\modHamNelS{\NelVec}$
with other operators and comparing it with corresponding correlators of
$\entDong{\entSurf}+2\pi \modHamMat$. To do so, however, we must first understand the role played by
the terms involving $\FaulknerCOf{\graviton}$ and $\FaulknerCOf{\qFieldsOther}$
appearing the second line of \eqref{eq:canal} which we previously set aside.

We will see later that those terms are merely artifacts of the diffeomorphism
introduced in \eqref{eq:samarium}. Thus, it is helpful to begin by introducing
the following operator:
\begin{align}
  \modHamNelSDiff{\NelVec}
  \equiv& \int_{\NelPatchPart{2}} \left( 
          \eomDensity{\graviton} \cdot \Lie{\NelVec} \graviton
          + \eomDensity{\qFieldsOther} \cdot \Lie{\NelVec} \qFieldsOther
          \right)
          \;.
          \label{eq:untimed}
\end{align} 
The effect of inserting this operator into a Euclidean correlation function is,
up to a transformation of the path integral measure, to apply the differential
operator $\Lie{\NelVec}$ to all other $(\graviton,\qFieldsOther)$ operators residing in $\NelPatchPart{2}$.
Naively, we have
\begin{align}
  \langle \modHamNelSDiff{\NelVec} \blkOp \rangle
  =&
     \left\langle
     (\varyWRT{\graviton,\Lie{\NelVec}\graviton} + \varyWRT{\qFieldsOther,\Lie{\NelVec}\qFieldsOther})
     \blkOp
     \right\rangle
     \;.
     \label{eq:bohrium}
\end{align}
This is a consequence of replacing as the path integration variables
\begin{align}
  \graviton \to& \graviton + \testFunc\Lie{\NelVec}\graviton \;,
  &
    \qFieldsOther \to& \qFieldsOther + \testFunc\Lie{\NelVec}\qFieldsOther \;,
                       \label{eq:foil}
\end{align}
with $\testFunc$ a smoothed indicator function that identically vanishes outside
$\NelPatchPart{2}$ while, inside $\NelPatchPart{2}$, very rapidly interpolates
to being a small constant (including on $\partial\NelPatchPart{2}\cap\bdyMan$;
recall the boundary terms \eqref{eq:wiring} cancel each other).
This change of variables should induce a nontrivial variation of the path
integral measure, giving an extra anomalous correction to \eqref{eq:bohrium} that one
ought to absorb into the definition \eqref{eq:untimed}. To avoid derailing the
current discussion however, we shall take \eqref{eq:untimed} and
\eqref{eq:bohrium} at face value, leaving the variation of the path integral measure to be discussed
in appendix \ref{sec:deflate}. 

To connect $\modHamNelSDiff{\NelVec}$, as introduced in \eqref{eq:untimed}, to
terms appearing in \eqref{eq:canal}, we take the
$\thTimeNelPatch{1}\to\thTimeNelPatch{2}^-$ limit appearing in \eqref{eq:canal},
using the identity \eqref{eq:feisty}:
\begin{align}
  \lim_{\thTimeNelPatch{1}\to\thTimeNelPatch{2}} \modHamNelSDiff{\NelVec}
  =& -2\pi\int_{\thTimeNelPatch{2}}
     \thVec^a \left(  
     \FaulknerCOf{\graviton}
     + \FaulknerCOf{\qFieldsOther}
     \right)_a
     \;.
     \label{eq:silliness}
\end{align}
We see that these are precisely (minus) the $\FaulknerCOf{\graviton}$ and
$\FaulknerCOf{\qFieldsOther}$ terms of \eqref{eq:canal}, and we thus
have 
\begin{align}
  \lim_{\thTimeNelPatch{1}\to\thTimeNelPatch{2}} (
  \modHamNelS{\NelVec}
  + \modHamNelSDiff{\NelVec}
  )
  =& - 2\pi \left(
     \int_{\entSurf} \charge{\thVec}
     + \int_{\thTimeNelPatch{2}} \current{\thVec}
     + \int_{\thTimeNelPatch{2}\cap\bdyMan} \thVec\cdot\LagDensityBdy
     \right) \;.
     \label{eq:wizard}
\end{align}
(In fact, though it won't be of much use to us, \eqref{eq:wizard} holds even
without the $\thTimeNelPatch{1}\to\thTimeNelPatch{2}^-$ limit\footnote{This can
  be shown, by a similar calculation to that presented in section
  \ref{sec:obliged}, using: the definitions \eqref{eq:francium} and
  \eqref{eq:untimed} for $\modHamNelS{\NelVec}$ and $\modHamNelSDiff{\NelVec}$;
  the definitions \eqref{eq:d4d5c4e6} and \eqref{eq:lutetium} for Noether charge
  and current; the identities \eqref{eq:jigsaw} and \eqref{eq:yarn} concerning
  the boundary Lagrangian $\LagDensityBdy$; and the following identity (see
  appendix B in \cite{Faulkner:2013ica}):
  \begin{align}
    -d(\egVec^a \FaulknerC_a)
    =& \sum_{\egField\in\{\metric,\graviton,\bgFieldsOther,\qFieldsOther\}}\eomDensity{\egField}
       \cdot \Lie{\egVec} \egField
       \;.
       \label{eq:osmium}
  \end{align}%
  \label{foot:anguished}
}%
.) Note, from \eqref{eq:bohrium} that
\begin{align}
  \langle \modHamNelSDiff{\NelVec} \rangle
  =& 0
     \label{eq:parasitic}
\end{align} 
so we were indeed justified previously in stating that the
$(\FaulknerCOf{\graviton}+\FaulknerCOf{\qFieldsOther})_a$ term of
\eqref{eq:canal} has vanishing expectation value and can thus be neglected when
evaluating the generalized entropy in \eqref{eq:radon}.

By comparing the correlation functions between $\modHamNelS{\NelVec}$ and other
operators with those of $\entDong{\entSurf}+2\pi \modHamMat$, we will now argue
using \eqref{eq:wizard} and \eqref{eq:bohrium} that the identification
\eqref{eq:mendelevium} of the matter modular Hamiltonian $\modHamMat$ with the
matter Noether current integral $-\int_{\thTimeNelPatch{2}}
\currentMat{\thVec}$, plus a boundary term, is indeed correct. For this purpose,
it is sufficient to consider an arbitrary operator $\blkOpInit$, which can be,
for example, a product of local operators, inserted on
the initial time slice $\entReg$ at $\thTime=0$. (We can evolve
more general operators, say a sequence of operators at arbitrary Euclidean or
Lorentzian times joined by some Schwinger-Keldysh contour, back to the $\entReg$
surface and consider the resulting `Heisenberg'-picture operator $\blkOpInit$.)
Reiterating \eqref{eq:panic} and \eqref{eq:aspirin}, for such an operator, we
have
\begin{align}
  \langle \blkOpInit \rangle
  =& \frac{
     \int [d\graviton] [d\qFieldsOther] \;
     e^{-\act[\metric,\graviton,\bgFieldsOther+\qFieldsOther]}
     \blkOpInit
     }{
     e^{-\eActMix[\metric]}
     }
     = \frac{\tr(\unStateMat \blkOpInit)}{\tr\unStateMat}
\end{align}
where $\unStateMat$ evolves time by path integration from $\thTime=0$ to
$\thTime=2\pi$. Now, we consider inserting $\modHamNelS{\NelVec}$ into this
expectation value. There are two cases to consider, depending on whether
$\entReg$ lies in the thermal time interval
$\timeOpenInterval{\thTimeNelPatch{1}}{\thTimeNelPatch{2}}$ in which
$\modHamNelSDiff{\NelVec}$ resides. Let us first consider the easier case where
$\entReg$ does not lie in
$\timeOpenInterval{\thTimeNelPatch{1}}{\thTimeNelPatch{2}}$. Then, the insertion
of $\modHamNelS{\NelVec}$ defined in \eqref{eq:francium} simply gives ---
\cf{}\eqref{eq:ytterbium} and \eqref{eq:thorium}:
\begin{align}
  \langle \modHamNelS{\NelVec} \blkOpInit \rangle
  =& - \frac{
     \partial_n\left.
     \int [d\graviton] [d\qFieldsOther] \;
     e^{-\act[\metricR{\repS{\entSurf}n},\graviton,\bgFieldsOtherR{\repS{\entSurf}n}+\qFieldsOther]}
     \blkOpInit
     \right|_{n=1}
     }{
     e^{-\eActMix[\metric]}
     }
     \qquad (\entReg \not\subseteq \timeOpenInterval{\thTimeNelPatch{1}}{\thTimeNelPatch{2}})
     \label{eq:lawrencium}
  \\
    =& \left\langle
       \left( 
       \entDong{\entSurf}[\metric,\graviton,\bgFieldsOther+\qFieldsOther]
       + \actGrav[\metric]
        \right) 
       \blkOpInit
       \right\rangle
     -\frac{
       \partial_n \left. \tr[(\unStateMat)^n \blkOpInit] \right|_{n=1}
       }{\tr\unStateMat}
  \\
  =& \left\langle
     \left(  
     \entDong{\entSurf}[\metric,\graviton,\bgFieldsOther+\qFieldsOther]
     + \actGrav[\metric]
     + 2\pi \modHamMat
     \right)
     \blkOpInit
     \right\rangle
     \;,
     \label{eq:mosaic}
\end{align} 
where in the second equality, we have used \eqref{eq:berkelium} and
\eqref{eq:handwork} as well as, for the gravitational parts of the action and
Wald-Dong entropy, \eqref{eq:catering}. On the other hand, from \eqref{eq:bohrium} and
\eqref{eq:wizard}\footnote{As noted below it, \eqref{eq:wizard} in fact holds
  without the $\thTimeNelPatch{1}\to\thTimeNelPatch{2}^-$ limit; consequently,
  \eqref{eq:labrador} does not actually require the limit, but merely that
  $\entReg$ lies outside
  $\timeOpenInterval{\thTimeNelPatch{1}}{\thTimeNelPatch{2}}$. This can be seen
  independently by consistency with \eqref{eq:mosaic}, which (subject to
  $\entReg\not\subseteq\timeOpenInterval{\thTimeNelPatch{1}}{\thTimeNelPatch{2}}$)
  is manifestly
  independent of $\thTimeNelPatch{1}$, $\thTimeNelPatch{2}$, and the precise profile of the $\NelFunc$
  function used to define $\NelVec$ in \eqref{eq:abamectine}.
  \label{foot:coliseum}
},
\begin{align}
  \lim_{\thTimeNelPatch{1}\to\thTimeNelPatch{2}^-}
  \langle \modHamNelS{\NelVec} \blkOpInit \rangle
  =& 2\pi \left\langle
     \left[ 
     -\int_\entSurf \charge{\thVec}
     - \int_{\thTimeNelPatch{2}} \current{\thVec}
     - \int_{\thTimeNelPatch{2}\cap\bdyMan} \thVec\cdot\LagDensityBdy
     \right]
     \blkOpInit
     \right\rangle
     \;.
     \qquad (
     \text{%
     $\entReg \not\subseteq \timeOpenInterval{\thTimeNelPatch{1}}{\thTimeNelPatch{2}}$
     in limit%
     }
     )
     \label{eq:labrador}
\end{align}
Comparing \eqref{eq:mosaic} and \eqref{eq:labrador} using \eqref{eq:famished}
and \eqref{eq:taps} suggests, in this case,
\begin{align}
  \entDong{\entSurf}[\metric,\graviton,\bgFieldsOther+\qFieldsOther]
  + 2\pi \modHamMat
  =& - 2\pi \left(
     \int_\entSurf \charge{\thVec}
     + \int_{\thTimeNelPatch{2}} \currentMat{\thVec}
     + \int_{\thTimeNelPatch{2}\cap\bdyMan} \thVec\cdot\LagDensityBdyMat
     \right) \;,
     \label{eq:fermium}
\end{align} 
which, upon identifying the Wald-Dong entropy and Noether charge terms, gives
the equality \eqref{eq:mendelevium} we are after. However, we have arrived at
\eqref{eq:fermium} here by considering the case where no operators reside in
$\timeOpenInterval{\thTimeNelPatch{1}}{\thTimeNelPatch{2}}$ as the limit of
vanishing interval size is taken; unfortunately, this is insufficient to
distinguish whether $\modHamNelSDiff{\NelVec}$ should also be included in
\eqref{eq:fermium} since
\begin{align}
  \langle \modHamNelSDiff{\NelVec} \blkOpInit \rangle
  =& 0 \;.
  &
    (\entReg \not\subseteq \timeOpenInterval{\thTimeNelPatch{1}}{\thTimeNelPatch{2}})
\end{align} 



To distinguish $\modHamNelS{\NelVec}+\modHamNelSDiff{\NelVec}$ from
$\modHamNelS{\NelVec}$ and to understand why the $\FaulknerCOf{\graviton}$ and
$\FaulknerCOf{\qFieldsOther}$ terms of \eqref{eq:canal} appeared in the
calculation, let us now consider the case\footnote{As already mentioned below
  \eqref{eq:wizard} and in footnote \ref{foot:coliseum}, much of the discussion
  in this section does not actually require us to take the limit of vanishing
  interval $\timeOpenInterval{\thTimeNelPatch{1}}{\thTimeNelPatch{2}}$ size. But
  even if we do take the limit, it is still possible to do so such that
  $\entReg$ stays within
  $\timeOpenInterval{\thTimeNelPatch{1}}{\thTimeNelPatch{2}}$, \eg{} by taking
  $\thTimeNelPatch{1}\to 0^-$ and $\thTimeNelPatch{2}\to 0^+$.} where the
initial time slice $\entReg$ intersects
$\timeOpenInterval{\thTimeNelPatch{1}}{\thTimeNelPatch{2}}$. Now,
\eqref{eq:lawrencium} is not quite right as written for the following reason.
Recall as explained below \eqref{eq:samarium} that, owing to the diffeomorphism invariance of
$\eActMix[\metricR{\repS{\entSurf}n}]$, we were able to choose coordinates for
$\metricR{\repS{\entSurf}n}$ such that the $\diffRel$ of \eqref{eq:samarium}
are treated as equalities; in particular, varying $n$ changes the metric, but keeps the
period of the $\thTime$ coordinate fixed. Thus, we must remember that
$\modHamNelS{\NelVec}$ not only acts to vary the replication number $n$, but
also to apply the diffeomorphism that realizes this. In contrast, in the RHS of
\eqref{eq:lawrencium}, when $n$ is varied, we have in mind that the period of
thermal time is proportionately increased while the background remains locally invariant. The insertion of $\blkOpInit$ spoils diffeomorphism invariance and
must be transformed accordingly:
\begin{align}
  \begin{split}
    \MoveEqLeft[3] \langle \modHamNelS{\NelVec} \blkOpInit \rangle
    \\
    =& - \frac{ \partial_n\left. \int [d\graviton] [d\qFieldsOther] \;
       e^{-\act[\metricR{\repS{\entSurf}n},\graviton,\bgFieldsOtherR{\repS{\entSurf}n}+\qFieldsOther]}
       \blkOpInit \right|_{n=1} }{ e^{-\eActMix[\metric]} } - \left\langle
       (\varyWRT{\graviton,\Lie{\NelVec}\graviton} + \varyWRT{\qFieldsOther,\Lie{\NelVec}\qFieldsOther})
       \blkOpInit \right\rangle
  \end{split}
       \label{eq:dubnium}
  \\
  =& \left\langle
     \left( 
     \entDong{\entSurf}[\metric,\graviton,\bgFieldsOther+\qFieldsOther]
     + \actGrav[\metric]
     + 2\pi \modHamMat
     -\modHamNelSDiff{\NelVec}
     \right)
     \blkOpInit
     \right\rangle
     \;,
     \label{eq:rutherfordium}
\end{align}
where, in the second equality, we have now used \eqref{eq:bohrium}. Outside of
$\timeOpenInterval{\thTimeNelPatch{1}}{\thTimeNelPatch{2}}$, the diffeomorphism
is trivial
; thus, in the case of the previous paragraph, we were justified in brushing
this subtlety under the rug. In the present case where $\entReg$ intersects
$\timeOpenInterval{\thTimeNelPatch{1}}{\thTimeNelPatch{2}}$, however,
$\modHamNelSDiff{\NelVec}$ can act nontrivially on $\blkOpInit$. Still, from
\eqref{eq:wizard}, we again recover \eqref{eq:fermium} and consequently
\eqref{eq:mendelevium}.

To summarize, we have shown that $\modHamNelSDiff{\NelVec}$ and the
$\FaulknerCOf{\graviton}$ and $\FaulknerCOf{\qFieldsOther}$ terms appearing in \eqref{eq:canal} capture the effect of the extra
diffeomorphism required to turn the $\diffRel$ in \eqref{eq:samarium} into
equalities. As such, we have argued that these contributions are artifacts of
the trick we introduced in section \ref{sec:velcro} as a proxy for extending the
period of thermal time and should therefore be excluded from the modular
Hamiltonian. Moreover, we have seen, by comparison of correlation functions with
other operators, that \eqref{eq:fermium}, and thus our relation
\eqref{eq:mendelevium} between the matter modular Hamiltonian and Noether
current, are indeed correct as operator equations.

Let us end by commenting that the arbitrariness of the choice of
$\thTimeNelPatch{2}$ in \eqref{eq:mendelevium} and \eqref{eq:fermium} is not at
all surprising --- the interpretation of this is simply that, for thermal
states, the modular Hamiltonian $\modHamMat$ is itself the generator of time
evolution and thus commutes with the time evolution operator
$e^{-\thTime\modHamMat}$. (For the remaining sections of this paper, it will be
simplest to take $\thTimeNelPatch{2}=0$, corresponding to the initial time slice
$\entReg$.) In appendix \ref{sec:immorally}, we derive the time evolution
generator for quantum fields on backgrounds which are not $\thVec$-symmetric ---
we find that the generator, modulo spacetime boundary, Gibbons-Hawking-like, and joint
terms, is still the usual canonical one (see \eqref{eq:snowfield}), but the
evolution operator is now a path-ordered exponential of an integral over a
foliation of time slices (see \eqref{eq:county}). Again, these more general
results cannot be expressed in terms of gravitational stress tensor components,
contrary to previous work \cite{Wong:2013gua}.


\subsection{Continuation to Lorentzian signature}
\label{sec:albatross}
In the discussion of the first law in section \ref{sec:neon}, we will consider
the first law of generalized entropy in Lorentzian signature, involving
variations which may not be easily described in terms of Euclidean preparation.
Thus, before proceeding, let us take a moment now to discuss how various
quantities considered so far in Euclidean signature are related to their
Lorentzian counterparts\footnote{In this paper, we use the words ``Euclidean''
  and ``Lorentzian'' rather loosely, specifying whether we are considering a
  spacetime with the time coordinate $\thTime$ or $\LorTime=-i\thTime$. Note
  that a configuration $(\metric,\graviton,\bgFieldsOther+\qFieldsOther)$ that
  has real Lorentzian components for all real $\LorTime$ might yet fail to have
  real ``Euclidean'' components for all real $\thTime$. To give an example, a
  Lorentzian black hole metric rotating in the angular coordinate $\varphi$ will
  typically have an imaginary $\metric_{\thTime \varphi}$ component (unless one
  also Wick-rotates the angular velocity parameter by hand, which one must undo
  in the end anyway \cite{Hawking:1998kw}). Thus, a path integral over real
  configurations in $\LorTime$ might continue to a path integral involving
  complex configurations in $\thTime$. The conversions between ``Euclidean'' and
  ``Lorentzian'' objects in this section can be regarded as being evaluated on
  any configurations that might appear in a given path integral, real or
  otherwise.}.

We begin with the usual relations
\begin{align}
  \thTime
  =& i \LorTime \;,
  &
    \Euc{\thVec}
    =&\partial_{\thTime}
       = -i \partial_{\LorTime}
       = -i \, \Lor{\thVec} \;,
  &
    d\thTime
    =& i d\LorTime \;,
  &
    \Euc{\volForm}
    =& i\, \Lor{\volForm}
       \label{eq:proud}
\end{align}
between Euclidean thermal time $\thTime$ and the Lorentzian physical time
$\LorTime$. Here, we have also written the relations between the thermal Killing
vector $\Euc{\thVec}$ and the Lorentzian Killing vector $\Lor{\thVec}$. More
generally, we shall use superscripts $\Euc{\bullet}$ and $\Lor{\bullet}$ to
distinguish Euclidean and Lorentzian objects, for example, the volume forms
$\Euc{\volForm}$ and $\Lor{\volForm}$ respectively in Euclidean and Lorentzian
spacetimes. From the above, we see that Euclidean and Lorentzian time
derivatives are related by
\begin{align}
  \Lie{\Euc{\thVec}}
  =& -i \Lie{\Lor{\thVec}} \;.
     \label{eq:risotto}
\end{align} 
We also have
\begin{align}
  \Euc{\LagDensity}
  =& -i\, \Lor{\LagDensity} \;,
  &
    \EucEomDensity{\egField}
    =& -i\, \LorEomDensity{\egField} \;,
  &
    \Euc{\sympPot}[\vary\egField]
    =& -i\, \Lor{\sympPot}[\vary\egField] \;,
  &
    (\egField\in\{\metric,\graviton,\bgFieldsOther,\qFieldsOther\})
    \label{eq:overbook}
\end{align}
where the latter equalities follow from the first, giving the standard relation
between Euclidean and Lorentzian Lagrangian densities. Note then that,
consistently taking the definition of the stress tensor to be
\eqref{eq:friendless}, we have from \eqref{eq:proud} and \eqref{eq:overbook} the
following relations between Euclidean and Lorentzian stress tensors\footnote{One
  must be careful when comparing time components of tensors between Euclidean
  and Lorentzian signatures. For instance, the time-time component of
  \eqref{eq:marshy} reads
  $\Euc{\stress}^{\thTime\thTime}=-\Lor{\stress}^{\thTime\thTime}=\Lor{\stress}^{\LorTime\LorTime}$,
where $\bullet^{\thTime}=(d\thTime)_a\, \bullet^a$ indicates contraction with
$d\thTime$ while
$\bullet^{\LorTime}=(d\LorTime)_a\, \bullet^a$ indicates contraction with $d\LorTime$.}:
\begin{align}
       \Euc{\stressDensity}^{ab}
       =& -i\,\Lor{\stressDensity}^{ab}
  &
    \Euc{\stress}^{ab}
    =& - \Lor{\stress}^{ab} \;.
       \label{eq:marshy}
\end{align} 

Corollary to \eqref{eq:proud}, \eqref{eq:risotto}, and \eqref{eq:overbook}
above, we have additionally
\begin{align}
  \Euc{\thVec}\cdot\Euc{\LagDensity}
  =& -\Lor{\thVec} \cdot \Lor{\LagDensity} \;,
  &
    \Euc{\sympPot}[\Lie{\Euc{\thVec}}\egField]
    =& -\Lor{\sympPot}[\Lie{\Lor{\thVec}}\egField] \;,
  &
    (\egField\in\{\metric,\graviton,\bgFieldsOther,\qFieldsOther\})
    \label{eq:pummel}
\end{align} 
In general, we will consistently use the same definitions to define quantities
in terms of the Lagrangian, regardless of the spacetime signature. Consequently,
by the above and the definitions \eqref{eq:d4d5c4e6} and \eqref{eq:lutetium},
\begin{align}
  \EucCurrent{\Euc{\thVec}}
  =& -\LorCurrent{\Lor{\thVec}} \;,
  &
    \EucCharge{\Euc{\thVec}}
    =& -\LorCharge{\Lor{\thVec}} \;.
       \label{eq:anxiety}
\end{align} 
With the minus signs in the above relations, we must be careful when rewriting
our results in Lorentzian signature; for example, the minus signs of the Noether
current and charge terms of \eqref{eq:radon}, \eqref{eq:mendelevium}, and
\eqref{eq:fermium} disappear:
\begin{align}
  \entDong{\entSurf}[\metric,\graviton,\bgFieldsOther+\qFieldsOther]
  =& -2\pi \int_{\entSurf} \EucCharge{\Euc{\thVec}}
     = 2\pi \int_{\entSurf} \LorCharge{\Lor{\thVec}} \;,
  \\
  \modHamMat
  =& -\int_{\entReg}\EucCurrentMat{\Euc{\thVec}} - \int_{\bdyReg} \Euc{\thVec}\cdot\EucLagDensityBdyMat
     = \int_{\entReg} \LorCurrentMat{\Lor{\thVec}} + \int_{\bdyReg} \Lor{\thVec}\cdot\LorLagDensityBdyMat \;,
     \label{eq:washbasin}
\end{align}
with the first line simply reiterating \eqref{eq:e4c5nf3d6d4cxd4nxd4nf6nc3a6}
for $\thVec$-symmetric background geometries $\metric$. As already noted in
footnote \ref{foot:cartridge}, we will also consistently stick with the
conventions \eqref{eq:stucco} and \eqref{eq:defame} for orienting surfaces,
regardless of the spacetime signature\footnote{The normal vector
  $\normalVecOf{\thTime}$ in \eqref{eq:stucco} and \eqref{eq:defame} is
  generalized in the obvious way to $\normalVecOf{\LorTime}$ in Lorentzian
  signature, the future-directed normal of constant-$\LorTime$ surfaces.}.


\section{First law of generalized entropy}
\label{sec:neon}
Here, we wish to calculate the variation of generalized entropy, given by
\eqref{eq:solitude}, due to perturbations of the bulk background fields
$\metric$ and $\bgFieldsOther$ and of the state $\stateMat$ of the quantum
matter fields $\graviton$ and $\qFieldsOther$. Taking the unperturbed setup to
be thermal, as described in section \ref{sec:rockBottom} above
\eqref{eq:e4c5nf3d6d4cxd4nxd4nf6nc3a6} and considered in section
\ref{sec:fluorine}, we aim to recover the first law of generalized entropy
\eqref{eq:fake} as well as its necessary and sufficient conditions in terms of
gravitational dynamics. As described in section \ref{sec:helium} (and appendix
\ref{sec:d4f5g3nf6bg2e6nf3be7o-oo-oc4d5}), we again work with a semiclassical
theory with a quantum graviton $\graviton$ and other arbitrary matter background
plus quantum tensor fields $\bgFieldsOther+\qFieldsOther$ on a background
geometry $\metric$. Our goal will be to evaluate the first variation of
generalized entropy \eqref{eq:solitude} under arbitrary variations
$(\vary\metric,\vary\bgFieldsOther)$ and $\vary\stateMat$ of the background
fields and the state\footnote{Perhaps we should comment on a previous work
  \cite{Costa:2019yoy,Costa:2020ees} which ostensibly derived the first law for
  the generalized entropy. However, the work there essentially consisted of
  wrapping Wald's original classical calculation \cite{Wald:1993nt} in
  expectation value brackets. (A summary of Wald's classical calculation is
  illustrated below in section \ref{sec:exhaust} for just the metric field; one
  can easily imagine including other classical fields.) The only significant
  novelty was the observation that the symplectic density $\langle
  \symp[\vary\metric,\vary\graviton,\vary\bgFieldsOther,\vary\qFieldsOther;\Lie{\thVec}\graviton,\Lie{\thVec}\qFieldsOther]
  \rangle$ (defined by \eqref{eq:placidly} below) does not vanish. Rather, its
  integral over the bulk spatial region $\entReg$ was identified (seemingly by
  classical intuition) as the variation of the modular Hamiltonian
  $\modHamMat$'s expectation
  value, 
  and subsequently by the first law of von Neumann entropy \eqref{eq:e4e6}, the
  variation $\vary\ent[\stateMat]$ of the bulk matter entropy. Of course, in
  QFT, the objects $\vary\graviton$ and $\vary\qFieldsOther$ require some
  explanation. In \cite{Costa:2019yoy,Costa:2020ees}, the variation of quantum
  operators is defined by absorbing the variation of the state $\stateMat$.
  Specifically, \cite{Costa:2019yoy,Costa:2020ees} considers a unitary
  transformation to the state $\stateMat+\vary\stateMat = \unitary^\dagger
  \stateMat \unitary$ and then defines the variation $\vary\blkOp$ of an
  operator $\blkOp$ according to $\blkOp+\vary\blkOp=\unitary \blkOp
  \unitary^\dagger$ --- see above (11) in \cite{Costa:2019yoy} and above (3.21)
  in \cite{Costa:2020ees}. But, von Neumann entropy is invariant under unitary
  transformations. So, if unitary variations are all that is considered, then
  any variation to the interesting bulk von Neumann part $\ent[\stateMat]$ of
  generalized entropy is precluded! Perhaps a resolution to this problem might
  be to allow the unitary transformation $\unitary$ to act on the full
  purification $|\stateMatPurified\rangle$ of
  $\stateMat=\tr_{\aux}|\stateMatPurified\rangle\langle \stateMatPurified |$,
  \ie{}$\unitary|\stateMatPurified\rangle$. Upon tracing out the purifying
  auxiliary system, it is then possible to obtain non-unitary transformations
  --- specifically, completely positive and trace preserving (CPTP) maps --- to
  $\stateMat$ which modify the matter entropy. Still, one would need to verify
  $\int_{\entReg}\langle
  \symp[\vary\metric,\vary\graviton,\vary\bgFieldsOther,\vary\qFieldsOther;\Lie{\thVec}\graviton,\Lie{\thVec}\qFieldsOther]
  \rangle\eqDubious\varyWRT{\stateMat}\langle \modHamMat \rangle$. Note that it
  is very important that it is the state variation $\varyWRT{\stateMat}\langle
  \modHamMat \rangle=\tr(\vary\stateMat\modHamMat)$ which appears here, since
  this is what appears in the first law of von Neumann entropy \eqref{eq:e4e6}.
  So, in a sense, a significant portion of our work in this section and in
  appendix \ref{sec:commodore} can be interpreted as demonstrating that this
  equality is morally true (modulo some terms $\int_{\entReg}\thVec\cdot\langle
  \eomDensity{\graviton}\cdot\vary\graviton+\eomDensity{\qFieldsOther}\cdot\vary\qFieldsOther\rangle$),
  with significant attention paid to distinguishing between variations of states
  and operators, as well as functional variations in background fields.} of the
quantum matter fields $(\graviton,\qFieldsOther)$. In the end, we shall find
that when the gravitational equations of motion are satisfied for the
unperturbed and perturbed metrics and states, a first law emerges for
generalized entropy. As in \cite{Faulkner:2013ica,Swingle:2014uza}, we can also
look in a converse direction: we can see precisely what is required of the
perturbed background and state for the satisfaction of the first law. As
mentioned in the introductory section \ref{sec:rockBottom}, our results
generalize the results of \cite{Swingle:2014uza} to address a multitude of
subtleties related to non-minimal coupling and incongruities between the
gravitational stress tensor and the matter modular Hamiltonian (discussed
previously for scalars in \cite{Jafferis:2015del}\footnote{See, in particular,
  section 4.1 in \cite{Jafferis:2015del}.}), background variations unsuppressed
in $\GNewton$, and possible contributions to asymptotic energy from the matter
sector. From here on, we shall work in Lorentzian signature, implicitly
suppressing the $\Lor{\bullet}$ superscripts used in section
\ref{sec:albatross}.

Our calculations here are organized as follows. It will be instructive for us to
calculate the variations of the classical and quantum parts of generalized
entropy \eqref{eq:solitude} separately. Thus, we split the Wald-Dong entropy
$\entDong{\entSurf}$ into a classical purely gravitational part, arising from
the gravitational action $\actGrav[\metric]$, and the remaining quantum matter
part arising from $\actMat[\metric,\graviton,\bgFieldsOther+\qFieldsOther]$:
\begin{align}
  \entDong{\entSurf}[\metric,\graviton,\bgFieldsOther+\qFieldsOther]
  =& \entDongGrav{\entSurf}[\metric]
     + \entDongMat{\entSurf}[\metric,\graviton,\bgFieldsOther+\qFieldsOther]
     \;.
\end{align}
In section \ref{sec:exhaust}, we start by calculating the variation of the
purely classical gravitational piece $\entDongGrav{\entSurf}$ --- this will
follow along the same lines as, and thus permits a review of, the derivation of
the classical first law of black hole thermodynamics
\cite{Wald:1993nt,Iyer:1994ys,Faulkner:2013ica,Jacobson:1993vj}. In section
\ref{sec:manatee}, we move onto the remaining quantum part of generalized
entropy,
\begin{align}
  \langle \entDongMat{\entSurf}[\metric] \rangle
  + \ent[\stateMat] \;.
  \label{eq:float}
\end{align}
Using the first law of von Neumann entropy \eqref{eq:e4e6}, the variation of the
above is calculated and we obtain an answer involving the operator variation
$2\pi\int \varyWRT{\blkOp}\currentMat{\thVec}$ of the matter Noether current,
induced by perturbations of the background fields --- such a contribution would
have been discarded in previous works merely interested in background
perturbations suppressed in $\GNewton$. As we explain in section
\ref{sec:bobtail}, the calculation of this contribution is somewhat subtle as it
is not merely given by the functional variation
$\varyWRT{\metric}+\varyWRT{\bgFieldsOther}$ of the expression for
$2\pi\int\currentMat{\thVec}$ written out in terms of the fields $\metric$,
$\graviton$, $\bgFieldsOther$, and $\qFieldsOther$. Having obtained the
variations of all requisite pieces, we finally collect our results together in
section \ref{sec:sponge} to recover the first law of generalized entropy.

\subsection{The classical calculation}
\label{sec:exhaust}
The calculation to determine the first variation of
$\entDongGrav{\entSurf}[\metric]$ is straightforward, as it simply involves
taking the functional variation $\varyWRT{\metric}$ resulting from the
perturbation $\vary{\metric}$ of the classical field $\metric$:
\begin{align}
  \vary\entDongGrav{\entSurf}[\metric]
  =& \varyWRT{\metric} \entDongGrav{\entSurf}[\metric]
     = 2\pi \varyWRT{\metric} \int_{\entSurf} \chargeGrav{\thVec} \;,
     \label{eq:weary}
\end{align}
where, in the last equality, we have used the fact that the unperturbed
background metric $\metric$ is boost-symmetric around $\entSurf$ so that the
corrections to \eqref{eq:e4c5nf3d6d4cxd4nxd4nf6nc3a6} vanish at linear order in
perturbations, as hinted below that equation\footnote{To see that
  \eqref{eq:e4c5nf3d6d4cxd4nxd4nf6nc3a6} persists classically at linear order in
  perturbations away from $\thVec$-symmetric fields, see Theorem 6.1 (in
  particular, (93) and (95)) in \cite{Iyer:1994ys} and notice that the
  corrections in (1.5) of \cite{Dong:2013qoa} are quadratic in the extrinsic
  curvature.\label{foot:sequester}}. The classical calculation relating the RHS
of \eqref{eq:weary} to the variation of an asymptotic energy proceeds much the
same as in \cite{Faulkner:2013ica}. One starts by calculating the variation of
the Noether current defined in \eqref{eq:d4d5c4e6}. Using the identity
\eqref{eq:struggle} for the Lie derivative, one has
\begin{align}
  \varyWRT{\metric}\currentGrav{\egVec}
  =& -\egVec\cdot \left(
     \eomDensityGrav{\metric}\cdot\vary\metric
     \right)
     + d(\egVec\cdot \sympPotGrav[
     \vary\metric
     ])
     + \sympGrav[
     \vary\metric
     ;
     \Lie{\egVec}\metric
     ]
     \;,
     \label{eq:resilience}
\end{align}
where the symplectic density is defined by
\begin{align}
  \symp[\varyWRT{1}\egField;\varyWRT{2}\egField]
  \equiv& \varyWRT{1}\sympPot[\varyWRT{2}\egField]
          - \varyWRT{2}\sympPot[\varyWRT{1}\egField] \;.
          \label{eq:placidly}
\end{align} 
For the calculation at hand, we will use this with the vector field
$\egVec=\thVec$. Using \eqref{eq:lutetium}, the above can be related to the
variation of the Noether charge $\chargeGrav{\thVec}$:
\begin{align}
  d \left( 
  \varyWRT{\metric}
  \chargeGrav{\thVec}
  - \thVec \cdot \sympPotGrav[
  \vary\metric
  ]
  \right)
  =& 
     \varyWRT{\metric}
     \left(
     d  \chargeGrav{\thVec}
     - \currentGrav{\thVec}
     \right)
     - \thVec\cdot\left(
     \eomDensityGrav{\metric}\cdot\vary\metric
     \right)
  \\
  =& - \thVec^a
     \varyWRT{\metric}
     \FaulknerCGrav_a
     - \thVec\cdot\left(
     \eomDensityGrav{\metric}\cdot\vary\metric
     \right)
     \label{eq:callous}
\end{align}
where the $\sympGrav$ of \eqref{eq:resilience} has vanished due to
$\Lie{\thVec}\metric=0$. Integrating \eqref{eq:callous} on $\entReg$, we then
find that the variation \eqref{eq:weary} of the purely gravitational part of the
Wald-Dong entropy is given by
\begin{align}
  \vary
  \entDongGrav{\entSurf}[\metric]
  =& 2\pi 
     \int_{\bdyReg}  \left(
     \varyWRT{\metric}\chargeGrav{\thVec}
     - \thVec \cdot \sympPotGrav[\vary\metric]
     \right)
     + 2\pi 
     \int_{\entReg} \left[
     \thVec^a \varyWRT{\metric} \FaulknerCGrav_a
     + \thVec\cdot\left(
     \eomDensityGrav{\metric}\cdot\vary\metric
     \right)
     \right]
     \;.
     \label{eq:charlie}
\end{align} 
If one were just considering a classical theory of the metric $\metric$, the
interpretation of \eqref{eq:charlie} as a first law then follows from
identifying the integral over the boundary region $\bdyReg$ as the variation of
an asymptotic energy \cite{Wald:1993nt,Iyer:1994ys}. It is straightforward to
include other classical fields in this calculation, if one so desires, and we
see that the classical first law of black hole thermodynamics \eqref{eq:hunger}
is manifest in \eqref{eq:charlie}: the vanishing of the latter terms in
\eqref{eq:charlie} involving the unperturbed and linearized equations of motion
(see \eqref{eq:roentgenium} and \eqref{eq:e4c5}) constitute the necessary and
sufficient conditions for the classical first law.

\subsection{Variation of the quantum part of generalized entropy}
\label{sec:manatee}
Let us move now to the remaining quantum part \eqref{eq:float} of the
generalized entropy \eqref{eq:solitude}. First of all, the first law of von
Neumann entropy \eqref{eq:e4e6} applied to \eqref{eq:washbasin} gives
\begin{align}
  \vary\ent[\stateMat]
  =& 2\pi \varyWRT{\stateMat} \langle \modHamMat \rangle
     = 2\pi \left(
     \int_{\entReg} \varyWRT{\stateMat}\langle \currentMat{\thVec} \rangle
     + \int_{\bdyReg} \thVec\cdot
     \varyWRT{\stateMat}\left\langle
     \LagDensityBdyMat
     \right\rangle
     \right)\;,
     \label{eq:sociopath}
\end{align}
where, for an operator $\blkOpInit$,
\begin{align}
  \varyWRT{\stateMat}\langle \blkOpInit \rangle
  \equiv \tr(\vary\stateMat \blkOpInit)
\end{align}
means the variation of an expectation value resulting only from the perturbation
of the state $\stateMat$, while fixing\footnote{Here, ``fixing'' means fixing
  the expression of the operator written in terms of the field operators
  $\graviton,\qFieldsOther$ and their \emph{conjugate momenta} on $\entReg$, as
  opposed to their time derivatives $\Lie{\thVec}$. It is the former that are
  truly operators on $\entReg$, while time derivatives require a definition of
  evolution away from $\entReg$. See section \ref{sec:bobtail} for further
  discussion of this point.} the operator $\blkOpInit$. In contrast, the
variation of an expectation value $\langle \blkOpInit \rangle$ such
as\footnote{As written in \eqref{eq:e4c5nf3d6d4cxd4nxd4nf6nc3a6} and explained
  in footnote \ref{foot:keg} and below \ref{eq:everyone}, the Wald-Dong entropy
  can be equated to the Noether charge for $\thVec$-symmetric geometries
  $\metric$; moreover, by Theorem 6.1 of \cite{Iyer:1994ys},
  \eqref{eq:e4c5nf3d6d4cxd4nxd4nf6nc3a6} continues to hold classically for
  linear perturbations from $\thVec$-symmetric fields as explained in footnote
  \ref{foot:sequester}. At the semiclassical level, the same proof for this
  theorem shows that the latter statement continues to hold as expectation
  values:
  \begin{align}
    \vary\langle \entDong{\entSurf} \rangle
    =& 2\pi \int_{\entSurf} \vary\langle \charge{\thVec} \rangle \;
  \end{align}
  with, in the proof, $\Lie{\thVec}\langle \WaldX \rangle=0$ following from the
  $\thVec$-symmetry of the unperturbed background and path integral. (Again, we
  should make the choice $\WaldY=0$ and $\WaldZ=0$ --- see footnote
  \ref{foot:keg}.)
  \label{foot:evoke}
}
\begin{align}
  \langle \entDongMat{\entSurf}[\metric] \rangle
  =& 2\pi \int_{\entSurf} \langle \chargeMat{\thVec} \rangle
     \label{eq:psychopath}
\end{align}
must account for both the state variation as well as the variation
$\varyWRT{\blkOp}$ of the operator, resulting from perturbations of the
background $\metric,\bgFieldsOther$:
\begin{align}
  \vary\langle \blkOpInit \rangle
  =& \varyWRT{\stateMat} \langle \blkOpInit \rangle
     + \langle \varyWRT{\blkOp} \blkOpInit \rangle
     \;,
\end{align}
for instance,
\begin{align}
  \vary\langle \chargeMat{\thVec} \rangle
  =& \varyWRT{\stateMat} \langle \chargeMat{\thVec} \rangle
     + \langle \varyWRT{\blkOp}\chargeMat{\thVec} \rangle \;.
\end{align}
Now, we use the definition \eqref{eq:lutetium} of $\chargeMat{\thVec}$ in terms
of $\currentMat{\thVec}$ to write
\begin{align}
  \int_{\entReg} \currentMat{\thVec}
  =& \int_{\bdyReg} \chargeMat{\thVec}
     - \int_{\entSurf} \chargeMat{\thVec}
     + \int_{\entReg} \thVec^a \FaulknerCMat_a
     \;,
     \label{eq:blinks}
\end{align}
allowing us to combine \eqref{eq:sociopath} with the variation of
\eqref{eq:psychopath} to obtain
\begin{align}
  \begin{split}
    \MoveEqLeft[3]
    \vary
    \langle \entDongMat{\entSurf}[\metric] \rangle
    + \vary\ent[\stateMat]
    \\
    =& 2\pi \left(\int_{\bdyReg} \left\{
       \vary\langle \chargeMat{\thVec} \rangle
       + \thVec\cdot
       \varyWRT{\stateMat}\left\langle
       \LagDensityBdyMat
       \right\rangle
       \right\}
           + \int_{\entReg}\left\{
           \thVec^a \vary\langle \FaulknerCMat_a \rangle
           - \langle \varyWRT{\blkOp} \currentMat{\thVec} \rangle
           \right\}
           \right)
           \;.
           \label{eq:king}
  \end{split}
\end{align}
We recognize the Noether charge and $\FaulknerCMat_a$ terms as elements that can
be readily combined with those of \eqref{eq:charlie} 
in the variation of the purely gravitational part
$\vary\entDongGrav{\entSurf}[\metric]$ of generalized entropy. In section
\ref{sec:bobtail} below, we shall clarify the meaning of and calculate the final
term in \eqref{eq:king} involving the operator variation of the matter Noether
current. When we combine all the pieces of the first law of generalized entropy
in section \ref{sec:sponge}, we shall also see how variations of the boundary
Lagrangian $\LagDensityBdyMat$ and symplectic potential $\sympPot$ terms, such
as that appearing in \eqref{eq:charlie}, can be cohesively combined in the
complete expression for asymptotic energy.

Before proceeding, however, let us mention that the $\FaulknerCMat$ of the
second term in \eqref{eq:king} can in fact be reduced to just the metric part
$\FaulknerCMatOf{\metric}$ (recall the definitions \eqref{eq:roentgenium} and
\eqref{eq:lutetium})\footnote{This ultimately traces back to our special
  treatment of the metric and graviton fields, with respect to which we have
  Legendre transformed to obtain an effective action --- see appendix
  \ref{sec:d4f5g3nf6bg2e6nf3be7o-oo-oc4d5}. If we had performed similar Legendre
  transformations on other non-scalar fields for which \eqref{eq:roentgenium} is
  nontrivial, we must also include their
  $\langle(\FaulknerCMatOf{\bgFieldsOther})_a\rangle$.}:
\begin{align}
  \int_{\entReg}
  \thVec^a \,\vary\langle \FaulknerCMat_a \rangle
  =& \int_{\entReg}
     \thVec^a \,\vary\langle (\FaulknerCMatOf{\metric})_a \rangle \;.
     \label{eq:upside}
\end{align}
For other background fields $\bgFieldsOther$, shift-invariance of the
$\qFieldsOther$ path integral results in trivial one-point functions for
$\eomDensity{\bgFieldsOther}$, as noted in \eqref{eq:uranium}, so by definition
\eqref{eq:roentgenium},
\begin{align}
  \langle (\FaulknerCOf{\bgFieldsOther})_a \rangle
  =& 0 \;.
     \label{eq:skewer}
\end{align}
By changing path integration variables according to the transformation
\eqref{eq:doze}, one can similarly argue
\begin{align}
  \langle (\FaulknerCOf{\graviton})_a \rangle
  =& 0 \;,
  &
    \langle (\FaulknerCOf{\qFieldsOther})_a \rangle
    =& 0 \;.
       \label{eq:surpass}
\end{align} 
Actually, these relations are corrected by variations of the path integration
measure, but we leave discussion of this to appendix \ref{sec:deflate} (see
\eqref{eq:harmonic}).

\subsection{Operator variation of Noether current}
\label{sec:bobtail}
Now, we turn to the calculation of the operator variation
$2\pi\int_{\entReg}\varyWRT{\blkOp}\currentMat{\thVec}$ encountered in the
variation \eqref{eq:king} of the quantum part of generalized entropy. Note that
this should not be confused\footnote{It is perhaps helpful to briefly summarize
  why it is $\int_{\entReg}\varyWRT{\blkOp} \currentMat{\thVec}$ rather than
  $\vary\modHamMat$ that appears in our derivation of the first law of
  generalized entropy. Recall from our calculation in section \ref{sec:manatee}
  that the first law of von Neumann entropy \eqref{eq:sociopath} removes the
  need to consider $\vary\modHamMat$, setting the variation of von Neumann
  entropy equal just to the variation
  $2\pi\varyWRT{\stateMat}\langle\modHamMat\rangle=2\pi\int_{\entReg}
  \varyWRT{\stateMat} \langle\currentMat{\thVec}\rangle$ of the expectation
  value of the \emph{unperturbed} $\modHamMat$ operator, resulting only from the
  perturbation $\vary\stateMat$ of the state. However, when reexpressing the
  variation $\vary\langle \entDong{\entSurf} \rangle=2\pi\int_{\entSurf}
  \vary\langle \charge{\thVec} \rangle$ of the Wald-Dong contribution to
  generalized entropy using the relation \eqref{eq:blinks}, one also obtains a
  term $-2\pi\int_{\entReg} \vary\langle \currentMat{\thVec} \rangle$. This term
  can be split into two contributions, resulting from the perturbation of the
  matter state $\stateMat$, and the perturbation of the operator itself due to
  changes in the background fields $\metric$ and $\bgFieldsOther$; the former
  cancels against the variation of von Neumann entropy, while the latter
  survives and must be evaluated.
  \label{foot:apache}
} with the modular Hamiltonian variation
$2\pi\vary\modHamMat=-\vary\log\unStateMat$. Rather, the Noether current
$\currentMat{\thVec}$ is always defined in terms of the Lagrangian and
symplectic potential by \eqref{eq:d4d5c4e6}, but can generally depend on the
background fields $\metric$ and $\bgFieldsOther$.
  
Thus, to calculate this contribution, one might naively calculate, in the same
manner as in \eqref{eq:resilience},
\begin{align}
  \begin{split}
    (\varyWRT{\metric}+\varyWRT{\bgFieldsOther})\currentMat{\thVec}
    =&
       -\thVec\cdot\left( \eomDensityMat{\metric}\cdot\vary\metric +
       \eomDensityMat{\bgFieldsOther}\cdot\vary\bgFieldsOther \right) +
       d(\thVec\cdot \sympPotMat[\vary\metric,\vary\bgFieldsOther])
    \\
     &+ \sympMat[ \vary\metric, \vary\bgFieldsOther; \Lie{\thVec}\graviton,
       \Lie{\thVec}\qFieldsOther ] \;.
  \end{split}
       \label{eq:hemlock}
\end{align}
To spell out the above calculation \eqref{eq:hemlock} explicitly, one writes
$\currentMat{\thVec}$ as an expression in terms of fields
$(\metric,\bgFieldsOther,\graviton,\qFieldsOther)$ on $\entReg$ as well as their
$\Lie{\thVec}$ time derivatives; then, the classical background fields
$(\metric,\bgFieldsOther)$ (and their derivatives) are perturbed, while doing
nothing to $(\graviton,\qFieldsOther)$ and their $\Lie{\thVec}$ time
derivatives. This, however, is not quite what we want for $\varyWRT{\blkOp}$.

Since time derivatives of quantum fields $(\graviton,\qFieldsOther)$ are not
truly operators on the initial surface $\entReg$ unless a time evolution is
prescribed, we should instead consider the conjugate momenta of the quantum
fields, which fundamentally are operators on $\entReg$. To properly calculate
$\varyWRT{\blkOp}\currentMat{\thVec}$, therefore, we should write
$\currentMat{\thVec}$ as an expression in terms of conjugate momenta of
$(\graviton,\qFieldsOther)$, rather than their time derivatives, and fix these
conjugate momenta as the background fields $(\metric,\bgFieldsOther)$ are
perturbed. From the definition \eqref{eq:d4d5c4e6} of the current
$\currentMat{\thVec}$,
\begin{align}
  \currentMat{\thVec}
  =& \sympPotMat[\Lie{\thVec}\metric,\Lie{\thVec}\bgFieldsOther]
     + \HamDensitySimpleMat{\thVec}
     \label{eq:reexamine}
  \\
  \HamDensitySimpleMat{\thVec}
  \equiv& \sympPot[\Lie{\thVec}\graviton,\Lie{\thVec}\qFieldsOther]
          - \thVec\cdot \LagDensityMat \;,
          \label{eq:dubbed}
\end{align}
we see that, modulo an extra
$\sympPotMat[\Lie{\thVec}\metric,\Lie{\thVec}\bgFieldsOther]$, it is related to
the symplectic potential
$\sympPotMat[\Lie{\thVec}\graviton,\Lie{\thVec}\qFieldsOther]$ and Lagrangian
$\thVec\cdot\LagDensityMat$ for the quantum fields $\graviton$ and
$\qFieldsOther$ in the typical way expected for a Hamiltonian. Now, since the
unperturbed background is $\thVec$-symmetric, satisfying $\Lie{\thVec}\metric=0$
and $\Lie{\thVec}\bgFieldsOther=0$, the only contribution to the operator
variation of the extra
$\sympPotMat[\Lie{\thVec}\metric,\Lie{\thVec}\bgFieldsOther]$ term can come from
the functional variation of the arguments $\Lie{\thVec}\metric$ and
$\Lie{\thVec}\bgFieldsOther$:
\begin{align}
  \varyWRT{\blkOp}\sympPotMat[\Lie{\thVec}\metric,\Lie{\thVec}\bgFieldsOther]
  =& \sympPotMat[\Lie{\thVec}\vary\metric, \Lie{\thVec}\vary\bgFieldsOther]
     = (\varyWRT{\metric} + \varyWRT{\bgFieldsOther})
     \sympPotMat[\Lie{\thVec}\metric,\Lie{\thVec}\bgFieldsOther]\;.
     \label{eq:commotion}
\end{align}
On the other hand, the variation of a Hamiltonian, such as
$\int_{\entReg}\HamDensitySimpleMat{\thVec}$, expressed in terms of conjugate
momenta, is equal to minus the variation of the corresponding Lagrangian, in
this case $\int_{\entReg} \thVec\cdot\LagDensityMat$, written in terms of time
derivatives. Also properly accounting for boundary contributions to the
Hamiltonian and Lagrangian gives
\begin{align}
  \left\langle
  \varyWRT{\blkOp}
  \left( 
  \int_{\entReg} 
  \HamDensitySimpleMat{\thVec}
  + \int_{\bdyReg} \thVec\cdot \LagDensityBdyMat
  \right)
  \right\rangle
  =& \left\langle
     (\varyWRT{\metric}+\varyWRT{\bgFieldsOther})\left(
     -\int_{\entReg} \thVec\cdot\LagDensityMat
     + \int_{\bdyReg} \thVec\cdot\LagDensityBdyMat
     \right)
     \right\rangle\;.
     \label{eq:renewal}
\end{align} 
In appendix \ref{sec:commodore}, we provide a more careful derivation for
\eqref{eq:renewal} which explains the presence of the terms on the boundary
region $\bdyReg=\entReg\cap\bdyMan$.

Putting \eqref{eq:commotion} together with \eqref{eq:renewal}, we have that the
operator variation of \eqref{eq:reexamine} is given by
\begin{align}
  \int_{\entReg} 
  \langle \varyWRT{\blkOp} \currentMat{\thVec}  \rangle
  =& \int_{\entReg} \left\langle 
     (\varyWRT{\metric} + \varyWRT{\bgFieldsOther})\left(
     \sympPotMat[\Lie{\thVec}\metric, \Lie{\thVec}\bgFieldsOther]
     - \thVec\cdot\LagDensityMat
     \right)
     \right\rangle
     + \int_{\bdyReg} \thVec
     \cdot \left\langle
     (\varyWRT{\metric}+\varyWRT{\bgFieldsOther}-\varyWRT{\blkOp}) \LagDensityBdyMat
     \right\rangle
  \\
  =& \int_{\entReg} \left\langle
     (\varyWRT{\metric}+\varyWRT{\bgFieldsOther})\left(
     \currentMat{\thVec} - \sympPot[\Lie{\thVec}\graviton,\Lie{\thVec}\qFieldsOther]
     \right)
     \right\rangle
     + \int_{\bdyReg} \thVec
     \cdot \left\langle
     (\varyWRT{\metric}+\varyWRT{\bgFieldsOther}-\varyWRT{\blkOp}) \LagDensityBdyMat
     \right\rangle
     \;.
     \label{eq:fenwick}
\end{align}
We have already calculated the functional variation of the Noether current
$\currentMat{\thVec}$ in \eqref{eq:hemlock} above. From \eqref{eq:commotion} and
the $\thVec$ boost-invariance of the unperturbed background and path integral,
we have for the symplectic potential $\sympPot$ term of \eqref{eq:fenwick},
\begin{align}
  \left\langle
  (\varyWRT{\metric}+\varyWRT{\bgFieldsOther})
  \sympPot[\Lie{\thVec}\graviton,\Lie{\thVec}\qFieldsOther]
  \right\rangle
  =& \left\langle
     (\varyWRT{\metric}+\varyWRT{\bgFieldsOther})
     \sympPotMat[
     \Lie{\thVec}\metric,
     \Lie{\thVec}\graviton,
     \Lie{\thVec}(\bgFieldsOther+\qFieldsOther)
     ]
     \right \rangle
     - \Lie{\thVec} \left\langle
     \sympPotMat[\vary\metric,\vary\bgFieldsOther]
     \right \rangle
  \\
  =& \left\langle
     \sympMat[
     \vary\metric,\vary\bgFieldsOther;
     \Lie{\thVec} \graviton,\Lie{\thVec}\qFieldsOther
     ]
     \right\rangle
     \;.
\end{align}
Thus, using \eqref{eq:hemlock}, with the vanishing \eqref{eq:uranium} of the
equation of motion's expectation value $\langle \eomDensity{\bgFieldsOther}
\rangle$, and the above, we find that \eqref{eq:fenwick} can be written as
\begin{align}
  \int_{\entReg} \left\langle
  \varyWRT{\blkOp} \currentMat{\thVec}
  \right\rangle
  =& -\int_{\entReg} 
     \thVec\cdot\left(
     \left\langle
     \eomDensityMat{\metric}
     \right \rangle
     \cdot\vary\metric
     \right)
         + \int_{\bdyReg}
         \thVec\cdot \left\langle 
         \sympPotMat[\vary\metric,\vary\bgFieldsOther]
         + (\varyWRT{\metric}+\varyWRT{\bgFieldsOther}-\varyWRT{\blkOp})
         \LagDensityBdyMat 
         \right \rangle 
         \;.
         \label{eq:e4e5ke2}
\end{align}

\subsection{Collecting all pieces}
\label{sec:sponge}
Finally, we can put together \eqref{eq:charlie} and \eqref{eq:king} with
\eqref{eq:e4e5ke2} to get the total variation of generalized entropy
\eqref{eq:solitude} around a thermal setup:
\begin{align}
  \begin{split}
    \vary \entGen{\entSurf}[\metric]
    =& 2\pi \int_{\bdyReg}  \left\{ 
       \vary
       \langle
       \charge{\thVec}
       \rangle
       - \thVec \cdot
       \langle
       \sympPot[\vary\metric, \vary\bgFieldsOther]
       \rangle
       + \thVec \cdot (\vary - \varyWRT{\metric} - \varyWRT{\bgFieldsOther})
       \langle \LagDensityBdyMat \rangle
       \right\}
    \\
     &+ 2\pi \int_{\entReg} \left[ 
       \thVec^a \vary\left\langle
       (\FaulknerCOf{\metric})_a
       \right\rangle
       + \thVec \cdot \left(
       \left\langle \eomDensity{\metric} \right\rangle
       \cdot \vary\metric
       \right)
       \right]
  \end{split}
       \label{eq:renounce}
\end{align}
Note that, for $\vary\metric$ and $\vary\bgFieldsOther$ which preserve the
boundary conditions\footnote{We are primarily interested in background
  variations which preserve the boundary conditions at the spacetime boundary
  $\bdyMan$, as it is only for such variations that it seems reasonable to
  ascribe a variation of asymptotic energy. So, for the remainder of this
  section, we will no longer reiterate this assumption. However, it seems from
  our derivation that the result \eqref{eq:renounce} holds more generally, even
  allowing for variations in the background fields $\metric$ and
  $\bgFieldsOther$ which alter the boundary conditions at $\bdyMan$. (In
  holography, such variations would correspond to varying sources in the CFT.)}
at the spacetime boundary $\bdyMan$, one has from \eqref{eq:wiring} the
simplification
\begin{align}
  -\langle
  \sympPot[\vary\metric, \vary\bgFieldsOther]
  \rangle
  |_{\bdyMan}
  + (\vary - \varyWRT{\metric} - \varyWRT{\bgFieldsOther})
  \langle \LagDensityBdyMat\rangle
  |_{\bdyMan}
  =& \vary\langle \LagDensityBdy \rangle |_{\bdyMan}\;.
\end{align}
so that we may write\footnote{One may try to compare our \eqref{eq:e4c5} and
  \eqref{eq:calamity} with (3.8) in \cite{Swingle:2014uza}, where the analogue
  of the last term in \eqref{eq:calamity} has already been dropped (taking the
  unperturbed metric to be on-shell). Note, however, that their ``$E^g_{ab}$''
  is obtained by varying the gravitational action with respect to the inverse
  metric $\metric^{ab}$, and so should be equated to our
  $-(\eomGrav{\metric})_{ab}=-\metric_{ac} \, \metric_{bd} \,
  (\eomGrav{\metric})^{cd}$ with $(\eomGrav{\metric})^{ab}$ being given by the
  variation of the gravitational action with respect to the metric
  $\metric_{ab}$. On the other hand, we both use the same standard definition of
  the stress tensor. Accounting for all this, there is still an extra sign when
  comparing our \eqref{eq:calamity} with their (3.8). This can likely be traced
  back to either a difference in defining the orientation of $\entReg$ --- see
  \eqref{eq:stucco} for our conventions --- or a sign error inherited by
  \cite{Swingle:2014uza} from \cite{Faulkner:2013ica}.

  In particular, we note the following apparent inconsistency in
  \cite{Faulkner:2013ica}. It is clear from the derivation of the expression
  (5.21) for the variation asymptotic energy that the integrand written there,
  to be called ``$\bm{\chi}$'' in (5.22), should be integrated using the
  orientation induced on ``$\partial\mathcal{C}$'' as the boundary the Cauchy
  surface ``$\mathcal{C}$''. Correspondingly, ``$\bm{\chi}$'' in (4.14) should
  be integrated using the orientation induced on ``$B$'' as part of the boundary
  $\partial\Sigma$ of $\Sigma$. However, the opposite orientation is used in
  writing the third equality in (4.17); it seems, to sensibly correct this
  inconsistency, one ought to flip the signs of everything on one side or the
  other of that equality. Note that the signs in (4.17) of
  \cite{Faulkner:2013ica} seems to have been directly inherited by (3.8) in
  \cite{Swingle:2014uza}.
  \label{foot:frail}
}
\begin{align}
  \vary \entGen{\entSurf}[\metric]
  =& 2\pi \vary \langle \HamBdy{\thVec} \rangle
     + 2\pi \int_{\entReg} \left[ 
     \thVec^a \vary\left\langle
     (\FaulknerCOf{\metric})_a
     \right\rangle
     + \thVec \cdot \left(
     \left\langle \eomDensity{\metric} \right\rangle
     \cdot \vary\metric
     \right)
     \right]
     \label{eq:calamity}
\end{align}
where we have identified the asymptotic energy (\cf{}(82) in
\cite{Iyer:1994ys}\footnote{Note that (80) in \cite{Iyer:1994ys} defining the
  ``$\WaldB$'' appearing in their (82) is analogous to our more locally written
  property \eqref{eq:wiring} for $-\LagDensityBdy$. Apart from some intermediate
  equations which were integrated at least over the boundary spacial directions
  anyway, the derivations of the main results of this paper do not make use of
  the spatial locality of \eqref{eq:wiring}. So, there is no obvious harm in
  weakening \eqref{eq:wiring} to:
  \begin{align}
    \int_{\LorTime\cap\bdyMan}
    \thVec\cdot\sympPot[\vary\egField]
    =& -\int_{\LorTime\cap\bdyMan} \thVec\cdot\varyWRT{\egField}\LagDensityBdy \;,
    &
      (
      \egField\in\{\metric,\graviton,\bgFieldsOther,\qFieldsOther\},
      \text{$\vary\egField$ preserves b.c.}
      )
      \label{eq:viral}
  \end{align}
  for all constant time surfaces, denoted here by the shorthand $\LorTime$, and
  all field variations that preserve the boundary conditions at the spacetime
  boundary. This is just (80) in \cite{Iyer:1994ys} with their ``$\WaldB$''
  replaced by $-\LagDensityBdy$.
  \label{foot:mosaic}
})
\begin{align}
  \HamBdy{\thVec}
  \equiv& \int_{\bdyReg}  \left(
          \charge{\thVec} + \thVec\cdot\LagDensityBdy
          \right) \;.
          \label{eq:cymbal}
\end{align}

Finally, let us comment now on the interpretation of our result
\eqref{eq:calamity}. From \eqref{eq:calamity}, we see that the first law of
generalized entropy is equivalent to
\begin{align}
  \vary \entGen{\entSurf}[\metric]
  =& 2\pi \vary \langle \HamBdy{\thVec} \rangle
  &
  &\iff
  &
    \int_{\entReg} \left[ 
    \thVec^a \vary\left\langle
    (\FaulknerCOf{\metric})_a
    \right\rangle
    + \thVec \cdot \left(
    \left\langle \eomDensity{\metric} \right\rangle
    \cdot \vary\metric
    \right)
    \right]
    =& 0 \;,
\end{align}
with $\eomDensity{\metric}^{ab}=\volForm \eom{\metric}^{ab}$ being the metric
equations of motion and we recall from \eqref{eq:e4c5} that
$\FaulknerCOf{\metric}^a = -2\volForm_b \eom{\metric}^{ab}$. In particular, we
see that if the gravitational equations of motion are satisfied by the
unperturbed thermal background and state as well as at linear order by the
perturbation, then the first law of generalized entropy holds:
\begin{align}
  0 =&
       \langle \eom{\metric}^{ab} \rangle
       = \vary \langle \eom{\metric}^{ab} \rangle 
  &
  &\implies
  &
    \vary \entGen{\entSurf}[\metric]
    =& 2\pi \vary \langle \HamBdy{\thVec} \rangle
       \;.
       \label{eq:finale}
\end{align}
Conversely, assuming the unperturbed thermal setup is a solution to the
gravitational equations of motion, then the first law implies an integral of the
linear order gravitational equations of motion for the perturbation vanishes:
\begin{align}
  \vary \entGen{\entSurf}[\metric]
  =& 2\pi \vary \langle \HamBdy{\thVec} \rangle
  &
  &\implies
  &
    0=& \int_{\entReg}  
        \thVec_a \volForm_b \;
        \vary
        \langle
        \eom{\metric}^{ab}
        \rangle
  &
    (\langle \eom{\metric}^{ab} \rangle =& 0) \;.
                                           \label{eq:stylist}
\end{align}
In a holographic setting, \cite{Swingle:2014uza} examined this converse
direction\footnote{The derivation of the analogue of \eqref{eq:calamity} in
  \cite{Swingle:2014uza} was in the simplified setting of Einstein gravity and
  it is assumed that the modular Hamiltonian is given by the integral of
  components of the stress tensor, as summarized in section \ref{sec:rockBottom}
  around \eqref{eq:e4e6d4d5e5}. Possible non-minimal couplings of matter to
  curvature, for example, would break that derivation. However, assuming
  \eqref{eq:calamity}, it seems the discussion of \cite{Swingle:2014uza} beyond
  that point continues to hold.}, in the particular case where the unperturbed
setup is the AdS vacuum with $\entReg$ being a half-ball centred on the boundary
$\bdyMan$. It was found that sufficient symmetry exists there to translate,
rotate, boost, and dilate the region $\entReg$ such that imposing the first law
of generalized entropy for all half-balls $\entReg$ implies that the linearized
equation $\vary\langle \eom{\metric}^{ab} \rangle = 0$ vanishes locally
everywhere.


\section{Discussion}
\label{sec:d4f5g3nf6bg2e6nf3o-oo-oc4}
In this work, we have considered a semiclassical gravitational theory, allowing
for arbitrary tensor\footnote{Although, we do not anticipate much difficulty in
  extending our calculations to allow for spinor fields --- it seems one would
  just need to work with vielbeins rather than a metric field.} field matter
that can couple non-minimally to curvature. We have seen in section
\ref{sec:helium}, how generalized entropy in such a theory can be written in
terms of the gravitational effective action in a Callan-Wilczek-like
\cite{Callan:1994py} form \eqref{eq:astatine}, involving backgrounds containing
regulated conical singularities. With this, we showed in section
\ref{sec:fluorine} that the modular Hamiltonian for matter fields in thermal
setups, which possess a rotation symmetry in thermal time, is generally given by
the sum \eqref{eq:mendelevium} (or, equivalently continued to Lorentzian
signature, \eqref{eq:washbasin}) of the integrated bulk Noether current and a
boundary term. In general, this does not coincide with components of the
gravitational stress tensor \eqref{eq:e4e6d4d5e5}, particularly in the presence
of non-minimal coupling of matter to curvature. While the relationship
\eqref{eq:e4e6d4d5e5} seemed to be an important assumption in the previous
discussion \cite{Swingle:2014uza} of the first law of generalized entropy, we
were nonetheless able to recover similar relations between the generalized first
law and gravitational dynamics in section \ref{sec:neon} using our general
expression \eqref{eq:mendelevium} for the modular Hamiltonian. Let us now
discuss some noteworthy features of our calculations, the generalizations they
afford relative to previous work, and possible future directions to pursue.

As overviewed in section \ref{sec:helium} and further expanded upon in appendix
\ref{sec:d4f5g3nf6bg2e6nf3be7o-oo-oc4d5}, we have chosen to work with the
gravitational effective action in this paper in order to relate gravitational
dynamics to the first law of generalized entropy. Because we were working with
the effective action, as opposed to simply the logarithm of the partition
function, it was sensible to consider off-shell backgrounds, as in our
generalized entropy formula \eqref{eq:astatine}. Of course, \eqref{eq:astatine},
evaluated on holographic RT surfaces, agrees the more standard holographic
computation of generalized entropy seen in literature involving smooth bulk
backgrounds, as shown in appendix \ref{sec:d4f5g3nf6bg2e6nf3be7o-oo-oc4d5}; for
example, as reviewed there, \cite{Dong:2017xht} evaluates the effective action
on smooth but off-shell backgrounds in order to take variations of the RT
surface when deriving its quantum extremality. Indeed, as manifest in
\eqref{eq:astatine}, there seems to be no obstruction\footnote{An objection may
  be to point out the difficulty in practically evaluating the analytic
  continuation in $n$ for arbitrary regions. But of course, this issue plagues
  any replica calculation of entropy. One can nonetheless argue formally for the
  existence of an analytic continuation as described in footnote
  \ref{foot:pretext}.} in using this Callan-Wilczek-like formula to define
generalized entropy for any bulk region, not just those enclosed by RT surfaces
in holography. Defining generalized entropy for arbitrary regions using the
standard formula \eqref{eq:solitude} is of course relatively straightforward for
simple matter fields, \eg{}scalar fields, but is made difficult when treating
gravity dynamically; it is difficult to even assign graviton fluctuations to a
particular region in a gauge-invariant manner\footnote{Therefore, the
  calculations in section \ref{sec:helium} connecting the standard formula
  \eqref{eq:solitude} to our Callan-Wilczek-like formula \eqref{eq:astatine} are
  somewhat dubious in light of issues of gauge-fixing, as admitted in footnote
  \ref{foot:d4nf6c4e6nf3b6g3}. However, with this loose connection between
  \eqref{eq:solitude} and \eqref{eq:astatine} in mind, we are suggesting here
  that perhaps \eqref{eq:astatine} should be taken as the definition of
  generalized entropy of arbitrary regions in gravitational theories, where
  \eqref{eq:solitude} might be difficult to define.}. Yet, there does not seem
to be any immediate difficulty in including gravitons in the evaluation of our
Callan-Wilczek-like formula \eqref{eq:astatine}.

In fact, in evaluating the gravitational effective action, at least to
one-loop-order in $\GNewton$, as described in section \ref{sec:helium} and
appendix \ref{sec:d4f5g3nf6bg2e6nf3be7o-oo-oc4d5}, it seems natural to include
graviton fluctuations as an extra matter field. We should admit, however, that
our treatment of the graviton is still not quite complete, because, as noted in
footnote \ref{foot:d4nf6c4e6nf3b6g3}, we have yet to properly treat issues of
gauge fixing. Naively, we do not anticipate much difficulty resulting from the
introduction of ghost fields into the path integral defining the effective
action. However, additional care must be taken if one wants the resulting
effective action to be invariant under gauge transformations of the background
field and independent of the gauge-fixing conditions for the graviton.
Fortunately, past work by Vilkovisky and DeWitt
\cite{Vilkovisky:1984st,Vilkovisky:1984tga}\footnote{In footnote
  \ref{foot:banker}, we very briefly outline the general idea.} has clarified
how this can be achieved. An interesting question for future work to examine
would be whether our Callan-Wilczek-like formula \eqref{eq:astatine} gives a
sensible notion of entropy for arbitrary regions in gravitational theories when
evaluated using the gauge-fixed Vilkovisky-DeWitt effective action; \eg{}does
the generalized entropy thus obtained satisfy a sensible first law?

Using \eqref{eq:astatine}, we showed in section \ref{sec:fluorine} that the
modular Hamiltonian for quantum matter fields is given by the sum of the bulk
Noether current plus a boundary term, as written in \eqref{eq:mendelevium} in
Euclidean signature, or equivalently, continued to Lorentzian signature,
\eqref{eq:washbasin}. By working carefully with this expression for the modular
Hamiltonian (as opposed to the supposed expression \eqref{eq:e4e6d4d5e5} in
terms of the stress tensor used in \cite{Swingle:2014uza}) and accounting for
matter contributions to Wald-Dong entropy, we derived the relation
\eqref{eq:calamity} between gravitational dynamics and the first law of
generalized entropy. As foreshadowed in the introductory section
\ref{sec:rockBottom}, beyond being valid for matter non-minimally coupled to
curvature where \eqref{eq:e4e6d4d5e5} generally fails, our approach extends
beyond that of \cite{Swingle:2014uza} in a couple other ways.

Firstly, while \cite{Swingle:2014uza} implicitly assumes that background
perturbations are suppressed in $\GNewton$ due to the subtleties pointed out in
section \ref{sec:rockBottom}, our calculations make no such assumption. The fact
that our results hold for arbitrary first order variations is not terribly
surprising. After all, the classical first law for Noether charge entropy
\cite{Wald:1993nt,Iyer:1994ys} doesn't even make reference to a $\GNewton$
parameter.

Secondly, our result \eqref{eq:calamity} fully accounts for contributions of
matter to the asymptotic energy $\HamBdy{\thVec}$ written in \eqref{eq:cymbal}.
This is to be contrasted with \cite{Swingle:2014uza} which dedicates a paragraph
to explaining that if matter is assumed to fall off sufficiently quickly at the
spacetime boundary, then the matter sector should not modify the expression for
asymptotic energy. Indeed, \cite{Swingle:2014uza}'s derivation, as summarized in
our section \ref{sec:rockBottom}, would seem to suggest that the only direct
appearances of matter in the relation between the generalized first law and
gravitational dynamics are in the von Neumann entropy and as a source for the
gravitational equations of motion. Again, our more general result echos
intuition from the classical first law \cite{Wald:1993nt,Iyer:1994ys} where
matter and gravitational fields alike appear in the expression for asymptotic
energy.

One need not be too creative to find an example where matter contributes
directly to the asymptotic energy \eqref{eq:cymbal}. The particularly simple but
instructive example of a minimally coupled Maxwell field $\formMax$ in an
asymptotically-AdS spacetime is considered in appendix \ref{sec:garnish}. There,
electrically charged black holes are considered with boundary conditions
corresponding to fixed potential at the AdS boundary $\bdyMan$ and to fixed
charge density, as evaluated by Gauss's law on boundary time slices. In either
case, it is shown that the Maxwell field makes a nonvanishing contribution
$\chargeMax{\thVec}|_{\bdyReg}$ to the Noether charge appearing in the
asymptotic energy \eqref{eq:cymbal}. For fixed potential boundary conditions,
the Maxwell contribution $\HamBdyMax{\thVec}$ to asymptotic energy is given by
the integral $\int_{\bdyReg}\chargeMax{\thVec}=-\potElecBH \totalChargeElec$ and
can be identified as the product of a fixed electric potential $\potElecBH$ and
a fluctuating electric charge $\totalChargeElec$. Thus, the first law of
generalized entropy \eqref{eq:calamity} with a quantum Maxwell field
appropriately recovers the expected `grand canonical' thermodynamic first law of
electrically charged black holes with the appearance of a $\vary\langle
\HamBdyMax{\thVec} \rangle=-\potElecBH\vary\langle \totalChargeElec \rangle$
term. Obviously, the Maxwell contribution to the asymptotic energy, in
particular the Noether charge $\charge{\thVec}|_{\bdyReg}$, is crucial to this
result. For the fixed charge density boundary conditions, one must include an
extra boundary action $\int_{\bdyMan} \LagDensityBdyMax$ so that the variational
problem in $\formMax$ is well-defined in the sense described around
\eqref{eq:wiring}. It is found, in this case, that the two terms in the Maxwell
part $\HamBdyMax{\thVec}=\int_{\bdyReg}(\chargeMax{\thVec}+\thVec
\cdot\LagDensityBdyMax)$ of asymptotic energy, though each nonvanishing, exactly
cancel each other. Thus, the expected `canonical' thermodynamic first law is
recovered with the absence of a charge variation term. These examples
demonstrate the importance of matter contributions to both the Noether charge
and boundary Lagrangian terms of the asymptotic energy \eqref{eq:cymbal}.

Let us conclude by pointing out some directions for future work. Firstly, as
already mentioned above, it would be interesting to carefully work out the
details of gauge-fixing in relation to our Callan-Wilczek-like generalized
entropy formula \eqref{eq:astatine}, including ghost fields and working with a
gauge-invariant (\eg{}Vilkovisky-DeWitt) effective action. The hope is that
\eqref{eq:astatine} might give a clean way to define generalized entropy for
arbitrary regions in gauge theories.

Another area that lacks in clarity relates to the path integration measure for
the quantum matter and how its variation might contribute to our calculations.
In appendix \ref{sec:deflate}, we have sketched out how we expect these
contributions to correct quantities encountered in this paper which have so far,
in the main text, been naively obtained only from considerations of the action.
We then argued that, while quantities like the matter stress tensor might be
corrected by variations of the path integration measure, the generalized first
law result \eqref{eq:calamity} is still expected to hold with the corrected
quantities. It may be worthwhile to verify the admittedly abstract claims made
in our appendix \ref{sec:deflate} by explicit calculation in some simple
examples.

Finally, one can try to extend our calculation of the variation of generalized
entropy to higher orders in perturbations, along the lines of
\cite{Faulkner:2017tkh}. In that paper, it was shown, for a classical bulk in
holography, that second order gravitational equations about the vacuum can be
derived by comparing the second variations of Wald-Dong entropy and asymptotic
energy with the second variation of relative entropy in the CFT. To lift these
results to the semiclassical level, one would need to evaluate the second
variation of the bulk von Neumann entropy which, in turn, seems to require one
to know how the bulk modular Hamiltonian varies away from thermal setups
(specifically, the vacuum in the considerations of \cite{Faulkner:2017tkh}). For
perturbations that can be easily described by a deformation of the path integral
preparation in the Euclidean past, the results in our appendix
\ref{sec:immorally} might be useful. We derive there the instantaneous generator
for time evolution on time-dependent backgrounds and write the modular
Hamiltonian for states prepared on such backgrounds as the logarithm of a
path-ordered exponential. (As in the time-independent case, we find that the
instantaneous time evolution generator does not generally correspond to
components of the gravitational stress tensor, contrary to previous work
\cite{Wong:2013gua}.) With this, one may be able to evaluate the second
variation of bulk von Neumann entropy for perturbations corresponding to
background variations in the Euclidean past, \eg{}perturbations to the boundary
conditions set by background field asymptotics at the Euclidean spacetime
boundary corresponding to turning on CFT sources.


\acknowledgments


The author would like to thank Robert Myers for his comments and invaluable
discussion on the drafts of this paper. Research at Perimeter Institute is
supported in part by the Government of Canada through the Department of
Innovation, Science, and Economic Development Canada and by the Province of
Ontario through the Ministry of Colleges and Universities. The author is also
supported by the Natural Sciences and Engineering Research Council of Canada
through a Doctoral Canadian Graduate Scholarship. The author's attendance at
various workshops and meetings were supported by the It-from-Qubit
collaboration.


\appendix

\section{Effective actions and entropy}
\label{sec:d4f5g3nf6bg2e6nf3be7o-oo-oc4d5}
In this appendix, we first review the basic relationship between connected
generating functionals and effective actions in quantum field theory. This
material can be found in standard quantum field theory texts, \eg{}in sections
11.3 and 11.4 of \cite{Peskin:1995ev}, but we review it here to make our
discussion self-contained. Then, we will describe how the standard holographic
calculation of bulk generalized entropy, involving quantum gravity partition
functions, relates to our Callan-Wilczek form \eqref{eq:astatine} of generalized
entropy computed using the effective action. A reference which makes use
effective actions, in the context of holographic entropy calculations is
\cite{Dong:2017xht}, though they are perhaps less explicit about this --- see
their first paragraph of section 4.1. We will aim to be more explicit and
comment on connections to \cite{Dong:2017xht} as we proceed. Having everything
setup in the end, as a brief aside, we present a parallel calculation to
\cite{Dong:2017xht}, deriving the quantum extremality of bulk entangling
surfaces in holography, using our Callan-Wilczek equation \eqref{eq:astatine}.
As elsewhere in this paper, we do not treat the particulars of gauge-fixing so
our calculations involving gravitons are somewhat formal in this sense. A far
more sophisticated discussion of how one can obtain
field-reparametrization-invariant and gauge-invariant effective actions can be
found in \cite{Vilkovisky:1984st,Vilkovisky:1984tga}; for simplicity, we shall
focus our discussion on the naive effective action introduced in standard QFT
and leave issues of gauge-fixing to be examined in future work.

Let us consider a quantum field theory with a quantum field $\graviton$ on a
background $\metric$, with an action $\actQG[\metric+\graviton]$. As suggested
by notation, the background metric and graviton in the main text of this paper
take the roles of $\metric$ and $\graviton$. Other background fields
$\bgFieldsOther$ and quantum fields $\qFieldsOther$ will be omitted in the
present discussion for simplicity. They merely go along for the ride as extra
path integrated fields $[d\qFieldsOther]$ and extra arguments
$\bgFieldsOther+\qFieldsOther$ for the action. (We will not introduce bulk
sources or Legendre transform these other fields, as we do for $\metric$ and
$\graviton$ below.) Further, the action $\actQG[\metric+\graviton]$ should be
distinguished from the semiclassical action $\act[\metric,\graviton]$ to be
introduced below and used throughout this paper, starting in \eqref{eq:alone}.
If $\metric$ and $\graviton$ are metric and graviton fields, then the meaning of
the superscript on the action $\actQG$ is that a path integral with this action
gives expectation values $\langle \bullet \rangleQG$ in the quantum gravity
theory, at least symbolically.

Introducing a bulk source $\srcGrav$ for the field $\metric+\graviton$, the
connected generating functional $\eActCon[\srcGrav]$ is given by
\begin{align}
  \eActCon[\srcGrav]
  \equiv& -\log \int [d\graviton] e^{-\actQG[\metric+\graviton]-\frac{1}{2}\int \srcGrav\cdot (\metric+\graviton)}
          \;.
\end{align} 
While $\eActCon[0]$ gives the partition function of the quantum gravity theory,
the derivatives of $\eActCon[\srcGrav]$, of course, give the connected
correlation functions of $\metric+\graviton$. In particular, the one-point
function is given by the first derivative:
\begin{align}
  2 \frac{\delta\eActCon[\srcGrav]}{\delta\srcGrav}
  =& \metric+\langle \graviton \rangleQG_\srcGrav \;,
     \label{eq:darkening}
\end{align}
where $\langle \bullet \rangleQG_\srcGrav$ denotes the expectation value in the
presence of the source $\srcGrav$. Due to the path-integration over
$\graviton$, $\eActCon[\srcGrav]$ is in fact independent of the bulk profile of
the background $\metric$ (though it can depend on the boundary conditions of
$\metric+\graviton$ at the spacetime boundary $\bdyMan$ which may be implicitly
specified by $\metric$).

The connected generating functional and effective action are related by a
Legendre transformation. To describe this, it is helpful to define
$\srcGrav[\metric]$ as the particular choice of the source $\srcGrav$ such that
$\metric$ fully captures the expectation value of the complete field
$\metric+\graviton$, that is,
\begin{align}
  \metric + \langle \graviton \rangleQG_{\srcGrav[\metric]}
  \equiv& \, \metric \;,
  &
  &\text{\ie,}
  &
    \langle \graviton\rangleQG_{\srcGrav[\metric]}
    =& \, 0
 \;,
          \label{eq:ragged}
\end{align}
or equivalently, from \eqref{eq:darkening},
\begin{align}
  2 \left.
  \frac{\delta\eActCon[\srcGrav]}{\delta\srcGrav}
  \right|_{\srcGrav[\metric]}
  =& \metric
     \;.
     \label{eq:e4c5d4cxd4c3dxc3}
\end{align} 
The effective action $\eActMix[\metric]$ is then given by\footnote{To give a
  brief description of Vilkovisky and de Witt's field-reparametrization-invariant
  and gauge-invariant effective action
  \cite{Vilkovisky:1984st,Vilkovisky:1984tga}, the rough idea is as follows.
  Firstly, one must gauge-fix the path integral (in the usual way, introducing
  ghost fields) using ``mean-field gauge conditions'', which vary as one applies
  gauge transformations to the background field. Further, one must work
  covariantly in the space of field configurations. Instead of sourcing the
  `difference of coordinates' $\graviton$ between the points $\metric$ and
  $\metric+\graviton$ in the defining equation \eqref{eq:e4c5d4cxd4c3} for the
  effective action, one should instead consider the `geodesic distance' between
  $\metric$ and $\metric+\graviton$ in the space of metric configurations; the
  source for this geodesic distance is then chosen to set the
  expectation value of the distance to
  zero.\label{foot:banker}}
\begin{align}
  \eActMix[\metric]
  \equiv& \eActCon[\srcGrav[\metric]]
          - \frac{1}{2} \int \srcGrav[\metric] \cdot\metric
          \label{eq:lavender}
  \\
  =& -\log \int [d\graviton] e^{-\actQG[\metric+\graviton] -\frac{1}{2} \int \srcGrav[\metric] \cdot\graviton}\;.
     \label{eq:e4c5d4cxd4c3}
\end{align} 
In \cite{Dong:2017xht} --- see the first paragraph of section
4.1 therein --- the analogue\footnote{The relative factor of $\frac{1}{2}$
  comes from the fact that \cite{Dong:2017xht} doesn't seem to use the
  conventional normalization \eqref{eq:friendless} for stress tensors, while we
  have normalized $\srcGrav$ so that its appearance in \eqref{eq:foothill} is
  similar to a conventional stress tensor. Note also that they are sourcing fluctuations in the \emph{inverse} metric, whereas our graviton
  $\graviton_{ab}$ gives fluctuations in the metric, so the comparison between
  our $\frac{1}{2}\srcGrav^{ab}[\metric]$ and their
  $-\volForm\,(\DongStressBkg[\metric])_{ab}$ is not exact.} of our
$\frac{1}{2}\srcGrav[\metric]$ would be called $-\volForm\,\DongStressBkg[\metric]$, and by defining the log of the partition function
``off-shell'', they mean calculate the effective action $\eActMix[\metric]$ ---
see footnote 10 in \cite{Dong:2017xht}. While higher derivatives of $\eActMix[\metric]$ give one-particle-irreducible
(1PI) correlation functions\footnote{For this reason, \eqref{eq:lavender} is
  also known as the 1PI generating function in QFT. Note, however, that the more
  sophisticated Vilkovisky-DeWitt effective action mentioned in footnote
  \ref{foot:banker} is not a 1PI generating functional \cite{Burgess:1987zi}.},
the first derivative returns the source $\srcGrav[\metric]$, as can be seen from
\eqref{eq:e4c5d4cxd4c3dxc3} and \eqref{eq:lavender}:
\begin{align}
  -2\frac{\delta\eActMix[\metric]}{\delta\metric}
  =& \srcGrav[\metric]
  &
  &(\text{$\vary\metric$ away from $\bdyMan$})\;.
    \label{eq:foothill}
\end{align}
Setting the RHS of this equation to zero gives the quantum analogue of an
equation of motion
\begin{align}
  -2 \frac{\delta\eActMix[\metric]}{\delta\metric}
  =& \srcGrav[\metric]
     = 0
  &
  &(\text{$\vary\metric$ away from $\bdyMan$, $\metric$ on-shell})
    \label{eq:tavern}
\end{align} 
with the meaning that a background profile $\metric$ satisfying this equation
fully captures the expectation value of the total field $\metric+\graviton$ in
the absence a source $\srcGrav$:
\begin{align}
  \metric + \langle \graviton \rangleQG
  =& \metric \;,
  &
  &\text{\ie,}
  &
    \langle \graviton\rangleQG
    =&0
       \;.
  &
  &(\text{$\metric$ on-shell})
    \label{eq:sequence}
\end{align}
The on-shell condition \eqref{eq:tavern} is precisely that written in
\eqref{eq:lonely}. In fact, it is not difficult to see by definition
\eqref{eq:lavender} that \eqref{eq:foothill} and the on-shell condition
\eqref{eq:tavern} are respectively equivalent
to
\begin{align}
  -2 \varyWRT{\metric} \eActMix[\metric]
  =& \int \srcGrav[\metric]\cdot\vary\metric
  &
    (\text{$\vary\metric$ preserving \bc})
  \\
  =& 0  & (\text{$\vary\metric$ preserving \bc, $\metric$ on-shell})
          \label{eq:escalate}
\end{align}
for all $\vary\metric$ that preserve the boundary conditions at the spacetime
boundary $\bdyMan$,
\ie{}can be absorbed as a shift in the path-integration variable $\graviton$,
but need not have compact support localized in the spacetime interior.

Throughout the majority of this paper, we have found it useful to think of
$\eActMix[\metric]$ as simply giving the logarithm of the partition function in
a local ``semiclassical'' theory of a quantum matter field $\graviton$
propagating on a classical background $\metric$. That is, we would like to think
of the combination
$\actQG[\metric+\graviton]+\frac{1}{2}\int\srcGrav[\metric]\cdot\graviton$
appearing in \eqref{eq:e4c5d4cxd4c3} as itself an action
$\act[\metric,\graviton]$ local in $\metric$ and $\graviton$. (We will clarify
what ``semiclassical'' means below \eqref{eq:covenant}.) But in general,
$\srcGrav[\metric]$ is not local in $\metric$ --- from its definition
\eqref{eq:ragged}, we see that $\frac{1}{2}\srcGrav[\metric]$ is precisely the
sum of all amputated $\graviton$ tadpoles. However, this is local at leading
loop order:
\begin{align}
  \srcGrav[\metric]
  =& -2\frac{\delta\actQG[\metric]}{\delta\metric} + O(\GNewton^0) \;,
     \label{eq:capitulate}
\end{align}
where we have supposed that $\actQG$ contains a constant prefactor $1/\GNewton$
such that loop expansion becomes synonymous with expanding\footnote{Obviously,
  for the gravitational case, $\GNewton$ is Newton's constant. In the main text,
  where we consider additional matter fields $\bgFieldsOther+\qFieldsOther$, we
  do not assume that their parts of the action contain factors of $1/\GNewton$.
  Then, the simple correspondence between expanding in $\GNewton$ and in loops
  is lost; nonetheless, we will say ``$n$-loops-in-$\GNewton$'' to mean an order
  in $\GNewton$ that would correspond to $n$-loop-order had the theory only
  contained $\metric$ and $\graviton$ fields.} in $\GNewton$. Thus, if we are
merely interested in evaluating \eqref{eq:e4c5d4cxd4c3} to one-loop order, that
is, to order $\GNewton^0$, we may throw out the nonlocal loop contributions to
$\srcGrav[\metric]$, keeping only the local leading piece \eqref{eq:capitulate}.
The semiclassical action $\act[\metric,\graviton]$ can then be taken to be the
zeroth-order and quadratic terms of $\actQG[\metric+\graviton]$ expanded in
$\graviton$:
\begin{align}
  \act[\metric,\graviton]
  =& \actQG[\metric]
     + \frac{1}{2} \iint \graviton \cdot \frac{\delta^2\actQG[\metric]}{\delta\metric^2} \cdot \graviton
     \label{eq:critter}
\end{align} 
Terms of higher power in $\graviton$ can be dropped as they would only
contribute to the effective action \eqref{eq:e4c5d4cxd4c3} at order $\GNewton$
and higher. Now, we have
\begin{align}
  \eActMix[\metric]
  =& -\log \int [d\graviton]
     e^{-\act[\metric,\graviton]}
     + O(\GNewton) \;,
     \label{eq:faucet}
\end{align}
where, by construction, $\act[\metric,\graviton]$ is local in $\metric$ and
$\graviton$. This allows for the second interpretation of $\eActMix[\metric]$,
as the logarithm of a partition function in a semiclassical theory with action
$\act$.

To connect further with the notation used in the main text from \eqref{eq:alone}
onward, in the gravitational case, we can categorize the terms in
\eqref{eq:critter} as:
\begin{align}
  \actGrav[\metric]
  =& \actQG[\metric]
  \\
  \actMat[\metric,\graviton]
  =& \frac{1}{2} \iint \graviton \cdot \frac{\delta^2\actQG[\metric]}{\delta\metric^2} \cdot \graviton
     \;,
     \label{eq:covenant}
\end{align}
with the action of any extra matter fields $\bgFieldsOther+\qFieldsOther$, if
present, also added to $\actMat$. Thus, the graviton contributes to the matter
stress tensor in the semiclassical theory, as described below \eqref{eq:jeep}.
Indeed, the on-shell condition \eqref{eq:tavern} for the background $\metric$
now takes the expected semiclassical form \eqref{eq:lonely} of an equality
between a classical (higher-derivative equivalent of an) Einstein tensor
$(\eomGrav{\metric})^{ab}$ and a quantum matter stress tensor $\langle
\stress^{ab} \rangle$.

In the main text, the locality of the action $\act[\metric,\graviton]$ in this
semiclassical theory is useful because it permits the application of Iyer-Wald
formalism when performing entropy calculations\footnote{It is conceivable that
  one can find a way to work with the full nonlocal action
  $\actQG[\metric+\graviton]+\frac{1}{2}\int\srcGrav[\metric]\cdot\graviton$
  appearing in the exact effective action $\eActMix[\metric]$ in
  \eqref{eq:e4c5d4cxd4c3}, leveraging the fact that the nonlocal
  $\srcGrav[\metric]$ multiplies $\graviton$, whose one-point function vanishes
  by the definition \eqref{eq:ragged} of the source $\srcGrav[\metric]$.
  One can argue, for instance in calculations involving the first variation of
  the effective action $\eActMixMat[\metric]$,
  the variation of the term $\frac{1}{2}\int\srcGrav[\metric]\cdot\graviton$ does not contribute
  since it involves $\langle \graviton\rangleQG_{\srcGrav[\metric]}=0$.}. While the
original action $\actQG[\metric+\graviton]$ was also local in the first place,
the benefit of performing the Legendre transform \eqref{eq:lavender} to obtain the
effective action when doing entropy calculations is twofold. Firstly, one can
much more concretely define a generalized entropy across an arbitrary entangling
surface $\entSurf$ in an arbitrary background via \eqref{eq:astatine}. In
contrast, as mentioned below \eqref{eq:darkening}, the (bulk profile of the)
background is meaningless in the quantum gravity partition function
$\eActCon[0]$. For instance, evaluating the partition function on the replicated
background $\metricR{\repS{\entSurf} n}$ will give an answer independent of
$\entSurf$ --- the graviton path integral effectively erases any signature of
$\entSurf$, for example, invariably setting $\langle \metricR{\repS{\entSurf} n}
+ \graviton \rangleQG$ to a smooth on-shell configuration independent of
$\entSurf$. For this reason, \cite{Dong:2017xht} found it necessary to work with
the effective action when deriving the quantum extremality of $\entSurf$ --- we
briefly review this further below as an aside. For the main text of this paper,
however, the primary motivation behind working with the generating functional
$\eActMix[\metric]$ is that it provides a nontrivial equation of motion for the
metric \eqref{eq:tavern}, or expressed in the semiclassical formulation,
\eqref{eq:lonely}. As we find in \eqref{eq:finale}, linearized equations of
motion for the metric enter as sufficient conditions for the first law of
generalized entropy. Moreover, as written in \eqref{eq:stylist}, the first law
implies the integral of certain components of the linearized equations vanishes;
as mentioned there, this can be strengthened in certain maximally-symmetric
setups to even locally set all components of the linearized equations to zero.
This is a nontrivial result regarding how an information-theoretic statement and
gravitational dynamics are mutually dependent.

Having listed some advantages of working with the effective action
$\eActMix[\metric]$, we should however be mindful that, while the original
quantum gravity theory, with action $\actQG$, and resulting modified theory,
with action $\actQG+\frac{1}{2}\int \srcGrav[\metric]\cdot\graviton$ which
semiclassically becomes $\act$, are related, they are not in general the same.
For example, the expectation values $\langle \bullet \rangleQG$ and $\langle
\bullet \rangle$, computed respectively with these actions can differ, with the
two related by $\langle \bullet \rangle = \langle \bullet
\rangleQG_{\srcGrav[\metric]}$. (Throughout the main text, we have referred
almost exclusively to the latter expectation values calculated semiclassically
--- see \eqref{eq:desperate}.) This is the case if $\metric$ is taken off-shell:
while $\langle \graviton \rangleQG$ will take whatever value required to put the
combination $\metric+\langle \graviton \rangleQG$ back on-shell, we will always
invariably have $\langle \graviton \rangle=0$ --- recall, by definition
\eqref{eq:ragged}, the role of the source $\srcGrav[\metric]$ is precisely to
ensure this. If, however, $\metric$ is on-shell \eqref{eq:tavern}, then the
vanishing of $\srcGrav[\metric]$ implies we can exactly take
\begin{align}
  \act[\metric,\graviton]
  =& \actQG[\metric+\graviton]
  &
  &(\text{$\metric$ on-shell})
    \label{eq:trifocal}
\end{align}
so that, for any operator $\blkOp$,
\begin{align}
  \langle \blkOp \rangleQG
  =& \langle \blkOp \rangle \;.
  &
  &(\text{$\metric$ on-shell})
\end{align}
Furthermore, the connected generating functional and effective action will also
agree:
\begin{align}
  \eActCon[0]
  =& \eActMix[\metric] \;.
  &
    (\text{$\metric$ on-shell})
    \label{eq:bakery}
\end{align}

Given that, when $\metric$ is taken off-shell, the semiclassical theory departs
from the original quantum gravity theory we started out with, one might worry
that the generalized entropy calculated in the former might be of little
relevance in the context of the latter, especially since the
$\metricR{\repS{\entSurf}n}$ appearing in our generalized entropy formula
\eqref{eq:astatine} is clearly off-shell at $n\ne 1$. However, we shall
argue that the generalized entropy calculated using the effective action
$\eActMix$ with \eqref{eq:astatine} is precisely equivalent to the more standard
entropy calculation in holography involving quantum gravity partition functions,
which we now review\footnote{The discussion after this point is independent of
  having a local semiclassical theory, so there is no need to loop-expand
  $\eActMix$. Our goal will simply be to relate the holographic entropy
  calculation to \eqref{eq:astatine}, which can be taken to be
  our definition of generalized entropy in terms of the exact effective action
  $\eActMix$.}.

Let us begin with some notation on the CFT side. We shall consider a CFT state
$\state$ of the spacetime boundary region $\bdyReg$ with boundary
$\partial\bdyReg=\bdyEntSurf$. Further, we suppose that $\state$ is prepared by
an appropriately normalized Euclidean path-integral over the boundary manifold
$\bdyMan$ with cuts on $\bdyReg$ --- for example, see Figure
\ref{fig:e4e5ke2ke7}, though there, it appears $\bdyEntSurf$ is empty or otherwise at
infinity in the suppressed spatial directions. It is useful to also consider the
unnormalized state
\begin{align}
  \unState=\tr(\unState) \state
\end{align}
obtained from the unnormalized path-integral. For instance, the CFT partition
function on $\bdyMan$ is given by $\tr\unState$. According to the replica trick
(reviewed for the bulk state $\unStateMat$ below \eqref{eq:lost}) higher moments
$\tr\unState^n$ correspond to path integrals, \ie partition functions, on $n$
replicas of $\bdyMan$. These replicas are `branch'-stitched together along
$\bdyReg$, but one should exclude any curvature singularities produced on
$\bdyEntSurf$ --- for later use, let us call the resulting manifold
$\bdyManR{n}$. (In the notation introduced above \eqref{eq:trustless},
$\bdyManR{n}=\bdyManR{\rep{\bdyEntSurf}n}$.) To calculate the entropy of the
state $\state$, we can then apply \eqref{eq:lost}, now to the boundary state:
\begin{align}
  \ent[\state]
  =& - \partial_n \left( \log\tr \state^n \right)_{n=1}
     = \partial_n \left(  -\log \tr \unState^n + n\log\tr\unState \right)_{n=1} \;.
     \label{eq:lake}
\end{align}

Holography equates the CFT partition function $\tr\unState^n$ on $\bdyManR{n}$
to the bulk quantum gravity partition function in an asymptotically-AdS
spacetime with boundary $\bdyManR{n}$, thus allowing \eqref{eq:lake} to be
transformed into a bulk calculation. In particular, (again, focusing on just the
gravitational fields $\metric$ and $\graviton$ for simplicity) the holographic
dictionary states that
\begin{align}
  - \log\tr\unState^n
  =& \eActConR{n}
     \equiv -\log\int [d\graviton] e^{-\actQG[\metricR{n}+\graviton]} \;,
     \label{eq:dislodge}
\end{align}
where $\metricR{n}$ is any bulk background with boundary $\bdyManR{n}$ --- note
that the bulk partition function $\eActConR{n}$ is simply the connected
generating functional $\eActCon[0]$ with sources turned off and evaluated on
$\metricR{n}$. As noted below \eqref{eq:darkening}, the bulk profile of
$\metricR{n}$ in path integrals of the form \eqref{eq:dislodge} is immaterial
due to the path integration over $\graviton$ (but we will make a choice for
$\metricR{n}$ momentarily). We therefore have, upon combining \eqref{eq:lake}
and \eqref{eq:dislodge}:
\begin{align}
  \ent[\state]
  =& \partial_n \left(
     \eActConR{n}
     - n\eActConR{1}
     \right)_{n=1} \;.
     \label{eq:nastily}
\end{align}
We can now state our question concretely: can \eqref{eq:nastily} be equated with
our Callan-Wilczek formula \eqref{eq:astatine} for generalized entropy across
some entangling surface $\entSurf$ on some background $\metric$? After all,
\eqref{eq:nastily} and \eqref{eq:astatine} are tantalizingly similar, save for
the use of $\eActConR{n}$ versus $\eActMix[\metricR{\repS{\entSurf}n}]$.

Given that the connected generating functional and effective action can be
equated as in \eqref{eq:bakery} if the background is put on-shell, it is natural
now to make the definitive choice for $\metricR{n}$ to be on-shell:
\begin{align}
  \varyWRT{\metric}
  \left.
  \eActMix[\metric]
  \right|_{\metric=\metricR{n}}
  \equiv&\, 0\;,
  &
  &(\text{$\vary\metric$ preserving \bc})
    \label{eq:cardiac}
\end{align}
for then
\begin{align}
  \eActConR{n}
  =& \eActMix[\metricR{n}] \;.
\end{align}
But this still does not quite get us from \eqref{eq:nastily} to the form of
\eqref{eq:astatine}, but rather
\begin{align}
  \ent[\state]
  =& \partial_n\left(
     \eActMix[\metricR{n}]
     - n\eActMix[\metricR{1}]
     \right)_{n=1} \;.
     \label{eq:curtly}
\end{align}
It is this formula that \cite{Dong:2017xht} works mostly with --- in particular,
the ``$\DongActOrbR{n}$'' introduced there in (4.1), and used to express generalized
entropy by (4.3) and (4.4), is what we would call $\eActMix[\metricR{n}]/n$.
Though \eqref{eq:curtly} does not explicitly feature an entangling surface
$\entSurf$, as we soon review below, one is implicitly specified by the sequence
of metrics $\metricR{n}$. By considering transformations $\vary\metricR{n}$ of
$\metricR{n}$ which continue to diffeomorphisms $\vary\metricR{1}$ on
$\metricR{1}$, \cite{Dong:2017xht} go on to argue that generalized entropy is
extremized by $\entSurf$, \ie{}$\entSurf$ is quantum extremal. With the
exception of $n=1$, the $\vary\metricR{n}$ perturbations are generally off-shell,
and it is ultimately the extremality of $\eActMix[\metricR{n}]$ that results in
the quantum extremality of $\entSurf$.

Our present goal, however, is to go from \eqref{eq:curtly} to the Callan-Wilczek
form \eqref{eq:astatine} of generalized entropy which explicitly displays the
entangling surface. Comparing the second terms of \eqref{eq:curtly} and
\eqref{eq:astatine}, we see that we should aim for a generalized entropy
evaluated on the background $\metricR{1}$. We now choose the entangling surface
on $\metricR{1}$ as follows: for each integer $n\ge 2$, we expect the on-shell
background $\metricR{n}$ to share the $\integers_n$-symmetry of the boundary
$\bdyManR{n}$ and thus possess a codimension-two surface $\entSurfR{n}$ that is
fixed by this symmetry; we then suppose that this sequence of surfaces
$\entSurfR{n}$ can be continued to give a well-defined surface $\entSurfR{1}$ on
$\metricR{1}$ --- it is this surface that we will take to be the entangling
surface $\entSurf$ appearing in the generalized entropy formula
\eqref{eq:astatine}. With this, we can write down the difference between the
entropy $\ent[\state]$ written in \eqref{eq:curtly} and the generalized entropy
$\entGen{\entSurfR{1}}[\metricR{1}]$ given by \eqref{eq:astatine},
\begin{align}
  \ent[\state]
  -\entGen{\entSurfR{1}}[\metricR{1}]
  =& \partial_n \left(
     \eActMix[\metricR{n}]
     - \eActMix[\metricR{1\repS{\entSurfR{1}}n}]
     \right)_{n=1} \;,
     \label{eq:renewable}
\end{align}
with the aim to argue that this vanishes. For this purpose, it is helpful to
recall the notation $\bullet_{\orbS{\entSurf}n}$ introduced below
\eqref{eq:molybdenum} for the inverse operation to $\bullet_{\repS{\entSurf}n}$.
Making use of this, \eqref{eq:renewable} can be shown to vanish
\begin{align}
  \ent[\state]
  -\entGen{\entSurfR{1}}[\metricR{1}]
  =& (\partial_m+\partial_n)\left(
     \eActMix[\metricR{n\orbS{\entSurfR{n}}n \repS{\entSurfR{n}} m}]
     - \eActMix[\metricR{1\repS{\entSurf{1}}m}]
     \right)_{\substack{m=1 \\ n=1}}
  \\
  =& \left. \partial_n \eActMix[\metricR{n\orbS{\entSurfR{n}}n}] \right|_{n=1}
  \\
  =& 0\;,
     \label{eq:subplot}
\end{align}
where, in the last equality, the on-shell condition \eqref{eq:cardiac} for
$\metricR{1}$ was applied. We thus have the desired equivalence
between the usual starting point \eqref{eq:nastily} for holographically deriving
generalized entropy, featuring the quantum gravity partition function, and our
Callan-Wilczek equation \eqref{eq:astatine} which involves the effective action
$\eActMix$ and explicitly displays the entangling surface.


Finally, as an aside, we conclude this appendix with a parallel calculation to
\cite{Dong:2017xht}, showing that $\entSurfR{1}$ extremizes our Callan-Wilczek
form \eqref{eq:astatine} of generalized entropy
$\entGen{\entSurfR{1}}[\metricR{1}]$. As in \cite{Dong:2017xht}, let us consider
perturbing the metric $\metricR{1}$ by a diffeomorphism
$\vary\metric[\vary\entSurf]$ which captures the effect of an arbitrary
perturbation $\vary\entSurf$ of the entangling surface away from $\entSurfR{1}$
--- that is, fixing the coordinate location of $\entSurf=\entSurfR{1}$ while
perturbing the metric by $\vary\metric[\vary\entSurf]$ is equivalent to
perturbing the entangling surface by $\vary\entSurf$ (while fixing the metric
$\metric=\metricR{1}$)\footnote{Note that, since our generalized entropy
  equation \eqref{eq:astatine} involves only $\metricR{1}$, there is no need to
  extend $\vary\metric$ to $\metricR{n\ne 1}$ as was done in \cite{Dong:2017xht}. However,
  it will still be useful in the following calculation to consider the on-shell
  metrics $\metricR{n}$.}. Now, we simply evaluate the variation of
\eqref{eq:astatine}:
\begin{align}
  \int_\entSurf \vary\entSurf \cdot
  \left.
  \frac{\delta \entGen{\entSurf}[\metricR{1}]}{\delta \entSurf}
  \right|_{\entSurf=\entSurfR{1}}
  =& \int \vary\metric[\vary\entSurf] \cdot \left(
     \partial_n
     \frac{\delta \eActMix[\metricR{\repS{\entSurfR{1}}n}]}{\delta \metric}
     \right)_{\substack{\metric=\metricR{1} \\ n=1}}
  \\
  =& \int \vary\metric[\vary\entSurf] \cdot
     \left[ 
     \partial_n
     \left(
     \frac{\delta \eActMix[\metricR{\repS{\entSurfR{n}}n}]}{\delta \metric}
     \right)_{\metric=\metricR{n\orbS{\entSurfR{n}}n}}
     \right]_{n=1}
  \\
  =& 0 \;,
\end{align}
where, in the latter two equalities, we have made use of the on-shell condition
\eqref{eq:cardiac} for $\metricR{n}$, in particular, \eqref{eq:subplot} and
\begin{align}
  \left.
  \frac{\delta \eActMix[\metricR{\repS{\entSurfR{n}}n}]}{\delta \metric}
  \right|_{\metric=\metricR{n\orbS{\entSurfR{n}}n}}
  =& 0 \;.
  &
    (\text{$\vary\metric$ away from $\bdyMan$})
\end{align}
Thus, we conclude that $\entSurfR{1}$ is indeed quantum extremal in that it
extremizes our Callan-Wilczek form \eqref{eq:astatine} of generalized entropy.


\section{Evolution on time-dependent backgrounds}
\label{sec:immorally}

We saw in section \ref{sec:kebab} that, for a thermal state produced by
evolution on background fields $(\metric,\bgFieldsOther)$ which are symmetric
under rotations in thermal time $\thTime$ generated by
$\thVec=\partial_{\thTime}$, the matter modular Hamiltonian is given by the
integral of a Noether current plus a boundary term, as written in
\eqref{eq:mendelevium}. This operator then must generate evolution in time: for
any pure states $|\pureState_0\rangle$ and $\langle \pureState_2|$ on slices of
constant time, we have
\begin{align}
  \begin{split}
    \left\langle \pureState_2 \left| e^{ (\thTimeNelPatch{2}-\thTimeNelPatch{0})
    \left( \int_\entReg\currentMat{\thVec} + \int_{\bdyReg}
    \thVec\cdot\LagDensityBdyMat \right) } \right| \pureState_0
    \right\rangle = \int_{|\pureState_0\rangle}^{\langle \pureState_2 |}
    [d\graviton, d\qFieldsOther] e^{ -
    \actTimeClosedIntervalMat{\thTimeNelPatch{0}}{\thTimeNelPatch{2}} } \;,
    \\
    (\Lie{\thVec}\metric=0,\, \Lie{\thVec}\bgFieldsOther=0)
  \end{split}
  \label{eq:chewing}
\end{align}
where on the LHS, $\thVec$-symmetry has allowed us to consider, without loss of
generality, the initial slice $\entReg$ at $\thTime=0$; on the RHS,
$\actTimeClosedIntervalMat{\thTimeNelPatch{0}}{\thTimeNelPatch{2}}$ denotes the
action (to be dissected further below) in the spacetime interval
$\timeClosedInterval{\thTimeNelPatch{0}}{\thTimeNelPatch{2}}$ over which the
path integral is evaluated, with the states $|\pureState_0\rangle$ and $\langle
\pureState_2 |$ giving boundary conditions on the $\thTimeNelPatch{0}$ and
$\thTimeNelPatch{2}$ surfaces respectively. (We intend on reserving the symbol
$\thTimeNelPatch{1}$ for another time between $\thTimeNelPatch{0}$ and
$\thTimeNelPatch{2}$, as discussed below. We shall work in Euclidean signature
until the very end of this appendix, but a straightforward generalization to
Lorentzian signature is possible.)

Our goal in this appendix will be to find the generalization to
\eqref{eq:chewing} in the case where the background fields $\metric$ and
$\bgFieldsOther$ are not necessarily symmetric with respect to rotations in
thermal time generated by $\thVec$. We will still take
$\thVec=\partial_{\thTime}$ to be a vector field generating rotations in a
`thermal time' coordinate $\thTime$ around a codimension-two surface $\entSurf$
--- see Figure \ref{fig:e4e5ke2ke7} --- but the background fields will no longer
be required to be symmetric under this rotation\footnote{Note that, for a given
  background and $\entSurf$, one can generally have many possible choices of
  $\thVec$, which correspond to different coordinatizations of the spacetime.}.
(We will continue to take $\thTime$ near $\entSurf$ such that it parametrizes
proper angles around $\entSurf$.) The general idea behind our calculations will
be to identify the instantaneous generator
$-\int_{\thTime}\HamDensityMat{\thVec} - \int_{\thTime\cap\bdyMan}
\HamDensityBdyMat{\thVec}$ (here, $\thTime$ being shorthand for a constant
$\thTime$ surface) for time evolution that satisfies
\begin{align}
  \left\langle \pureState_2 \left| 
  \timeOrder e^{
  \int_{\thTimeNelPatch{0}}^{\thTimeNelPatch{2}} d\thTime\, \left( 
  \int_\thTime \HamDensityMat{\thVec}
  + \int_{\thTime\cap\bdyMan} \HamDensityBdyMat{\thVec}
  \right)
  }
  \right| \pureState_0 \right\rangle
  =& \int_{|\pureState_0\rangle}^{\langle \pureState_2 |}
     [d\graviton, d\qFieldsOther]
     e^{
     - \actTimeClosedIntervalMat{\thTimeNelPatch{0}}{\thTimeNelPatch{2}}
     } \;,
     \label{eq:county}
\end{align}
where $\timeOrder$ indicates time-ordering. (One may recognize some resemblance
of $\HamDensityMat{\thVec}$ with the symbol $\HamDensitySimpleMat{\thVec}$ from
\eqref{eq:dubbed} in section \ref{sec:bobtail} --- indeed, we shall find that
\eqref{eq:dubbed} gives part of the correct expression for
$\HamDensityMat{\thVec}$.) This will subsequently lead us to an expression for
the modular Hamiltonian of states prepared on arbitrary backgrounds. Restricting
our results to the $\thVec$-symmetric case offers a consistency check of
\eqref{eq:chewing} and our derivation of the thermal modular Hamiltonian in
section \ref{sec:fluorine}.

Before proceeding however, we should discuss precisely what is meant by the
action $\actTimeClosedIntervalMat{\thTimeNelPatch{0}}{\thTimeNelPatch{2}}$. In
addition to the bulk and spacetime boundary elements seen in
\eqref{eq:treadmill}, we must further introduce here additional
Gibbons-Hawking-like contributions\footnote{For Einstein gravity, one starts out
  with the usual Gibbons Hawking \cite{Gibbons:1976ue} Lagrangian density
  $-\frac{1}{8\pi\GNewton}\normalVec\cdot\volForm \extrK$ on codimension-1
  surfaces, where $\normalVec$ and $\extrK$ are the unit normal vector and trace
  of the extrinsic curvature (in the $\normalVec$ direction) of the surface. To
  Legendre transform to the semiclassical theory, as described in appendix
  \ref{sec:d4f5g3nf6bg2e6nf3be7o-oo-oc4d5}, one should keep as part of
  $\LagDensityGHMat$ any terms in
  $-\frac{1}{8\pi\GNewton}(\normalVec\cdot\volForm \extrK)[\metric+\graviton]$
  quadratic in $\graviton$. More generally, one can expect contributions to
  $\LagDensityGHMat$ whenever the bulk matter action $\actMat$ includes
  higher-than-first derivatives of matter fields.} to the action on the time
slices $\thTimeNelPatch{0}$ and $\thTimeNelPatch{2}$ bounding the path
integration spacetime region, where we fix the quantum field configurations
according to the states $|\pureState_0\rangle$ and $\langle \pureState_2|$.
Furthermore, we will also allow for the presence of ``joint'' terms at the
corners $\thTimeNelPatch{0}\cap\bdyMan$ and $\thTimeNelPatch{2}\cap\bdyMan$
(here, and elsewhere when easily understood, we will use $\thTime$ as shorthand
for a constant $\thTime$ surface). In all\footnote{It will be convenient in
  later discussions to suppose that the joint terms $\LagDensityJtMat$ are
  chosen so that, when evaluating the norm of a state
  $|\pureState_0\rangle=|\pureState_2\rangle$ by taking the
  $\thTimeNelPatch{0}\to\thTimeNelPatch{2}$ limit of \eqref{eq:county}, the
  joint terms vanish.
  \label{foot:litigator}},
\begin{align}
  \begin{split}
    \actTimeClosedIntervalMat{\thTimeNelPatch{0}}{\thTimeNelPatch{2}}
    =&
       \int_{\timeClosedInterval{\thTimeNelPatch{0}}{\thTimeNelPatch{2}}}
       \LagDensityMat + \left(
       \int_{\thTimeNelPatch{2}}-\int_{\thTimeNelPatch{0}}\right)
       \LagDensityGHMat
           + \left( \int_{\thTimeNelPatch{2}\cap\bdyMan} -
           \int_{\thTimeNelPatch{0}\cap\bdyMan}\right) \LagDensityJtMat\;.
    \\
     & + \int_{ \partial\timeClosedInterval{\thTimeNelPatch{0}}{\thTimeNelPatch{2}}
       \cap\bdyMan } \LagDensityBdyMat
  \end{split}
         \label{eq:statute}
\end{align}
Similar to $\LagDensityBdyMat$, whose role is described around
\eqref{eq:wiring}, the Gibbons-Hawking-like contribution $\LagDensityGHMat$ is
required to allow the fixing of field configurations on codimension-1 surfaces
to be consistent with a well-defined variational principle --- namely, that the
variation of the action, among the set of quantum field configurations permitted
by boundary conditions, is given solely by bulk equations of motion. This is
achieved by the identity\footnote{Suppose one views the difference of
  $\int_{\egReg}\LagDensityGH$ on either side of a codimension-one surface
  $\egReg$ as giving the bulk Lagrangian $\LagDensity$ integrated
  across a thin shell (\ie{}a neighbourhood) around $\egReg$, across which
  fields are smoothed. (This is easily seen in the example of Einstein gravity.) Then,
  \eqref{eq:sitcom} can be thought of as a consequence of considering the effect
  of field perturbations, which have support on one side of $\egReg$, on the
  bulk Lagrangian in the shell. \label{foot:sitcom}}
\begin{align}
  \left.\sympPot[\vary\graviton,\vary\qFieldsOther]\right|_{\egReg}
  =&
     -(\varyWRT{\graviton}+\varyWRT{\qFieldsOther})\LagDensityGHMat|_{\egReg}
     \;. 
  &(\text{codim.-1 surfaces $\egReg$ where $\vary\graviton,\vary\qFieldsOther=0$})
    \label{eq:sitcom}
\end{align}
This is analogous to \eqref{eq:wiring}\footnote{In \eqref{eq:wiring}, we
  considered also variations in the background fields $(\metric,\bgFieldsOther)$
  essentially arguing that relations for these follow from those of the quantum
  fields $(\graviton,\qFieldsOther)$ since $\bgFieldsOther$ and, prior to the
  Legendre transformation of the gravitational fields described in appendix
  \ref{sec:d4f5g3nf6bg2e6nf3be7o-oo-oc4d5}, $\metric$ merely shift
  $\qFieldsOther$ and $\graviton$. In this appendix, we will have no need to
  consider the analogue of \eqref{eq:sitcom} for background field variations
  (though we expect it to hold, if $(\metric,\bgFieldsOther)$ do indeed just
  shift $(\graviton,\qFieldsOther)$ prior to the Legendre transformation) and
  consider $(\metric,\bgFieldsOther)$ simply as nondynamical fields on which the
  action $\actMat$ depends.}. (Indeed, as mentioned below \eqref{eq:wiring}, one
expects Gibbons-Hawking-like terms to be included in $\LagDensityBdy$; as
described there, the $\LagDensityBdy$ typically requires even more terms to
implement \eqref{eq:wiring} at the spacetime boundary.) The Gibbons-Hawking-like
terms are supplemented here by the joint terms $\LagDensityJtMat$ where the
bounding surface turns sharply at the spacetime boundary.
Typically, 
joint terms are obtained simply from Gibbons-Hawking like terms in the limit
where the surface acquires a sharp edge, \ie{}``joint'', between two faces
\cite{Hayward:1993my}; on the spacetime boundary, it may be that the contents of
$\LagDensityJtMat$ are more complicated, just as $\LagDensityBdyMat$ can contain
more terms than $\LagDensityGHMat$. At any rate, analogous to
  \eqref{eq:wiring} and \eqref{eq:sitcom}, for any $\vary\graviton$ and
  $\vary\qFieldsOther$ that preserve boundary conditions on $\bdyMan$ and vanish
  on a codimension-one surface $\egReg$ (\eg{}a constant time surface) that
  intersects $\bdyMan$ at a joint $\egReg\cap\bdyMan$, we expect the
variation
  $(\varyWRT{\graviton}+\varyWRT{\qFieldsOther})\LagDensityJtMat|_{\egReg\cap\bdyMan}$ to cancel
  against any localized contributions that
  $\sympPot[\vary\graviton,\vary\qFieldsOther]$ might produce at the joint when integrated across
  it from $\egReg$ to $\bdyMan$. We will not speculate further on the precise forms of
$\LagDensityGHMat$, $\LagDensityJtMat$, and $\LagDensityBdyMat$ as they mostly
just get carried along in the following calculations.

One should note in \eqref{eq:statute}, however, the absence\footnote{Such terms
  near $\entSurf$ would obstruct the interpretation of the path integral in
  \eqref{eq:county}, when $\thTimeNelPatch{0}=0$ and $\thTimeNelPatch{2}=2\pi$,
  as giving matrix elements of an unnormalized state $\unStateMat$ on the
  $\thTime=0$ surface $\entReg$. Stitching together $n$ replicas of the path
  integral by identifying and integrating over the states $|\pureState_0\rangle$
  and $|\pureState_2\rangle$ between the replicas should allow one to calculate
  $\tr[(\unStateMat)^n]$, as described around \eqref{eq:rename}; since the
  quantum field values at $\entSurf$ should be integrated over in such traces,
  one does not expect their path integral representations to have any joint or
  Gibbons-Hawking-like terms at $\entSurf$.

  One can compare this with how, in the stitching process, the
  Gibbons-Hawking-like terms on the constant $\thTime$ surfaces included in
  \eqref{eq:statute} cancel between replicas or, more generally, reproduce the
  bulk contributions to the action localized in an infinitesimally thin shell around the $\thTime=0$
  surface $\entReg$; further, the joint terms at the spacetime boundary should
  similarly cancel between each other (once the normalization of the states
  $|\pureState_0\rangle$ and $|\pureState_2\rangle$ are accounted for --- see
  footnote \ref{foot:litigator}) or, more generally, reproduce contributions to
  $\LagDensityBdyMat$ localized on $\bdyReg=\entReg\cap\bdyMan$. If one were to
  apply a similar analysis to joint terms at or Gibbons-Hawking-like terms
  around $\entSurf$, one would find an extraneous leftover piece when attempting
  to evaluate $\tr[(\unStateMat)^n]$. \label{foot:backboard}} of a joint term at
the entangling surface, or equivalently, a Gibbons-Hawking-like term on an
infinitesimal surface
$\partial\neigh{0}{\entSurf}\cap\timeClosedInterval{\thTimeNelPatch{1}}{\thTimeNelPatch{2}}$
cutting off a zero-sized neighbourhood $\neigh{0}{\entSurf}$ of $\entSurf$ from
the remainder of the spacetime region
$\timeClosedInterval{\thTimeNelPatch{0}}{\thTimeNelPatch{2}}$. Instead, we shall
impose, as an additional boundary condition at $\entSurf$, that the set of
configurations for the quantum fields permitted in the path integral
\eqref{eq:county} be smooth and not conically-singular with respect to the
prescribed angle $\thTimeNelPatch{2}-\thTimeNelPatch{0}$ between the
$\thTimeNelPatch{0}$ and $\thTimeNelPatch{2}$ surfaces, \ie{}they behave near
$\entSurf$ as expected for smooth tensor fields restricted to a corner region
with opening angle $\thTimeNelPatch{2}-\thTimeNelPatch{0}$. \Eg{}graviton
fluctuations $\graviton$ are limited to those which preserve this proper
angle. 

We are now ready to proceed with our calculations to determine the
$\HamDensityMat{\thVec}$ and $\HamDensityBdyMat{\thVec}$ appearing in
\eqref{eq:county}. We will follow \cite{Wong:2013gua} closely, though we will
point out the terms missed by the authors there which led to the
misidentification of the modular Hamiltonian as the integral of stress tensor
components. The idea will be to differentiate the RHS of \eqref{eq:county} in
$\thTimeNelPatch{2}$ in order to identify
$\int_{\thTimeNelPatch{2}}\HamDensityMat{\thVec} +
\int_{\thTimeNelPatch{2}\cap\bdyMan} \HamDensityBdyMat{\thVec}$ as the resulting
extra insertion in
\begin{align}
  \begin{split}
    \MoveEqLeft[3]\partial_{\thTimeNelPatch{2}} \left\langle \pureState_2 \left|
    \timeOrder e^{ \int_{\thTimeNelPatch{0}}^{\thTimeNelPatch{2}} d\thTime
    \left( \int_\thTime \HamDensityMat{\thVec} + \int_{\thTime\cap\bdyMan}
    \HamDensityBdyMat{\thVec} \right) }
    \right| \pureState_0 \right\rangle \\
    =& \left\langle \pureState_2 \left| \timeOrder \left(
       \int_{\thTimeNelPatch{2}}\HamDensityMat{\thVec}
       +\int_{\thTimeNelPatch{2}\cap\bdyMan} \HamDensityBdyMat{\thVec}
       \right) \, e^{ \int_{\thTimeNelPatch{0}}^{\thTimeNelPatch{2}} d\thTime
       \left( \int_\thTime \HamDensityMat{\thVec} + \int_{\thTime\cap\bdyMan}
       \HamDensityBdyMat{\thVec} \right) } \right| \pureState_0
       \right\rangle
            \;.
  \end{split}
            \label{eq:impish}
\end{align}
We should emphasize in this manipulation that $\langle \pureState_2|$ should be
taken to be a fixed state (and does not evolve with $\thTimeNelPatch{2}$ as the
derivative is taken).

Now, by diffeomorphism invariance, differentiating the RHS of \eqref{eq:county}
in $\thTimeNelPatch{2}$, with fixed background fields $\metric$ and
$\bgFieldsOther$, is equivalent to fixing the coordinate interval
$\timeOpenInterval{\thTimeNelPatch{0}}{\thTimeNelPatch{2}}$ but applying an
infinitesimal diffeomorphism $\Lie{\NelVec}$ to the background fields which
drags more spacetime into the interval. In fact, in section \ref{sec:velcro}, we
constructed precisely such a diffeomorphism, generated by the vector field
$\NelVec$ defined in \eqref{eq:abamectine} interpolating between zero at
$\thTimeNelPatch{1}$, which we take to be between $\thTimeNelPatch{0}$ and
$\thTimeNelPatch{2}$, and $2\pi\thVec$ at $\thTimeNelPatch{2}$. This is
illustrated in Figure \ref{fig:e4e5f4exf4bc4}, though here wish to apply this
diffeomorphism everywhere between $\thTimeNelPatch{1}$ and $\thTimeNelPatch{2}$,
all the way up to $\entSurf$, and we are not assuming that the background fields
$\metric$ and $\bgFieldsOther$ are $\thVec$-symmetric. Thus, we are led to
consider the perturbation of background fields
\begin{align}
  \metric \to& \metric + \frac{\vary\thTimeNelPatch{2}}{2\pi} \Lie{\NelVec}\metric \;,
  &
    \graviton \to& \graviton + \frac{\vary\thTimeNelPatch{2}}{2\pi} \Lie{\NelVec}\graviton \;.
                   \label{eq:cytoplasm}
\end{align}
In order to mimic
$\thTimeNelPatch{2}\to\thTimeNelPatch{2}+\vary\thTimeNelPatch{2}$ using this
trick, we must also transform the path integration variables so that the quantum
fields satisfy appropriately perturbed boundary conditions around $\entSurf$
(recall the discussion in the paragraph preceding \eqref{eq:impish}). This is
can be achieved by
\begin{align}
  \graviton \to& \graviton + \testFunc\Lie{\NelVec}\graviton \;,
  &
    \qFieldsOther \to& \qFieldsOther + \testFunc\Lie{\NelVec}\qFieldsOther \;,
                       \label{eq:abiding}
\end{align} 
where now $\testFunc(\thTime)$ is a smooth
function over
  $\timeClosedInterval{\thTimeNelPatch{1}}{\thTimeNelPatch{2}}$
that vanishes at $\thTimeNelPatch{1}$ and
  $\thTimeNelPatch{2}$, but takes the small constant value
  $\vary\thTimeNelPatch{2}/(2\pi)$ almost everywhere in
  $\timeOpenInterval{\thTimeNelPatch{1}}{\thTimeNelPatch{2}}$. (We want to leave the
quantum field configurations on the $\thTimeNelPatch{0}$ and
$\thTimeNelPatch{2}$ surfaces fixed by $|\pureState_0\rangle$ and $\langle
\pureState_2 |$.)

Altogether, we then have
\begin{align}
  \begin{split}
    \MoveEqLeft[3] -2\pi\partial_{\thTimeNelPatch{2}}
    \int_{|\pureState_0\rangle}^{\langle \pureState_2 |} [d\graviton,
    d\qFieldsOther] e^{ -
    \actTimeClosedIntervalMat{\thTimeNelPatch{0}}{\thTimeNelPatch{2}} }
    \\
    =& \left\langle \pureState_2 \left| \timeOrder \, ( \modHamNelMat{\NelVec} +
       \modHamNelMatDiff{\NelVec} )\, e^{
       \int_{\thTimeNelPatch{0}}^{\thTimeNelPatch{2}} d\thTime \left(
       \int_\thTime \HamDensityMat{\thVec} + \int_{\thTime\cap\bdyMan}
       \HamDensityBdyMat{\thVec} \right) } \right| \pureState_0
       \right\rangle \;,
  \end{split}
       \label{eq:impurity} 
\end{align}
where, in analogy with \eqref{eq:francium} and \eqref{eq:untimed},
\begin{align}
  \begin{split}
    \modHamNelMat{\NelVec} 
    \equiv& \int_{\NelPatchPart{2}} \left(
            \eomDensityMat{\metric}
            \cdot \Lie{\NelVec} \metric + \eomDensity{\bgFieldsOther} \cdot
            \Lie{\NelVec} \bgFieldsOther \right) + \int_{\partial\NelPatchPart{2}}
            \sympPotMat[ \Lie{\NelVec} \metric,\Lie{\NelVec} \bgFieldsOther ]
    \\
          & + 2\pi\int_{\thTimeNelPatch{2}} (
            \varyWRT{\metric,\Lie{\thVec}\metric}
            + \varyWRT{\bgFieldsOther,\Lie{\thVec}\bgFieldsOther}
            ) \LagDensityGHMat
            +
            2\pi \int_{\thTimeNelPatch{2}\cap\bdyMan} (
            \varyWRT{\metric,\Lie{\thVec}\metric}
            + \varyWRT{\bgFieldsOther,\Lie{\thVec}\bgFieldsOther}
            ) \LagDensityJtMat
    \\
          & + \int_{\partial\NelPatchPart{2}\cap\bdyMan} (
            \varyWRT{\metric,\Lie{\NelVec}\metric}
            + \varyWRT{\bgFieldsOther,\Lie{\NelVec}\bgFieldsOther}
            ) \LagDensityBdyMat \;,
  \end{split}
            \label{eq:stimuli}
  \\
  \modHamNelDiff{\NelVec}
  \equiv&
          \int_{\NelPatchPart{2}} \left( 
          \eomDensity{\graviton} \cdot \Lie{\NelVec} \graviton
          + \eomDensity{\qFieldsOther} \cdot \Lie{\NelVec} \qFieldsOther
          \right)
          + \int_{\partial\NelPatchPart{2} \cap \partial\NelPatchPart{1}}
          \sympPot[\Lie{\NelVec}\graviton, \Lie{\NelVec}\qFieldsOther]
          \;,
          \label{eq:bubble}
\end{align}
and $\NelPatchPart{1}$ and $\NelPatchPart{2}$ are the spacetime region
illustrated in Figure \ref{fig:e4e5f4exf4bc4}, with a zero size $a\to 0$ for the
neighbourhood
$\NelPatchPart{1}=\neigh{0}{\entSurf}\cap\timeOpenInterval{\thTimeNelPatch{1}}{\thTimeNelPatch{2}}$
of $\entSurf$ in $\timeOpenInterval{\thTimeNelPatch{1}}{\thTimeNelPatch{2}}$.
Put differently, $\NelPatchPart{2}$ covers the thermal time interval
$\timeOpenInterval{\thTimeNelPatch{1}}{\thTimeNelPatch{2}}$, but we should be
mindful that its boundary $\partial\NelPatchPart{2}$ includes an infinitesimal
arc near $\entSurf$ which would be identified by
$\partial\NelPatchPart{2}\cap\partial\NelPatchPart{1}$ in Figure
\ref{fig:e4e5f4exf4bc4} --- we explain this in the following paragraph. The
interpretations of $\modHamNelMat{\NelVec}$ and $\modHamNelDiff{\NelVec}$ are
analogous to those of $\modHamNelS{\NelVec}$ and $\modHamNelSDiff{\NelVec}$
introduced in \eqref{eq:francium} and \eqref{eq:untimed} in section
\ref{sec:fluorine}: here, they implement the transformations
\eqref{eq:cytoplasm} and \eqref{eq:abiding} of the background and path
integration variables. (Strictly speaking, \eqref{eq:stimuli} and
\eqref{eq:bubble} should also have anomalous corrections arising from the
variations of the path integral measure in response to these transformations,
but we leave discussion of such terms to appendix \ref{sec:deflate}.)

Before, proceeding, we should comment on some differences between the
calculation here and that of section \ref{sec:velcro}. Recall there that we were
interested in evaluating the generalized entropy using our Callan-Wilczek
formula \eqref{eq:astatine} which required consideration of the deformed
replicated backgrounds
$(\metricR{\repS{\entSurf}n},\bgFieldsOtherR{\repS{\entSurf}n})$ containing
regulated conical singularities. To reproduce the variation of these backgrounds
from $n=1$, we transformed the background fields with $\Lie{\NelVec}$ in the
region we called $\NelPatchPart{2}$ outside a tiny neighbourhood
$\neigh{\neighSize}{\entSurf}$ of the entangling surface $\entSurf$; but within
the region
$\NelPatchPart{1}=\neigh{\neighSize}{\entSurf}\cap\timeOpenInterval{\thTimeNelPatch{1}}{\thTimeNelPatch{2}}$
inside the neighbourhood, we applied a transformation which interpolated to
doing nothing exactly at $\entSurf$. This is the situation illustrated in Figure
\ref{fig:e4e5f4exf4bc4}. Here, however, we simply wish to evaluate the effect of
the $\Lie{\NelVec}$ transformation on the background fields within the thermal
time interval $\timeOpenInterval{\thTimeNelPatch{1}}{\thTimeNelPatch{2}}$. The
action in this interval is the $a\to 0$ limit of the action over
$\NelPatchPart{2}$ and we see from the variation of the latter that we must be
careful to include symplectic potential $\sympPotMat$ terms on
$\partial\NelPatchPart{2}$, in particular, the arc
$\partial\NelPatchPart{2}\cap\partial\NelPatchPart{1}$ near $\entSurf$ which
becomes infinitesimal as $a\to 0$. Moreover, because the background fields
$(\metric,\bgFieldsOther)$ are no longer assumed to be $\thVec$-symmetric, the
$\sympPotMat[\Lie{\NelVec}\metric,\Lie{\NelVec}\bgFieldsOther]=\sympPotMat[\Lie{\thVec}\metric,\Lie{\thVec}\bgFieldsOther]$
term on the $\thTimeNelPatch{2}$ surface where $\NelVec=2\pi\thVec$ is now
nontrivial. In all, the $\sympPotMat$ boundary term of \eqref{eq:stimuli}
contains the following pieces\footnote{For simplicity, we have assumed that the
  foliation of the spacetime by constant time $\thTime$ slices is such that
  there are no finite contributions due to
  $\sympPotMat[\Lie{\NelVec}\metric,\Lie{\NelVec}\bgFieldsOther]$ integrated
  around the joints at the corners of $\partial\NelPatchPart{2}$ (see Figure
  \ref{fig:e4e5f4exf4bc4}). One expects this to be the case so long as the
  proper angles that constant time $\thTime$ slices make with the spacetime
  boundary $\bdyMan$ and the arc
  $\partial\NelPatchPart{2}\cap\partial\NelPatchPart{1}$, as measured in the
  background $(\metric,\bgFieldsOther)$, are invariant in time. (This should
  automatically be true for the joints bounding
  $\partial\NelPatchPart{2}\cap\partial\NelPatchPart{1}$, which should always
  have opening angle $\pi/2$.) Otherwise, one must carry such contributions
  along in the calculation, much like the $\LagDensityJt$ term in
  \eqref{eq:stimuli}.}:
\begin{align}
  \begin{split}
    \MoveEqLeft[3]\int_{\partial\NelPatchPart{2}} \sympPotMat[ \Lie{\NelVec}
    \metric,\Lie{\NelVec} \bgFieldsOther ]
    \\
    =& \int_{\partial\NelPatchPart{2}\cap\partial\NelPatchPart{1}} \sympPotMat[
       \Lie{\NelVec} \metric,\Lie{\NelVec} \bgFieldsOther ] +
       2\pi\int_{\thTimeNelPatch{2}} \sympPotMat[ \Lie{\thVec} \metric,\Lie{\thVec}
       \bgFieldsOther ] +\int_{\partial\NelPatchPart{2}\cap\bdyMan} \sympPotMat[
       \Lie{\NelVec} \metric,\Lie{\NelVec} \bgFieldsOther ] \;,
  \end{split}
       \label{eq:reprint}
\end{align} 
whereas only the analogue of the last piece appeared in \eqref{eq:francium}.
Obviously, there are also the Gibbons-Hawking-like and joint terms which
additionally appear in \eqref{eq:stimuli} due to the fixing of quantum field
configurations on the $\thTimeNelPatch{2}$ surface. In \cite{Wong:2013gua}, all
boundary terms are neglected, which as we will see, leads to the false
identification of the time translation generator $\HamDensityMat{\thVec}$ with
components of the stress tensor.

One can also compare the $\modHamNelDiff{\NelVec}$ introduced here in
\eqref{eq:bubble} to the $\modHamNelSDiff{\NelVec}$ introduced in
\eqref{eq:untimed} of section \ref{sec:kebab}. Just as
$\modHamNelSDiff{\NelVec}$ appeared as a byproduct of the way we modelled the
extension of the proper time period using a diffeomorphism on background fields,
$\modHamNelDiff{\NelVec}$ has appeared here resulting from the way we have
similarly modelled the variation of the final thermal time $\thTimeNelPatch{2}$.
In either case, the need to add these operators can be verified by the
consideration of their effect on other operators inserted in the thermal time
interval $\timeOpenInterval{\thTimeNelPatch{1}}{\thTimeNelPatch{2}}$, as
described for $\modHamNelSDiff{\NelVec}$ in section \ref{sec:kebab}. (We shall
not repeat the nearly identical analysis for $\modHamNelDiff{\NelVec}$ here.)
Indeed, the need to include $\modHamNelDiff{\NelVec}$ in our present calculation
\eqref{eq:impurity} is made even more obvious by the variation in boundary
conditions around $\entSurf$, as described around \eqref{eq:abiding} --- this is
captured by the $\sympPot[\Lie{\NelVec}\graviton,\Lie{\NelVec}\qFieldsOther]$
term included in \eqref{eq:bubble}. Note that, as in
\eqref{eq:untimed}, the symplectic potential and boundary Lagrangian variation
terms in response to \eqref{eq:abiding} vanish on the spacetime boundary
$\bdyMan$ owing to \eqref{eq:wiring}.
  Moreover, there are no terms on the $\thTimeNelPatch{1}$ and
  $\thTimeNelPatch{2}$ surfaces owing to \eqref{eq:sitcom}, nor are there contributions
  localized at the joints $\thTimeNelPatch{1}\cap\bdyMan$ and
  $\thTimeNelPatch{2}\cap\bdyMan$, given the description of the joint terms below
that equation.

Let us now proceed with our goal of extracting the instantaneous generator of
time evolution. Considering \eqref{eq:impish}, we see that we must extract the
generator $-\int_\thTime \HamDensityMat{\thVec} -
\int_{\thTime\cap\bdyMan}\HamDensityBdyMat{\thVec}$ for time translations as an
operator inserted to the left of the evolution operator, at the latest time
$\thTimeNelPatch{2}$. Thus, similar to our calculation in sections
\ref{sec:obliged} and \ref{sec:kebab}, we will now to take the
$\thTimeNelPatch{1}\to\thTimeNelPatch{2}^-$ limit\footnote{Actually, taking this
  limit is not necessary but simplifies the calculation. Just as for
  \eqref{eq:wizard}, the expression \eqref{eq:overfeed} we eventually obtain for
  $\modHamNelMat{\NelVec}+\modHamNelDiff{\NelVec}$ in fact continues to hold
  generally without the limit, as can be shown by a calculation analogous to the
  one mentioned in footnote \ref{foot:anguished} using additionally
  \eqref{eq:everyone} (restricting now to just objects constructed from the
  matter action).
} of \eqref{eq:impurity}, \eqref{eq:stimuli}, and \eqref{eq:bubble}. Using
\eqref{eq:feisty}, we obtain equations similar to \eqref{eq:guzzler} and
\eqref{eq:silliness}:
\begin{align}
  \begin{split}
    \lim_{\thTimeNelPatch{1}\to\thTimeNelPatch{2}^-} \modHamNelMat{\NelVec}
    =&
       -2\pi \int_{\thTimeNelPatch{2}} \thVec^a \left(
       \FaulknerCMatOf{\metric} +
       \FaulknerCOf{\bgFieldsOther}
       \right)_a
    \\
     &+ 2\pi \int_{\thTimeNelPatch{2}} (
       \varyWRT{\metric,\Lie{\thVec}\metric} +
       \varyWRT{\bgFieldsOther,\Lie{\thVec}\bgFieldsOther}
       ) \LagDensityGHMat +
       2\pi \int_{\thTimeNelPatch{2}\cap\bdyMan} (
       \varyWRT{\metric,\Lie{\thVec}\metric} +
       \varyWRT{\bgFieldsOther,\Lie{\thVec}\bgFieldsOther}
       ) \LagDensityJtMat
    \\
     & + \lim_{\thTimeNelPatch{1}\to\thTimeNelPatch{2}^-} \Bigg\{
       \int_{\partial\NelPatchPart{2}} \sympPotMat[ \Lie{\NelVec}\metric,
       \Lie{\NelVec} \bgFieldsOther ]
           + \int_{\partial\NelPatchPart{2}\cap\bdyMan} (
           \varyWRT{\metric,\Lie{\NelVec}\metric} +
           \varyWRT{\bgFieldsOther,\Lie{\NelVec}\bgFieldsOther}
           ) \LagDensityBdyMat
           \Bigg\} \;.
  \end{split}
           \label{eq:bok}
  \\
  \lim_{\thTimeNelPatch{1}\to\thTimeNelPatch{2}} \modHamNelDiff{\NelVec}
  =& -2\pi\int_{\thTimeNelPatch{2}}
     \thVec^a \left(  
     \FaulknerCOf{\graviton}
     + \FaulknerCOf{\qFieldsOther}
     \right)_a
     + \lim_{\thTimeNelPatch{1}\to\thTimeNelPatch{2}^-}
     \int_{\partial\NelPatchPart{2} \cap \partial\NelPatchPart{1}}
     \sympPot[
     \Lie{\NelVec} \graviton,
     \Lie{\NelVec} \qFieldsOther
     ]
     \;.
     \label{eq:quake}
\end{align}
If one ignores $\modHamNelDiff{\NelVec}$ as well as all terms in the latter
lines in \eqref{eq:bok}, then since the expectation value of the
$(\FaulknerCOf{\bgFieldsOther})_a$ vanishes as written in \eqref{eq:skewer}, one
would be tempted to identify the generator for time translations as
$-\HamDensityMat{\thVec}\eqDubious -\thVec_b \, (\FaulknerCMatOf{\metric})^b =
\volForm_a \thVec_b \stress^{ab}$ upon comparison of \eqref{eq:impish} with
\eqref{eq:impurity} and \eqref{eq:bok}. This is the conclusion that
\cite{Wong:2013gua} draws\footnote{One can also compare this with the
  often-written expression \eqref{eq:e4e6d4d5e5} for the thermal modular
  Hamiltonian in terms of the stress tensor. (Note that the stress tensor
  written there is the Lorentzian stress tensor, differing by a sign from the
  Euclidean stress tensor written here --- see \eqref{eq:marshy}.)}. But, of
course, $\modHamNelDiff{\NelVec}$ and the other terms in \eqref{eq:bok} cannot
simply be neglected --- \eg{} even if the action only consists of a bulk part
$\int\LagDensityMat$, with $\LagDensityGHMat$, $\LagDensityJtMat$, and
$\LagDensityBdyMat$ all vanishing, the symplectic potential $\sympPotMat$ terms
can still be nontrivial, particularly if there is non-minimal coupling between
matter and curvature.

To proceed, we next apply the definitions \eqref{eq:d4d5c4e6} and
\eqref{eq:lutetium} of Noether charge and current. It will be simplest now to
consider the sum of \eqref{eq:bok} and \eqref{eq:quake}, rather than treating
them separately as we did for $\modHamNelS{\NelVec}$ and
$\modHamNelSDiff{\NelVec}$ in sections \ref{sec:obliged} and \ref{sec:kebab}.
Rewriting \eqref{eq:bok} and \eqref{eq:quake} in terms of Noether current and
charge, we obtain
\begin{align}
  \begin{split}
    \lim_{\thTimeNelPatch{1}\to\thTimeNelPatch{2}^-} \left( \modHamNelMat{\NelVec}
    + \modHamNelDiff{\NelVec} \right)
    =& 2\pi \Bigg\{ \int_{\thTimeNelPatch{2}} \left(
       \sympPotMat[ \Lie{\thVec}\metric,\Lie{\thVec}\bgFieldsOther, ]
       -\currentMat{\thVec} + ( \varyWRT{\metric,\Lie{\thVec}\metric} +
       \varyWRT{\bgFieldsOther,\Lie{\thVec}\bgFieldsOther} ) \LagDensityGHMat
       \right)
    \\
     & + \int_{\thTimeNelPatch{2}\cap\bdyMan} (
       \varyWRT{\metric,\Lie{\thVec}\metric} +
       \varyWRT{\bgFieldsOther,\Lie{\thVec}\bgFieldsOther} ) \LagDensityJtMat +
       \int_{\partial\thTimeNelPatch{2}} \chargeMat{\thVec} \Bigg\}
    \\
     &+\lim_{\thTimeNelPatch{1}\to\thTimeNelPatch{2}^-} \Bigg\{
       \int_{\partial\NelPatchPart{2} \cap \partial\NelPatchPart{1}} \sympPotMat[
       \Lie{\NelVec} \metric, \Lie{\NelVec} \bgFieldsOther, \Lie{\NelVec}
       \graviton, \Lie{\NelVec} \qFieldsOther ]
    \\
     & + \int_{\partial\NelPatchPart{2}\cap\bdyMan} \left( \sympPotMat[
       \Lie{\NelVec} \metric, \Lie{\NelVec} \bgFieldsOther ] +(
       \varyWRT{\metric,\Lie{\NelVec}\metric} +
       \varyWRT{\bgFieldsOther,\Lie{\NelVec}\bgFieldsOther} ) \LagDensityBdyMat
       \right) \Bigg\} \;,
  \end{split}
       \label{eq:tummy}
\end{align}
where we have used \eqref{eq:reprint} to split up the $\sympPotMat$ term of
\eqref{eq:bok}, and $\partial\thTimeNelPatch{2}$ is shorthand for the boundaries
of the $\thTimeNelPatch{2}$ surface. Similar to the manipulations leading to
\eqref{eq:canal} in section \ref{sec:obliged}, one can use \eqref{eq:lilac} on
$\bdyMan$ and its analogue \eqref{eq:everyone}\footnote{Recall our conventions
  for defining the orientations of
  $\partial\NelPatchPart{1}\cap\partial\NelPatchPart{2}$ versus
  $\partial\NelPatchPart{2}\cap\partial\NelPatchPart{1}$ as stated after
  \eqref{eq:gallon}.} near $\entSurf$,
together with the identities \eqref{eq:wiring} and \eqref{eq:yarn} for the
variation of the boundary Lagrangian (all now applied to quantities obtained
just from the matter action), to greatly simplify the sum of the Noether charge
term with the latter two lines of \eqref{eq:tummy}. One thus obtains
\begin{align}
  \begin{split}
    \lim_{\thTimeNelPatch{1}\to\thTimeNelPatch{2}^-} \left( \modHamNelMat{\NelVec}
    + \modHamNelDiff{\NelVec} \right) =& 2\pi \Bigg\{
                                         \int_{\thTimeNelPatch{2}} \left(
                                         \sympPotMat[ \Lie{\thVec}\metric,\Lie{\thVec}\bgFieldsOther, ]
                                         -\currentMat{\thVec} + ( \varyWRT{\metric,\Lie{\thVec}\metric} +
                                         \varyWRT{\bgFieldsOther,\Lie{\thVec}\bgFieldsOther} ) \LagDensityGHMat
                                         \right)
    \\
                                       &+\int_{\thTimeNelPatch{2}\cap\bdyMan} \left( - \thVec\cdot\LagDensityBdyMat
                                         +( \varyWRT{\metric,\Lie{\thVec}\metric} +
                                         \varyWRT{\bgFieldsOther,\Lie{\thVec}\bgFieldsOther} ) \LagDensityJtMat
                                         \right) \Bigg\} \;.
  \end{split}
                                         \label{eq:overfeed}
\end{align}

Finally, comparing \eqref{eq:impish} with \eqref{eq:impurity} and using
\eqref{eq:overfeed}, we find that the instantaneous generator
$-\int_{\thTime}\HamDensityMat{\thVec} - \int_{\thTime\cap\bdyMan}
\HamDensityBdyMat{\thVec}$ for time evolution satisfying \eqref{eq:county} is
given by
\begin{align}
  \HamDensityMat{\thVec}
  =& \currentMat{\thVec}
     -\sympPotMat[ \Lie{\thVec}\metric,\Lie{\thVec}\bgFieldsOther, ]
     -(
     \varyWRT{\metric,\Lie{\thVec}\metric}
     + \varyWRT{\bgFieldsOther,\Lie{\thVec}\bgFieldsOther}
     ) \LagDensityGHMat
     \label{eq:snowfield}
  \\
  \HamDensityBdyMat{\thVec}
  =& \thVec\cdot\LagDensityBdyMat
     -(
     \varyWRT{\metric,\Lie{\thVec}\metric} +
     \varyWRT{\bgFieldsOther,\Lie{\thVec}\bgFieldsOther}
     ) \LagDensityJtMat \;.
     \label{eq:treble}
\end{align} 
As a consistency check, we see that when the background fields $\metric$ and
$\bgFieldsOther$ are $\thVec$-symmetric, $\HamDensityMat{\thVec}$ and
$\currentMat{\thVec}$ agree:
\begin{align}
      \HamDensityMat{\thVec}
      =& \currentMat{\thVec} \;,
  &
    \HamDensityBdyMat{\thVec}
    =& \thVec\cdot\LagDensityBdyMat \;,
  &
    (\Lie{\thVec}\metric=0,\, \Lie{\thVec}\bgFieldsOther=0)
\end{align} 
and indeed the evolution equations \eqref{eq:chewing} and \eqref{eq:county}
agree. The modular Hamiltonian for the thermal state prepared on such a
$\thVec$-symmetric background is given by the spatial integral $-\int_\entReg
\currentMat{\thVec}$ of the Noether current $\currentMat{\thVec}$ plus a
boundary term $-\int_{\bdyReg}\thVec\cdot\LagDensityBdyMat$, as written in
\eqref{eq:mendelevium}; more generally, for (unnormalized) states $\unStateMat=
e^{-2\pi \modHamMat}$ prepared by Euclidean path integration on backgrounds
which are not symmetric in time $\thTime$, the modular Hamiltonian $\modHamMat$
is given by
\begin{align}
  2\pi\modHamMat
  =& -\log \timeOrder e^{
     \int_0^{2\pi} d\thTime\;
     \left( 
     \int_\thTime \HamDensityMat{\thVec}
     + \int_{\thTime\cap\bdyMan} \HamDensityBdyMat{\thVec} 
     \right)
     } \;.
  &
    (\text{$\unStateMat$ prepared on $(\metric,\bgFieldsOther)$})
    \label{eq:barracuda}
\end{align}
with $\HamDensityMat{\thVec}$ and $\HamDensityBdyMat{\thVec}$ given by
\eqref{eq:snowfield} and \eqref{eq:treble}.

We end this appendix by commenting that, when moving between Euclidean and
Lorentzian signatures, the operators $\HamDensityMat{\thVec}$ and
$\HamDensityBdyMat{\thVec}$, as defined by \eqref{eq:snowfield} and
\eqref{eq:treble} in both signatures, is expected to acquire extra
signs.
This can be seen from \eqref{eq:risotto}, \eqref{eq:overbook} (with a similar
analytic continuation expected for the Lagrangian in each piece of the action
\eqref{eq:statute}), \eqref{eq:pummel}, and \eqref{eq:anxiety}:
\begin{align}
  \EucHamDensityMat{\Euc{\thVec}}
  =& -\LorHamDensityMat{\Lor{\thVec}} \;,
  &
    \EucHamDensityBdyMat{\Euc{\thVec}}
    =& - \LorHamDensityMat{\Lor{\thVec}}\;,
\end{align} 
where we are using the notation of section \ref{sec:albatross}. For instance,
\eqref{eq:county} reads in Lorentzian signature:
\begin{align}
  \left\langle \pureState_2 \left| 
  \timeOrder e^{
  -i \int_{\LorTime_0}^{\LorTime_2} d\LorTime
  \left(
  \int_\LorTime \LorHamDensityMat{\Lor{\thVec}}
  + \int_{\LorTime\cap\bdyMan} \LorHamDensityBdyMat{\Lor{\thVec}}
  \right)
  }
  \right| \pureState_0 \right\rangle
  =& \int_{|\pureState_0\rangle}^{\langle \pureState_2 |}
     [d\graviton, d\qFieldsOther] \,
     e^{
     i \, \LorActTimeClosedIntervalMat{\LorTime_{0}}{\LorTime_{2}}
     } \;.
     \label{eq:perenial}
\end{align} 
Alternatively, one can go through our derivation starting in Lorentzian
signature in the first place, finding that the $\LorHamDensityMat{\Lor{\thVec}}$
and $\LorHamDensityBdyMat{\Lor{\thVec}}$ satisfying \eqref{eq:perenial} are
given by \eqref{eq:snowfield} and \eqref{eq:treble}, with everything replaced by
their Lorentzian counterparts.


\section{Operator variation of the Noether current integral}
\label{sec:commodore}

Recall, as part of our derivation of the first law of generalized entropy, we
considered in section \ref{sec:bobtail} the operator variation $\int_{\entReg}
\varyWRT{\blkOp} \currentMat{\thVec}$ of the Noether current integral over the
initial surface $\entReg$. 
As we discuss around \eqref{eq:reexamine} to \eqref{eq:renewal}, evaluation of
the operator variation $\varyWRT{\blkOp}\currentMat{\thVec}$ is more subtle than
simply taking the functional variation
$\varyWRT{\metric}+\varyWRT{\bgFieldsOther}$ of \eqref{eq:d4d5c4e6} while fixing
in this expression the quantum fields $\graviton$ and $\qFieldsOther$ and their
time derivatives on $\entReg$. Instead, one must be careful to fix the conjugate
momenta of the quantum fields, rather than their time derivatives, as it is the
former which are truly operators on the initial surface $\entReg$.

As written in \eqref{eq:reexamine}, we found that the Noether current
$\currentMat{\thVec}$ can be split 
into a piece $\sympPotMat[\Lie{\thVec}\metric,\Lie{\thVec}\bgFieldsOther]$,
whose operator variation is easily calculated in \eqref{eq:commotion} due to the
$\thVec$-symmetry of the unperturbed background fields, and a piece
$\HamDensitySimpleMat{\thVec}$, given by \eqref{eq:dubbed}, whose spatial
integral $\int_{\entReg}\HamDensitySimpleMat{\thVec}$ resembles a Hamiltonian
with corresponding Lagrangian $\int_{\entReg}\thVec\cdot\LagDensityMat$. This
led us to speculate in \eqref{eq:renewal} that, up to boundary terms, the
operator variation of the Hamiltonian
$\int_{\entReg}\HamDensitySimpleMat{\thVec}$, with the conjugate momenta of the
quantum fields $\graviton$ and $\qFieldsOther$ fixed, is given simply by the
functional variation $\varyWRT{\metric}+\varyWRT{\bgFieldsOther}$ of the
Lagrangian $\int_{\entReg}\thVec\cdot\LagDensityMat$, with the time derivatives
of the quantum fields fixed.
In this appendix, we point out that the argument for \eqref{eq:renewal},
in the case where the unperturbed background is stationary
(\ie{}$\Lie{\thVec}\metric=0$ and
$\Lie{\thVec}\bgFieldsOther=0$) 
can be made precise from the (Lorentzian versions of the) results
\eqref{eq:snowfield} and \eqref{eq:treble}, together with
the evolution equation \eqref{eq:perenial}, of appendix \ref{sec:immorally} --- indeed notice that
the $\HamDensitySimpleMat{\thVec}$ defined in \eqref{eq:dubbed} and the boundary
term appearing in \eqref{eq:renewal} match the expressions of
$\HamDensityMat{\thVec}$ and $\HamDensityBdyMat{\thVec}$ found in
\eqref{eq:snowfield} and \eqref{eq:treble} for the instantaneous generator of
time evolution, modulo Gibbons-Hawking-like and joint terms.

Let us start by considering \eqref{eq:perenial} in the general case without any
assumption of symmetry for the background fields. Notice on the LHS of
\eqref{eq:perenial} that $\HamDensityMat{\thVec}$ and
$\HamDensityBdyMat{\thVec}$ should be interpreted as operators constructed from
the quantum fields and their conjugate momenta. In contrast, on the RHS, the
action $\actTimeClosedIntervalMat{\LorTime_0}{\LorTime_2}$ (whose components are
given in \eqref{eq:statute}) is written in terms of their time derivatives.
Thus, taking the variation of \eqref{eq:perenial} with respect to the background
fields $\metric$ and $\bgFieldsOther$ gives
\begin{align}
  \begin{split}
    \MoveEqLeft[3]
    \left\langle \pureState_2 \left| 
    \timeOrder
    \int_{\LorTime_0}^{\LorTime_2} d\LorTime \,
    \varyWRT{\blkOp}
    \left(
    \int_\LorTime \HamDensityMat{\thVec}
    + \int_{\LorTime\cap\bdyMan} \HamDensityBdyMat{\thVec}
    \right)
    e^{
    -i \int_{\LorTime_0}^{\LorTime_2} d\LorTime
    \left(
    \int_\LorTime \HamDensityMat{\thVec}
    + \int_{\LorTime\cap\bdyMan} \HamDensityBdyMat{\thVec}
    \right)
    }
    \right| \pureState_0 \right\rangle
    \\
    =& -\int_{|\pureState_0\rangle}^{\langle \pureState_2 |}
       [d\graviton, d\qFieldsOther] \,
       e^{
       i \actTimeClosedIntervalMat{\LorTime_{0}}{\LorTime_{2}}
       }
       \,
       (\varyWRT{\metric}+\varyWRT{\bgFieldsOther})
       \actTimeClosedIntervalMat{\LorTime_0}{\LorTime_2}
       \;.
  \end{split}
            \label{eq:bullpen}
\end{align}
(Actually, the RHS should also receive contributions from the variation of the
path integral measure; we leave consideration of such anomalous contributions to
appendix \ref{sec:deflate}.) Dividing by $\LorTime_2-\LorTime_0$ and taking the $\LorTime_0\to 0$ and
$\LorTime_2 \to 0$ limit gives
\begin{align}
  \varyWRT{\blkOp} 
  \left(
  \int_{\entReg} \HamDensityMat{\thVec}
  + \int_{\bdyReg} \HamDensityBdyMat{\thVec}
  \right)
  =&
     -
     \lim_{\LorTime_0,\LorTime_2\to 0}
  \frac{1}{\LorTime_2-\LorTime_0}
  (\varyWRT{\metric}+\varyWRT{\bgFieldsOther})
  \actTimeClosedIntervalMat{\LorTime_0}{\LorTime_2} \;,
  \label{eq:astonish}
\end{align} 
where, upon recalling that $|\pureState_0\rangle$ and $|\pureState_2\rangle$ can
be any pure states, we have written \eqref{eq:astonish} as an operator equation.
From the expression \eqref{eq:statute} for
$\actTimeClosedIntervalMat{\LorTime_0}{\LorTime_2}$, we see that the limit on
the RHS of \eqref{eq:astonish} reduces it to integrals on the initial time
slices $\entReg$ and $\bdyReg$ in the bulk and spacetime boundary\footnote{Note
  that the sign on the $\LagDensityBdyMat$ term arises from our conventions for
  orienting $\partial\timeClosedInterval{\LorTime_1}{\LorTime_2}\cap\bdyMan$ and
  $\bdyReg$ as stated around \eqref{eq:gallon} and \eqref{eq:defame}.}:
\begin{align}
     \lim_{\LorTime_0,\LorTime_2\to 0}
  \frac{\actTimeClosedIntervalMat{\LorTime_0}{\LorTime_2}}{\LorTime_2-\LorTime_0}
  =& \int_{\entReg} \left(
     \thVec\cdot\LagDensityMat + \Lie{\thVec}\LagDensityGHMat
     \right)
     + \int_{\bdyReg} \left(
     -\thVec\cdot\LagDensityBdyMat
     + \Lie{\thVec}\LagDensityJtMat 
     \right)
     \label{eq:rising}
\end{align}
Notice that \eqref{eq:astonish} with \eqref{eq:rising} nearly gives us the
equality \eqref{eq:renewal} we are after --- what remains is to treat the
Gibbons-Hawking-like and joint terms in \eqref{eq:rising} and
$\HamDensityMat{\thVec}$ and $\HamDensityBdyMat{\thVec}$ in \eqref{eq:snowfield}
and \eqref{eq:treble}. To do so, we now specialize to the case where the
unperturbed background fields and path integral (with which the expectation
value in \eqref{eq:renewal} is computed) are stationary:
\begin{align}
  \left\langle
  \varyWRT{\blkOp}
  (\varyWRT{\metric,\Lie{\thVec}\metric}+\varyWRT{\bgFieldsOther,\Lie{\thVec}\bgFieldsOther})
  \LagDensityGHMat
  \right\rangle
  =& 
     \left\langle
     (\varyWRT{\metric,\Lie{\thVec}\vary\metric}+\varyWRT{\bgFieldsOther,\Lie{\thVec}\vary\bgFieldsOther})
     \LagDensityGHMat
     \right\rangle
     =\left\langle
     (\varyWRT{\metric}+\varyWRT{\bgFieldsOther})
     \Lie{\thVec}\LagDensityGHMat
     \right\rangle
  \\
  \left\langle
  \varyWRT{\blkOp}
  (\varyWRT{\metric,\Lie{\thVec}\metric}+\varyWRT{\bgFieldsOther,\Lie{\thVec}\bgFieldsOther})
  \LagDensityJtMat
  \right\rangle 
  =& 
     \left\langle
     (\varyWRT{\metric,\Lie{\thVec}\vary\metric}+\varyWRT{\bgFieldsOther,\Lie{\thVec}\vary\bgFieldsOther})
     \LagDensityJtMat
     \right\rangle
     =\left\langle
     (\varyWRT{\metric}+\varyWRT{\bgFieldsOther})
     \Lie{\thVec}\LagDensityJtMat
     \right\rangle \;.
\end{align}
Thus, when the expectation value of \eqref{eq:astonish} is taken with a path
integral over a $\thVec$-symmetric background, the operator variations of the
Gibbons-Hawking-like and joint terms of the $\HamDensityMat{\thVec}$ and
$\HamDensityBdyMat{\thVec}$ (given by \eqref{eq:snowfield} and
\eqref{eq:treble}) on the LHS cancel the functional variations of the
corresponding terms on the RHS (given by \eqref{eq:rising}). Cancelling these
terms, we immediately recover \eqref{eq:renewal}.


\section{Maxwell contributions to asymptotic quantities}
\label{sec:garnish}

The point of this appendix will be to consider how quantum matter can contribute
nontrivially to the asymptotic energy \eqref{eq:cymbal} appearing in the first
law \eqref{eq:calamity} of generalized entropy. We will consider the
particularly simple but instructive example of a minimally coupled Maxwell field
in asymptotically AdS spacetime. In the end, we will demonstrate how the Maxwell
contributions to the Noether charge and boundary Lagrangian terms of the
asymptotic energy \eqref{eq:cymbal} are important for recovering, from the
generalized first law \eqref{eq:calamity}, the expected semiclassical
thermodynamics of electrically charged black holes at fixed potential and fixed
charge.

The Maxwell field is described by a (Lorentzian) bulk action
\begin{align}
  \int \LagDensityMax
  =& -\frac{1}{4} \int \fieldStrMax^{ab} \fieldStrMax_{ab} \volForm \;,
  &
    \formStrMax
    =& d\formMax
       \;.
       \label{eq:taunt}
\end{align}
It is straightforward to show that this Maxwell action leads to the extra
contributions
\begin{align}
  \sympPotMax[\vary\fieldMax]
  =& \vary\fieldMax_a\fieldStrMax^{ab} \volForm_b
     \;,
  &
    \chargeMax{\egVec}
    =& -\frac{1}{2} \egVec^c \fieldMax_c \fieldStrMax^{ab} \volForm_{ab}
       \label{eq:blip}
\end{align}
to the symplectic potential and Noether charge associated to an arbitrary vector
field $\egVec$. Another useful quantity for our discussion is the electric
charge
\begin{align}
  \totalChargeElec =& \int_{\LorTime\cap\bdyMan} \chargeElec
  &
    \chargeElec =& \frac{1}{2}\left( \fieldStrMax^{ab} \volForm_{ab}\right)_{\LorTime\cap\bdyMan}
                   \label{eq:drapery}
\end{align}
computed from Gauss's law evaluated on a time slice (denoted with the shorthand
$\LorTime\cap\bdyMan$) at the spacetime boundary $\bdyMan$. From the CFT
perspective, $\chargeElec$ gives the charge density of the current to which the
bulk Maxwell field is dual \cite{Marolf:2006nd}.

The asymptotics of the background geometry and of the Maxwell field are known in
an asymptotically-AdS spacetime. In asymptotically AdS spacetimes, the leading
behaviour of metric and volume form components near the AdS boundary is given by
\begin{align}
  \metric_{ab} dx^a dx^b
  \sim& \frac{dr^2}{r^2} + r^2 \metricCFT_{\mu \nu} dy^\mu dy^\nu \;,
  &
    \volComp_{r\mu_1\cdots\mu_d}
    \sim& r^{d-1} \;,
  &
    (r\to\infty)
    \label{eq:overcoat}
\end{align}
where $x^a=(r,y^\mu)$ and $y^\mu$ are respectively coordinates of the bulk and
AdS boundary at $r=\infty$, and $\metricCFT_{\mu\nu}$ is the (finite) CFT
metric. When studying the Maxwell contributions to asymptotic quantities, we
will find that expressions only depend on the leading asymptotic behaviour of
the metric and volume form \eqref{eq:overcoat}, so we need not specify the
geometry further. (If one so desires, for concreteness, one can fix the
background geometry to that of the classical solution for an electrically
charged black hole in the following discussions.)

It has been shown \cite{Marolf:2006nd} that the allowed asymptotics for the
Maxwell field in an asymptotically AdS spacetime is given by\footnote{Note that
  $d\fieldMaxGauge$ may not necessarily be produced by a local gauge
  transformation. This is evident in the electrically charged black hole example
  to be considered shortly --- analytically continuing to Euclidean signature,
  different values of $\potElecBH$ in \eqref{eq:strep} give different values for
  the Wilson loop running around the thermal time $\thTime=i\LorTime$ circle.}
\begin{align}
  \formMax
  \sim& d\fieldMaxGauge + \formMaxP + \formMaxM
        \;,
  &
    (r\to\infty)
\end{align}
where
\begin{align}
  \fieldMaxP_\mu
  \sim& O(r^{2-d}) \;,
  &
    \fieldMaxM_\mu
    \sim& O(r^0) \;,
  \\
  \fieldMaxP_r
  \sim& O(r^{1-d}) \;,
  &
    \fieldMaxM_r
    \sim& \begin{cases}
      O(r^{-3}) & d\ne 4 \\
      O(r^{-3}\log r) & d=4
    \end{cases} \;,
                        \label{eq:banister}
\end{align}
and correspondingly
\begin{align}
  \formStrMax
  \sim& \formStrMaxP+\formStrMaxM \;,
  &
    (r\to\infty)
\end{align}
with
\begin{align}
  (\fieldStrMaxP)^{r\mu}
  \sim& 
        O(r^{1-d}) 
        \;,
  &
    (\fieldStrMaxM)^{r\mu}
    \sim& \begin{cases}
      O(r^{-3}) & d\ne 4 \\
      O(r^{-3}\log r) & d=4
    \end{cases} \;.
\end{align}
As noted in \cite{Witten:2003ya,Marolf:2006nd}, the choice of boundary
conditions at the AdS boundary is not necessarily unique for the Maxwell field,
\eg: in $d\ge 3$, one may choose to fix $\formMax|_{\bdyMan}$
at\footnote{Recall, as specified in footnote \ref{foot:kinfolk}, this notation
  means also projection onto $\bdyMan$, unless indices are otherwise specified.}
the AdS boundary $\bdyMan$ (\ie{}fix $d\fieldMaxGauge|_{\bdyMan}$ and
$\formMaxM|_{\bdyMan}$), and consider field fluctuations falling off like
$\formMaxP$; in $d\le 3$, one can instead fix
$r^{d-1}\fieldStrMax^{r\mu}|_{\bdyMan}$ (\ie{}fix $\formMaxP$) while considering
fluctuations with the asymptotics of $d\fieldMaxGauge+\formMaxM$. Let us
decompose the gauge field $\formMax$ into a background part $\bgFormMax$ and a
fluctuating quantum part $\qFormMax$
\begin{align}
  \formMax
  =& \bgFormMax + \qFormMax \;,
  &
    (\bgFieldsOther_a=\bgFieldMax_a, \qFieldsOther_a=\qFieldMax_a)
\end{align}
where, in the more general notation used elsewhere in this paper,
$\bgFieldMax_a$ and $\qFieldMax_a$ would have been denoted as $\bgFieldsOther_a$
and $\qFieldsOther_a$. When $\formMax|_{\bdyMan}$ is fixed in $d\ge 3$, the
quantum fluctuations $\qFormMax$ fall off like
\begin{align}
  \qFormMax
  \sim& \qFormMaxP\;,
  &(\text{fixed $\formMax|_{\bdyMan}$ \bc}, r\to\infty)
    \label{eq:jubilance}
\end{align}
while, when $r^{d-1}\fieldStrMax^{r\mu}|_{\bdyMan}$ is fixed in $d\le 3$, they
behave like
\begin{align}
  \qFormMax
  \sim& d\qFieldMaxGauge + \qFormMaxM \;.
  &
    (\text{fixed $r^{d-1}\fieldStrMax^{r\mu}|_{\bdyMan}$ \bc}, r\to\infty)
    \label{eq:dock}
\end{align}
In $d=3$, various hybrid boundary conditions are possible, one choice of which
we will consider later. Restrictions must also be placed on the asymptotics of
$\fieldMaxGauge$, but all we will need is the condition that $\fieldMaxGauge$
remain finite at the asymptotic boundary $\bdyMan$ in $d=3$, as we later
consider fluctuations in $\fieldMaxGauge$. (Our discussion of boundary
conditions is not comprehensive; yet more possibilities can be found in
\cite{Marolf:2006nd}.)

It will be instructive to consider the asymptotic energy involved in the first
law of generalized entropy for electrically charged black holes. For fixed
boundary conditions, one can always absorb ambiguities in the choice of Maxwell
field background $\bgFormMax$ into shifts of the quantum field $\qFormMax$. So,
without loss of generality, let us fix $\bgFormMax=\formMaxBH$ to be the
classical solution for an electrically charged (possibly rotating) black hole
--- see, \eg{}\cite{Chamblin:1999tk,Caldarelli:1999xj} --- which behaves
asymptotically as
\begin{align}
  \formMaxBH \sim& d\fieldMaxGaugeBH+\formMaxBHP \;,
  &
    \formStrMaxBH
    \sim& \formStrMaxBHP \;,
  &
    \fieldMaxGaugeBH =& \potElecBH \LorTime \;,
  &
    (r\to\infty)
    \label{eq:strep}
\end{align}
where, $\LorTime$ is the time coordinate that is periodically identified
$\LorTime\sim\LorTime-2\pi i$ in the imaginary direction\footnote{In this paper,
  we have always taken the Euclidean time coordinate $\thTime=i\LorTime$ to be
  periodic, rescaling as needed so that the period (\ie{}inverse temperature) is
  $2\pi$. Since the Killing vector $\partial_{\LorTime}=-i\partial_{\thTime}$ of
  a thermal setup vanishes on the bifurcation surface $\entSurf$,
  $\partial_{\LorTime}$ must be tangent to the null geodesic generators of the
  horizon emanating from $\entSurf$. For a rotating black hole, this means that
  we have chosen somewhat nonstandard angular coordinates which rotate along
  with the black hole horizon. Thus, what we have been calling the asymptotic
  ``energy'' $\HamBdy{\thVec}$, associated with our
  $\thVec=\partial_{\LorTime}$, would also include what is called angular
  momentum in more standard language. (See the discussion around (90)-(91) in
  \cite{Iyer:1994ys}, whose $\thVec$ is proportional to ours.) Note however,
  that $\LorTime$ is still (at least proportional to) the standard time
  coordinate and the one-form $d\LorTime$, \eg{}appearing in $\formMaxBH$ in
  \eqref{eq:strep}, is unambiguous under time-dependent reparametrizations of
  other coordinates.}, and the constant $\potElecBH$ gives the electric
potential of the black hole\footnote{Specifically, $\potElecBH$ should be
  thought of as the difference in electric potential between the black hole
  horizon and the AdS boundary. By regularity of $\formMaxBH$,
  $\thVec\cdot\formMaxBH$ vanishes at the bifurcation surface, so $\potElecBH$
  is given by minus the electric potential at the AdS boundary. (Note that our
  convention is always to take $\formMax$ to be smooth in the bulk, whereas some
  references, \eg{}\cite{Caldarelli:1999xj} sometimes exclude $\potElecBH
  d\LorTime$ and keep only the inexact part of $\formMax$, which by itself would
  be singular at bifurcation surface.)}. 
When the background $\bgFormMax$ has the asymptotics of \eqref{eq:strep}, we
will refer to \eqref{eq:jubilance} as fixed potential boundary conditions,
since, where applicable ($d\ge 3$), they fix the electric potential to its
background value in \eqref{eq:strep}:
\begin{align}
  \formMax|_{\bdyMan}
  =& \potElecBH d\LorTime
  &
    (\text{fixed $\potElec$ \bc})
\end{align} 
as quantum fluctuations \eqref{eq:jubilance} in $\formMax$ are suppressed near
the AdS boundary. As hinted below \eqref{eq:dock}, in $d=3$ another interesting
boundary condition to consider is given by a hybrid of \eqref{eq:jubilance} and
\eqref{eq:dock}, where the electric charge density $\chargeElec$ defined in
\eqref{eq:drapery} is fixed but the electric potential
$\potElec(y^\mu)=\potElecBH+\qPotElec(y^\mu)$ is allowed to
fluctuate\footnote{Alternatively, one might fix the total charge
  $\totalChargeElec$ given in \eqref{eq:drapery} but allow the electric
  potential $\potElec(\LorTime)$ to fluctuate only as a function of time:
  \begin{align}
    \qFormMax
    \sim& \qFormMaxP + \qPotElec(\LorTime) d\LorTime \;,
    &
      \qTotalChargeElec
      =& \frac{1}{2}\int_{\LorTime\cap\bdyMan} (\qFieldStrMaxP)^{a b} \volForm_{a b}
         = 0 \;.
    &
      (\text{fixed $\totalChargeElec$ \bc}, r\to\infty)
      \label{eq:rosy}
  \end{align}
  (Note, however, that this boundary condition is not local in space.)
  \label{foot:coaster}
}:
\begin{align}
  \qFormMax
  \sim& \qFormMaxP + \qPotElec(y^\mu) d\LorTime \;,
  &
    \qChargeElec
    =& \frac{1}{2} \left[ (\qFieldStrMaxP)^{a b} \volForm_{a b}  \right]_{\LorTime\cap\bdyMan}
       = 0 \;.
  &
    (\text{fixed $\chargeElec$ \bc}, r\to\infty)
    \label{eq:grafted}
\end{align} 
We shall refer to \eqref{eq:grafted} as fixed charge density boundary conditions
for the obvious reason that the charge density is fixed to its background value:
\begin{align}
  \chargeElec
  =& \chargeElecBH
     = \frac{1}{2} \left[ (\fieldStrMaxBHP)^{ab}\volForm_{ab}  \right]_{\LorTime\cap\bdyMan}
     \;.
  &
    (\text{fixed $\chargeElec$ \bc})
\end{align} 

Let us consider fixed potential boundary conditions first. It follows, from
examination of the scaling \eqref{eq:overcoat} of asymptotic AdS and the
scalings \eqref{eq:banister} of the Maxwell background \eqref{eq:strep} and
permitted field variations \eqref{eq:jubilance}, that the Maxwell symplectic
potential \eqref{eq:blip} vanishes when evaluated on the AdS boundary $\bdyMan$
(in the admissible dimensions $d\ge 3$):
\begin{align}
  \sympPotMax[\vary\fieldMax] |_{\bdyMan}
  =& 0 \;,
  &
    \LagDensityBdyMax
    =& 0 \;
  &
    (\text{fixed potential \bc})
\end{align}
and consequently, no extra boundary Lagrangian $\LagDensityBdyMax$ must be added
to satisfy \eqref{eq:wiring}. Thus, the Maxwell contribution to the asymptotic
energy \eqref{eq:cymbal} can be written as
\begin{align}
  \HamBdyMax{\thVec}
  =& \int_{\bdyReg} \chargeMax{\thVec}
     = -\potElecBH \totalChargeElec \;,
  &
    (\text{fixed potential \bc})
    \label{eq:unpiloted}
        \;,
\end{align}
where, in the second equality, we have extracted the nonvanishing part of the
Noether charge \eqref{eq:blip} (given the asymptotics of the metric background
\eqref{eq:overcoat}, the Maxwell background \eqref{eq:strep}, and quantum
fluctuations \eqref{eq:jubilance}) and identified the electric charge
$\totalChargeElec$ defined by \eqref{eq:drapery}. Note that, in the last
expression of \eqref{eq:unpiloted}, only $\totalChargeElec$ depends on the
quantum field $\qFormMax$. Thus, for fixed potential boundary conditions, the
first law of generalized entropy \eqref{eq:calamity} for perturbations from (the
Hartle-Hawking state of) an electrically charged black hole receives the extra
contribution
\begin{align}
  \vary\langle \HamBdyMax{\thVec} \rangle
  =& -\potElecBH \vary\langle \totalChargeElec \rangle \;
     \label{eq:manifesto}
\end{align}
from the Maxwell field. We see that including the Maxwell field in the Noether
charge part of the asymptotic energy is crucial for recovering the familiar
thermodynamics of a fixed potential (\ie{}grand canonical) ensemble.

Finally, let us consider the fixed charge density boundary conditions
\eqref{eq:grafted}\footnote{For the fixed total charge boundary conditions
  described by \eqref{eq:rosy} in footnote \ref{foot:coaster}, the discussion is
  essentially the same. Since those boundary conditions are non-local in space,
  one should consider the correspondingly non-local analogue \eqref{eq:viral},
  written in footnote \ref{foot:mosaic}, of the condition \eqref{eq:wiring}. The
  Maxwell part of the more general condition \eqref{eq:viral} is satisfied by
  again the choice of boundary Lagrangian $\LagDensityBdyMax$ written in
  \eqref{eq:swiftness}. Once various quantities are integrated over the boundary
  spatial directions, the discussion in the text goes through just as well for
  fixed total charge boundary conditions \eqref{eq:rosy}.} in $d= 3$. Now, the
Maxwell symplectic potential \eqref{eq:blip} no longer vanishes at the AdS
boundary $\bdyMan$:
\begin{align}
  \sympPotMax[\vary\fieldMax]|_{\bdyMan}
  =& \vary\potElec\, (\fieldStrMaxBHP)^{t r}\volForm_r \;,
  &
    (\text{fixed $\chargeElec$ \bc})
    \label{eq:quote}
\end{align}
so we must supplement the bulk Maxwell action \eqref{eq:taunt} with a boundary
term \cite{Marolf:2006nd}
\begin{align}
  \int_{\bdyMan} \LagDensityBdyMax
  = -\int_{\bdyMan} \fieldMax_a \fieldStrMax^{a b} \volForm_b
  =& -\int_{\bdyMan} \potElec\, (\fieldStrMaxBHP)^{\LorTime r} \volForm_r \;,
  & (\text{fixed $\chargeElec$ \bc})
    \label{eq:swiftness}
\end{align}
so that \eqref{eq:wiring} is satisfied. Note that appending this term to the
action and integrating over the electric potential $\qPotElec$ introduced in
\eqref{eq:grafted} is tantamount to performing a Laplace transform on the bulk
path integral\footnote{See \cite{Witten:2003ya} for a general discussion about
  switching between the fixed $\formMax|_{\bdyMan}$ and fixed
  $r^{d-1}\fieldStrMax^{r\mu}|_{\bdyMan}$ boundary conditions described by
  \eqref{eq:jubilance} and \eqref{eq:dock}, including the interpretation from
  the CFT perspective. Note that what they call a ``Legendre transform'' is
  actually a Laplace transform. (When describing thermodynamics, physicists are
  often not careful in distinguishing Legendre and Laplace transforms as the two
  become equivalent upon taking the saddle-point approximation of partition
  functions.)}. As we will now describe, switching between fixed potential and
fixed charge density boundary conditions corresponds precisely to the analogous
Laplace transform between grand canonical and canonical thermodynamics.
Extracting the finite part of the Noether charge \eqref{eq:blip} appearing in
the asymptotic energy \eqref{eq:cymbal}, one finds
\begin{align}
  \chargeMax{\thVec}|_{\bdyReg}
  =& -\potElec\, (\fieldStrMaxBHP)^{\LorTime r} \volForm_{\LorTime r} \;.
  & (\text{fixed $\chargeElec$ \bc})
    \label{eq:matted}
\end{align} 
Comparing \eqref{eq:matted} to \eqref{eq:swiftness}, we thus find that the
Maxwell contribution to the asymptotic energy \eqref{eq:cymbal} now vanishes
\begin{align}
  \HamBdyMax{\thVec}
  =& 0 \;.
  & (\text{fixed $\chargeElec$ \bc})
    \label{eq:pledge}
\end{align}
So, for fixed charge density boundary conditions, the Maxwell field does not
contribute an extra term to the asymptotic energy appearing in the generalized
first law \eqref{eq:calamity}. It is, of course, unsurprising from the
thermodynamics standpoint that no term like \eqref{eq:manifesto} appears in the
first law, as we are working with fixed charge (\ie{}with boundary conditions
\eqref{eq:grafted}, the black hole is now more akin to a system in the canonical
rather than grand canonical ensemble). Note that, to see this, it was important
to consider both the Noether charge and boundary Lagrangian parts of asymptotic
energy \eqref{eq:cymbal}, as the Maxwell contribution to each part, by itself,
would be non-vanishing.



\section{UV-divergences and anomalies}
\label{sec:deflate}
In this appendix, we shall describe how various `anomalous' corrections, resulting
from variations of the path integral measure, fit into the narrative of this
paper. We will open, in appendix \ref{sec:moonrise}, with a brief discussion of the renormalization of UV
divergences in the effective action and generalized entropy. This will naturally
lead us to consider anomalous terms resulting from background field variations.
In appendix \ref{sec:affix}, we discuss such variations together with anomalous changes of path integration
variables. There, we establish notation for and define the anomalous
contributions to various quantities encountered throughout this paper, \eg{}equations of
motion, the symplectic potential, the stress tensor, Wald-Dong entropy, and
Noether current and charge. With relevant notation established, we shall then catalogue all the
appearances of possible anomalous corrections to the calculations of this paper:
the derivations of the modular Hamiltonian from section \ref{sec:fluorine} and
the first law of generalized entropy from section \ref{sec:neon} are
covered in appendices \ref{sec:campfire} and \ref{sec:dexterous} respectively.
Ultimately, we shall argue that, while the Noether charge and modular
Hamiltonian discussed in section \ref{sec:fluorine} can generally receive
anomalous contributions, the relationship between the first law of generalized
entropy and the effective gravitational equations of motion derived in section
\ref{sec:neon} continues to hold with the appropriately corrected quantities.

\subsection{UV-divergences of the effective action}
\label{sec:moonrise}
Recall, for the purposes of evaluating the quantum gravity effective action to
one-loop-order in $\GNewton$, the graviton can be treated in a semiclassical
theory as just another matter field propagating on a classical background
geometry $\metric$ --- see appendix \ref{sec:d4f5g3nf6bg2e6nf3be7o-oo-oc4d5}. As
reviewed in chapter 6 of \cite{Birrell:1982ix}, the bare matter part of the
effective action in semiclassical theories
\begin{align}
  \eActMixMatBare[\metric]
  =& -\log \int [d\graviton\, d\qFieldsOther]_{\bare}
     \, e^{-\actMat[\metric,\graviton,\bgFieldsOther+\qFieldsOther]} 
\end{align}
typically contains UV divergences, even in the simplest case of free quantum
matter fields. These divergences are inherited, for example, by the bare stress
tensor
\begin{align}
  2\frac{\delta \eActMixMatBare[\metric]}{\delta \metric_{ab}}
  =& \langle \stressBareDensity^{ab} \rangle
     = \langle \stressBare^{ab}  \rangle\volForm \;.
  &
  &(\text{$\vary\metric$ away from $\bdyMan$})
    \label{eq:diffusive}
\end{align}
However, in semiclassical theories, the divergences of
$\eActMixMatBare[\metric]$ can be extracted as a local action
$\actMatDiver[\metric]$ containing divergent coefficients. Thus, the complete
effective action
\begin{align}
  \eActMix[\metric]
  =& \actGravBare[\metric] + \eActMixMatBare[\metric]
\end{align}
can nonetheless be rendered UV-finite if $\actGravBare[\metric]$ contains
similarly divergent terms which cancel against $\actMatDiver[\metric]$. Then, we
see that generalized entropy $\entGen{\entSurf}[\metric]$ defined in
\eqref{eq:solitude}, also expressible in the Callan-Wilczek form
\eqref{eq:astatine} in terms of the effective action, must similarly be
UV-finite\footnote{See, \eg{}\cite{Solodukhin:1995ak,Cooperman:2013iqr}.
  However, unlike what is suggested in these references, note that the quantity
  given by the Callan-Wilczek formula \eqref{eq:astatine} should generally not
  be confused with a von Neumann entropy --- see footnotes \ref{foot:f3} and
  \ref{foot:elm}. For example, one can apply \eqref{eq:astatine} to just
  $\eActMixMatBare[\metric]$ to get the bare matter Wald-Dong plus von Neumann
  entropy, to $\eActMixMatRen[\metric]$ introduced below to get the renormalized
  matter Wald-Dong plus von Neumann entropy, or to $\eActMix[\metric]$ to get
  the full generalized entropy.}, even though its constituent bare quantities,
\eg{}the classical Wald-Dong entropy $\entDongGravBare{\entSurf}[\metric]$,
arising from $\actGravBare[\metric]$, and the von Neumann entropy
$\ent[\stateMat]$, are individually UV-divergent. (As we shall explain further
below, taking the zero limit of the regulator $\neighSize$ for the smoothed
conical singularities produced by the $\bullet_{\repS{\entSurf}n}$ operation in
\eqref{eq:astatine} is not expected to introduce any new divergences in
generalized entropy absent in the effective action $\eActMix[\metric]$ evaluated
on smooth $\metric$.)

Rather than working with divergent quantities, however, one can choose to
renormalize by explicitly absorbing the divergences $\actMatDiver[\metric]$ of
the bare matter part $\eActMixMatBare[\metric]$ of the effective action into the
bare gravitational action $\actGravBare[\metric]$:
\begin{align}
  \actGravRen[\metric]
  =& \actGravBare[\metric] + \actMatDiver[\metric]
     \label{eq:exorcist}
  \\
  \eActMixMatRen[\metric]
  =& \eActMixMatBare[\metric] - \actMatDiver[\metric] \;,
     \label{eq:groggily}
\end{align}
The above renormalized combinations are UV-finite, so that, for example, the
correspondingly renormalized stress tensor
\begin{align}
  2\frac{\delta \eActMixMatRen[\metric]}{\delta \metric_{ab}}
  =& \langle \stressRenDensity^{ab} \rangle
     = \langle \stressRen^{ab}  \rangle\volForm 
  &
  &(\text{$\vary\metric$ away from $\bdyMan$})
    \label{eq:frigidly}
\end{align}
is finite. Moreover, the classical Wald-Dong entropy
$\entDongGravRen{\entSurf}[\metric]$ arising from $\actGravRen[\metric]$ is also
finite. The divergences of $\entDongGravBare{\entSurf}[\metric]$ will have been
transferred by this renormalization process into the quantum part $\langle
\entDongMatRen{\entSurf}[\metric,\graviton,\bgFieldsOther+\qFieldsOther]
\rangle$ of the Wald-Dong entropy\footnote{If one absorbs 
  $\actMatDiver[\metric]$ into the path integration measure, as written below in
\eqref{eq:donated}, then the divergent part of
$\entDongGravBare{\entSurf}[\metric]$ is specifically transferred to the
anomalous contribution to $\langle
\entDongMatRen{\entSurf}[\metric,\graviton,\bgFieldsOther+\qFieldsOther]
\rangle$, as we will describe further below \eqref{eq:sift} in appendix \ref{sec:affix}.}, which is now responsible for cancelling the
UV divergences of $\ent[\stateMat]$ in the generalized entropy
\eqref{eq:solitude}.

One way to absorb $-\actMatDiver[\metric]$ into the bare matter part of the
effective action as in \eqref{eq:groggily} is to define the renormalized path
integral measure
\begin{align}
  [d\graviton\, d\qFieldsOther]_\ren
  =& e^{\actMatDiver[\metric]}[d\graviton\, d\qFieldsOther]_\bare \;.
     \label{eq:donated}
\end{align}
so that
\begin{align}
  \eActMixMatRen[\metric]
  =&  -\log \int [d\graviton\, d\qFieldsOther]_\ren
     e^{-\actMat[\metric,\graviton,\bgFieldsOther+\qFieldsOther]} \;.
     \label{eq:duress}
\end{align} 
Note that this explicitly introduces to the path integral measure an extra
dependence on the metric via $\actMatDiver[\metric]$. However, even before this
renormalization, there may already have been metric dependence intrinsic to the
bare path integral measure $[d\graviton\, d\qFieldsOther]_{\bare}$, due for
example to the metric-dependence of modes used to define path integration in the
first place. This naturally segues into our discussion in the next section
regarding the anomalous terms arising from the background-dependence of the path
integration measure.

Before moving on, however, let us also alleviate some concerns regarding another
possible source of UV-divergences in the generalized entropy formula
\eqref{eq:astatine} that might arise even if the effective action
$\eActMix[\metric]$ evaluated on smooth backgrounds is finite. Recall that,
implicit in the definition of the deformed replication
$\bullet_{\repS{\entSurf}n}$ appearing in the RHS of \eqref{eq:astatine}, is a
regulator $\neighSize$ describing, for smooth $\metric$, the size of the
neighbourhood $\neigh{\neighSize}{\entSurf}$ of $\entSurf$ over which the
would-be conical singularity of $\metricR{\repS{\entSurf}n}$ is smoothed. Even
while the effective action evaluated on smooth backgrounds is taken to be
UV-finite, a concern that one might raise is whether the expression
\eqref{eq:astatine} for generalized entropy blows up as the limit $\neighSize\to
0$ is taken. For RT surfaces in holography, we certainly expect a UV-finite bulk
effective action to lead to a UV-finite generalized entropy since one can
alternatively write generalized entropy in terms of the effective action
evaluated on smooth backgrounds $\metricR{n}$, as in \eqref{eq:curtly}, where
there is obviously no additional regulator $\neighSize$ to worry about. In fact,
the calculation leading to \eqref{eq:subplot} relating the two expressions
\eqref{eq:astatine} and \eqref{eq:curtly} seems to suggest a more general
argument for the finiteness of \eqref{eq:astatine} under $\neighSize\to 0$
perhaps extending beyond holography. As in appendix
\ref{sec:d4f5g3nf6bg2e6nf3be7o-oo-oc4d5}, let us denote by
$\bdyManR{n}=\bdyManR{\rep{\bdyEntSurf}n}$ the spacetime boundary manifold
replicated around $\bdyEntSurf=\entSurf\cap\bdyMan$. Then, for a general
$\metric$ and $\entSurf$, suppose that one can construct a sequence of smooth
backgrounds $\metricR{n}$ that satisfy boundary conditions on $\bdyManR{n}$ and
are $\integers_n$-symmetric around codimension-two surfaces $\entSurfR{n}$ such
that $\metricR{1}=\metric$, $\entSurfR{1}=\entSurf$, and the orbifolds
$\metricR{n\orb{\entSurf}n}$ (and hence $\metricR{n\orbS{\entSurf}n}$) can be
analytically continued to real $n$ near $n=1$. (For example, \cite{Dong:2017xht}
must have implicitly made this assumption for bulk solutions $\metric$ and
non-RT $\entSurf$ when verifying the quantum extremality of RT surfaces as
described below \eqref{eq:curtly}.) Then, the calculation around
\eqref{eq:subplot} shows that the generalized entropy
$\entGen{\entSurf}[\metric]$ defined by \eqref{eq:astatine} is given by
\begin{align}
  \entGen{\entSurf}[\metric]
  =& \partial_n\left(
     \eActMix[\metricR{n}]
     - n\eActMix[\metricR{1}]
     - \eActMix[\metricR{n\orbS{\entSurfR{n}}n}]
     \right)_{n=1}
     \;,
     \label{eq:duke}
\end{align}
where the first two terms are manifestly independent of $\neighSize$ while the
last term can be expressed as
\begin{align}
  \partial_n \left. \eActMix[\metricR{n\orbS{\entSurfR{n}}n}] \right|_{n=1}
  =& \int \frac{\delta \eActMix[\metric]}{\delta \metric} \cdot
     \partial_n \left.\metricR{n\orbS{\entSurfR{n}}n} \right|_{n=1} \;.
\end{align}
Even for off-shell $\metric$, we expect this to be finite in the $\neighSize\to
0$ limit as it only involves the boundary-condition-preserving variation
$\partial_n \left.\metricR{n\orbS{\entSurfR{n}}n} \right|_{n=1}$ of the metric
(as opposed to, \eg{}higher powers of curvature which would become dangerously
divergent in $\neigh{\neighSize}{\entSurfR{n}}$ at $n\ne 1$). We conclude
therefore that the generalized entropy given by \eqref{eq:duke}, or equivalently
\eqref{eq:astatine}, is UV-finite even in the $\neighSize\to 0$ limit, provided
the effective action $\eActMix[\metric]$ evaluated on smooth backgrounds is
UV-finite, either by renormalization in the semiclassical theory, as described
earlier in this appendix, or for some other reason.

\subsection{Variations of the path integral measure}
\label{sec:affix}
Let us begin by very formally describing the bare path integration measure. One
can think of this as a `volume form' on the space of UV-regulated field
configurations $(\metric+\graviton,\bgFieldsOther+\qFieldsOther)$, which can be
constructed out of `one-forms' on this space, \ie{}some basis of field
fluctuation modes, which we schematically denote $d\graviton$ and
$d\qFieldsOther$, and some `metric determinant' on the space of field
configurations, which can generally vary with
$(\metric+\graviton,\bgFieldsOther+\qFieldsOther)$. In general, the UV
regulation of fluctuations can also depend on
$(\metric+\graviton,\bgFieldsOther+\qFieldsOther)$. Thus, to display these
dependences explicitly, we can formally write the bare path integration measure
as
\begin{align}
  [d\graviton, d\qFieldsOther; \metric+\graviton, \bgFieldsOther+\qFieldsOther]_{\bare} \;.
  \label{eq:unrevised}
\end{align}
Considering the semiclassical effective action, however, the background metric
$\metric$ itself plays a distinguished role. For instance, we might choose to
evaluate the renormalized matter part \eqref{eq:duress} using the renormalized
measure \eqref{eq:donated} --- notice here that the absorbed counterterm action
$\actMatDiver[\metric]$ depends on the background metric $\metric$ alone. Thus,
we will allow for a slight generalization to the form of the path integration
measure\footnote{In fact, if we are only interested in using our semiclassical
  theory to compute the quantum gravity effective action up to $O(\GNewton)$
  corrections, as described around \eqref{eq:faucet}, it may suffice to simply
  consider the measure
  \begin{align}
    [d\graviton, d\qFieldsOther; \metric, \bgFieldsOther+\qFieldsOther]\;,
    \label{eq:floss}
  \end{align}
  as graviton fluctuations are suppressed in $\GNewton$. We would then
  essentially be zooming in to the space of small metric fluctuations around
  $\metric$, which can be approximated by the `tangent space' at the `point'
  $\metric$. \label{foot:scowling}}:
\begin{align}
  [d\graviton, d\qFieldsOther; \metric, \graviton, \bgFieldsOther+\qFieldsOther]_{\ren} \;.
  \label{eq:overgrown}
\end{align}
From here on, we shall adopt this more general form and discard subscripts
$\bullet_\bare$ and $\bullet_\ren$; unless otherwise specified, one is free to
interpret the discussions after this point
as referring either to bare or renormalized quantities.

It is perhaps helpful to pause briefly to connect our description of the path
integration measure above with the simplified example of a quantum scalar field propagating
on a classical background
spacetime \cite{Hawking:1976ja,Fujikawa:1980vr}. Here, it is standard to define
the path integration as being over the expansion coefficients of scalar
field fluctuations with respect to eigenmodes of a differential operator,
typically extracted from the terms in the action quadratic in the
fluctuations. The orthonormalization of the eigenmodes is taken with respect to
the usual inner product $(f_1,f_2) = \int \volForm f_1^* f_2$ of scalar
functions $f_1, f_2$ on a curved manifold. To connect with the language of the
previous paragraph, this orthonormalization condition is what determines the
`metric' on the space of field configurations and it obviously depends on the
spacetime metric. The UV-regulation of the bare path integration can be applied
as a restriction on the eigenvalues of the differential operator, which is
metric dependent. Furthermore, if, for example, the scalar is charged under a gauge symmetry
carried by a Maxwell field, then the differential operator should also depend on
the Maxwell field. Thus, we see that it is plausible for the path integration
measure to depend on both the spacetime metric and matter fields.

Let us proceed now to discuss the anomalous terms that can arise from variations
of the path integration measure. For this, we must appeal to the locality of the
response of the path integral measure to variations in the background and to
changes of path integration variables. By this, we mean that if we perturb the
background $(\metric,\bgFieldsOther)$ or apply an infinitesimal transformation
to the quantum fields $(\graviton,\qFieldsOther)$ in a restricted spacetime
neighbourhood in $(\metric+\graviton,\bgFieldsOther+\qFieldsOther)$, then only
the path integration measure over quantum field fluctuations
$(d\graviton,d\qFieldsOther)$ in that neighbourhood ought to be affected.
Obviously, the metric-dependence of the $e^{\actMatDiver[\metric]}$ factor in
the renormalized measure \eqref{eq:donated} is local, since the action
$\actMatDiver[\metric]$ is local. To see that the response of the bare path
integration measure should also be local to begin with, one can choose to define
the path integration using quantum field fluctuation
$(d\graviton,d\qFieldsOther)$ modes given by localized wavepackets\footnote{Note
  that the path integration measure does not depend on the specific choice of
  modes \perse, but rather the `volume form' on the space of configurations,
  \eg{}implied by an orthonormalization condition of the modes. Thus, the path
  integral does not care whether one chooses to consider, say, the eigenmodes of
  some differential operator, or a basis of localized wavepackets, so long as the modes satisfy
  the same orthonormalization condition and span the same space of UV-regulated
  fluctuations.} --- it is then natural for the orthonormalization condition and
UV-regulation of the wavepackets to be dependent only on the field profiles
$(\metric+\graviton,\bgFieldsOther+\qFieldsOther)$ in the wavepackets' regions
of support. Moreover, if we reparameterize the quantum field fluctuations
$(d\graviton,d\qFieldsOther)$ in a restricted spacetime neighbourhood, then we
expect fluctuations corresponding to orthonormal wavepackets localized elsewhere
in the spacetime to remain orthonormal.

Expressing this locality mathematically, we shall write the variation of the
path integration measure resulting from a background perturbation
$(\vary\metric,\vary\bgFieldsOther)$ as the insertion of operators living in the
supports of $\vary\metric$ and $\vary\bgFieldsOther$; by linearity in
$\vary\metric$ and $\vary\bgFieldsOther$ and using integrating by parts, we can
generally write this as
\begin{align}
  (\varyWRT{\metric}+\varyWRT{\bgFieldsOther})
  [d\graviton, d\qFieldsOther; \metric, \graviton, \bgFieldsOther+\qFieldsOther]
  =& -[d\graviton, d\qFieldsOther;\metric, \graviton, \bgFieldsOther+\qFieldsOther] \int \left(
     \eomDensityAnom{\metric}\cdot\vary\metric
     + \eomDensityAnom{\bgFieldsOther}\cdot\vary\bgFieldsOther
     + d\sympPotAnom[\vary\metric, \vary\bgFieldsOther]
     \right) \;,
     \label{eq:illusive}
\end{align}
where $\sympPotAnom$ is linear in its arguments. In contrast to their
counterparts $\eomDensity{\metric}$, $\eomDensity{\bgFieldsOther}$, and
$\sympPot[\vary\metric,\vary\bgFieldsOther]$ obtained from the Lagrangian
$\LagDensity$, the operators $\eomDensityAnom{\metric}$,
$\eomDensityAnom{\bgFieldsOther}$, and
$\sympPotAnom[\vary\metric,\vary\bgFieldsOther]$ might not have simple
expressions in terms of field operators; we shall simply take
\eqref{eq:illusive} to be their definition.

We can develop similar notation for changes
$(\vary\graviton,\vary\qFieldsOther)$ of path integration variables. Here,
$\vary\graviton$ and $\vary\qFieldsOther$ denote linearized transformations
acting on the space field configurations and can generally vary over this space,
\ie{}$(\vary\graviton,\vary\qFieldsOther)$ may not necessarily be c-numbers. In
the same way that $(\vary\metric,\vary\bgFieldsOther)$ play the roles of
linearization parameters in \eqref{eq:illusive}, it is helpful here to introduce
a scalar c-number function $\testFunc$ over spacetime and consider the effect of
the change of variables $(\testFunc \vary\graviton, \testFunc
\vary\qFieldsOther)$ on the path integration measure, linearized in $\testFunc$.
Then, analogous to \eqref{eq:illusive}, we write
\begin{align}
  \begin{split}
    \MoveEqLeft[3] \varyWRT{\testFunc} [ d(\graviton+\testFunc\vary\graviton),
    d(\qFieldsOther+\testFunc\vary\qFieldsOther); \metric,
    \graviton+\testFunc\vary\graviton,
    \bgFieldsOther+\qFieldsOther+\testFunc\vary\qFieldsOther ]
    \\
    =& -[d\graviton, d\qFieldsOther;\metric, \graviton,
       \bgFieldsOther+\qFieldsOther] \int \left\{ \left(
       \eomDensityAnom{\graviton,\vary\graviton} +
       \eomDensityAnom{\qFieldsOther, \vary\qFieldsOther} \right)\testFunc +
       d\left( \sympPotAnomQ{\graviton,\vary\graviton}[\testFunc] +
       \sympPotAnomQ{\qFieldsOther,\vary\qFieldsOther}[\testFunc] \right)
       \right\} \;,
  \end{split}
       \label{eq:barricade}
\end{align}
where $\sympPotAnomQ{\egField,\vary\egField}[\testFunc]$ is linear in
$\testFunc$ (and its derivatives). Furthermore, note that, for any function
$\testFunc$, we have by definition that
\begin{align}
  \eomDensityAnom{\egField,\vary\egField}\, \testFunc
  =& \eomDensityAnom{\egField,\testFunc\vary\egField}
     \;,
  &
    \sympPotAnomQ{\egField,\vary\egField}[\testFunc]
    =& \sympPotAnomQ{\egField,\testFunc\vary\egField}[1] \;.
  &
  &(\egField\in\{\graviton,\qFieldsOther\})
    \label{eq:hardwired}
\end{align}
It will be helpful, therefore, to abbreviate
\begin{align}
  \sympPotAnomQ{\egField,\vary\egField}
  =& \sympPotAnomQ{\egField,\vary\egField}[1] \;.
\end{align} 
Note that, owing to the shift-invariance of the `one-forms' in the path
integration measure,
\begin{align}
  [d(\graviton+\vary\graviton), d(\qFieldsOther+\vary\bgFieldsOther);
  \metric, \graviton, \bgFieldsOther+\qFieldsOther]
  =&
     [d\graviton, d\qFieldsOther; \metric, \graviton, \bgFieldsOther+\qFieldsOther]
  &
    (\text{c-number $\vary\graviton,\vary\qFieldsOther$})
    \;,
    \label{eq:symphonic}
\end{align}
we have for c-number variations,
\begin{align}
  \eomDensityAnom{\qFieldsOther,\vary\bgFieldsOther}
  =& \eomDensityAnom{\bgFieldsOther} \cdot \vary\bgFieldsOther \;,
  &
    \sympPotAnomQ{\qFieldsOther,\vary\bgFieldsOther}
    =& \sympPotAnom[\vary\bgFieldsOther]\;,
  &
    (\text{c-number $\vary\bgFieldsOther$})
    \label{eq:nappy}
\end{align}
where $\eomDensityAnom{\bgFieldsOther}$ and $\sympPotAnom[\vary\bgFieldsOther]$
are the same objects defined previously in \eqref{eq:illusive}; similarly, we
can write\footnote{Note that if we had used the form \eqref{eq:unrevised} of the
  path integration measure instead of the more general \eqref{eq:overgrown}, we
  would find that $\eomDensityAnom{\graviton}=\eomDensityAnom{\metric}$ and
  $\sympPotAnom[\vary\graviton]=\sympPotAnom[\vary\metric=\vary\graviton]$ in
  \eqref{eq:spiffy} are also the same objects defined previously in
  \eqref{eq:illusive}. Moreover, if we neglect the graviton dependence of the
  measure, as suggested in \eqref{eq:floss} of footnote \ref{foot:scowling}, we
  would have found
  \begin{align}
    \eomDensityAnom{\graviton,\vary\graviton}
    =& 0 \;,
    &
      \sympPotAnomQ{\graviton,\vary\graviton} =& 0 \;.
    &
      (\text{c-number $\vary\graviton$})
      \label{eq:ultimate}
  \end{align} 
  This is just \eqref{eq:spiffy} with the $\vary\graviton$ on the RHS set to be
  negligibly small (to be consistent with the assumption of suppressed graviton
  fluctuations that led to \eqref{eq:floss}).}, for some
$\eomDensityAnom{\graviton}$ and some $\sympPotAnom[\vary\graviton]$ linear in
$\vary\graviton$,
\begin{align}
  \eomDensityAnom{\graviton,\vary\graviton}
  =& \eomDensityAnom{\graviton}\cdot\vary\graviton \;,
  &
    \sympPotAnomQ{\graviton,\vary\graviton}
    =& \sympPotAnom[\vary\graviton] \;.
  &
    (\text{c-number $\vary\graviton$})
    \label{eq:spiffy}
\end{align}

Although the $\eomDensityAnom{}$ and $\sympPotAnom$ objects defined by
\eqref{eq:illusive} and \eqref{eq:barricade} might not have obvious expressions
in terms of fields and aren't obtained simply from some Lagrangian, it is
natural to include these anomalous terms as corrections to various operators.
Perhaps the most obvious example is the matter stress tensor $\stress^{ab}$,
whose expectation value is given by the first variation of the matter part
$\eActMixMat[\metric]$ of the effective action, as in the first equality of
\eqref{eq:friendless}. Accounting for the variations of both the action and the
path integral measure, we see that the full stress tensor can be expressed as
\begin{align}
  (\eomDensityMat{\metric} + \eomDensityAnom{\metric})^{ab}
  =& \stressDensity^{ab}
     = \stress^{ab} \volForm \;,
     \label{eq:tightwad}
\end{align}
so that
\begin{align}
  0
  =& \frac{\delta \eActMix[\metric]}{\delta\metric_{ab}}
     = \left\langle
     \left(
     \eomDensity{\metric}
     + \eomDensityAnom{\metric}
     \right)^{ab}
     \right\rangle
     = (\eomDensityGrav{\metric})^{ab} + \frac{1}{2}\left\langle\stressDensity^{ab}\right\rangle
  &
  &(\text{$\vary\metric$ away from $\bdyMan$, $\metric$ on-shell})
    \label{eq:residual}
\end{align} 
gives the anomaly-corrected semiclassical equations of motion \eqref{eq:lonely}
for the metric $\metric$. Moreover, the property \eqref{eq:symphonic} of the
path integration measure and the fact that $\bgFieldsOther$ only appears as a
shift of $\qFieldsOther$ in both the action
$\act[\metric,\graviton,\bgFieldsOther+\qFieldsOther]$ and the measure
$[d\graviton, d\qFieldsOther; \metric, \graviton, \bgFieldsOther+\qFieldsOther]$
gives
\begin{align}
  0
  =& \frac{\delta \eActMix[\metric]}{\delta \bgFieldsOther}
     = \langle \eomDensity{\bgFieldsOther} + \eomDensityAnom{\bgFieldsOther}\rangle \;,
  &
  &(\text{$\vary\bgFieldsOther$ away from $\bdyMan$})
    \label{eq:litigate}
\end{align}
correcting the more naively written \eqref{eq:uranium}. For the quantum fields
$\egField\in\{\graviton,\qFieldsOther\}$, the anomalous corrections
$\eomDensityAnom{\egField,\vary\egField}$ and
$\sympPotAnomQ{\egField,\vary\egField}$ give the variations of the path
integration measure in response to reparametrizations of the integration
variables by $\vary\egField$; thus,
we have
\begin{align}
  \langle
  \eomDensity{\graviton} \cdot \vary\graviton
  + \eomDensityAnom{\graviton,\vary\graviton}
  \rangle
  =& 0 \;,
  &
    \langle
    \eomDensity{\qFieldsOther}\cdot\vary\qFieldsOther
    +\eomDensityAnom{\qFieldsOther,\vary\qFieldsOther}
    \rangle
    =& 0 \;,
       \label{eq:sandlot}
\end{align}
of which \eqref{eq:litigate} is a special case, as seen from \eqref{eq:nappy}.
(The above holds locally at any point in the interior of the spacetime region of
path integration, because we can consider in \eqref{eq:barricade} the
perturbations $\testFunc\vary\graviton$ and $\testFunc\vary\qFieldsOther$ for
any scalar function $\testFunc$ with arbitrary support away from the boundary of
the region--- recall also \eqref{eq:hardwired}.)

It will be helpful also to define the anomalous corrections to other operators
encountered in this paper, such as the Wald-Dong entropy, Noether charge, and
Noether current. Let us begin with the Wald-Dong entropy which was naively given
by \eqref{eq:handwork} or equivalently
\eqref{eq:scone}.
We shall analogously define the anomalous correction to the Wald-Dong entropy to
be\footnote{Recall, from the discussion below \eqref{eq:fondue} and
  \eqref{eq:crummiest} that $\bullet_{\rep{\entSurf}n\orbS{\entSurf}n}$ has the
  interpretation, at any real $n>0$ as a deformation localized in the tiny
  neighbourhood $\neigh{\neighSize}{\entSurf}$ of $\entSurf$ which multiplies
  the opening angle around exactly $\entSurf$ by $n$.}
\begin{align}
  \begin{split}
    \MoveEqLeft[3]
    [d\graviton,d\qFieldsOther;\metric,\graviton,\bgFieldsOther+\qFieldsOther]\,
    \entDongAnom{\entSurf}[\metric,\graviton,\bgFieldsOther+\qFieldsOther] \\
    =& \partial_n \left( [ d(\gravitonR{\rep{\entSurf}n\orbS{\entSurf}n}),
       d(\qFieldsOtherR{\rep{\entSurf}n\orbS{\entSurf}n});
       \metricR{\rep{\entSurf}n\orbS{\entSurf}n},
       \gravitonR{\rep{\entSurf}n\orbS{\entSurf}n},
       \bgFieldsOtherR{\rep{\entSurf}n\orbS{\entSurf}n} +
       \qFieldsOtherR{\rep{\entSurf}n\orbS{\entSurf}n} ] \right)_{n=1}
  \end{split}
       \label{eq:glance}
  \\
  =& \partial_n \left(
     [
     d(\gravitonR{\rep{\entSurf}n\orbS{\entSurf}n}),
     d(\qFieldsOtherR{\rep{\entSurf}n\orbS{\entSurf}n});
     \metricR{\rep{\entSurf}n},
     \gravitonR{\rep{\entSurf}n\orbS{\entSurf}n},
     \bgFieldsOtherR{\rep{\entSurf}n} + \qFieldsOtherR{\rep{\entSurf}n\orbS{\entSurf}n}
     ]
     - [
     d\graviton,
     d\qFieldsOther;
     \metricR{\repS{\entSurf}n},
     \graviton,
     \bgFieldsOtherR{\repS{\entSurf}n} + \qFieldsOther
     ]
     \right)_{n=1} \;,
     \label{eq:retread}
\end{align}
so that the parenthesized difference in measures of \eqref{eq:retread} is
precisely the anomalous correction of \eqref{eq:berkelium} mentioned below that
equation. Consequently, if we correct the Wald-Dong entropy to be given by the
naive \eqref{eq:handwork}, derived from the action, plus the anomalous
correction \eqref{eq:glance} so that
\begin{align}
  \entDong{\entSurf}[
  \metric,\graviton,\bgFieldsOther+\qFieldsOther
  ]
  =& \partial_n \left.
     \actEntSurf{\entSurf,n}[
     \metric,\graviton,\bgFieldsOther+\qFieldsOther
     ]
     \right|_{n=1}
     + \entDongAnom{\entSurf}[
     \metric,\graviton,\bgFieldsOther+\qFieldsOther
     ] \;,
     \label{eq:sift}
\end{align}
then the generalized entropy \eqref{eq:solitude} is still given in terms of the
effective action by \eqref{eq:astatine}, even after accounting for variations of
the path integration measure\footnote{To clarify our conventions for notation in
  this section, we should state that, with the exception of the stress tensor
  $\stress^{ab}$ defined by \eqref{eq:tightwad} and the Wald-Dong entropy
  defined by \eqref{eq:sift}, the symbols for all other quantities shall retain
  their original meanings defined in the main text of this paper in terms of the
  action $\act$. The addition of anomalous corrections to those other quantities
  shall be written explicitly.}. We note that, through the renormalization
\eqref{eq:donated} of the path integration measure, the divergent contribution
to Wald-Dong entropy from the divergent action $\actMatDiver[\metric]$ is
transferred between the two terms in \eqref{eq:sift}, but \eqref{eq:sift} as a
whole remains invariant.

As in \eqref{eq:crummiest}, we can try to express the anomalous contribution
\eqref{eq:glance} to the Wald-Dong entropy in terms of the anomalous symplectic
potential $\sympPotAnom$:
\begin{align}
  \begin{split}
    \MoveEqLeft[3]\entDongAnom{\entSurf}[
    \metric,\graviton,\bgFieldsOther+\qFieldsOther
    ]
    \\
    \eqDubious& \int_{\partial\neigh{0}{\entSurf}}
                \left(
                \sympPotAnom[
                \partial_n\metricR{\rep{\entSurf}n\orbS{\entSurf}n}|_{n=1},
                \partial_n \bgFieldsOtherR{\rep{\entSurf}n\orbS{\entSurf}n}|_{n=1}
                ]
                + \sympPotAnomQ{\graviton,\partial_n \gravitonR{\rep{\entSurf}n\orbS{\entSurf}n}|_{n=1}}
                + \sympPotAnomQ{\qFieldsOther,\partial_n \qFieldsOtherR{\rep{\entSurf}n\orbS{\entSurf}n}|_{n=1}}
                \right)
                \;.
  \end{split}
                \label{eq:rocker}
\end{align}
However, we expect similar subtleties as for \eqref{eq:crummiest} to plague
\eqref{eq:rocker} as well. (This is perhaps most obvious in the case where one
considers the renormalized path integration measure \eqref{eq:donated} which
absorbs the factor $e^{\actMatDiver[\metric]}$; the resulting extra contribution
to the anomalous Wald-Dong entropy \eqref{eq:glance} from this factor can be
calculated by applying the same procedure as in \cite{Dong:2013qoa} to the
action $-\actMatDiver[\metric]$.) However, one can hope that when the extrinsic
curvatures of the entangling surface $\entSurf$ vanish\footnote{One might, for
  example, try to argue, along the lines of the discussion after
  \eqref{eq:everyone}, that in these cases the action of
  $\bullet_{\rep{\entSurf}n\orbS{\entSurf}n}$, for $n$ close to $1$ and finite
  $\neighSize$, produces a change in the path integration measure that is not
  particularly pathological on approach to $\entSurf$. \label{foot:strobe}}, as
in thermal setups, corrections to \eqref{eq:rocker} vanish, just as the
corrections found by \cite{Dong:2013qoa} to \eqref{eq:crummiest} vanish in these
cases. Indeed, it will be interesting to compare \eqref{eq:rocker} to a very
similar expression for the anomalous contribution to a Noether charge integrated
along $\entSurf$, as we will discuss in appendix \ref{sec:campfire}.

Let us now consider the anomalous corrections to the $\FaulknerC_a$ forms and
Noether charge, defined by \eqref{eq:roentgenium} and \eqref{eq:lutetium}, and
the Noether current given by \eqref{eq:d4d5c4e6}. Let us start with the
$(\FaulknerCOf{\egField})_a$ forms originally given by \eqref{eq:roentgenium}
for each field $\egField\in\{\metric,\graviton,\bgFieldsOther,\qFieldsOther\}$.
For the background fields, the generalization is obvious: tacking on
superscripts $\bullet^{\anom}$ to \eqref{eq:roentgenium}, we have
\begin{align}
  (\FaulknerCAnomOf{\egField})_a
  \equiv& \sum_{i=1}^r
          (\eomAnom{\egField})^{b_1 \cdots b_s}_{c_1 \cdots a \cdots c_r}
          \egField^{c_1 \cdots c_i \cdots c_r}_{b_1 \cdots b_s}
          \volForm_{c_i}
          - \sum_{i=1}^s
          (\eomAnom{\egField})^{b_1 \cdots b_i \cdots b_s}_{c_1 \cdots c_r}
          \egField^{c_1 \cdots c_r}_{b_1 \cdots a \cdots b_s}
          \volForm_{b_i}
          \;.
  & (\egField\in\{\metric,\bgFieldsOther\})
    \label{eq:atrophy}
\end{align}
where
\begin{align}
  \eomDensityAnom{\egField}
  =& \volForm\eomAnom{\egField} \;.
\end{align} 
To define $(\FaulknerCAnomOf{\egField})_a$ for the quantum fields $\graviton$
and $\qFieldsOther$, it is perhaps better to consider an alternative to the
definition \eqref{eq:roentgenium}. For any vector field $\egVec$, it can be
shown simply by writing $\Lie{\egVec}$ out in terms of covariant derivatives
that $(\FaulknerCOf{\egField})_a$ and $(\FaulknerCOf{\egField})_a$ defined by
\eqref{eq:roentgenium} and \eqref{eq:atrophy} satisfy \footnote{See (B.2) in
  \cite{Faulkner:2013ica}.}
\begin{align}
  \eomDensity{\egField} \cdot \Lie{\egVec} \egField
  =& \egVec^a (\FaulknerBOf{\egField})_a - d\{\egVec^a (\FaulknerCOf{\egField})_a\} \;,
     \label{eq:grime}
  \\
  \eomDensityAnom{\egField} \cdot \Lie{\egVec} \egField
  =& \egVec^a (\FaulknerBAnomOf{\egField})_a - d\{\egVec^a (\FaulknerCAnomOf{\egField})_a\}
     \;,
   & (\egField\in\{\metric,\bgFieldsOther\})
     \label{eq:speckled} 
\end{align}
for some top-dimensional forms $(\FaulknerBOf{\egField})_a$ and
$(\FaulknerBAnomOf{\egField})_a$; these serve as an alternative to
\eqref{eq:roentgenium} and \eqref{eq:atrophy} for defining
$(\FaulknerCOf{\egField})_a$ and $(\FaulknerCAnomOf{\egField})_a$. Thus, a
reasonable definition for the $(\FaulknerCAnomOf{\egField})_a$ of a quantum
field $\egField\in\{\graviton,\qFieldsOther\}$ is given by writing
\begin{align}
  \eomDensityAnom{\egField,\Lie{\egVec}\egField}
  \equiv& \egVec^a (\FaulknerBAnomOf{\egField})_a - d\{\egVec^a (\FaulknerCAnomOf{\egField})_a\} \;,
  & (\egField\in\{\graviton,\qFieldsOther\})
    \label{eq:onset}
\end{align}
for some top-dimensional form $(\FaulknerBAnomOf{\egField})_a$. We note that
$\eomDensityAnom{\egField,\Lie{\egVec}\egField}$ should be expressible in this
form, because the change of integration variables
$\vary\egField=\Lie{\egVec}\egField$ is linear in $\egVec$ and depends on no
higher derivatives of $\egVec$ than its first, so the same statements should be
true of $\eomDensityAnom{\egField,\Lie{\egVec}\egField}$ as defined in
\eqref{eq:barracuda}\footnote{In particular, if we expand
  \begin{align}
    \Lie{\egVec}\egField
    =&\egVec^a \,(\vary_1)_a \egField + \nabla_b \,
       \egVec^a \,
       \tensor{(\vary_2)}{_a^b} \egField \;,
    & (\egField\in\{\graviton,\qFieldsOther\})
      \label{eq:tattoo}
  \end{align} 
  where the variations $(\vary_1)_a \egField$ and $\tensor{(\vary_2)}{_a^b}
  \egField$ are independent of $\egVec$, then we expect
  \begin{align}
    \eomDensityAnom{\egField,\Lie{\egVec}\egField}
    =&\egVec^a
       \eomDensityAnom{\egField,(\vary_1)_a \egField}
       + \nabla_b \egVec^a
       \eomDensityAnom{\egField,\tensor{(\vary_2)}{_a^b}\egField}
       \;.
    & (\egField\in\{\graviton,\qFieldsOther\})
      \label{eq:dilute}
  \end{align}
  In \eqref{eq:tattoo} and \eqref{eq:dilute}, $\nabla$ can be any derivative
  operator independent of the quantum fields $(\graviton,\qFieldsOther)$, but
  taking it to be the derivative operator preserving some volume form, \eg{}the
  volume form $\volForm$ of the background metric, makes it easy to see
  \eqref{eq:onset} is satisfied for
  \begin{align}
    (\FaulknerBAnomOf{\egField})_a
    =&
       \eomDensityAnom{\egField,(\vary_1)_a \egField}
       -\nabla_b \eomDensityAnom{\egField,\tensor{(\vary_2)}{_a^b}\egField} \;,
    &
      (\FaulknerCAnomOf{\egField})_a
      =&
         -\eomAnom{\egField,\tensor{(\vary_2)}{_a^b}\egField} \volForm_b
         \;,
    & (\egField\in\{\graviton,\qFieldsOther\})
      \label{eq:buffoon}
  \end{align}
  where
  \begin{align}
    \eomDensityAnom{\egField,\tensor{(\vary_2)}{_a^b}\egField}
    =& \volForm\eomAnom{\egField,\tensor{(\vary_2)}{_a^b}\egField} \;.
    & (\egField\in\{\graviton,\qFieldsOther\})
  \end{align} 
  (These expressions for $(\FaulknerBAnomOf{\egField})_a$ and
  $(\FaulknerCAnomOf{\egField})_a$, with $(\vary_1)_a \egField$ and
  $\tensor{(\vary_2)}{_a^b}\egField$ defined by \eqref{eq:tattoo}, again do not
  depend on the choice of the $(\graviton,\qFieldsOther)$-independent derivative
  operator $\nabla$.)}. Obviously, we can define, in analogy with
\eqref{eq:lutetium}, the total anomalous $\FaulknerCAnom_a$ form:
\begin{align}
  \FaulknerCAnom_a
  \equiv& \sum_{\egField\in\{\metric,\graviton,\bgFieldsOther,\qFieldsOther\}}
          (\FaulknerCAnomOf{\egField})_a \;.
\end{align} 

Before proceeding to discuss Noether charges and currents, let us pause briefly
to examine the expectation values of the anomaly-corrected
$\FaulknerC_a+\FaulknerCAnom_a$ forms. Let us begin with the metric part. Notice
from \eqref{eq:roentgenium} and \eqref{eq:atrophy} that, in analogy with
\eqref{eq:e4c5}, we have
\begin{align}
  (\FaulknerCOf{\metric} + \FaulknerCAnomOf{\metric})^a
  =& -2\volForm_b (\eom{\metric} + \eomAnom{\metric})^{ab}
     = -\volForm_b \left(2\eomGrav{\metric} + \stress\right)^{ab} \;,
     \label{eq:praising}
\end{align} 
involving the full stress tensor \eqref{eq:tightwad}, accounting for metric
variations in the path integration measure. Setting the expectation value of
\eqref{eq:praising} to zero yields the semiclassical equations of motion
\eqref{eq:residual} for the metric $\metric$. On the other hand, from
\eqref{eq:roentgenium}, \eqref{eq:atrophy}, and \eqref{eq:onset}, one can argue
that the expectation value of the anomaly-corrected
$(\FaulknerCOf{\egField}+\FaulknerCAnomOf{\egField})_a$ forms for the other
background fields $\bgFieldsOther$ and the quantum fields
$(\graviton,\qFieldsOther)$ vanish by virtue of \eqref{eq:litigate} and
\eqref{eq:sandlot}:
\begin{align}
  \left\langle 
  (\FaulknerCOf{\egField}+\FaulknerCAnomOf{\egField})_a
  \right\rangle
  =& 0 \;.
  &
    (\egField\in\{\bgFieldsOther,\graviton,\qFieldsOther\})
    \label{eq:harmonic}
\end{align}
To show the above for the case of the quantum fields
$\egField\in\{\graviton,\qFieldsOther\}$, note from \eqref{eq:grime} and
\eqref{eq:onset} that the integral of
$-\egVec^a(\FaulknerCOf{\egField}+\FaulknerCAnomOf{\egField})_a$ over any
codimension-one region $\egReg$ is expressible as the integral of
$\eomDensity{\egField}\cdot\Lie{\testFunc\egVec}\egField+\eomDensityAnom{\egField,\Lie{\testFunc\egVec}\egField}$
over an infinitesimally thin slab with $\egReg$ as one face, where $\testFunc$
is a smooth scalar function interpolating from one on $\egReg$ to zero on the
other face of the slab\footnote{Alternatively, one can directly write
  $(\FaulknerCOf{\egField}+\FaulknerCAnomOf{\egField})_a$, for
  $\egField\in\{\graviton,\qFieldsOther\}$ in terms of equations of motion using
  \eqref{eq:roentgenium} and \eqref{eq:buffoon}.}. Thus, \eqref{eq:harmonic}
gives the anomaly-corrected versions of \eqref{eq:skewer} and
\eqref{eq:surpass}.

Now, to define the anomalous contribution to the Noether charge, we can employ a
similar strategy as for \eqref{eq:onset}. Notice from \eqref{eq:d4d5c4e6} and
\eqref{eq:lutetium} that, in analogy to \eqref{eq:grime}, the Noether charge can
be defined as the total derivative part of the symplectic potential when using
the product rule to strip derivatives from $\egVec$:
\begin{align}
  \sympPot[
  \Lie{\egVec}\metric,
  \Lie{\egVec}\graviton,
  \Lie{\egVec}(\bgFieldsOther+\qFieldsOther)
  ]
  =& \egVec^a(\FaulknerC_a+\volForm_a\Lag) + d\charge{\egVec}
\end{align}
Thus, let us similarly define the anomalous correction $\chargeAnom{\egVec}$
such that
\begin{align}
  \sympPotAnom[\Lie{\egVec}\metric,\Lie{\egVec}\bgFieldsOther]
  + \sympPotAnomQ{\graviton,\Lie{\egVec}\graviton}
  + \sympPotAnomQ{\qFieldsOther,\Lie{\egVec}\qFieldsOther}
  =& \egVec^a \FaulknerAAnom_a + d\chargeAnom{\egVec}
     \label{eq:dumpling}
\end{align} 
for some $\FaulknerAAnom_a$. With $\FaulknerCAnom_a$ and $\chargeAnom{\egVec}$
now defined, we can proceed to also define the anomalous correction
$\currentAnom{\egVec}$ to the Noether current by analogy to \eqref{eq:lutetium}:
\begin{align}
  \currentAnom{\egVec}
  =& d\chargeAnom{\egVec} + \egVec^a \FaulknerCAnom_a \;.
     \label{eq:unplanted}
\end{align} 
Correcting \eqref{eq:lutetium} with this, we see from \eqref{eq:praising} and
\eqref{eq:harmonic} that the exterior derivative
\begin{align}
  d\left\langle
  \current{\egVec}+\currentAnom{\egVec}
  \right\rangle
  =& -2\, d\left\langle
     \volForm_a \egVec_b \left(
     \eom{\metric} + \eomAnom{\metric}
     \right)^{ab}
     \right\rangle 
\end{align} 
vanishes when the anomaly-corrected semiclassical equations of motion
\eqref{eq:residual} for the metric are satisfied; this expresses conservation of
the Noether current $\langle \current{\egVec}+\currentAnom{\egVec} \rangle$.

We end this section by introducing some final anomalous ingredients relevant for
considering perturbations which reach the boundary of spacetime regions over
which path integration is carried out --- \eg{} at the spacetime boundary
$\bdyMan$ or, when one is interested in path integrals over only a portion of
the bulk (as we were in appendix \ref{sec:immorally}), the bounding
codimension-one surfaces on which the quantum field configurations are fixed.
Let us begin by treating the latter types of surfaces. Suppose one starts out,
with a path integral, with measure
$[d\graviton,d\qFieldsOther;\metric,\graviton,\bgFieldsOther+\qFieldsOther]$,
covering, for simplicity, a compact spacetime without boundary. Now, we
bipartition the spacetime into two complementary regions $\egPatch$ and
$\egPatch'$. We shall take, as a fundamental property, the fact that path
integrals over spacetime regions can be joined together or split apart along
codimension-one surfaces. (Indeed, this is what permits the replica calculation
of R\'enyi entropies, as described after \eqref{eq:lost} in section
\ref{sec:helium}, and the interpretation of path integration as giving matrix
elements of an evolution operator, as described in appendix
\ref{sec:immorally}.) To be specific, suppose we fix, in our example, the
quantum field configuration\footnote{Without loss of generality, we can focus on
  quantum field states corresponding to fixed field configurations on
  $\partial\egPatch$. More general wave functionals on $\partial\egPatch$ can
  then be constructed from the basis formed by these states.} on
$\partial\egPatch=\partial\egPatch'$ to what we will denote with a state
$|\pureState\rangle$ --- this can be accomplished by inserting into the path
integral a delta functional
$\delta_{\partial\egPatch,|\pureState\rangle}[\graviton,\qFieldsOther]$ which
constrains the quantum fields $(\graviton,\qFieldsOther)$ on $\partial\egPatch$
to match $|\pureState\rangle$. We would like this delta functional to have the
property that, when
$\delta_{\partial\egPatch,|\pureState\rangle}[\graviton,\qFieldsOther]$ is
integrated with some measure over the space of $|\pureState\rangle$, the
identity is recovered. Then, we might naively hope to be able to split
\begin{align}
  \begin{split}
    \MoveEqLeft[3]\int[d\graviton,d\qFieldsOther;\metric,\graviton,\bgFieldsOther+\qFieldsOther]\,
    \delta_{\partial\egPatch,|\pureState\rangle}[\graviton,\qFieldsOther]\, \bullet
    \\
    \eqDubious& \int
                [
                d\graviton,d\qFieldsOther;
                \metric,\graviton,\bgFieldsOther+\qFieldsOther
                ]_{\egPatch,|\pureState\rangle} \,
                [
                d\graviton,d\qFieldsOther;
                \metric,\graviton,\bgFieldsOther+\qFieldsOther
                ]_{\egPatch',\langle \pureState|} \, \bullet \;,
  \end{split}
                \label{eq:bakeshop}
\end{align}
where
$[d\graviton,d\qFieldsOther;\metric,\graviton,\bgFieldsOther+\qFieldsOther]_{\egPatch,|\pureState\rangle}$
is the induced measure over quantum field fluctuations in $\egPatch$ which leave
the configuration on $\partial\egPatch$ fixed to $|\pureState\rangle$, and we
similarly define
$[d\graviton,d\qFieldsOther;\metric,\graviton,\bgFieldsOther+\qFieldsOther]_{\egPatch',\langle
  \pureState|}$. But, of course, we know that integrals (even over finite
dimensional spaces) do not generally factorize in this way. For specificity, let
us consider using localized wavepackets for modes, which approximately
diagonalize the metric over the space of quantum field fluctuations. Then,
\eqref{eq:bakeshop} is still problematic as it is possible for the metric in the
directions `transverse' to the `surface' (in the space of field configurations)
specified by the constraints of $|\pureState\rangle$ to be dependent on the
location along the `surface'. In particular, it may be that the
orthonormalization of the wavepackets centred on $\partial\egPatch$ depends on
the derivatives of fields across $\partial\egPatch$ (and thus on the field
configurations in the interiors of $\egPatch$ and $\egPatch'$). To more
generally allow for such dependences while still maintaining some notion of
factorizability of the path integral, we thus write, instead of
\eqref{eq:bakeshop},
\begin{align}
  \begin{split}
    \MoveEqLeft[3]
    \int[d\graviton,d\qFieldsOther;\metric,\graviton,\bgFieldsOther+\qFieldsOther]\,
    \delta_{\partial\egPatch,|\pureState\rangle}[\graviton,\qFieldsOther]\,
    \bullet
    \\
    =& \int e^{-\int_{\partial\egPatch}\LagDensityGHAnom} [
       d\graviton,d\qFieldsOther; \metric,\graviton,\bgFieldsOther+\qFieldsOther
       ]_{\egPatch,|\pureState\rangle} \,
       e^{-\int_{\partial\egPatch'}\LagDensityGHAnom} [ d\graviton,d\qFieldsOther;
       \metric,\graviton,\bgFieldsOther+\qFieldsOther ]_{\egPatch',\langle
       \pureState|} \, \bullet \;,
  \end{split}
\end{align}
where $\LagDensityGHAnom$ is some local operator. (Note that $\LagDensityGHAnom$
and the Gibbons-Hawking-like term $\LagDensityGH$ contributing to the action
play analogous roles in that they capture the effects that derivatives of fields
across a codimension-one surface have on the path integration measure and action
when configurations are joined along the surface.) Thus, for bounded spacetime
regions $\egPatch$ where field configurations are fixed on its codimension-one
boundary, it is most natural to consider path integrals where the measure
implicitly includes a factor of $e^{-\int_{\partial\egPatch}\LagDensityGHAnom}$;
explicitly
\begin{align}
            \int e^{-\int_{\partial\egPatch}\LagDensityGHAnom}
            [
            d\graviton,d\qFieldsOther;
            \metric,\graviton,\bgFieldsOther+\qFieldsOther
            ]_{\egPatch,|\pureState\rangle} \,
            \bullet
            \;.
            \label{eq:cosigner}
\end{align}

It follows that when we perturb the background fields, or when we perform a
change of path integration variables (which preserves the quantum field
configurations on the codimension-one surface set by $|\pureState\rangle$), we
should account for variations
\begin{align}
  -\int_{\partial\egPatch} \vary \LagDensityGHAnom
\end{align} 
of $\LagDensityGHAnom$ in addition to the terms displayed in \eqref{eq:illusive}
and \eqref{eq:barricade}. In particular, we note that, for changes of path
integration variables $(\vary\graviton,\vary\qFieldsOther)$, we expect
$(\varyWRT{\graviton}+\varyWRT{\qFieldsOther})\LagDensityGHAnom$ and
$\sympPotAnomQ{\graviton,\vary\graviton}+\sympPotAnomQ{\qFieldsOther,\vary\qFieldsOther}$
to cancel against each other (analogous to \eqref{eq:sitcom}):
\begin{align}
  \left(
  \sympPotAnomQ{\graviton,\vary\graviton}+\sympPotAnomQ{\qFieldsOther,\vary\qFieldsOther}
  \right)_{\egReg}
  =&
     -(\varyWRT{\graviton}+\varyWRT{\qFieldsOther})\LagDensityGHAnom |_{\egReg}
     \;.
  &(\text{bulk codim.-1 $\egReg$ where $\vary\graviton,\vary\qFieldsOther=0$})
    \label{eq:provolone}
\end{align} 
We expect this identity to hold, because for any codimension-one surface
$\egReg$, by design,
$(\varyWRT{\graviton}+\varyWRT{\qFieldsOther})e^{-\int_{\egReg}\LagDensityGHAnom}$
should recover the variation of the induced metric determinant over the space of
quantum fluctuations localized around $\egReg$, in response to a change of
variables $(\vary \graviton,\vary \qFieldsOther)$ with support on one side of
$\egReg$ (\cf{}footnote \ref{foot:sitcom}).

There is, further, a more general set of considerations to support
\eqref{eq:sitcom} and \eqref{eq:provolone}. If one considers classical field
theory, then a desirable property of the full action over a spacetime region
$\egPatch$, including Gibbons-Hawking-like boundary contributions on
$\partial\egPatch$, is that it is extremized by solutions of the bulk equations
of motion (as also mentioned around \eqref{eq:sitcom}). This is ensured if, for
an arbitrary variation $(\vary\graviton,\vary\qFieldsOther)$ of the fields which
preserve the boundary conditions at $\partial\egPatch$, the boundary integrals
$\int_{\partial\egPatch}$ cancel in the response
$\int_{\egPatch}(\eomDensity{\graviton}\cdot\vary\graviton +
\eomDensity{\qFieldsOther}\cdot\vary\qFieldsOther) +
\int_{\partial\egPatch}\sympPot[\vary\graviton,\vary\qFieldsOther] +
\int_{\partial\egPatch}
(\varyWRT{\graviton}+\varyWRT{\qFieldsOther})\LagDensityGH$ of the action to
$(\vary\graviton,\vary\qFieldsOther)$ --- hence one requires \eqref{eq:sitcom}.
To put this in a more cumbersome way, we can say that, from the perspective of
the action, the responses to $(\vary\graviton,\vary\qFieldsOther)$ and
$(\testFunc\vary\graviton,\testFunc\vary\qFieldsOther)$ are indistinguishable,
where $\testFunc$ is a smooth test function which is $1$ everywhere in
$\egPatch$ except an infinitesimally small neighbourhood of $\partial\egPatch$
where it interpolates to zero, \ie{}$\testFunc$ is a smoothed indicator function
with support in $\egPatch$. (The latter response consists of just the equations
of motion part
$\int_{\egPatch}\testFunc(\eomDensity{\graviton}\cdot\vary\graviton +
\eomDensity{\qFieldsOther}\cdot\vary\qFieldsOther) \to
\int_{\egPatch}(\eomDensity{\graviton}\cdot\vary\graviton +
\eomDensity{\qFieldsOther}\cdot\vary\qFieldsOther)$.) Now, moving to our
semiclassical theory, we can similarly demand the path integral measure's
response to $(\vary\graviton,\vary\qFieldsOther)$ and
$(\testFunc\vary\graviton,\testFunc\vary\qFieldsOther)$ be identical. This is
reasonable, given that we expect the norm of the difference between the two
vectors $(\vary\graviton,\vary\qFieldsOther)$ and
$(\testFunc\vary\graviton,\testFunc\vary\qFieldsOther)$ in the space of field
configurations to be vanishingly small in the limit where $\testFunc$ becomes an
indicator function for $\egPatch$ --- otherwise, the difference between
$(\vary\graviton,\vary\qFieldsOther)$ and
$(\testFunc\vary\graviton,\testFunc\vary\qFieldsOther)$ in the limit would seem
to correspond to a direction in field space that is unconstrained by the action.
Thus, it is natural to require \eqref{eq:provolone} and interpret the role of
$\LagDensityGHAnom$ in \eqref{eq:cosigner} as ensuring that the measure $
e^{-\int_{\partial\egPatch}\LagDensityGHAnom} [ d\graviton,d\qFieldsOther;
\metric,\graviton,\bgFieldsOther+\qFieldsOther ]_{\egPatch,|\pureState\rangle}$
varies smoothly over the space of field configurations over distances in the
space deemed infinitesimal by consideration of variations of the action.

Thus far, we have considered codimension-one surfaces in the interior of the
spacetime on which field configurations are fixed in a path integral. One can
also consider the case where the path integral covers a full spacetime with
boundary $\bdyMan$, where boundary conditions require the quantum fields
$(\graviton,\qFieldsOther)$ to scale appropriately on approach to $\bdyMan$ such
that $(\metric+\graviton,\bgFieldsOther+\qFieldsOther)$ have certain desired
asymptotics. One can try to obtain a metric over the space of all field
fluctuations, which can in general reach the spacetime boundary $\bdyMan$, by
naive extension of the same metric one would apply to bounded fluctuations ---
\eg{} for fluctuations of scalar fields, one can continue trying to apply the
metric $(f_1,f_2) = \int \volForm f_1^* f_2$ to scalar functions $f_1,f_2$ even
if their support extends to the spacetime boundary $\bdyMan$. Then, one might
naively try to integrate using the path integral measure
$[d\graviton,d\qFieldsOther;\metric,\graviton,\bgFieldsOther+\qFieldsOther]$
induced by this metric, on the space of field configurations satisfying the
boundary conditions. This can, however, be problematic as such fluctuations
extending to the spacetime boundary generally have infinite norms, as in the
case of non-normalizable modes in holography. (It may often be the case that
infinite-norm fluctuations are disallowed by the boundary conditions at the
spacetime boundary $\bdyMan$. Even so, one might however still be concerned when
seeking to perturb the boundary conditions, \eg{}to compute CFT correlation
functions in holography, being worried that the configuration space metric
varies infinitely under the perturbation.) Moreover, even if we concern
ourselves only with fluctuations of non-infinite norm, we should still require,
as described in the previous paragraph, consistency between variations of the
action and the path integration measure. In particular, one can again ask
whether the action and path integral measure can tell the difference between
fluctuations $(\vary\graviton,\vary\qFieldsOther)$ and
$(\testFunc\vary\graviton,\testFunc\vary\qFieldsOther)$, where now
$(\vary\graviton,\vary\qFieldsOther)$ can reach but preserve boundary conditions
at $\bdyMan$ and $\testFunc$ is a smoothed indicator function for a bounded
region which covers the spacetime in a limit. Now due to \eqref{eq:wiring}, the
action \eqref{eq:treadmill} including the boundary contribution
$\int_{\bdyMan}\LagDensityBdy$ cannot tell such a difference. It is again
natural to require that the difference between the two vectors
$(\vary\graviton,\vary\qFieldsOther)$ and
$(\testFunc\vary\graviton,\testFunc\vary\qFieldsOther)$ in the space of field
configurations be vanishing. However, we see in \eqref{eq:barricade} that the
response of the path integral measure
$[d\graviton,d\qFieldsOther;\metric,\graviton,\bgFieldsOther+\qFieldsOther]$
includes an extra $-\int_\bdyMan ( \sympPotAnomQ{\graviton,\vary\graviton} +
\sympPotAnomQ{\qFieldsOther,\vary\qFieldsOther} )$ if
$(\vary\graviton,\vary\qFieldsOther)$ is considered rather than
$(\testFunc\vary\graviton,\testFunc\vary\qFieldsOther)$.

To resolve the issues raised in the previous paragraph, let us therefore include
an extra factor in path integrals that reach the spacetime boundary $\bdyMan$;
for those path integrals with boundary $\bdyMan$ for instance, analogous to
\eqref{eq:cosigner}, we shall integrate using
\begin{align}
  \int e^{-\int_{\bdyMan}\LagDensityBdyAnom}
  [
  d\graviton,d\qFieldsOther;
  \metric,\graviton,\bgFieldsOther+\qFieldsOther
  ]\,
  \bullet
  \;.
\end{align}
Here, the role of the factor $e^{-\int_{\bdyMan}\LagDensityBdyAnom}$ can be
interpreted as renormalizing the configuration space metric on those field
fluctuations which extend to the spacetime boundary $\bdyMan$, so that the
vanishing, finiteness, or infiniteness of their norms is consistent with what
one expects from consideration of the action. In particular, the discussion of
the previous paragraph suggests
\begin{align}
  \left(
  \sympPotAnomQ{\graviton,\vary\graviton}+\sympPotAnomQ{\qFieldsOther,\vary\qFieldsOther}
  \right)_{\bdyMan}
  =&
     -(\varyWRT{\graviton}+\varyWRT{\qFieldsOther})\LagDensityBdyAnom |_{\bdyMan}
     \;,
     \label{eq:goggles}
\end{align}
analogous to \eqref{eq:provolone}. Moreover, since $\bgFieldsOther$ and, when
considering the form \eqref{eq:unrevised} of the measure, $\metric$ merely shift
$\qFieldsOther$ and $\graviton$, we also expect
\begin{align}
  \left.
  \sympPotAnom[\vary\metric,\vary\bgFieldsOther]
  \right|_{\bdyMan}
  =&
     -(\varyWRT{\metric}+\varyWRT{\bgFieldsOther})\LagDensityBdyAnom |_{\bdyMan}
     \;.
     \label{eq:confound}
\end{align}
This should also hold for the more general form \eqref{eq:overgrown} of the path
integration measure if it is obtained from \eqref{eq:unrevised} via
\eqref{eq:donated} with $\actMatDiver[\metric]$ a local action of the form
\eqref{eq:treadmill}, satisfying \eqref{eq:wiring}.

More generally, the boundary of the path integral region can include both
portions in the interior and portions on the boundary of the spacetime; examples
are the path integrals considered in appendix \ref{sec:immorally} over finite
time intervals. As a general guide, one should incorporate $\LagDensityGHAnom$
and $\LagDensityBdyAnom$ terms into the path integral measure when the
corresponding $\LagDensityGH$ and $\LagDensityBdy$ terms appear in the action.
In appendix \ref{sec:immorally}, we also considered the possibility of joint
contributions $\LagDensityJtMat$ to the action where constant time slices and
the spacetime boundary $\bdyMan$ met. One might similarly anticipate
corresponding $\LagDensityJtAnom$ contributions to the path integral measure
which have the property\footnote{One might also expect that, when the path
  integrals over two adjacent time intervals
  $\timeOpenInterval{\thTimeNelPatch{0}}{\thTimeNelPatch{1}}$ and
  $\timeOpenInterval{\thTimeNelPatch{1}}{\thTimeNelPatch{2}}$ are stitched
  together along $\thTimeNelPatch{1}$, the $\LagDensityJtAnom$ terms at
  $\thTimeNelPatch{1}\cap\bdyMan$ account for any contributions to
  $\LagDensityBdyAnom$ in the resulting joined path integral that are localized
  on $\thTimeNelPatch{1}\cap\bdyMan$. (\Cf{}the second paragraph of footnote
  \ref{foot:backboard}.)} that their variations cancel against any localized
contributions the anomalous symplectic potential
$\sympPotAnomQ{\graviton,\vary\graviton}+\sympPotAnomQ{\qFieldsOther,\vary\qFieldsOther}$
might produce when integrated across a joint, for $\vary\graviton$ and
$\vary\qFieldsOther$ which preserve boundary conditions on the surfaces meeting
at the joint.

\subsection{Anomalous correction to the modular Hamiltonian}
\label{sec:campfire}
With the notation set up in appendix \ref{sec:affix}, we now proceed to show how
various objects in our calculations of section \ref{sec:fluorine}, relating
generalized entropy to Noether charge and current in thermal setups, are
affected by anomalous variations of the path integral measure. We also comment
on how the closely related calculations of appendix \ref{sec:immorally} are
corrected.

Let us begin with the thermal case considered in section \ref{sec:fluorine}. The
first appearance of anomalous corrections occurs in \eqref{eq:thorium} and
\eqref{eq:sprain}. As mentioned above these equations, one must additionally
account for the variation of the path integral measure in response to the
variations of the background fields; consequently, the full
$\modHamNelS{\NelVec}$ operator reads
\begin{align}
  \begin{split}
    \modHamNelS{\NelVec}
    =& \int_{\NelPatchPart{2}}
       \left[
       (
       \eomDensity{\metric}
       + \eomDensityAnom{\metric}
       )
       \cdot \Lie{\NelVec} \metric
       + (
       \eomDensity{\bgFieldsOther}
       +\eomDensityAnom{\bgFieldsOther}
       ) 
       \cdot \Lie{\NelVec} \bgFieldsOther
       \right]
    \\
     &+ \int_{\partial\NelPatchPart{2} \cap \bdyMan} \left\{ 
       \sympPot[
       \Lie{\NelVec} \metric,\Lie{\NelVec} \bgFieldsOther
       ] 
       + \sympPotAnom[
       \Lie{\NelVec} \metric,\Lie{\NelVec} \bgFieldsOther
       ]
       + \left(
       \varyWRT{\metric,\Lie{\NelVec}\metric} +
       \varyWRT{\bgFieldsOther,\Lie{\NelVec}\bgFieldsOther}
       \right)
       \left(
       \LagDensityBdy + \LagDensityBdyAnom
       \right)
       \right\}
       \;.
  \end{split}
       \label{eq:shed}
\end{align} 
From \eqref{eq:atrophy} (or \eqref{eq:speckled}) and \eqref{eq:onset}, one finds
relations between the anomalous corrections to the equations of motion
$\eomDensityAnom{}$ and the $\FaulknerCAnom$ forms analogous to
\eqref{eq:feisty}:
\begin{align}
  \lim_{\thTimeNelPatch{1}\to\thTimeNelPatch{2}^-}
  \int_{\NelPatchPart{2}} \eomDensityAnom{\egField} \cdot \Lie{\NelVec} \egField
  =& -2\pi \int_{\thTimeNelPatch{2}} \thVec^a(\FaulknerCAnomOf{\egField})_a \,,
  & (\egField\in\{\metric,\bgFieldsOther\})
    \label{eq:violate}
  \\
  \lim_{\thTimeNelPatch{1}\to\thTimeNelPatch{2}^-}
  \int_{\NelPatchPart{2}} \eomDensityAnom{\egField,\Lie{\NelVec}\egField} 
  =& -2\pi \int_{\thTimeNelPatch{2}} \thVec^a(\FaulknerCAnomOf{\egField})_a \,.
  & (\egField\in\{\graviton,\qFieldsOther\})
    \label{eq:casually}
\end{align} 
Using \eqref{eq:violate}, we can take the
$\thTimeNelPatch{1}\to\thTimeNelPatch{2}$ limit of \eqref{eq:shed} as in
\eqref{eq:guzzler} to find
\begin{align}
  \begin{split}
    \MoveEqLeft[3]
    \lim_{\thTimeNelPatch{1}\to\thTimeNelPatch{2}^-} \modHamNelS{\NelVec}
    \\
    =&
       -2\pi \int_{\thTimeNelPatch{2}} \thVec^a \left( \FaulknerCOf{\metric} +
       \FaulknerCAnomOf{\metric} + \FaulknerCOf{\bgFieldsOther} +
       \FaulknerCAnomOf{\bgFieldsOther} \right)_a
    \\
     &+
       \lim_{\thTimeNelPatch{1}\to\thTimeNelPatch{2}^-}
       \int_{\partial\NelPatchPart{2} \cap \bdyMan} \left\{ 
       \sympPot[
       \Lie{\NelVec} \metric,\Lie{\NelVec} \bgFieldsOther
       ] 
       + \sympPotAnom[
       \Lie{\NelVec} \metric,\Lie{\NelVec} \bgFieldsOther
       ]
       + \left(
       \varyWRT{\metric,\Lie{\NelVec}\metric} +
       \varyWRT{\bgFieldsOther,\Lie{\NelVec}\bgFieldsOther}
       \right)
       \left(
       \LagDensityBdy + \LagDensityBdyAnom
       \right)
       \right\} \;.
  \end{split}
\end{align} 
Since our definition of $\currentAnom{\egVec}$ in \eqref{eq:unplanted} is simply
\eqref{eq:lutetium} with $\bullet^{\anom}$ superscripts tacked on, we can
express the above in terms of anomaly-corrected currents and charges in much the
same way as \eqref{eq:campsite}. Moreover, we have from \eqref{eq:dumpling}
the relation
\begin{align}
  2\pi\int_{\thTimeNelPatch{2}\cap\bdyMan} \chargeAnom{\thVec}
  =& - \lim_{\thTimeNelPatch{1}\to\thTimeNelPatch{2}^-}
     \int_{\partial\NelPatchPart{2}\cap\bdyMan} \left(
     \sympPotAnom[
     \Lie{\NelVec}\metric,
     \Lie{\NelVec}\bgFieldsOther]
     + \sympPotAnomQ{\graviton,\Lie{\NelVec}\graviton}
     + \sympPotAnomQ{\qFieldsOther,\Lie{\NelVec}\qFieldsOther}
     \right) \;,
\end{align} 
analogous to \eqref{eq:lilac}. Using these equations, together with
\eqref{eq:wiring} and \eqref{eq:goggles}, we can also carry out the analogous
simplification of spacetime boundary terms that led to \eqref{eq:canal}. The
result can be written as the corrected version of \eqref{eq:wizard}:
\begin{align}
  \lim_{\thTimeNelPatch{1}\to\thTimeNelPatch{2}} (
  \modHamNelS{\NelVec}
  + \modHamNelSDiff{\NelVec}
  )
  =& - 2\pi \left[ 
     \int_{\entSurf} \left(
     \charge{\thVec}
     + \chargeAnom{\thVec}
     \right)
     + \int_{\thTimeNelPatch{2}} \left(
     \current{\thVec}
     + \currentAnom{\thVec}
     \right)
     + \int_{\thTimeNelPatch{2}\cap\bdyMan} \thVec\cdot\left(
     \LagDensityBdy + \LagDensityBdyAnom
     \right)
     \right] \;,
\end{align}
where now, by \eqref{eq:feisty} and \eqref{eq:casually},
\begin{align}
  \modHamNelSDiff{\NelVec}
  \equiv& \int_{\NelPatchPart{2}} \left(
          \eomDensity{\graviton} \cdot \Lie{\NelVec} \graviton +
          \eomDensityAnom{\graviton,\Lie{\NelVec}} + \eomDensity{\qFieldsOther}
          \cdot \Lie{\NelVec} \qFieldsOther +
          \eomDensityAnom{\qFieldsOther,\Lie{\qFieldsOther}} \right)
            \label{eq:saddling}
  \\
  \lim_{\thTimeNelPatch{1}\to\thTimeNelPatch{2}} \modHamNelSDiff{\NelVec}
  =&
     -2\pi\int_{\thTimeNelPatch{2}} \thVec^a \left( \FaulknerCOf{\graviton} +
     \FaulknerCAnomOf{\graviton} + \FaulknerCOf{\qFieldsOther} +
     \FaulknerCAnomOf{\qFieldsOther} \right)_a
\end{align} 
similar to \eqref{eq:untimed} and \eqref{eq:silliness}.

The interpretation of $\modHamNelSDiff{\NelVec}$ is precisely that described by
\eqref{eq:bohrium} which now holds exactly: it implements the variation of path
integration variables described around \eqref{eq:foil}. Indeed, the definition
\eqref{eq:saddling} of $\modHamNelSDiff{\NelVec}$ now contains both the response
of the action and the path integration measure to such a variation --- the
latter is exactly the anomalous correction mentioned below \eqref{eq:foil}.
Thus, by the same argument as given for \eqref{eq:fermium} in section
\ref{sec:kebab}, we find that we should identify, for thermal setups,
\begin{align}
  \begin{split}
    \MoveEqLeft[3]
    \entDong{\entSurf}[\metric,\graviton,\bgFieldsOther+\qFieldsOther]
    + 2\pi\modHamMat
    \\
    =& \lim_{\thTimeNelPatch{1}\to\thTimeNelPatch{2}} (
       \modHamNelS{\NelVec}
       + \modHamNelSDiff{\NelVec}
       )
       -\actGrav[\metric]
    \\
    =& -2\pi\left[
       \int_{\entSurf}\left( \charge{\thVec}+\chargeAnom{\thVec} \right)
       + \int_{\thTimeNelPatch{2}} \left( \currentMat{\thVec}+\currentAnom{\thVec} \right)
       + \int_{\thTimeNelPatch{2}\cap\bdyMan} \thVec\cdot\left(
       \LagDensityBdyMat + \LagDensityBdyAnom
       \right)
       \right] \;,
  \end{split}
       \label{eq:squint}
\end{align}
where \eqref{eq:famished} and \eqref{eq:taps} were used in the last equality.
Just as we equated the contributions \eqref{eq:crummiest} and
\eqref{eq:everyone} to the Wald-Dong and Noether charge entropies coming from
the action, so too can we cast, using \eqref{eq:dumpling}, the anomalous part to
the Noether charge entropy in a form
\begin{align}
  - 2\pi\int_{\entSurf} \chargeAnom{\thVec}
  =& 
     \int_{\partial\NelPatchPart{1}\cap\partial\NelPatchPart{2}} \left(
     \sympPotAnom[
     \Lie{\NelVec}\metric,
     \Lie{\NelVec}\bgFieldsOther]
     + \sympPotAnomQ{\graviton,\Lie{\NelVec}\graviton}
     + \sympPotAnomQ{\qFieldsOther,\Lie{\NelVec}\qFieldsOther}
     \right) \;,
\end{align} 
resembling the anomalous part \eqref{eq:rocker} of the Wald-Dong entropy. One
may hope\footnote{See footnote \ref{foot:strobe}.} that, in the thermal case,
\eqref{eq:rocker} is indeed a correct expression for the anomalous part of
Wald-Dong entropy and that the corrections analogous to those found by
\cite{Dong:2013qoa} vanish just as they do for \eqref{eq:crummiest}. Then,
arguing as below \eqref{eq:everyone}, we expect to also have
\begin{align}
  \entDongAnom{\entSurf}[\metric,\graviton,\bgFieldsOther+\qFieldsOther]
  =& -2\pi\int_{\entSurf} \chargeAnom{\thVec} \;,
  &
    \entDong{\entSurf}[\metric,\graviton,\bgFieldsOther+\qFieldsOther]
    =& -2\pi\int_{\entSurf} (\charge{\thVec}+\chargeAnom{\thVec}) \;,
       \label{eq:domain}
\end{align} 
where the latter is the full Wald-Dong entropy \eqref{eq:sift}, including the
anomalous part $\entDongAnom{\entSurf}$. It then follows from \eqref{eq:squint}
and \eqref{eq:domain} that the thermal modular Hamiltonian is given by
\begin{align}
  \modHamMat
  =& -\int_{\thTimeNelPatch{2}} \left(
     \currentMat{\thVec}
     + \currentAnom{\thVec}
     \right) 
     - \int_{\thTimeNelPatch{2}\cap\bdyMan} \thVec\cdot\left(
     \LagDensityBdyMat + \LagDensityBdyAnom
     \right)
     \;,
     \label{eq:outplayed}
\end{align} 
which is also the generator for time evolution in this case.

More generally, one can consider time evolution on backgrounds which are not
$\thVec$-symmetric, as in appendix \ref{sec:immorally}. Since the calculations
there mirror those of section \ref{sec:fluorine} and anomalous corrections can
be incorporated in nearly the same manner as above, we will not redo the
calculations in detail. Instead, let us state the final result for the
instantaneous generator
$-\int_{\thTime}(\HamDensityMat{\thVec}+\HamDensityAnom{\thVec}) -
\int_{\thTime\cap\bdyMan}
(\HamDensityBdyMat{\thVec}+\HamDensityBdyAnom{\thVec})$ of time evolution. The
anomalous corrections are given by the obvious analogues of \eqref{eq:snowfield}
and \eqref{eq:treble}:
\begin{align}
  \HamDensityAnom{\thVec}
  =& \currentAnom{\thVec} - \sympPotAnom[\Lie{\thVec}\metric,\Lie{\thVec}\bgFieldsOther] 
     -(
     \varyWRT{\metric,\Lie{\thVec}\metric}
     + \varyWRT{\bgFieldsOther,\Lie{\thVec}\bgFieldsOther}
     ) \LagDensityGHAnom
     \label{eq:wake}
  \\
  \HamDensityBdyAnom{\thVec}
  =& \thVec\cdot\LagDensityBdyAnom
     -(
     \varyWRT{\metric,\Lie{\thVec}\metric} +
     \varyWRT{\bgFieldsOther,\Lie{\thVec}\bgFieldsOther}
     ) \LagDensityJtAnom 
     \;.
     \label{eq:fox}
\end{align}

\subsection{Anomalies and the first law of generalized entropy}
\label{sec:dexterous}

Let us proceed now to discuss how anomalous corrections affect our results
\eqref{eq:renounce} and
\eqref{eq:calamity} derived in section \ref{sec:neon} relating the generalized
first law and gravitational dynamics. As in section \ref{sec:neon}, we now move
to Lorentzian signature --- see section \ref{sec:albatross}. Of course, the
completely classical calculations of section \ref{sec:exhaust} are unaffected by
variations of the path integral measure.

The first encounter with anomalous corrections is in \eqref{eq:sociopath}; the
first law of von Neumann entropy (the first equality) continues to hold, but the
modular Hamiltonian is corrected as we found in \eqref{eq:outplayed}:
\begin{align}
  \vary\ent[\stateMat]
  =&
     2\pi \left(
     \int_{\entReg} \varyWRT{\stateMat}\left\langle
     \currentMat{\thVec}
     + \currentAnom{\thVec}
     \right\rangle 
     + \int_{\bdyReg} \thVec\cdot
     \varyWRT{\stateMat}\left\langle
     \LagDensityBdyMat
     + \LagDensityBdyAnom
     \right\rangle
     \right)\;,
\end{align} 
By the definition \eqref{eq:unplanted} of the anomalous correction to Noether
current, we have the analogue of \eqref{eq:blinks}:
\begin{align}
  \int_{\entReg} \currentAnom{\thVec}
  =& \int_{\bdyReg} \chargeAnom{\thVec}
     - \int_{\entSurf} \chargeAnom{\thVec}
     + \int_{\entReg} \thVec^a \FaulknerCAnom_a
\end{align}
so the variation \eqref{eq:king} of the quantum part of generalized entropy now
reads\footnote{We are assuming here that the relation \eqref{eq:domain} between
  the anomalous corrections to the Wald-Dong and Noether charge entropies
  continues to hold, at least as expectation values, at linear order in
  perturbations. (\Cf{}footnote \ref{foot:evoke}.)}
\begin{align}
  \begin{split} 
    \vary \langle \entDongMat{\entSurf}[\metric] \rangle + \vary\ent[\stateMat]
    =& 2\pi \Bigg( \int_{\bdyReg}\left\{ \vary\langle \chargeMat{\thVec} +
      \chargeAnom{\thVec}\rangle + \thVec\cdot \varyWRT{\stateMat}\left\langle
        \LagDensityBdyMat + \LagDensityBdyAnom \right\rangle \right\}
    \\
    &+ \int_{\entReg}\left\{ \thVec^a \vary\langle \FaulknerCMat_a +
      \FaulknerCAnom_a \rangle - \left\langle
        \varyWRT{\blkOp}(\currentMat{\thVec} + \currentAnom{\thVec})
      \right\rangle \right\} \Bigg)\;.
  \end{split}
      \label{eq:endeared}
\end{align} 
Using \eqref{eq:harmonic}, giving the anomaly-corrected versions of
\eqref{eq:skewer} and \eqref{eq:surpass}, we find again that it is actually only
the metric part $(\FaulknerCMatOf{\metric})_a + (\FaulknerCAnomOf{\metric})_a$
of the $\FaulknerC_a$ forms that contributes to the above RHS.

The calculation of the operator variation of the Noether current proceeds much
in the same manner as in section \ref{sec:bobtail} and appendix
\ref{sec:commodore}. We can repeat the argument of appendix
\ref{sec:commodore}, but now using the time evolution generator corrected by
\eqref{eq:wake} and \eqref{eq:fox} and also accounting for variations of the
path integral measure correcting \eqref{eq:bullpen}. The result is, in place of
\eqref{eq:renewal},
\begin{align}
  \begin{split}
    \MoveEqLeft[3]
    \left\langle
    \varyWRT{\blkOp} \left[
    \int_{\entReg} \left(
    \HamDensitySimpleMat{\thVec} +
    \HamDensitySimpleAnom{\thVec}
    \right) 
    + \int_{\bdyReg} \thVec\cdot \left( \LagDensityBdyMat + \LagDensityBdyAnom \right)
    \right]
    \right\rangle
    \\
    =& -\int_{\entReg}\thVec\cdot\left\langle
       (\eomDensityMat{\metric} + \eomDensityAnom{\metric}) \cdot \vary\metric
       + d\left(
        \sympPotMat[\vary\metric,\vary\bgFieldsOther] +
       \sympPotAnom[\vary\metric,\vary\bgFieldsOther]
       \right)
       \right\rangle
    \\
    &+ \int_{\bdyReg} \thVec\cdot\left\langle
      (\varyWRT{\metric}+\varyWRT{\bgFieldsOther})\left( \LagDensityBdyMat +
        \LagDensityBdyAnom \right) \right\rangle \;,
  \end{split}
      \label{eq:hankering}
\end{align}
where, analogous to \eqref{eq:reexamine},
\begin{align}
  \HamDensitySimpleAnom{\thVec}
  \equiv& \currentAnom{\thVec} - \sympPotAnom[\Lie{\thVec}\metric,\Lie{\thVec}\bgFieldsOther] 
          \;,
          \label{eq:bonehead}
\end{align}
and we have made use of the vanishing expectation value \eqref{eq:litigate} of
the corrected equations of motion for $\bgFieldsOther$.
Now, using \eqref{eq:reexamine} and \eqref{eq:bonehead}, together with the invariance of the
unperturbed background fields and path integral under $\thVec$, we have,
similar to \eqref{eq:commotion},
\begin{align}
  \varyWRT{\blkOp}\left(
  \currentMat{\thVec} + \currentAnom{\thVec}
  - \HamDensitySimpleMat{\blkOp} - \HamDensitySimpleAnom{\blkOp}
  \right)
  =& \varyWRT{\blkOp}\left(
     \sympPotMat[\Lie{\thVec}\metric,\Lie{\thVec}\bgFieldsOther]
     + \sympPotAnom[\Lie{\thVec}\metric,\Lie{\thVec}\bgFieldsOther]
     \right)
  \\
  =& 
     \sympPotMat[\Lie{\thVec}\vary\metric,\Lie{\thVec}\vary\bgFieldsOther]
     + \sympPotAnom[\Lie{\thVec}\vary\metric,\Lie{\thVec}\vary\bgFieldsOther]
  \\
  \left\langle 
  \varyWRT{\blkOp}\left(
  \currentMat{\thVec} + \currentAnom{\thVec}
  - \HamDensitySimpleMat{\blkOp} - \HamDensitySimpleAnom{\blkOp}
  \right)
  \right\rangle
  =& \Lie{\thVec} \left\langle
     \sympPotMat[\vary\metric,\vary\bgFieldsOther]
     + \sympPotAnom[\vary\metric,\vary\bgFieldsOther]
     \right\rangle \;.
     \label{eq:phoenix}
\end{align} 
Applying the Lie derivative identity \eqref{eq:struggle} to \eqref{eq:phoenix}
and combining with \eqref{eq:hankering}, we then have the analogue of
\eqref{eq:e4e5ke2}:
\begin{align}
  \begin{split}
    \MoveEqLeft[3] \int_{\entReg} \left\langle \varyWRT{\blkOp} (
      \currentMat{\thVec} + \currentAnom{\thVec} ) \right\rangle
    \\
    =& - \int_{\entReg} \thVec\cdot\left( \left\langle \eomDensityMat{\metric} +
        \eomDensityAnom{\metric} \right \rangle \cdot\vary\metric \right) +
    \int_{\bdyReg}  \thVec\cdot \left\langle
        \sympPotMat[\vary\metric,\vary\bgFieldsOther] +
        \sympPotAnom[\vary\metric,\vary\bgFieldsOther] \right \rangle 
    \\
    &+ \int_{\bdyReg} \thVec\cdot\left\langle ( \varyWRT{\metric}
      +\varyWRT{\bgFieldsOther} - \varyWRT{\blkOp} ) \left( \LagDensityBdyMat +
        \LagDensityBdyAnom \right) \right\rangle \;.
  \end{split}
      \label{eq:cabana}
\end{align}

At last, putting together the variations of the classical \eqref{eq:charlie} and
quantum \eqref{eq:endeared} parts of generalized entropy, and applying
\eqref{eq:cabana}, we obtain the final result
\begin{align}
  \begin{split}
  \vary \entGen{\entSurf}[\metric]
    =&
     2\pi\int_{\bdyReg}  \Big\{
     \vary
     \langle
     \charge{\thVec} + \chargeAnom{\thVec}
     \rangle
     - \thVec \cdot 
     \langle
     \sympPot[\vary\metric, \vary\bgFieldsOther]
     + \sympPotAnom[\vary\metric,\vary\bgFieldsOther]
     \rangle
     \\
     &+ \thVec \cdot (\vary - \varyWRT{\metric} - \varyWRT{\bgFieldsOther})
     \langle 
     \LagDensityBdyMat + \LagDensityBdyAnom
     \rangle
     \Big\}
       \\
     &+ 2\pi \int_{\entReg} \left[ 
     \thVec^a \vary\left\langle
     (\FaulknerCOf{\metric} + \FaulknerCAnomOf{\metric})_a
     \right\rangle
     + \thVec \cdot \left(
     \left\langle \eomDensity{\metric} + \eomDensityAnom{\metric}\right\rangle
     \cdot \vary\metric
     \right)
     \right]
  \end{split}
     \label{eq:reanalyze}
\end{align} 
in place of \eqref{eq:renounce}. As in \eqref{eq:calamity} and
\eqref{eq:cymbal}, for variations $\vary\metric$ and
$\vary\bgFieldsOther$ that preserve boundary conditions at the spacetime
boundary $\bdyMan$, we can use \eqref{eq:wiring} and \eqref{eq:confound} to
identify the first integral in \eqref{eq:reanalyze} as the variation of an
asymptotic energy:
\begin{align}
  \vary \entGen{\entSurf}[\metric]
    =&
     2\pi \vary \langle \HamBdy{\thVec} \rangle
     + 2\pi \int_{\entReg} \left[ 
     \thVec^a \vary\left\langle
     (\FaulknerCOf{\metric} + \FaulknerCAnomOf{\metric})_a
     \right\rangle
     + \thVec \cdot \left(
     \left\langle \eomDensity{\metric} + \eomDensityAnom{\metric}\right\rangle
     \cdot \vary\metric
     \right)
       \right]
\label{eq:slobbery}
  \\
  \HamBdy{\thVec}
  \equiv& \int_{\bdyReg} \left\{ 
     \charge{\thVec} + \chargeAnom{\thVec}
     + \thVec \cdot (
     \LagDensityBdy + \LagDensityBdyAnom
     )
     \right\} \;.
\end{align}
The interpretations of \eqref{eq:reanalyze} and \eqref{eq:slobbery} are the same
as for \eqref{eq:renounce} and \eqref{eq:calamity} of
course, except now all anomalous corrections have been accounted for. In
particular, notice that the equations of motion
\begin{align}
  \left(
  \FaulknerCOf{\metric}
  + \FaulknerCAnomOf{\metric}
  \right)^a
  =& -2 \volForm_b \left(
     \eom{\metric}
     + \eomAnom{\metric}
     \right)^{ab}
  \\
  \left(
  \eom{\metric}
  + \eomAnom{\metric}
  \right)^{ab}
  =& \left(
     \eomGrav{\metric}
     + \frac{1}{2} \stress
     \right)^{ab}
\end{align}
involve the full stress tensor \eqref{eq:tightwad} including the response of the
path integral measure to metric variations.










\bibliographystyle{JHEP}
\bibliography{references}

\providecommand{\href}[2]{#2}\begingroup\raggedright\begin{thebibliography}{10}

\bibitem{Ryu:2006bv}
S.~Ryu and T.~Takayanagi, \emph{{Holographic derivation of entanglement entropy
  from AdS/CFT}},
  \href{https://doi.org/10.1103/PhysRevLett.96.181602}{\emph{Phys. Rev. Lett.}
  {\bfseries 96} (2006) 181602}
  [\href{https://arxiv.org/abs/hep-th/0603001}{{\ttfamily hep-th/0603001}}].

\bibitem{Ryu:2006ef}
S.~Ryu and T.~Takayanagi, \emph{{Aspects of Holographic Entanglement Entropy}},
  \href{https://doi.org/10.1088/1126-6708/2006/08/045}{\emph{JHEP} {\bfseries
  08} (2006) 045} [\href{https://arxiv.org/abs/hep-th/0605073}{{\ttfamily
  hep-th/0605073}}].

\bibitem{Lewkowycz:2013nqa}
A.~Lewkowycz and J.~Maldacena, \emph{{Generalized gravitational entropy}},
  \href{https://doi.org/10.1007/JHEP08(2013)090}{\emph{JHEP} {\bfseries 08}
  (2013) 090} [\href{https://arxiv.org/abs/1304.4926}{{\ttfamily 1304.4926}}].

\bibitem{Jacobson:1993xs}
T.~Jacobson and R.C.~Myers, \emph{{Black hole entropy and higher curvature
  interactions}},
  \href{https://doi.org/10.1103/PhysRevLett.70.3684}{\emph{Phys. Rev. Lett.}
  {\bfseries 70} (1993) 3684}
  [\href{https://arxiv.org/abs/hep-th/9305016}{{\ttfamily hep-th/9305016}}].

\bibitem{Hung:2011xb}
L.-Y.~Hung, R.C.~Myers and M.~Smolkin, \emph{{On Holographic Entanglement
  Entropy and Higher Curvature Gravity}},
  \href{https://doi.org/10.1007/JHEP04(2011)025}{\emph{JHEP} {\bfseries 04}
  (2011) 025} [\href{https://arxiv.org/abs/1101.5813}{{\ttfamily 1101.5813}}].

\bibitem{Dong:2013qoa}
X.~Dong, \emph{{Holographic Entanglement Entropy for General Higher Derivative
  Gravity}}, \href{https://doi.org/10.1007/JHEP01(2014)044}{\emph{JHEP}
  {\bfseries 01} (2014) 044} [\href{https://arxiv.org/abs/1310.5713}{{\ttfamily
  1310.5713}}].

\bibitem{Faulkner:2013ana}
T.~Faulkner, A.~Lewkowycz and J.~Maldacena, \emph{{Quantum corrections to
  holographic entanglement entropy}},
  \href{https://doi.org/10.1007/JHEP11(2013)074}{\emph{JHEP} {\bfseries 11}
  (2013) 074} [\href{https://arxiv.org/abs/1307.2892}{{\ttfamily 1307.2892}}].

\bibitem{Engelhardt:2014gca}
N.~Engelhardt and A.C.~Wall, \emph{{Quantum Extremal Surfaces: Holographic
  Entanglement Entropy beyond the Classical Regime}},
  \href{https://doi.org/10.1007/JHEP01(2015)073}{\emph{JHEP} {\bfseries 01}
  (2015) 073} [\href{https://arxiv.org/abs/1408.3203}{{\ttfamily 1408.3203}}].

\bibitem{Dong:2017xht}
X.~Dong and A.~Lewkowycz, \emph{{Entropy, Extremality, Euclidean Variations,
  and the Equations of Motion}},
  \href{https://doi.org/10.1007/JHEP01(2018)081}{\emph{JHEP} {\bfseries 01}
  (2018) 081} [\href{https://arxiv.org/abs/1705.08453}{{\ttfamily
  1705.08453}}].

\bibitem{Wald:1993nt}
R.M.~Wald, \emph{{Black hole entropy is the Noether charge}},
  \href{https://doi.org/10.1103/PhysRevD.48.R3427}{\emph{Phys. Rev. D}
  {\bfseries 48} (1993) 3427}
  [\href{https://arxiv.org/abs/gr-qc/9307038}{{\ttfamily gr-qc/9307038}}].

\bibitem{Iyer:1994ys}
V.~Iyer and R.M.~Wald, \emph{{Some properties of Noether charge and a proposal
  for dynamical black hole entropy}},
  \href{https://doi.org/10.1103/PhysRevD.50.846}{\emph{Phys. Rev. D} {\bfseries
  50} (1994) 846} [\href{https://arxiv.org/abs/gr-qc/9403028}{{\ttfamily
  gr-qc/9403028}}].

\bibitem{Bekenstein:1973ur}
J.D.~Bekenstein, \emph{{Black holes and entropy}},
  \href{https://doi.org/10.1103/PhysRevD.7.2333}{\emph{Phys. Rev. D} {\bfseries
  7} (1973) 2333}.

\bibitem{Wall:2011hj}
A.C.~Wall, \emph{{A proof of the generalized second law for rapidly changing
  fields and arbitrary horizon slices}},
  \href{https://doi.org/10.1103/PhysRevD.85.104049}{\emph{Phys. Rev. D}
  {\bfseries 85} (2012) 104049}
  [\href{https://arxiv.org/abs/1105.3445}{{\ttfamily 1105.3445}}].

\bibitem{Bousso:2015eda}
R.~Bousso and N.~Engelhardt, \emph{{Generalized Second Law for Cosmology}},
  \href{https://doi.org/10.1103/PhysRevD.93.024025}{\emph{Phys. Rev. D}
  {\bfseries 93} (2016) 024025}
  [\href{https://arxiv.org/abs/1510.02099}{{\ttfamily 1510.02099}}].

\bibitem{Bousso:2015mna}
R.~Bousso, Z.~Fisher, S.~Leichenauer and A.C.~Wall, \emph{{Quantum focusing
  conjecture}}, \href{https://doi.org/10.1103/PhysRevD.93.064044}{\emph{Phys.
  Rev. D} {\bfseries 93} (2016) 064044}
  [\href{https://arxiv.org/abs/1506.02669}{{\ttfamily 1506.02669}}].

\bibitem{Faulkner:2013ica}
T.~Faulkner, M.~Guica, T.~Hartman, R.C.~Myers and M.~Van~Raamsdonk,
  \emph{{Gravitation from Entanglement in Holographic CFTs}},
  \href{https://doi.org/10.1007/JHEP03(2014)051}{\emph{JHEP} {\bfseries 03}
  (2014) 051} [\href{https://arxiv.org/abs/1312.7856}{{\ttfamily 1312.7856}}].

\bibitem{Swingle:2014uza}
B.~Swingle and M.~Van~Raamsdonk, \emph{{Universality of Gravity from
  Entanglement}},  \href{https://arxiv.org/abs/1405.2933}{{\ttfamily
  1405.2933}}.

\bibitem{Jacobson:2015hqa}
T.~Jacobson, \emph{{Entanglement Equilibrium and the Einstein Equation}},
  \href{https://doi.org/10.1103/PhysRevLett.116.201101}{\emph{Phys. Rev. Lett.}
  {\bfseries 116} (2016) 201101}
  [\href{https://arxiv.org/abs/1505.04753}{{\ttfamily 1505.04753}}].

\bibitem{Jacobson:2018ahi}
T.~Jacobson and M.~Visser, \emph{{Gravitational Thermodynamics of Causal
  Diamonds in (A)dS}},
  \href{https://doi.org/10.21468/SciPostPhys.7.6.079}{\emph{SciPost Phys.}
  {\bfseries 7} (2019) 079} [\href{https://arxiv.org/abs/1812.01596}{{\ttfamily
  1812.01596}}].

\bibitem{Casini:2014yca}
H.~Casini, F.D.~Mazzitelli and E.~Test\'e, \emph{{Area terms in entanglement
  entropy}}, \href{https://doi.org/10.1103/PhysRevD.91.104035}{\emph{Phys. Rev.
  D} {\bfseries 91} (2015) 104035}
  [\href{https://arxiv.org/abs/1412.6522}{{\ttfamily 1412.6522}}].

\bibitem{Casini:2019qst}
H.~Casini, S.~Grillo and D.~Pontello, \emph{{Relative entropy for coherent
  states from Araki formula}},
  \href{https://doi.org/10.1103/PhysRevD.99.125020}{\emph{Phys. Rev. D}
  {\bfseries 99} (2019) 125020}
  [\href{https://arxiv.org/abs/1903.00109}{{\ttfamily 1903.00109}}].

\bibitem{Jafferis:2015del}
D.L.~Jafferis, A.~Lewkowycz, J.~Maldacena and S.J.~Suh, \emph{{Relative entropy
  equals bulk relative entropy}},
  \href{https://doi.org/10.1007/JHEP06(2016)004}{\emph{JHEP} {\bfseries 06}
  (2016) 004} [\href{https://arxiv.org/abs/1512.06431}{{\ttfamily
  1512.06431}}].

\bibitem{Callan:1994py}
C.G.~Callan, Jr. and F.~Wilczek, \emph{{On geometric entropy}},
  \href{https://doi.org/10.1016/0370-2693(94)91007-3}{\emph{Phys. Lett. B}
  {\bfseries 333} (1994) 55}
  [\href{https://arxiv.org/abs/hep-th/9401072}{{\ttfamily hep-th/9401072}}].

\bibitem{Ammon:2015wua}
M.~Ammon and J.~Erdmenger, \emph{{Gauge/gravity duality}: {Foundations and
  applications}}, Cambridge University Press, Cambridge (4, 2015).

\bibitem{Marolf:2006nd}
D.~Marolf and S.F.~Ross, \emph{{Boundary Conditions and New Dualities: Vector
  Fields in AdS/CFT}},
  \href{https://doi.org/10.1088/1126-6708/2006/11/085}{\emph{JHEP} {\bfseries
  11} (2006) 085} [\href{https://arxiv.org/abs/hep-th/0606113}{{\ttfamily
  hep-th/0606113}}].

\bibitem{Klebanov:1999tb}
I.R.~Klebanov and E.~Witten, \emph{{AdS / CFT correspondence and symmetry
  breaking}}, \href{https://doi.org/10.1016/S0550-3213(99)00387-9}{\emph{Nucl.
  Phys. B} {\bfseries 556} (1999) 89}
  [\href{https://arxiv.org/abs/hep-th/9905104}{{\ttfamily hep-th/9905104}}].

\bibitem{Dong:2016hjy}
X.~Dong, A.~Lewkowycz and M.~Rangamani, \emph{{Deriving covariant holographic
  entanglement}}, \href{https://doi.org/10.1007/JHEP11(2016)028}{\emph{JHEP}
  {\bfseries 11} (2016) 028}
  [\href{https://arxiv.org/abs/1607.07506}{{\ttfamily 1607.07506}}].

\bibitem{Solodukhin:1995ak}
S.N.~Solodukhin, \emph{{One loop renormalization of black hole entropy due to
  nonminimally coupled matter}},
  \href{https://doi.org/10.1103/PhysRevD.52.7046}{\emph{Phys. Rev. D}
  {\bfseries 52} (1995) 7046}
  [\href{https://arxiv.org/abs/hep-th/9504022}{{\ttfamily hep-th/9504022}}].

\bibitem{Cooperman:2013iqr}
J.H.~Cooperman and M.A.~Luty, \emph{{Renormalization of Entanglement Entropy
  and the Gravitational Effective Action}},
  \href{https://doi.org/10.1007/JHEP12(2014)045}{\emph{JHEP} {\bfseries 12}
  (2014) 045} [\href{https://arxiv.org/abs/1302.1878}{{\ttfamily 1302.1878}}].

\bibitem{Nelson:1994na}
W.~Nelson, \emph{{A Comment on black hole entropy in string theory}},
  \href{https://doi.org/10.1103/PhysRevD.50.7400}{\emph{Phys. Rev. D}
  {\bfseries 50} (1994) 7400}
  [\href{https://arxiv.org/abs/hep-th/9406011}{{\ttfamily hep-th/9406011}}].

\bibitem{Wong:2013gua}
G.~Wong, I.~Klich, L.A.~Pando~Zayas and D.~Vaman, \emph{{Entanglement
  Temperature and Entanglement Entropy of Excited States}},
  \href{https://doi.org/10.1007/JHEP12(2013)020}{\emph{JHEP} {\bfseries 12}
  (2013) 020} [\href{https://arxiv.org/abs/1305.3291}{{\ttfamily 1305.3291}}].

\bibitem{Hawking:1998kw}
S.W.~Hawking, C.J.~Hunter and M.~Taylor, \emph{{Rotation and the AdS / CFT
  correspondence}},
  \href{https://doi.org/10.1103/PhysRevD.59.064005}{\emph{Phys. Rev. D}
  {\bfseries 59} (1999) 064005}
  [\href{https://arxiv.org/abs/hep-th/9811056}{{\ttfamily hep-th/9811056}}].

\bibitem{Costa:2019yoy}
B.A.~Costa, \emph{{Laws of black hole thermodynamics in semiclassical
  gravity}}, \href{https://doi.org/10.1088/1361-6382/abb638}{\emph{Class.
  Quant. Grav.} {\bfseries 37} (2020) 225004}
  [\href{https://arxiv.org/abs/1905.10823}{{\ttfamily 1905.10823}}].

\bibitem{Costa:2020ees}
B.A.~Costa, \emph{Energy, entropy, and spacetime: lessons from semiclassical
  black holes}, Ph.D. thesis, University of British Columbia, 2020.

\bibitem{Jacobson:1993vj}
T.~Jacobson, G.~Kang and R.C.~Myers, \emph{{On black hole entropy}},
  \href{https://doi.org/10.1103/PhysRevD.49.6587}{\emph{Phys. Rev. D}
  {\bfseries 49} (1994) 6587}
  [\href{https://arxiv.org/abs/gr-qc/9312023}{{\ttfamily gr-qc/9312023}}].

\bibitem{Vilkovisky:1984st}
G.A.~Vilkovisky, \emph{{The Unique Effective Action in Quantum Field Theory}},
  \href{https://doi.org/10.1016/0550-3213(84)90228-1}{\emph{Nucl. Phys. B}
  {\bfseries 234} (1984) 125}.

\bibitem{Vilkovisky:1984tga}
G.A.~Vilkovisky, \emph{{The Gospel according to DeWitt}},  in \emph{{Quantum
  Theory of Gravity}: {Essays in honor of the 60th birthday of Bryce S
  DeWitt}}, S.M.~Christensen, ed., (Bristol), pp.~169--209, Adam Hilger (1984).

\bibitem{Faulkner:2017tkh}
T.~Faulkner, F.M.~Haehl, E.~Hijano, O.~Parrikar, C.~Rabideau and
  M.~Van~Raamsdonk, \emph{{Nonlinear Gravity from Entanglement in Conformal
  Field Theories}}, \href{https://doi.org/10.1007/JHEP08(2017)057}{\emph{JHEP}
  {\bfseries 08} (2017) 057}
  [\href{https://arxiv.org/abs/1705.03026}{{\ttfamily 1705.03026}}].

\bibitem{Peskin:1995ev}
M.E.~Peskin and D.V.~Schroeder, \emph{{An Introduction to quantum field
  theory}}, Addison-Wesley, Reading, USA (1995).

\bibitem{Burgess:1987zi}
C.P.~Burgess and G.~Kunstatter, \emph{{On the Physical Interpretation of the
  Vilkovisky-de Witt Effective Action}},
  \href{https://doi.org/10.1142/S0217732387001117}{\emph{Mod. Phys. Lett. A}
  {\bfseries 2} (1987) 875}.

\bibitem{Gibbons:1976ue}
G.W.~Gibbons and S.W.~Hawking, \emph{{Action Integrals and Partition Functions
  in Quantum Gravity}},
  \href{https://doi.org/10.1103/PhysRevD.15.2752}{\emph{Phys. Rev. D}
  {\bfseries 15} (1977) 2752}.

\bibitem{Hayward:1993my}
G.~Hayward, \emph{{Gravitational action for space-times with nonsmooth
  boundaries}}, \href{https://doi.org/10.1103/PhysRevD.47.3275}{\emph{Phys.
  Rev. D} {\bfseries 47} (1993) 3275}.

\bibitem{Witten:2003ya}
E.~Witten, \emph{{SL(2,Z) action on three-dimensional conformal field theories
  with Abelian symmetry}},  in \emph{{From Fields to Strings: Circumnavigating
  Theoretical Physics: A Conference in Tribute to Ian Kogan}}, pp.~1173--1200,
  7, 2003 [\href{https://arxiv.org/abs/hep-th/0307041}{{\ttfamily
  hep-th/0307041}}].

\bibitem{Chamblin:1999tk}
A.~Chamblin, R.~Emparan, C.V.~Johnson and R.C.~Myers, \emph{{Charged AdS black
  holes and catastrophic holography}},
  \href{https://doi.org/10.1103/PhysRevD.60.064018}{\emph{Phys. Rev. D}
  {\bfseries 60} (1999) 064018}
  [\href{https://arxiv.org/abs/hep-th/9902170}{{\ttfamily hep-th/9902170}}].

\bibitem{Caldarelli:1999xj}
M.M.~Caldarelli, G.~Cognola and D.~Klemm, \emph{{Thermodynamics of
  Kerr-Newman-AdS black holes and conformal field theories}},
  \href{https://doi.org/10.1088/0264-9381/17/2/310}{\emph{Class. Quant. Grav.}
  {\bfseries 17} (2000) 399}
  [\href{https://arxiv.org/abs/hep-th/9908022}{{\ttfamily hep-th/9908022}}].

\bibitem{Birrell:1982ix}
N.D.~Birrell and P.C.W.~Davies, \emph{{Quantum Fields in Curved Space}},
  Cambridge Monographs on Mathematical Physics, Cambridge Univ. Press,
  Cambridge, UK (2, 1984),
  \href{https://doi.org/10.1017/CBO9780511622632}{10.1017/CBO9780511622632}.

\bibitem{Hawking:1976ja}
S.W.~Hawking, \emph{{Zeta Function Regularization of Path Integrals in Curved
  Space-Time}}, \href{https://doi.org/10.1007/BF01626516}{\emph{Commun. Math.
  Phys.} {\bfseries 55} (1977) 133}.

\bibitem{Fujikawa:1980vr}
K.~Fujikawa, \emph{{Comment on Chiral and Conformal Anomalies}},
  \href{https://doi.org/10.1103/PhysRevLett.44.1733}{\emph{Phys. Rev. Lett.}
  {\bfseries 44} (1980) 1733}.

\end{thebibliography}\endgroup

\end{document}